# Entropy-Driven Phase Transitions
## in
## Colloidal Systems

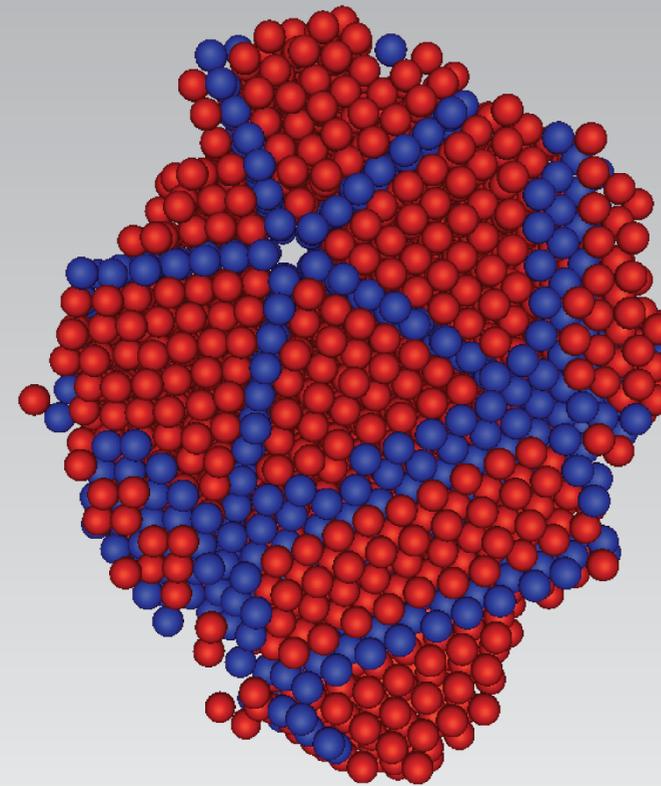

## Ran Ni

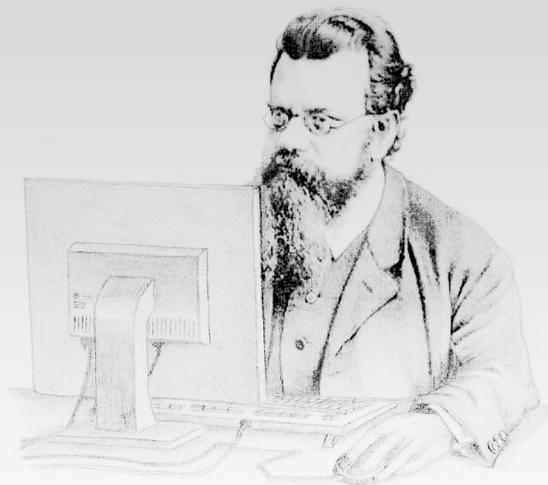

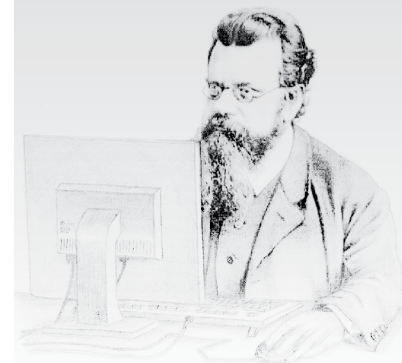

# Entropy-Driven Phase Transitions in Colloidal Systems

**Cover:**
front: the snapshot of a nucleus possessing a five-fold symmetry consisting of cyclic colloidal polymers with length of 10, where the red and blue spheres are fcc- and hcp-like spheres, respectively.
back: a portrait of Ludwig Boltzmann doing computer simulations.

# Entropy-Driven Phase Transitions in Colloidal Systems

Entropie-gedreven fase-overgangen in colloïdale systemen

## 胶体体系的熵驱动相变

Proefschrift

**Promotor:**      Prof. dr. ir. M. Dijkstra

This research was supported by an NWO-VICI grant.

$$S = k \log W$$

# Contents

# 1

## Introduction

### 1.1 Colloids

Colloids are particles within a size range of a nanometer to several micrometers that are moving around in a solvent. We make use of colloids almost every day. For instance, milk, latex paint, mayonnaise, ice creams, etc., by definition, are all consisting of colloidal particles. Therefore, it is clear that a better understanding of colloidal systems will improve these products and contribute to industrial applications. Besides applications, colloids are also good model systems to study in order to gain insights in the behavior of atomic and molecular systems. More than hundred years ago, Perrin pointed out that the motion of colloids is caused by the collisions with the molecules of the solvent, which causes the colloids to move in an irregular fashion, i.e., Brownian motion [1]. Due to the Brownian motion, colloidal particles can explore phase space, and self-assemble similarly to atomic and molecular systems. Moreover, colloidal particles are several orders of magnitude larger than atoms or molecules, and they move much slower. This makes it possible to track the motion of the particles in real time using an optical microscopy and study the dynamics of physical processes, such as nucleation, which are very difficult to investigate in atomic or molecular systems.

Moreover, the interactions between colloidal particles can be tuned relatively easily, which opens up the possibility of fabricating functional materials with rational designed properties. For instance, colloids are normally charged particles, since the surface of a colloid always adsorbs ions and releases some ions to the solvent. The ions in the solvent screen the bare Coulomb interactions between charged colloids. When the concentration of ions is low, the electrostatic repulsion dominates the phase behavior of colloidal systems, and makes the spherical colloidal particles form a body-centered cubic (bcc) crystal when the density of colloids is high. By adding salt into the solution, one can tune the colloid-colloid interaction into a hard-sphere like repulsion, which drives hard-sphere like particles to crystallize into a face-centered cubic (fcc) arrangement at high density [2]. This means that one can control the resulting crystal structure in colloidal systems by changing the properties of the solution.

Therefore, the colloidal self-assembly is important for both applications in material science and as model systems for studying physical phenomena such as nucleation, glasses, gels, etc.

## 1.2   Interactions

Recent breakthroughs in particle synthesis have resulted in a spectacular variety of building blocks with anisotropic interactions [3]. For example, rod-like [4], platelet-like [5] and polyhedral particles [6] have been successfully synthesized in experiments. In addition, as mentioned previously, one can also tune the interaction between colloidal particles by changing the properties of solvent. External fields can be employed to drive the self-assembly of colloidal systems as well [7].

In this thesis, we mainly focus on interactions induced by excluded volume effects, i.e., hard-core interactions. Although it is a simplification of the colloid-colloid interactions, there are many systems which are well modeled by hard interactions. For instance, in typical colloidal systems consisting of silica or PMMA particles coated with polymer brushes, the interaction between the colloidal particles include a hard-core repulsion, a steric repulsion associated with the polymer brushes, van der Waals forces, and electrostatic interactions induced by charges adsorbed on the surface of the colloids. In most experimental systems, the colloids are suspended in a medium with a similar refractive index that can minimize the van der Waals interactions. The steric stabilizer on the surface of colloids helps to prevent the particles to aggregate as it basically masks the van der Waals interactions. The electrostatic interactions can be screened by adding salt into the solution. Therefore, the resulting effective interaction between particles can be well modeled as hard-sphere repulsion, and the calculated equilibrium phase diagram of hard spheres indeed agrees well with experiments [8].

## 1.3   Simulation techniques

In this thesis, we perform computer simulations to study the phase behavior in colloidal systems. Two main simulation methods are employed: Monte Carlo (MC) simulations and molecular dynamics (MD) simulations. For example, to determine the nucleation rate in computer simulations, we first calculate the Gibbs free energy as a function of cluster size by using umbrella sampling MC simulations, and then compute the attachment rate by performing MD simulations starting from top of the nucleation barrier. Moreover, in the calculation of the equilibrium phase diagram of colloidal systems, we employ MC simulations to calculate the equation of state (EOS) and the free energies of various phases. In this section, we will give a brief explanation of these two methods.

### 1.3.1   Monte Carlo simulations

Monte Carlo simulations are usually employed to measure the equilibrium properties of a system by sampling the phase space with a Boltzmann distribution. Instead of following the real dynamic evolution of the system, a Monte Carlo simulation simply samples the most relevant states of the system, which is called importance sampling or Metropolis sampling [9].

Consider a system of $N$ interacting particles in a volume $V$ at temperature $T$. The probability of the system to be in a state with particle positions $\mathbf{r}^N$ and momenta $\mathbf{p}^N$ is proportional to the Boltzmann weight with respect to the Hamiltonian $\mathcal{H}$ of the system:

$$\mathcal{P}(\mathbf{r}^N, \mathbf{p}^N) \propto \exp\left[-\beta \mathcal{H}(\mathbf{r}^N, \mathbf{p}^N)\right] \tag{1.1}$$

with $\beta = 1/k_B T$ and $k_B$ Boltzmann constant. Then the mathematical expectation of a quantity $A$ of the system is the weighted average of $A$ over all possible states:

$$\langle A \rangle = \frac{\int \mathrm{d}\mathbf{r}^N \int \mathbf{p}^N A(\mathbf{r}^N, \mathbf{p}^N) \mathcal{P}(\mathbf{r}^N, \mathbf{p}^N)}{\int \mathrm{d}\mathbf{r}^N \int \mathrm{d}\mathbf{p}^N \mathcal{P}(\mathbf{r}^N, \mathbf{p}^N)}, \tag{1.2}$$

where the partition function for this system is given by

$$Q = \int \mathrm{d}\mathbf{r}^N \int \mathrm{d}\mathbf{p}^N \mathcal{P}(\mathbf{r}^N, \mathbf{p}^N). \tag{1.3}$$

The Hamiltonian can be written as

$$\mathcal{H}(\mathbf{r}^N, \mathbf{p}^N) = \mathcal{U}(\mathbf{r}^N) + \mathcal{K}(\mathbf{p}^N) \tag{1.4}$$

where $\mathcal{U}(\mathbf{r}^N)$ and $\mathcal{K}(\mathbf{p}^N)$ are the total potential energy and kinetic energy of the system, respectively. If the quantity $A$ is independent of the velocities of the particles, Eq. 1.2 can be re-written as:

$$\langle A \rangle = \frac{\int \mathrm{d}\mathbf{r}^N A(\mathbf{r}^N) \exp\left[-\beta \mathcal{U}(\mathbf{r}^N)\right]}{\int \mathrm{d}\mathbf{r}^N \exp\left[-\beta \mathcal{U}(\mathbf{r}^N)\right]} \tag{1.5}$$

A simple Monte Carlo method of evaluating Eq. 1.5 is to randomly generate a number of configurations, and determine the value $A$ and the weight of $\exp\left[-\beta \mathcal{U}(\mathbf{r}^N)\right]$ for each configuration. However, for any system of more than a few particles, this method is very inefficient, since it has a big probability of selecting configurations with very small Boltzmann weight $\exp\left[-\beta \mathcal{U}(\mathbf{r}^N)\right]$. Metropolis *et al.* introduced an ingenious method, in which $\langle A \rangle$ is calculated by averaging $A(\mathbf{r}^N)$ selected according to a Boltzmann distribution of $\exp\left[-\beta \mathcal{U}(\mathbf{r}^N)\right]$. Thus, if we randomly generate $M$ configurations according to the Boltzmann distribution, then

$$\langle A \rangle = \frac{1}{M} \sum_{i=1}^{M} A(\mathbf{r}_i^N). \tag{1.6}$$

Therefore, the Metropolis method suggests a way of sampling the most relevant configurations in phase space according to the Boltzmann distribution, which is also called importance sampling. In practice, one can generate a Markov chain of configurations according to the Boltzmann distribution, and simply average the quantity $A$ over the configurations [10].

However, in this thesis we study nucleation in colloidal systems. Nucleation is a rare event, which means it is not regarded as the "important" part of phase space in the conventional Metropolis method. The statistics of calculating the free energy barrier of nucleation in Metropolis Monte Carlo simulations are very poor. Thus, we need to bias the simulation to sample the nucleation barrier, which is called umbrella sampling. Basically,

we re-write Eq. 1.5 as

$$
\begin{aligned}
\langle A \rangle &= \frac{\int \mathrm{d}\mathbf{r}^N A(\mathbf{r}^N) \exp\left[-\beta \mathcal{U}(\mathbf{r}^N)\right]}{\int \mathrm{d}\mathbf{r}^N \exp\left[-\beta \mathcal{U}(\mathbf{r}^N)\right]} \\
&= \frac{\int \mathrm{d}\mathbf{r}^N A(\mathbf{r}^N) \exp\left[\beta \mathcal{W}(\mathbf{r}^N)\right] \exp\left\{-\beta\left[\mathcal{U}(\mathbf{r}^N) + \mathcal{W}(\mathbf{r}^N)\right]\right\}}{\int \mathrm{d}\mathbf{r}^N \exp\left[\beta \mathcal{W}(\mathbf{r}^N)\right] \exp\left\{-\beta\left[\mathcal{U}(\mathbf{r}^N) + \mathcal{W}(\mathbf{r}^N)\right]\right\}} \\
&= \frac{\left\langle A \exp\left[\beta \mathcal{W}(\mathbf{r}^N)\right]\right\rangle_{\mathcal{W}}}{\left\langle \exp\left[\beta \mathcal{W}(\mathbf{r}^N)\right]\right\rangle_{\mathcal{W}}},
\end{aligned}
\tag{1.7}
$$

where $\mathcal{W}(\mathbf{r}^N)$ is the biasing potential which is used to bias the simulation to sample the specific part of the phase space. In order to calculate the free energy barrier around cluster size $n_0$, we use

$$
\beta \mathcal{W}(\mathbf{r}^N) = \frac{1}{2} k \left[n(\mathbf{r}^N) - n_0\right]^2,
\tag{1.8}
$$

where $n(\mathbf{r}^N)$ is the biggest cluster size in the configuration $\mathbf{r}^N$ and $k = 0.2$ is the spring constant controlling the window width of the sampling. Here we described the Monte Carlo method in the canonical ensemble ($NVT$), and it can easily be adapted to other ensemble by modifying the partition function in Eq. 1.3. For instance, in the isothermal-isobaric ensemble ($NPT$), there is one more $PV$ term in the statistical weight, where $P$ and $V$ are the pressure and volume of the system, respectively.

## 1.3.2 Molecular Dynamics simulations

In molecular dynamics (MD) simulations, the dynamics of the system are explicitly taken into account, and the particles move according to Newton's Law which defines the equations of motion of the particles in the system. In conventional MD simulations, where the potential continuously changes as a function of the distance between particles, the equations of motion of the particles can be integrated with fixed time steps. The evolution of the system is then driven by time steps. However, this time-driven scheme cannot work for systems where the forces between the particles are instantaneous which is due to the discontinuity of the potential in the system. For instance, in MD simulations of hard particles, the particles only feel the interactions with other particles when they collide, but the collision cannot be detected before the overlap of particles in the time-driven MD simulations. In this thesis, we employ a scheme called event-driven molecular dynamics (EDMD) simulations to study a system of hard particles. In EDMD simulations, we first predict the times of all possible collisions explicitly, and put them into an events tree. We then evolve the system to the earliest event selected from the events tree and process the event and modify the events tree. Combined with using a cell list to speed up the simulation, the outline of the EDMD simulation for a system of hard particles is described below:

1. Initialize the velocities and positions of the particles in the system.

2. Initialize the cell list of the system, and predict and store next cell cross event for each particle.

3. Predict and store the collision events between the particles in neighboring cells.

4. Select the first predicted event, and evolve the system to the time of event.

5. In the case of a cell crossing event, update the cell list, and predict and store the collisions with the particles in the new neighboring cells.

6. In the case of a collision event, update the velocities of the particles; delete all events associated with the two colliding particles, and then predict and store the new collisions for the colliding particles.

7. In both cases, delete the old cell crossing event, and predict the new cell crossing for the particles involved in the event.

8. Repeat steps $4 \sim 7$.

In this thesis, we employ EDMD simulations mainly to study systems of anisotropic particles with rotational symmetry along one axis, i.e., hard dumbbells and hard spherocylinders. In the following, we take hard spherocylinders as an example to describe the free-flight dynamics and the collision dynamics [11].

Consider at time $t$, a hard spherocylinder located at $\mathbf{r}(t)$ with an orientation $\mathbf{u}(t)$ which is the unit vector along the axis of the cylindric part of the particle. The velocity and angular velocity of the particle are $\mathbf{v}(t)$ and $\omega(t)$, respectively, and by ignoring the spinning of the particle around $\mathbf{u}(t)$, we have $\omega(t) \cdot \mathbf{u}(t) = 0$. We assume that the particles move freely between collisions. The position and orientation of the particle at $t + \Delta t$ can be easily described by

$$
\begin{align}
\mathbf{r}(t + \Delta t) &= \mathbf{r}(t) + \mathbf{v}(t)\Delta t, \tag{1.9} \\
\mathbf{u}(t + \Delta t) &= \cos(|\omega(t)|\Delta t)\,\mathbf{u}(t) + \sin(|\omega(t)|\Delta t)\,\hat{\omega}(t) \times \mathbf{u}(t). \tag{1.10}
\end{align}
$$

where $\hat{\omega}(t) = \omega(t)/|\omega(t)|$.

To find the time at which the collision between two particles $i$ and $j$ occurs, we need to employ an overlap function which is a continuous function $f(\mathbf{r}_i, \mathbf{u}_i, \mathbf{r}_j, \mathbf{u}_j)$ depending on the location $\mathbf{r}_{i,j}$ and orientation $\mathbf{u}_{i,j}$ of the two particles. The overlap function is defined such that $f(\mathbf{r}_i, \mathbf{u}_i, \mathbf{r}_j, \mathbf{u}_j) < 0$, when the two particles are overlapped, and larger than zero, otherwise. Thus, to predict the time of collision, we essentially solve the equation:

$$
f(\mathbf{r}_i(t_{ij}), \mathbf{u}_i(t_{ij}), \mathbf{r}_j(t_{ij}), \mathbf{u}_j(t_{ij})) = 0. \tag{1.11}
$$

For a system of hard spherocylinders with a diameter $\sigma$ and a cylindrical segment of length $L$, we use $f(\mathbf{r}_i, \mathbf{u}_i, \mathbf{r}_j, \mathbf{u}_j) = d(\mathbf{r}_i, \mathbf{u}_i, \mathbf{r}_j, \mathbf{u}_j) - \sigma$, where $d(\mathbf{r}_i, \mathbf{u}_i, \mathbf{r}_j, \mathbf{u}_j)$ is the shortest distance between two line segments at $\mathbf{r}_{i,j}$ and orientation $\mathbf{u}_{i,j}$ [12]. Taking into account the free-flight dynamics of hard spherocylinders, one needs to solve a transcendental equation to predict the time of collision. Although it is hard to have analytical solutions for those equations, it can still be solved numerically [13].

As long as we have found the times of collisions in the system, we can select the first collision event and evolve the system to the time of the collision. We assume that particle $i$ and $j$ at positions $\mathbf{r}_i$, $\mathbf{r}_j$ with orientations $\mathbf{u}_i$, $\mathbf{u}_j$, momenta $\mathbf{p}_i$, $\mathbf{p}_j$ and angular velocities $\omega_i$, $\omega_j$, respectively, collide at the contact point $\mathbf{r}_c$ as shown in Fig. 1.1. The moment of inertia of the particle is $I$, thus the angular momentum of particle $i$ is $\mathbf{J}_i = I\omega_i$.

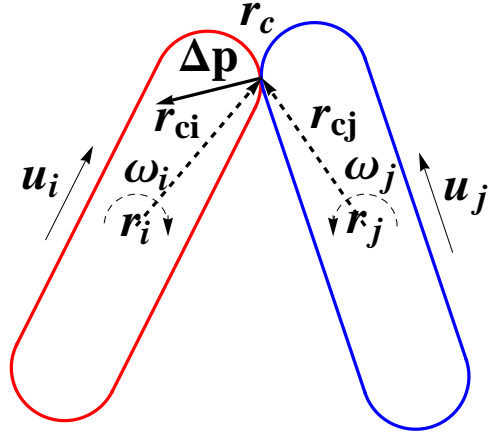

**Figure 1.1**: Illustration of the collision between two spherocylinders $i$ and $j$.

The post-collisional values of the linear and angular momenta of the particles, denoted by primes, are then given in terms of the collision impulse $\Delta \mathbf{p}$ at the contact point:

$$
\begin{aligned}
\mathbf{p}'_i &= \mathbf{p}_i + \Delta\mathbf{p}, & (1.12)\\
\mathbf{p}'_j &= \mathbf{p}_j - \Delta\mathbf{p}, & (1.13)\\
\mathbf{J}'_i &= \mathbf{J}_i + \mathbf{r}_{ci} \times \Delta\mathbf{p}, & (1.14)\\
\mathbf{J}'_j &= \mathbf{J}_j - \mathbf{r}_{cj} \times \Delta\mathbf{p}, & (1.15)
\end{aligned}
$$

where $\mathbf{r}_{ci,cj}$ is the position of point of collision with respect to the center of the particle $i, j$. Supposing that the collision is elastic and the two particles have the same mass $m$, we find

$$
\frac{|\Delta\mathbf{p}|^2}{m} + \mathbf{g}_{ij} \cdot \Delta\mathbf{p} + \frac{1}{2I}\left(|\mathbf{r}_{ci} \times \Delta\mathbf{p}|^2 + |\mathbf{r}_{cj} \times \Delta\mathbf{p}|^2\right) = 0, \tag{1.16}
$$

where

$$
\mathbf{g}_{ij} = \mathbf{v}_i + \omega_i \times \mathbf{r}_{ci} - \mathbf{v}_j - \omega_j \times \mathbf{r}_{cj} \tag{1.17}
$$

is the relative velocity of the two colliding points. Moreover, since the impulse is perpendicular to the surface, $\Delta\mathbf{p} = \mathbf{n}|\Delta\mathbf{p}|$, where $\mathbf{n}$ is the unit vector perpendicular to the surface of collision, and this gives

$$
|\Delta\mathbf{p}| = \frac{-\mathbf{g}_{ij} \cdot \mathbf{n}}{(1/m) + \left(|\mathbf{r}_{ci} \times \mathbf{n}|^2 + |\mathbf{r}_{cj} \times \mathbf{n}|^2\right)/2I}. \tag{1.18}
$$

This equation is generally applicable to hard particles with rotational symmetry.

## 1.4   Outline of this thesis

The remainder of this thesis is organized as follows. This thesis can be divided into two independent parts. In the first part of this thesis, we focus on studying the kinetic pathways of nucleation in colloidal systems. In Chapter 2, we briefly introduce the relevant theory of nucleation, i.e., classic nucleation theory. Then in Chapter 3, we investigate the crystal nucleation in the "simplest" model system for colloids, i.e., the monodisperse hard-sphere system, by using three different simulation methods, i.e., molecular dynamics, forward flux sampling and umbrella sampling simulations. Subsequently, we apply our simulation methods to a more realistic system of colloidal hard spheres in Chapter 4. Furthermore, we study the nucleation in a variety of systems consisting of hard particles, i.e., hard dumbbells (Chapter 5), hard rods (Chapter 6), hard colloidal polymers (Chapter 7) and binary hard-sphere mixtures (Chapter 8). In the second part of this thesis, we study the phase behavior of several colloidal systems. In Chapter 9, we study the equilibrium phase diagram of colloidal hard superballs whose shape interpolates from cubes to octahedra via spheres. We investigate the micellization of asymmetric patchy dumbbells induced by the depletion attraction in Chapter 10.

## 1.5  Appendix: $NPT$ Monte Carlo simulations

In this section, we formulate the partition function for the $NPT$ ensemble in a deformable box and the corresponding acceptance rule for volume moves in Monte Carlo simulations.

We start with the partition function for the $NVT$ ensemble:

$$Q(N, V, T) = C \int \mathrm{d}\mathbf{r}^N \exp\left[-\beta \mathcal{U}(\mathbf{r}^N)\right],\qquad(1.19)$$

where $C$ is a constant. If we use a matrix $\mathbf{h} = (\mathbf{e}_1, \mathbf{e}_2, \mathbf{e}_3)$ to fix the box shape, where $\mathbf{e}_i$ is the $i$th box length vector, the spatial coordinate of particle $k$ can be written as

$$\mathbf{r}_k = \mathbf{h}\mathbf{s}_k,\qquad(1.20)$$

where $\mathbf{s}_k$ is the fractional coordinate of particle $k$. Then the partition function for the $N\mathbf{h}T$ ensemble reads:

$$Q(N, \mathbf{h}, T) = CV^N \int \mathrm{d}\mathbf{s}^N \exp\left[-\beta \mathcal{U}(\mathbf{s}^N)\right],\qquad(1.21)$$

where $V = \det(\mathbf{h})$ is the volume of the system.

For the $NPT$ ensemble, we decompose the box into two terms: the volume of the box and the shape of the box. If we use the matrix $\mathbf{h}_0 = \mathbf{h}/V^{1/3}$ to denote the shape of the box, the partition function for the $NPT$ ensemble can be written as

$$Q(N, P, T) = \frac{1}{V_0} \int_0^\infty \mathrm{d}V \int \mathrm{d}\mathbf{h}_0 \exp(-\beta PV) Q(N, \mathbf{h}, T)\delta\left[\det(\mathbf{h}_0) - 1\right],\qquad(1.22)$$

where $V_0$ is the unit of volume, and $\delta(*)$ is the Dirac delta function. We note that $\det(\mathbf{h}_0) = \det(\mathbf{h})/V$, and we find

$$Q(N, P, T) = \frac{1}{V_0} \int_0^\infty \mathrm{d}V \int \mathrm{d}\mathbf{h} \det\left[J\left(\frac{\partial \mathbf{h}_0}{\partial \mathbf{h}}\right)\right] \exp(-\beta PV) Q(N, \mathbf{h}, T)\delta\left[\frac{\det(\mathbf{h})}{V} - 1\right],$$
$$(1.23)$$

where $J\left(\frac{\partial \mathbf{h}_0}{\partial \mathbf{h}}\right)$ is the Jacobian of $\mathbf{h}_0$ with respect to $\mathbf{h}$. We note that $\mathbf{h}$ has nine, i.e., $3 \times 3$, independent elements, and $\frac{\partial \mathbf{h}_{0,i}}{\partial \mathbf{h}_j} = V^{-1/3}\delta_K(i,j)$, where $\delta_K(i,j)$ is the Kronecker delta function. We arrive at

$$J\left(\frac{\partial \mathbf{h}_0}{\partial \mathbf{h}}\right) = \begin{pmatrix} V^{-1/3} & 0 & 0 & 0 & 0 & 0 & 0 & 0 & 0 \\ 0 & V^{-1/3} & 0 & 0 & 0 & 0 & 0 & 0 & 0 \\ 0 & 0 & V^{-1/3} & 0 & 0 & 0 & 0 & 0 & 0 \\ 0 & 0 & 0 & V^{-1/3} & 0 & 0 & 0 & 0 & 0 \\ 0 & 0 & 0 & 0 & V^{-1/3} & 0 & 0 & 0 & 0 \\ 0 & 0 & 0 & 0 & 0 & V^{-1/3} & 0 & 0 & 0 \\ 0 & 0 & 0 & 0 & 0 & 0 & V^{-1/3} & 0 & 0 \\ 0 & 0 & 0 & 0 & 0 & 0 & 0 & V^{-1/3} & 0 \\ 0 & 0 & 0 & 0 & 0 & 0 & 0 & 0 & V^{-1/3} \end{pmatrix},$$

and

$$\det\left[J\left(\frac{\partial \mathbf{h}_0}{\partial \mathbf{h}}\right)\right] = V^{-3}. \tag{1.24}$$

By utilizing the properties of the Dirac delta function and implying $\delta\left[\det(\mathbf{h})/V - 1\right] = V\delta\left[\det(\mathbf{h}) - V\right]$, the partition function can be re-written as

$$Q(N,P,T) = \frac{C}{V_0}\int \mathrm{d}\mathbf{h}[\det(\mathbf{h})]^{N-2}\exp\left[-\beta P \det(\mathbf{h})\right]\int \mathrm{d}\mathbf{s}^N \exp\left[-\beta\mathcal{U}(\mathbf{s}^N,\mathbf{h})\right]. \tag{1.25}$$

We can now construct a Monte Carlo scheme to simulate the system in the $NPT$ ensemble. The acceptance rule for a trial move from $\mathbf{h}$ to $\mathbf{h}'$ is

$$\begin{aligned} acc(o \to n) = & \min\left(1, \exp\left\{-\beta\left[\mathcal{U}(\mathbf{s}^N,\mathbf{h}') - \mathcal{U}(\mathbf{s}^N,\mathbf{h})\right.\right.\right. \\ & \left.\left.\left. + P\left[\det(\mathbf{h}') - \det(\mathbf{h})\right] - (N-2)\beta^{-1}\ln\left(\frac{\det(\mathbf{h}')}{\det(\mathbf{h})}\right)\right]\right\}\right). \end{aligned} \tag{1.26}$$

More generally, for the $NPT$ ensemble in $d$ dimensional systems, the partition function is

$$Q(N,P,T) = \frac{C}{V_0}\int \mathrm{d}\mathbf{h}[\det(\mathbf{h})]^{N+1-d}\exp\left[-\beta P \det(\mathbf{h})\right]\int \mathrm{d}\mathbf{s}^N \exp\left[-\beta\mathcal{U}(\mathbf{s}^N,\mathbf{h})\right]. \tag{1.27}$$

In the following, we derive the partition function for the $NPT$ ensemble in three specific cases, which are commonly used.

### 1.5.1   Cubic box

The most commonly used $NPT$- MC simulations are performed in a cubic simulation box without shape deformation. When we fix the shape of the cubic box with length $L$, the box matrix becomes

$$\mathbf{h} = \begin{pmatrix} L & 0 & 0 \\ 0 & L & 0 \\ 0 & 0 & L \end{pmatrix}. \tag{1.28}$$

Then the tensor $\mathrm{d}\mathbf{h} = \mathrm{d}L$, and the Jacobian in Eq. 1.23 becomes $\det\left[J\left(\frac{\partial \mathbf{h_0}}{\partial \mathbf{h}}\right)\right] = 1/L = V^{-1/3}$. The resulting partition function becomes

$$Q(N, P, T) = \frac{C}{3V_0} \int \mathrm{d}V V^N \exp(-\beta PV) \int \mathrm{d}\mathbf{s}^N \exp\left[-\beta \mathcal{U}(\mathbf{s}^N)\right], \qquad (1.29)$$

which recovers the partition function for $NPT$ ensemble without box shape deformation in Ref. [10].

### 1.5.2 Rectangular box

$NPT$ simulations in rectangular simulation boxes are also frequently employed, since many crystals cannot be simulated in a cubic box, e.g., hexagonal-close-packed crystals. The box matrix for a rectangular box is

$$\mathbf{h} = \begin{pmatrix} L_x & 0 & 0 \\ 0 & L_y & 0 \\ 0 & 0 & L_z \end{pmatrix}, \qquad (1.30)$$

where $L_x$, $L_y$ and $L_z$ are three independent variables denoting the box length in the $x$, $y$ and $z$ directions, respectively. The tensor $\mathrm{d}\mathbf{h}$ reduces to $\mathrm{d}L_x \mathrm{d}L_y \mathrm{d}L_z$, and the Jacobian in Eq. 1.24 becomes $\det\left[J\left(\frac{\partial \mathbf{h_0}}{\partial \mathbf{h}}\right)\right] = V^{-1}$. Subsequently, the partition function is

$$Q(N, P, T) = \frac{C}{V_0} \int \mathrm{d}L_x \mathrm{d}L_y \mathrm{d}L_z (L_x L_y L_z)^N \exp(-\beta P L_x L_y L_z) \int \mathrm{d}\mathbf{s}^N \exp\left[-\beta \mathcal{U}(\mathbf{s}^N)\right], \qquad (1.31)$$

which has the same acceptance rule for volume moves as for cubic simulation boxes. However, we note that the $NPT$ ensembles with a cubic box and a rectangular box are two different ensembles, and that the system in a rectangular box has more entropy associated with the box shape.

### 1.5.3 Floppy box without box rotation

The partition function for the $NPT$ ensemble in a general floppy box is described in Eq. 1.26, and one can perform Monte Carlo simulations with random walks in the nine elements of $\mathbf{h}$. However, the random walk in all the nine elements of $\mathbf{h}$ will result in a rotating simulation box. To prevent the rotation of the simulation box, a common trick is to fix the first axis of the box along the $x$ axis and the second axis of the box in the $x - y$ plane. The third axis of the box can freely rotate. The box matrix becomes an upper triangular matrix:

$$\mathbf{h} = \begin{pmatrix} h_1 & h_6 & h_5 \\ 0 & h_2 & h_4 \\ 0 & 0 & h_3 \end{pmatrix}, \qquad (1.32)$$

which uses the Voigt notation: $11 = 1$, $22 = 2$, $33 = 3$, $23(= 32) = 4$, $13(= 31) = 5$ and $12(= 21) = 6$. Then the Jacobian in Eq. 1.24 becomes $\det\left[J\left(\frac{\partial \mathbf{h_0}}{\partial \mathbf{h}}\right)\right] = V^{-2}$. Therefore, the partition function is

$$Q(N, P, T) = \frac{C}{V_0} \int \mathrm{d}\mathbf{h}[\det(\mathbf{h})]^{N-1} \exp\left[-\beta P \det(\mathbf{h})\right] \int \mathrm{d}\mathbf{s}^N \exp\left[-\beta \mathcal{U}(\mathbf{s}^N)\right]. \qquad (1.33)$$

And the corresponding acceptance rule for the trial move from $\mathbf{h}$ to $\mathbf{h}'$ is

$$
\begin{aligned}
acc(o \rightarrow n) \ = \ & \min\Big(1, \exp\Big\{-\beta\Big[\mathcal{U}(\mathbf{s}^N, \mathbf{h}') - \mathcal{U}(\mathbf{s}^N, \mathbf{h}) \\
& + P\left[\det(\mathbf{h}') - \det(\mathbf{h})\right] - (N-1)\beta^{-1}\ln\left(\frac{\det(\mathbf{h}')}{\det(\mathbf{h})}\right)\Big]\Big\}\Big). \quad (1.34)
\end{aligned}
$$

The $NPT$ ensembles in cubic, rectangular and floppy boxes are, in principle, equivalent when the system is infinitely large. However, for a finite system, they are different.

# Part I

# Nucleation in colloidal systems

# 2

---

# Theory of nucleation

---

In this chapter, we briefly describe the physical background of nucleation and a classic theoretical interpretation, i.e., classical nucleation theory (CNT). We shortly derive CNT for estimating the free-energy barrier of nucleation and the resulting nucleation rate.

## 2.1 Nucleation

Nucleation is the onset of a first order phase transition, which happens when the first order derivative of the free energy with respect to a thermodynamic variable is discontinuous. For example, the gas-liquid phase transition is a typical first order transition, which has a discontinuity in density $\rho$. Let us take the van der Waals fluid as example. In his thesis of 1873, Van der Waals proposed two correction terms to the ideal gas law $p = Nk_BT/V$ [14]. Firstly, he argued that the actual volume available to a molecule is smaller than the total volume $V$ of the container due to the excluded volume effect between molecules. Secondly, he argued that the attractions between the molecules reduce the pressure $p$ by an amount $-a\rho^2$, where $a > 0$ is a measure for the attraction between the molecules. Therefore, Van der Waals wrote

$$p = \frac{Nk_BT}{V - Nb} - a\rho^2 = \frac{\rho k_BT}{1 - \rho b} - a\rho^2, \tag{2.1}$$

with two phenomenological parameters $a$ and $b$. He also found that there is a critical temperature $T_c = 8a/27bk_B$, below which the system undergoes a gas-liquid phase transition upon increasing the density. A typical equation of state for a system of a van der Waals fluid at temperature $T < T_c$ is shown in Fig. 2.1.

By applying a common tangent construction on the free energy density curve as shown in Fig. 2.1, one finds that the gas-liquid phase coexistence densities are $\rho_{gas}$ and $\rho_{liquid}$, respectively, also called the *binodal* points. When the density of the system $\rho$ lies well inside the region $\rho_{gas} < \rho < \rho_{liquid}$, the system is metastable (supersaturated or undersaturated), and in equilibrium it phase separates into a gas phase with density $\rho_{gas}$ and a liquid phase with density $\rho_{liquid}$. The phase transition is triggered by fluctuations in the system. Moreover, for each isotherm $T < T_c$, there are two points where $\partial p/\partial \rho = 0$ equivalent to $\partial \rho/\partial p = \infty$, which means that a tiny fluctuation in the pressure can induce huge density fluctuations. These two points are called *spinodal* points, which are the stability limits of the metastable phases. Thus in the shadow region of Fig. 2.1, the gas-liquid phase transition needs to be generated by finite and localized fluctuations, which is called *nucleation*. During nucleation, the metastable phase has to overcome a finite free energy barrier, after which the nuclei can grow spontaneously.

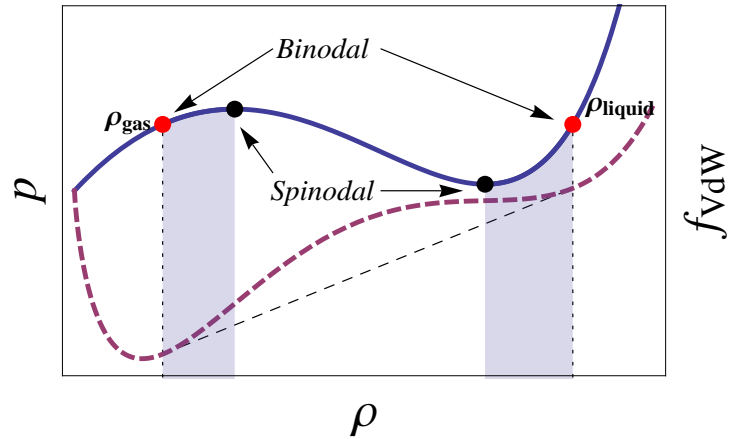

**Figure 2.1**: The equation of state (solid line) and the corresponding conveniently shifted and scaled Helmholtz free energy density $f_{\mathrm{vdW}} = F/V$ (thick dashed line) as a function of density $\rho$ for a system of van der Waals fluid at temperature $T < T_c$. The gas-liquid phase coexistence densities are $\rho_{\mathrm{gas}}$ and $\rho_{\mathrm{liquid}}$, respectively.

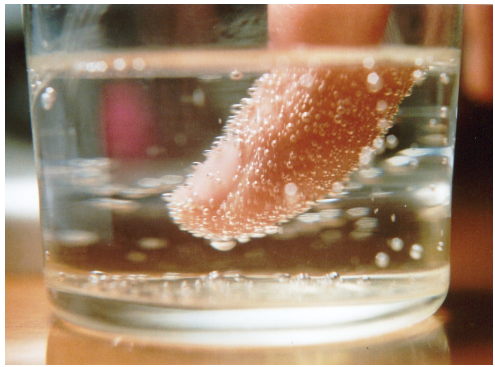

**Figure 2.2**: Nucleation of carbon dioxide bubbles around a finger.

Moreover, when we talk about nucleation, it is important to distinguish between *heterogeneous* and *homogeneous* nucleation. Heterogeneous nucleation happens when it can be assisted by a heterogeneity in the system, such as walls or impurities. For instance, when we put a finger into a cup of Sprite as shown in Fig. 2.2, there are more bubbles of carbon dioxide on the finger compared to the bulk phase, as nucleation of carbon dioxide bubbles is promoted by the interface created by the finger, i.e., heterogeneous nucleation. In contrast, homogeneous nucleation occurs due to spontaneous fluctuations in the bulk phase. Although in our real life, heterogeneous nucleation is more likely to happen, homo-geneous nucleation is not just a theoretical simplification for studying the physics of phase transitions. There are still a lot of situations in the real world where homogeneous nucleation dominates, such as condensation in supersonic nozzles [15], explosions which occurs when a cold liquid contacts a much hotter one [16], formation of heavily microcrystallized ceramics [17], (nano) particle synthesis, etc. Furthermore, experimental techniques, have advanced significantly in recent years, which makes it possible to study homogeneous nucleation in experiments. Nevertheless, the mechanism of homogeneous nucleation is still an open question, and even for a system of hard spheres, probably the simplest model to describe colloidal systems, there is still an ongoing debate on the discrepancy in the measured nucleation rates between experimental and theoretical methods [18–21]. This makes homogeneous nucleation an interesting and challenging topic. In the following, we briefly derive a commonly used theory to describe homogeneous nucleation, i.e., classical nucleation theory (CNT), which was first formulated by Volmer and Weber [22].

## 2.2   Free energy barrier

We consider a metastable phase $A$, e.g., a gas phase, at the thermodynamic condition where phase $B$, e.g., a liquid phase, is the stable phase. As shown in Fig. 2.3, droplets of phase $B$ exist in the phase $A$, and they may grow and shrink due to thermal fluctuations.

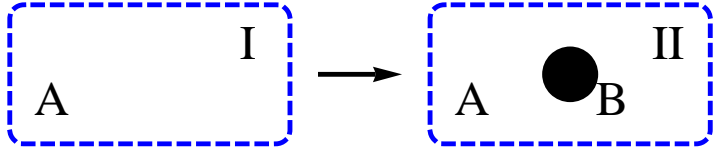

**Figure 2.3**:  Illustration of the nucleation of phase $B$ in a metastable phase $A$.

In the following, we first consider a homogeneous system containing only the metastable phase $A$ denoted as system I (left in Fig. 2.3) and a system II containing the metastable phase $A$ and a droplet (or a cluster) of phase $B$ (right in Fig. 2.3). The temperature of both systems is fixed and equal to $T$.

The internal energy of the homogeneous system I can be written as

$$U^{\mathrm{I}} = TS^{\mathrm{I}} - P^{\mathrm{I}}V^{\mathrm{I}} + \mu_A^{\mathrm{I}}N, \qquad (2.2)$$

where $T$ is the temperature of the system, $S^{\mathrm{I}}$ the total entropy, $P^{\mathrm{I}}$ the pressure, $V^{\mathrm{I}}$ the total volume, $\mu_A^{\mathrm{I}}$ the chemical potential and $N$ the total number of particles. The internal energy of system II, which contains the metastable phase $A$ and a cluster of phase $B$, is

$$U^{\mathrm{II}} = TS^{\mathrm{II}} - P_A^{\mathrm{II}}V_A^{\mathrm{II}} - P_B^{\mathrm{II}}V_B^{\mathrm{II}} + \mathcal{A}\gamma + \mu_A^{\mathrm{II}}N_A + \mu_B^{\mathrm{II}}N_B, \qquad (2.3)$$

where $P_{A,B}^{\mathrm{II}}$, $V_{A,B}^{\mathrm{II}}$ and $\mu_{A,B}^{\mathrm{II}}$ are the pressure, volume and chemical potential of phase $A$, $B$, respectively, with the superscripts indicating the values in system II, $\gamma$ the interfacial tension between $A$ and $B$ with $\mathcal{A}$ the surface area of cluster $B$, and $N_A$ and $N_B$ are the number of particles of phase $A$ and $B$ in system II, respectively.

If we keep the pressure in the metastable phase $A$ fixed, i.e., $P^{\mathrm{I}} = P_A^{\mathrm{II}} = P$, and the total number of particles constrain, i.e., $N = N_A + N_B$, then the Gibbs free energy in system I and II are

$$
\begin{aligned}
G^{\mathrm{I}} &= U^{\mathrm{I}} - TS^{\mathrm{I}} + P^{\mathrm{I}}V^{\mathrm{I}} = \mu_A^{\mathrm{I}} N & (2.4) \\
G^{\mathrm{II}} &= U^{\mathrm{II}} - TS^{\mathrm{II}} + P_A^{\mathrm{II}}(V_A^{\mathrm{II}} + V_B^{\mathrm{II}}) \\
&= (P - P_B^{\mathrm{II}})V_B^{\mathrm{II}} + \gamma\mathcal{A} + \mu_A^{\mathrm{II}} N_A + \mu_B^{\mathrm{II}} N_B & (2.5)
\end{aligned}
$$

Furthermore, as the pressure and temperature of the metastable phase $A$ are fixed, the chemical potential is constant, i.e., $\mu_A^{\mathrm{I}} = \mu_A^{\mathrm{II}}$, and the Gibbs free energy difference can be written as

$$
\begin{aligned}
\Delta G &= G^{\mathrm{II}} - G^{\mathrm{I}} \\
&= (P - P_B^{\mathrm{II}})V_B^{\mathrm{II}} + \mathcal{A}\gamma + \left[\mu_B^{\mathrm{II}}(P_B^{\mathrm{II}}) - \mu_A^{\mathrm{I}}(P_A)\right] N_B. & (2.6)
\end{aligned}
$$

To obtain the expression of the free energy in CNT, we make a few assumptions below:

1. The interfacial tension $\gamma$ is independent of the size of the cluster, or $\gamma = \gamma_\infty$, with $\gamma_\infty$ the interfacial tension of an infinite cluster or the planar interface.

2. The cluster is incompressible, meaning that its density $\rho$ does not change with pressure. This assumption is valid for the nucleation of a denser phase from a dilute phase, e.g. gas-liquid or fluid-crystal nucleation. Using the Gibbs-Duhem equation, the chemical potential of phase $B$ can be written as

$$
\begin{aligned}
\mu_B^{\mathrm{II}}(P_B^{\mathrm{II}}) &= \mu_B^{\mathrm{II}}(P) + \int_P^{P_B^{\mathrm{II}}} \frac{1}{\rho(P')}\mathrm{d}P' \\
&= \mu_B^{\mathrm{II}}(P) + \frac{P_B^{\mathrm{II}} - P}{\rho_B}, & (2.7)
\end{aligned}
$$

where $\rho_B$ is the density of phase $B$ at pressure $P$.

Therefore, we can re-write Eq. 2.6 into

$$
\begin{aligned}
\Delta G(N_B) &= \mathcal{A}(N_B)\gamma_\infty + \left[\mu_B^{\mathrm{II}}(P) - \mu_A^{\mathrm{I}}(P)\right] N_B \\
&= \mathcal{A}(N_B)\gamma_\infty - |\Delta\mu| N_B, & (2.8)
\end{aligned}
$$

where $\Delta\mu = \mu_B^{\mathrm{II}}(P) - \mu_A^{\mathrm{II}}(P)$ is the chemical potential difference between the two phases. Given phase $B$ is more stable than phase $A$, we have $\Delta\mu < 0$. In Eq. 2.8, the surface area of the cluster $\mathcal{A}$ depends on the cluster size $N_B$ and its shape. CNT assumes that the shape of the cluster is roughly spherical, and hence $\mathcal{A} = 4\pi R^2$ where $R$ is the radius of the spherical cluster. The Gibbs free energy of a spherical cluster with radius $R$ is given by

$$
\Delta G(R) = 4\pi R^2 \gamma_\infty - \frac{4}{3}\pi R^3 \rho_B |\Delta\mu|, \qquad (2.9)
$$

which contains two terms:

- a "surface" term $4\pi R^2 \gamma_\infty$, that takes into account the free energy cost of creating an interface between phase $A$ and $B$;

- a "volume" term $-\frac{4}{3}\pi R^3 \rho_B |\Delta\mu|$, that indicates the fact that phase $B$ is more stable than phase $A$, and can be interpreted as the driving force for the formation of phase $B$.

A typical representation of $\Delta G$ as a function of the cluster radius $R$ is shown in Fig. 2.4.

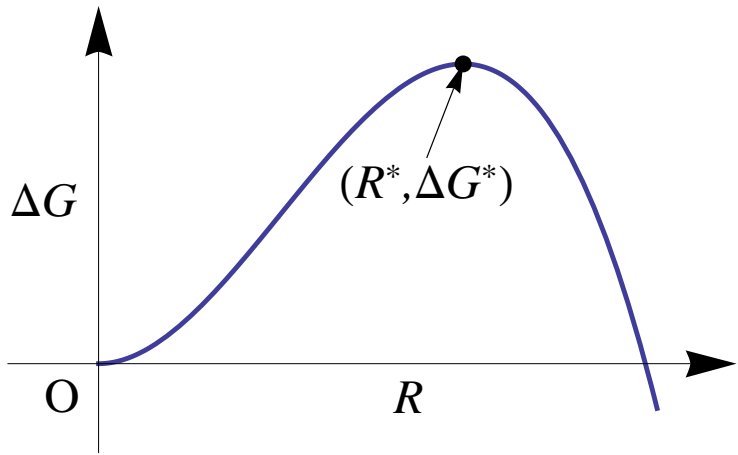

**Figure 2.4**: Gibbs free energy $\Delta G$ of a cluster as function of the cluster radius $R$ according to the classic nucleation theory where $R^*$ and $\Delta G^*$ are the critical cluster and the height of free energy barrier, respectively.

One finds that the free energy $\Delta G$ goes through a maximum at the critical cluster size

$$R^* = \frac{2\gamma_\infty}{\rho_B |\Delta\mu|}, \tag{2.10}$$

beyond which the nuclei can spontaneously grow. The height of the free energy barrier reads

$$\Delta G^* = \frac{16\pi}{3} \frac{\gamma_\infty^3}{(\rho_B |\Delta\mu|)^2}. \tag{2.11}$$

We note that CNT always predicts a finite free energy barrier, but the region where nucleation and growth can happen is bounded by the spinodal as shown in Fig. 2.1. In Chapter 6, we show that the nucleation of the stable smectic phase out of a supersaturated isotropic phase of hard rods is suppressed by an isotropic-nematic spinodal instability.

## 2.3   Kinetics of nucleation

The kinetics of nucleation is generally interpreted via a phenomenological reaction rate theory, which was first formulated by Volmer and Weber [22].

Volmer and Weber assumed that clusters of phase $B$ slowly grow or shrink via attachment or detachment of single particles:

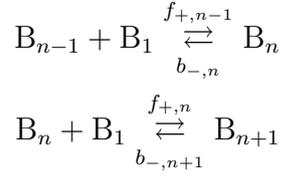

where $B_{n-1}$ is a cluster of $n-1$ particles, $B_1$ a cluster of one particle (monomer), $f_{+,n-1}$ and $b_{-,n}$ are the attachment and detachment rate of a single monomer to a cluster of $n-1$ and from a cluster of $n$ particles, respectively. This assumption is proposed on the basis of that the concentration of monomers is much higher than that of dimers, trimers, etc., and the interaction between the growing clusters is ideal-gas like [23]. Hence, reactions of clusters with dimers, trimers, etc., are infrequent compared with single particle attachment and detachment. The time-dependent cluster distribution $N_n(t)$ obeys the following Master equation:

$$\frac{\mathrm{d}N_n(t)}{\mathrm{d}t} = N_{n-1}(t)f_{+,n-1} + N_{n+1}(t)b_{-,n+1} - [N_n(t)f_{+,n} + N_n(t)b_{-,n}]. \tag{2.12}$$

The nucleation rate for a cluster size $n$ is the time-dependent flux of clusters that reaches $n$:

$$I_{n,t} = N_n(t)f_{+,n} - N_{n+1}(t)b_{-,n+1}. \tag{2.13}$$

We now assume that the system is in a steady state, in which the cluster size distribution does not change in time the nucleation rate is a constant.

$$I = N_n^s f_{+,n} - N_{n+1}^s b_{-,n+1}. \tag{2.14}$$

The equations can be solved by recurrence (see Ref. [24])

$$I = N_1^s \left[ \sum_{n=1}^{\infty} \frac{1}{f_{+,n}\xi_n} \right]^{-1}, \tag{2.15}$$

where

$$\xi_n = \prod_{i=1}^{n-1} \frac{f_{+,i}}{b_{-,i+1}} \quad \text{for} \quad n > 1. \tag{2.16}$$

The fluid is, due to the very small steady state flux, in a kind of quasi equilibrium, where for small cluster sizes $n \ll n^*$, the steady state cluster size distribution is almost equal to the equilibrium distribution. Since the clusters of size $n$ only interact with monomers, we assume that clusters of size $n$ are in equilibrium with respect to monomers, we then find

$$nN_1 \overset{K}{\rightleftharpoons} N_n, \tag{2.17}$$

where $N_1$ and $N_n$ are the equilibrium distribution of cluster size 1 and $n$, respectively, and $K$ the equilibrium constant. The ratio of products $\xi_n$ in Eq. 2.16 is just the equilibrium constant $K$ [24]. The equilibrium constant $K$ for this reaction is simply given by the free energy of forming a cluster with size $n$, and we find

$$\xi_n = K = \exp\left[-\beta \Delta G(n)\right] \tag{2.18}$$

Thus Eq. 2.15 can be re-written as

$$I = N_1 \left\{ \sum_{n=1}^{\infty} \frac{1}{f_{+,n} \exp\left[-\beta \Delta G(n)\right]} \right\}^{-1}. \tag{2.19}$$

In order to calculate the nucleation rate $I$, Becker and Döring made several approximations [25]:

1. the terms corresponding to the clusters near the top of the free energy barrier dominate the summation in Eq. 2.19;

2. the shape of $\Delta G(n)$ around $n^*$ can be approximated by a Taylor expansion at the top of free energy barrier

   $$\Delta G(n) = \Delta G(n^*) + \frac{1}{2} \Delta G(n^*)''(n - n^*)^2;$$

3. $f_{+,n}$ is further replaced by $f_{+,n^*}$;

4. the sum gets replaced by a integral from $n - n^* = -\infty$ and $n - n^* = \infty$, by considering $N_n$ to be a continuous function of $n$.

We then find the final expression of the steady-state nucleation rate

$$I = N_1 f_{+,n^*} \left( \frac{|\Delta G''(n^*)|}{2\pi k_B T} \right)^{1/2} \exp\left[-\beta \Delta G(n^*)\right], \tag{2.20}$$

where $\Delta G''(n^*)$ is the second order derivative of the Gibbs free energy with respect to the cluster size at the top of the free energy barrier $n^*$, and

$$Z = \left( \frac{|\Delta G''(n^*)|}{2\pi k_B T} \right)^{1/2} \tag{2.21}$$

is called the Zeldovitch factor. On the basis of this theory, Bennett [26] and Chandler [27] proposed a two step scheme to calculate the nucleation rate in simulations: the free energy barrier $\Delta G(n)$ can be obtained by performing Monte Carlo simulations using the umbrella sampling technique, and molecular dynamics simulations starting from configurations on the top of the free energy barrier can be employed to compute the attachment rate $f_{+,n^*}$.

The most important assumption in this theory is that when the cluster size is smaller than the critical cluster size, the system is in quasi-equilibrium, which could be problematic in some cases. For instance, in systems of binary mixtures, dynamical heterogeneities may make the kinetic pathways of nucleation out of equilibrium [28], and it may be influenced by the order parameter used in the umbrella sampling simulations. In Chapter 8, we will discuss the effect of the order parameter on the nucleation of binary hard-sphere mixtures.

## 2.4   Equilibrium distribution of cluster sizes

In order to determine the nucleation rate, we need to calculate the probability of finding a critical cluster, see Eq. 2.20. To this end, we consider a (meta) stable fluid in a cubic box consisting of $N$ particles at pressure $P$ and temperature $T$. The partition function is given by

$$Q(N, P, T) = \frac{\beta P}{\Lambda^{3N} N!} \int dV V^N \exp(-\beta PV) \int ds^N \exp(-\beta \mathcal{U}), \qquad (2.22)$$

where $\Lambda$ is the de Broglie wavelength, and $\mathbf{s}^N$ are the factional coordinates of the particles. We define an order parameter of the configuration $(\mathbf{s}^N, V)$ as $f(\mathbf{s}^N, V) = n$, where $n$ is the size of the largest cluster. This order parameter can be used to calculate the nucleation rate according to transition state theory, when it can distinguish the two different states, i.e., fluid and crystal. The partition function of a system where the largest cluster size is $n$ can be written as

$$Q_n(N, P, T) = \frac{\beta P}{\Lambda^{3N} N!} \int dV V^N \exp(-\beta PV) \int ds^N \exp(-\beta \mathcal{U}) \delta_K \left[ f(\mathbf{s}^N, V) - n \right], \quad (2.23)$$

where $\delta_K$ is the Kronecker delta function. Then the free energy difference between a system of $N$ particles containing a largest cluster of size $n$ and the (meta) stable fluid phase is

$$\begin{aligned}
\Delta G(n)_{NPT} &= -k_B T \ln \left[ \frac{Q_n(N, P, T)}{Q(N, P, T)} \right] \\
&= -k_B T \ln \left\langle \delta_K \left[ f(\mathbf{s}^N, V) - n \right] \right\rangle_{NPT}.
\end{aligned} \qquad (2.24)$$

The formation of a large cluster in the fluid phase is a rare event. When the cluster size is around the critical size, there is usually one large cluster present in the system. Thus

$$\Delta G(n)_{NPT} = -k_B T \ln \left\langle \delta_K \left[ f(\mathbf{s}^N, V) - n \right] \right\rangle_{NPT} = -k_B T \ln \left\langle N_n \right\rangle_{NPT}, \qquad (2.25)$$

where $N_n$ is the number of clusters of size $n$. This is the free energy of finding a cluster of size $n$ in a system of $N$ particles, and we scale this to a unit system by adding a $k_B T \ln N$ term in the free energy, since the nucleation rate is system size independent. The free energy barrier in a unit system is

$$\Delta G(n) = \Delta G(n)_{NPT} + k_B T \ln N = -k_B T \ln \left\langle \frac{N_n}{N} \right\rangle_{NPT}. \qquad (2.26)$$

We note that we can approximate $N \simeq \sum_{k=0}^{\infty} N_k$, as the number of fluid-like particles $N_0$ is dominant. We re-write Eq. 2.26

$$\begin{aligned}
\Delta G(n) &= -k_B T \ln \left\langle \frac{N_n}{\sum_{k=0}^{\infty} N_k} \right\rangle_{NPT} \\
&= -k_B T \ln P(n),
\end{aligned} \qquad (2.27)$$

where $P(n)$ is the probability distribution function of finding a cluster of size $n$. This can be calculated by Monte Carlo simulations with umbrella sampling technique. However, we should note that in our actual calculation, we still use Eq. 2.26 to compute the free energy barrier.

Moreover, when calculating the free energy barrier of nucleation, one should make sure that there is only one critical cluster in the simulation box. Thus the system size should be large but *not* too large that multiple critical clusters may exist.

# 3

## Crystal nucleation of hard spheres using molecular dynamics, umbrella sampling and forward flux sampling: A comparison of simulation techniques

Over the last number of years several simulation methods have been introduced to study rare events such as nucleation. In this chapter we examine the crystal nucleation rate of hard spheres using three such numerical techniques: molecular dynamics, forward flux sampling and a Bennett-Chandler type theory where the nucleation barrier is determined using umbrella sampling simulations. The resulting nucleation rates are compared with the experimental rates of Harland and Van Megen [J. L. Harland and W. van Megen, Phys. Rev. E **55**, 3054 (1997)], Sinn *et al.* [C. Sinn *et al.*, Prog. Colloid Polym. Sci. **118**, 266 (2001)] and Schätzel and Ackerson [K. Schätzel and B.J. Ackerson, Phys. Rev. E, **48**, 3766 (1993)] and the predicted rates for monodisperse and 5% polydisperse hard spheres of Auer and Frenkel [S. Auer and D. Frenkel, Nature **409**, 1020 (2001)]. When the rates are examined in long-time diffusion units, we find agreement between all the theoretically predicted nucleation rates, however, the experimental results display a markedly different behavior for low supersaturation. Additionally, we examined the pre-critical nuclei arising in the molecular dynamics, forward flux sampling, and umbrella sampling simulations. The structure of the nuclei appear independent of the simulation method, and in all cases, the nuclei contain on average significantly more face-centered-cubic ordered particles than hexagonal-close-packed ordered particles.

## 3.1 Introduction

Nucleation processes are ubiquitous in both natural and artificially-synthesized systems. However, the occurrence of a nucleation event is often rare and difficult to examine both experimentally and theoretically.

Colloidal systems are almost ideal model systems for studying nucleation phenomena. Nucleation and the proceeding crystallization in such systems often take place on experimentally accessible time scales, and due to the size of the particles, they are accessible to a wide variety of scattering and imaging techniques, such as (confocal) microscopy [29], holography [30], and light and x-ray scattering. Additionally, progress in particle synthesis [3], solvent manipulation, and the application of external fields [31] allows for significant control over the interparticle interactions, allowing for the study of a large variety of nucleation processes [32].

One such colloidal system is the experimental realization of "hard" spheres comprised of sterically stabilized polymethylmethacrylate (PMMA) particles suspended in a liquid mixture of decaline and carbon disulfide [33]. Experimentally, the phase behaviour of such a system has been examined by Pusey and Van Megen [8] and maps well onto the phase behaviour predicted for hard spheres. Specifically when the effective volume fraction of their system is scaled to reproduce the freezing volume fraction of hard spheres ($\phi = 0.495$) the resulting melting volume fraction is $\phi = 0.545 \pm 0.003$ [8] which is in good agreement with that predicted for hard spheres [34]. The nucleation rates have been measured using light scattering by Harland and Van Megen [33], Sinn *et al.* [35], Schätzel and Ackerson [36] and predicted theoretically by Auer and Frenkel [18].

On the theoretical side, hard-sphere systems are one of the simplest systems which can be applied to the study of colloidal and nanoparticle systems, and generally, towards the nucleation process itself. As such, it is an ideal system to examine various computational methods for studying nucleation, and comparing the results with experimental data. Such methods include, but are not limited to, molecular dynamics (MD) simulations, umbrella sampling (US), forward flux sampling (FFS), and transition path sampling (TPS). It is worth noting here that Auer and Frenkel [18] used umbrella sampling simulations to study crystal nucleation of hard spheres and found a significant difference between their predicted rates and the experimental rates of Refs. [33, 35, 36]. However, it was unclear where this difference originated. In this chapter we compare the nucleation rates for the hard-sphere system from MD, US and FFS simulations with the experimental results of Refs. [33, 35, 36]. We demonstrate that the three simulation techniques are consistent in their prediction of the nucleation rates, despite the fact that they treat the dynamics differently. Thus we conclude that the difference between the experimental and theoretical nucleation rates identified by Auer and Frenkel is not due to the simulation method.

A nucleation event occurs when a statistical fluctuation in a supersaturated liquid results in the formation of a crystal nucleus large enough to grow out and continue crystallizing the surrounding fluid. In general, small crystal nuclei are continuously being formed and melting back in a liquid. However, while most of these small nuclei will quickly melt, in a supersaturated liquid a fraction of these nuclei will grow out. Classical nucleation theory (CNT) is the simplest theory available for describing this process. In CNT it is assumed that the free energy for making a small nucleus is given by a surface free energy cost which is proportional to the surface area of the nucleus and a bulk free energy gain proportional to its volume. More specifically, according to CNT the Gibbs free energy difference between a homogeneous bulk fluid and a system containing a spherical nucleus of radius $R$ is given by

$$\Delta G(R) = 4\pi\gamma R^2 - \frac{4}{3}\pi \left|\Delta\mu\right| \rho_s R^3 \tag{3.1}$$

where $\left|\Delta\mu\right|$ is the difference in chemical potential between the fluid and solid phases, $\rho_s$ is the density of the solid, and $\gamma$ is the interfacial free energy density of the fluid-solid interface. This free energy difference is usually referred to as the nucleation barrier. From this expression, the radius of the critical cluster is found to be $R^* = 2\gamma/\left|\Delta\mu\right|\rho_s$ and the barrier height is $\Delta G^* = 16\pi\gamma^3/3\rho_s^2\left|\Delta\mu\right|^2$.

Umbrella sampling [37, 38] is a method to examine the nucleation process from which the nucleation barrier is easily obtained. The predicted barrier can then be used in combination with kinetic Monte Carlo (KMC) or MD simulations to determine the nucleation rate [18]. In US an order parameter for the system is chosen and configuration averages for sequential values of the order parameter are taken. In order to facilitate such averaging, the system is biased towards particular regions in configuration space. The success of the method is expected to depend largely on the choice of order parameter and biasing potential. Note that the free energy barrier is only defined in equilibrium, and thus is only applicable to systems which are in (quasi-) equilibrium.

Forward flux sampling [39–41] is a method of studying rare events, such as nucleation, in both equilibrium and non-equilibrium systems. Using FFS, the transition rate constants (e.g. the nucleation rate) for rare events can be determined when brute force simulations are difficult or even not possible. In FFS, a reaction coordinate $Q$ (similar to the order parameter in US) is introduced which follows the rare event. The transition rate between phase A and B is then expressed as a product of the flux ($\Phi_{A\lambda_0}$) of trajectories crossing the A state boundary, typically denoted $\lambda_0$, and the probability ($P(\lambda_B|\lambda_0)$) that a trajectory which has crossed this boundary will reach state B before returning to state A. Thus the transition rate constant is written as

$$k_{AB} = \Phi_{A\lambda_0} P(\lambda_B|\lambda_0). \tag{3.2}$$

Forward flux sampling facilitates the calculation of probability $P(\lambda_B|\lambda_0)$ by breaking it up into a set of probabilities between sequential values of the reaction coordinate. Little information regarding the details of the nucleation process is required in advance, and the choice of reaction coordinate is expected to be less important than the order parameter in US. Additionally, unlike US, FFS utilizes dynamical simulations and hence this technique does not assume that the system is in (quasi-)equilibrium.

Molecular dynamics and Brownian dynamics (BD) simulations are ideal for studying the time evolution of systems, and, when possible, they are the natural technique to study dynamical processes such as nucleation. Unfortunately, however, available the computational time often limits the types of systems which can be effectively studied by these dynamical techniques. Brownian dynamics simulations, which would be the natural choice to use for colloidal systems, are very slow due to the small time steps required to handle the steep potential used to approximate the hard-sphere potential. Event driven

| $\phi$ | $\beta p\sigma^3$ | $\beta\,\lvert\Delta\mu\rvert$ | $\rho_s\sigma^3$ |
|--------|-------------------|-------------------------------|------------------|
| 0.5214 | 15.0 | 0.34 | 1.107 |
| 0.5284 | 16.0 | 0.44 | 1.122 |
| 0.5316 | 16.4 | 0.48 | 1.128 |
| 0.5348 | 16.9 | 0.53 | 1.135 |
| 0.5352 | 17.0 | 0.54 | 1.136 |
| 0.5381 | 17.5 | 0.58 | 1.142 |
| 0.5414 | 18.0 | 0.63 | 1.148 |
| 0.5478 | 19.1 | 0.74 | 1.161 |
| 0.5572 | 20.8 | 0.90 | 1.178 |

**Table 3.1**: Packing fraction ($\phi = \pi\sigma^3 N/6V$) , reduced pressure ($\beta p\sigma^3$) and reduced chemical potential difference between the fluid and solid phases ($\beta\,\lvert\Delta\mu\rvert$) and reduced number density of the solid phase $\rho_s$ of the state points studied in this chapter. The chemical potential difference was determined using thermodynamic integration[10], and the equations of state for the fluid and solid are from Refs. [43, 44] respectively.

MD simulations are much more efficient to simulate hard spheres and enable us to study spontaneous nucleation of hard-sphere mixtures over a range of volume fractions. The main difference between the two simulation methods lies in how they treat the short-time motion of the particles. Fortunately, the nucleation rate is only dependent on the long-time dynamics which are not sensitive to the details of the short-time dynamics of the system [42].

In this chapter we study in detail the application of US and FFS techniques to crystal nucleation of hard spheres, and predict the associated nucleation rates. Combining these nucleation rates with results from MD simulations, we make predictions for the nucleation rates over a wide range of packing fractions $\phi = 0.5214 - 0.5572$, with corresponding pressures and supersaturations shown in Table 3.1. We compare these theoretical nucleation rates with the rates measured experimentally by Refs. [33, 35, 36].

This chapter is organized as follows: in section 3.2 we discuss the model, in section 3.3 we describe and examine the order parameter used to distinguish between solid- and fluid-like particles throughout this chapter, in section 3.4 we calculate essentially the "exact" nucleation rates using MD simulations, in sections 3.5 and 3.6 we calculate the nucleation rates of hard spheres using US and FFS respectively, and discuss difficulties in the application of these techniques, in section 3.7 we summarize the theoretical results and compare the predicted nucleation rates with the measured experimental rates of Harland and Van Megen [33], Sinn *et al.* [35], and Schätzel and Ackerson [36] and section 3.8 contains our conclusions.

## 3.2   Model

In this chapter we examine the nucleation rate between spheres with diameter $\sigma$ which
interact via a hard-sphere pair potential given by

$$\beta U^{\text{HS}}(r_{ij}) = \begin{cases} 0 & r_{ij} \geq \sigma \\ \infty & r_{ij} < \sigma. \end{cases} \tag{3.3}$$

where $r_{ij}$ is the center-to-center distance between particles $i$ and $j$ and $\beta = 1/k_B T$ with
$k_B$ Boltzmann's constant and $T$ the temperature. This is in contrast to several studies on
"hard" spheres where the hard sphere potential is approximated by a slightly soft potential
(e. g., Refs. [20, 45]) so that Brownian dynamics simulations or traditional molecular
dynamics simulations (i. e., molecular dynamics which is not event driven), which require
a continuous potential, can be used. We would like to emphasize this distinction here
as the hardness of the interaction has previously been shown to play a significant role in
nucleation rates [46, 47], see also Chapter 4 for a discussion.

## 3.3   Order Parameter

In this chapter, an order parameter is used to differentiate between liquid-like and solid-
like particles and a cluster algorithm is used to identify the solid clusters. For this study we
have chosen to use the local bond-order parameter introduced by Ten Wolde *et al.* [48, 49]
in the study of crystal nucleation in a Lennard-Jones system. This order parameter has
been used in many crystal nucleation studies, including a previous study of hard-sphere
nucleation by Auer and Frenkel [18].

In the calculation of the local bond order parameter a list of "neighbours" is determined
for each particle. The neighbours of particle $i$ include all particles within a radial distance
$r_c$ of particle $i$, and the total number of neighbours is denoted $N_b(i)$. A bond orientational
order parameter $q_{l,m}(i)$ for each particle is then defined as

$$q_{l,m}(i) = \frac{1}{N_b(i)} \sum_{j=1}^{N_b(i)} \Upsilon_{l,m}(\theta_{i,j}, \phi_{i,j}) \tag{3.4}$$

where $\Upsilon_{l,m}(\theta, \phi)$ are the spherical harmonics, $m \in [-l, l]$ and $\theta_{i,j}$ and $\phi_{i,j}$ are the polar and
azimuthal angles of the center-of-mass distance vector $\mathbf{r}_{ij} = \mathbf{r}_j - \mathbf{r}_i$ with $\mathbf{r}_i$ the position
vector of particle $i$. Solid-like particles are identified as particles for which the number of
connections per particle $\xi(i)$ is at least $\xi_c$ and where

$$\xi(i) = \sum_{j=1}^{N_b(i)} H\left[d_l(i,j) - d_c\right], \tag{3.5}$$

$H$ is the Heaviside step function, $d_c$ is the dot-product cutoff, and

$$d_l(i,j) = \frac{\sum\limits_{m=-l}^{l} q_{l,m}(i) q_{l,m}^*(j)}{\left(\sum\limits_{m=-l}^{l} |q_{l,m}(i)|^2\right)^{1/2} \left(\sum\limits_{m=-l}^{l} |q_{l,m}(j)|^2\right)^{1/2}}. \tag{3.6}$$

A cluster contains all solid-like particles which have a solid-like neighbour in the same cluster. Thus each particle can be a member of only a single cluster.

The parameters contained in this algorithm include the neighbour cutoff $r_c$, the dot-product cutoff $d_c$, the critical value for the number of solid-like neighbours $\xi_c$, and the symmetry index for the bond orientational order parameter $l$. The solid nucleus of a hard-sphere crystal is expected to have randomly stacked hexagonal order, thus the symmetry index is chosen to be 6 in all cases in this study. Note that this order parameter does not distinguish between FCC and HCP ordered particles.

To investigate the effect of the choice of $\xi_c$, we examined the number of correlated bonds per particle at the liquid-solid interface. To this end, we constructed a configuration in the coexistence region in an elongated box by attaching a box containing an equilibrated random-hexagonal-close-packed (RHCP) crystal to a box containing an equilibrated fluid. Note that the RHCP crystal was placed in the box such that the hexagonal layers were parallel to the interface. The new box was then equilibrated in an $NpT$ MC simulation. We then examined the density profile of solid-like particles as determined by our order parameter using $r_c = 1.4\sigma$, $d_c = 0.7$ and $\xi_c = 5, 7$ and 9. As shown in Fig. 3.1, for all values of $\xi_c$ that we examined if the order parameter appears to consistently identify the particles belonging to the bulk fluid and solid regions. For comparison we also show a typical configuration of the RHCP crystal in coexistence with the fluid phase. The solid-like particles as defined by the order parameter are labelled according to the number of solid-like neighbours while the fluid-like particles are denoted by dots. The main difference between these order parameters relates to distinguishing between fluid- and solid-like particles at the fluid-solid interface. Unsurprisingly, the location of the interface seems to shift in the direction of the bulk solid as $\xi_c$ is increased. We note that the dips in the density profile correspond to HCP stacked layers which are more pronounced for higher values of $\xi_c$.

## 3.4   Molecular Dynamics

### 3.4.1   Nucleation Rates

In MD simulations the equations of motion are integrated to follow the time evolution of the system. Since the hard-sphere potential is discontinuous the interactions only take place when particles collide. Thus the particles move in straight lines (ballistic) until they encounter another particle with which they perform an elastic collision [50]. These collision events are identified and handled in order of occurrence using an event driven simulation.

In theory, using an MD simulation to determine nucleation rates is quite simple. Starting with an equilibrated fluid configuration, an MD simulation is used to evolve the system until the largest cluster in the system exceeds the critical nucleus size. The MD time associated with such an event is then measured and averaged over many initial configurations. The nucleation rate is given by

$$I = \frac{1}{\langle t \rangle V} \tag{3.7}$$

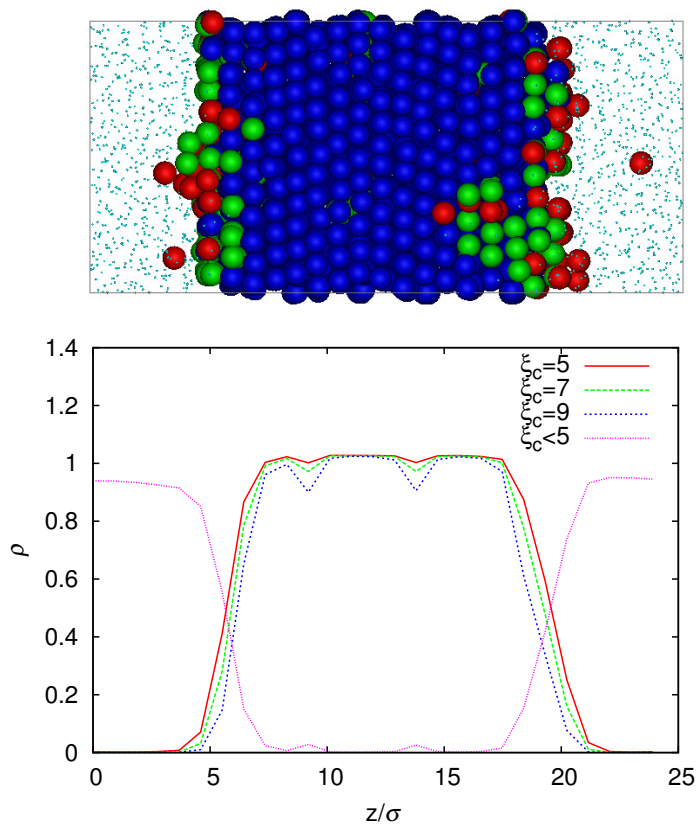

**Figure 3.1**: Top: A typical configuration of an equilibrated random-hexagonal-close-packed (RHCP) crystal in coexistence with an equilibrated fluid. The crystalline particles are labelled according to three different crystallinity criteria: the red particle have between $\xi = 5$ and 6 crystalline bonds, the green particles have between $\xi = 7$ and 8 crystalline bonds and the blue particles have $\xi \geq 9$ or more crystalline bonds. The fluid-like particles ($\xi < 5$) are denoted by dots. Bottom: The density profile of particles with a minimum number of neighbours $\xi$ as labelled. Note that the dips in the density profile correspond to HCP stacked layers. This implies that near the interface, the order parameter is slightly more sensitive to FCC ordered particles than to HCP ordered particles.

where $V$ is the volume of the system and $\langle t \rangle$ is the average time to form a critical nucleus. Measuring this time is relatively easy for low supersaturations where the nucleation times are relatively long compared to the nucleation event itself, which corresponds with a steep increase in the crystalline fraction of the system. However, for high supersaturations pinpointing the time of a nucleation event is more difficult. Often many nuclei form immediately and the critical nucleus sizes must be estimated from CNT or US simulations. Additionally, the precise details of the initial configuration can play a role at high supersaturations since the equilibration time of the fluid is of the same order of magnitude as the nucleation time. Hence, for each individual MD simulation we used a new initial configuration which was created by quenching the system very quickly.

For the results in this chapter, we performed MD simulations with up to 100,000 particles in a cubic box with periodic boundary conditions in an NVE ensemble. Time was measured in MD units $\sigma\sqrt{m/k_BT}$. The order parameter was measured every 10 time units and when the largest cluster exceeded the critical size by 100 percent we estimated the time $\tau_{\mathrm{nucl}}$ at which the critical nucleus was formed using stored previous configurations. We performed up to 20 runs for every density and averaged the nucleation times.

| Volume fraction | Average nucleation time | Rate |
|---|---|---|
| $\phi$ | $t\sqrt{k_BT/(m\sigma^2)}$ | $I\sigma^5/(6D_l)$ |
| 0.5316 | $1\cdot10^6$ | $5\cdot10^{-9}$ |
| 0.5348 | $1.7\cdot10^4$ | $3.6\cdot10^{-7}$ |
| 0.5381 | $1.4\cdot10^3$ | $5.3\cdot10^{-6}$ |
| 0.5414 | $2.0\cdot10^2$ | $4.3\cdot10^{-5}$ |
| 0.5478 | 42 | $3.0\cdot10^{-4}$ |
| 0.5572 | 10 | $2.4\cdot10^{-3}$ |

**Table 3.2**: The average nucleation time, obtained from MD simulations, to form a critical cluster that grew out and filled the box. The last column contains the rate ($I$) in units of $(6D_l)/\sigma^5$.

The results are shown in Table 3.2. The nucleation times shown here are for a system of $2.0\cdot10^4$ particles and in MD time units. To compare with other data we convert the MD time units to units of $\sigma^2/(6D_l)$ with $D_l$ the long-time self diffusion coefficient measured in the same MD simulations. We were not able to measure the long-time self diffusion coefficients for high densities because our measurements were influenced by crystallization. We used the fit obtained by Zaccarelli *et al.* [51] who used polydisperse particles to prevent crystallization. For $\phi < 0.54$, we find good agreement between our data for $D_l$ and this fit.

## 3.5   Umbrella Sampling

### 3.5.1   Gibbs Free-Energy Barriers

Umbrella sampling is a technique developed by Torrie and Valleau to study systems where Boltzmann-weighted sampling is inefficient [37]. This method has been applied frequently to study rare events, such as nucleation [38], and specifically has been applied in the past to study the nucleation of hard spheres [18]. In general, umbrella sampling is used to examine parts of configurational space which are inaccessible by traditional schemes, eg. Metropolis Monte Carlo simulations. Typically, a biasing potential is added to the true interaction potential causing the system to oversample a region of configuration space. The biasing potential, however, is added in a manner such that is it easy to "un"-bias the measurables.

In the case of nucleation, while it is simple to sample the fluid, crystalline clusters of larger sizes will be rare, and as such, impossible to sample on reasonable time scales. The typical biasing potential for studying nucleation is given by [48, 52]

$$U_{\mathrm{bias}}[n(\mathbf{r}^N)] = \frac{\lambda}{2}\left[n(\mathbf{r}^N) - n_C\right]^2 \qquad (3.8)$$

where $\lambda$ is a coupling parameter, $n(\mathbf{r}^N)$ is the size of the largest cluster associated with configuration $\mathbf{r}^N$, and $n_C$ is the targeted cluster size. By choosing $\lambda$ carefully, the simulation will fluctuate around the part of configurational space with $n(\mathbf{r}^N)$ in the vicinity of $n_C$. The expectation value of an observable $A$ is then given by

$$\langle A \rangle = \frac{\left\langle A/W(n(\mathbf{r}^N)) \right\rangle_{\text{bias}}}{\langle 1/W(n(\mathbf{r}^N)) \rangle_{\text{bias}}} \qquad (3.9)$$

where

$$W(n) = e^{-\beta U_{\text{bias}}(n)}. \qquad (3.10)$$

Using this scheme to measure the probability distribution $P(n)$ for clusters of size $n$, the Gibbs free energy barrier can be determined by [53]

$$\beta \Delta G(n) = \text{constant} - \ln[P(n)]. \qquad (3.11)$$

Many more details on this method are given elsewhere [10, 53].

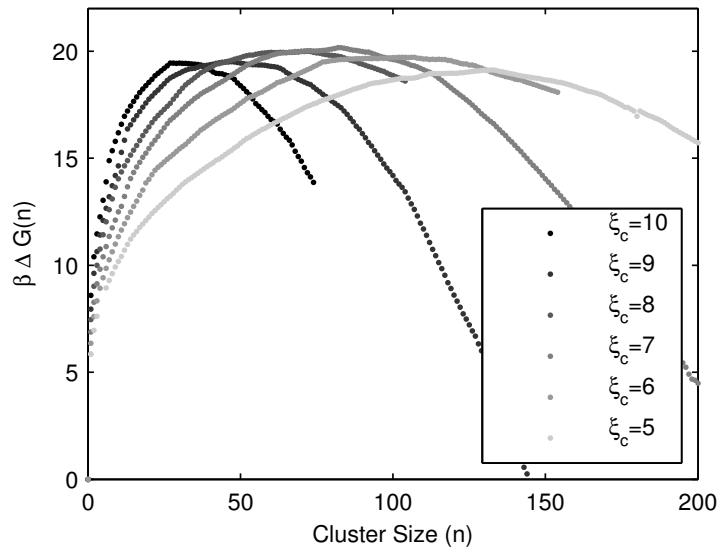

**Figure 3.2**: Gibbs free energy barriers $\beta\Delta G(n)$ as a function of cluster-size $n$ as obtained from umbrella sampling simulations at a reduced pressure of $\beta p\sigma^3 = 17$ for varying critical number of solid-like neighbours $\xi_c$ as labelled. For $\xi_c = 5, 7, 9$, the neighbour cutoff is $r_c = 1.4$ and for $\xi_c = 6, 8, 10$, $r_c = 1.3$. In all cases the dot product cutoff is $d_c = 0.7$.

For a pressure of $\beta p\sigma^3 = 17$, corresponding to a supersaturation of $\beta |\Delta\mu| = 0.54$, we examine the effect of one of the order parameter variables, namely $\xi_c$, on the prediction of the nucleation barriers. The barriers predicted by US using $\xi_c = 5, 6, 7, 8, 9$ and 10 are shown in Fig. 3.2. Note that the height of the barriers does not depend on $\xi_c$ within error bars. In general, for larger values of $\xi_c$ more particles are identified as fluid as compared with smaller values of $\xi_c$. This is consistent with the differences between these order parameters as demonstrated in Fig. 3.1. Thus, the radius measured in our simulation will depend on the definition of the order parameter. However, from classical nucleation

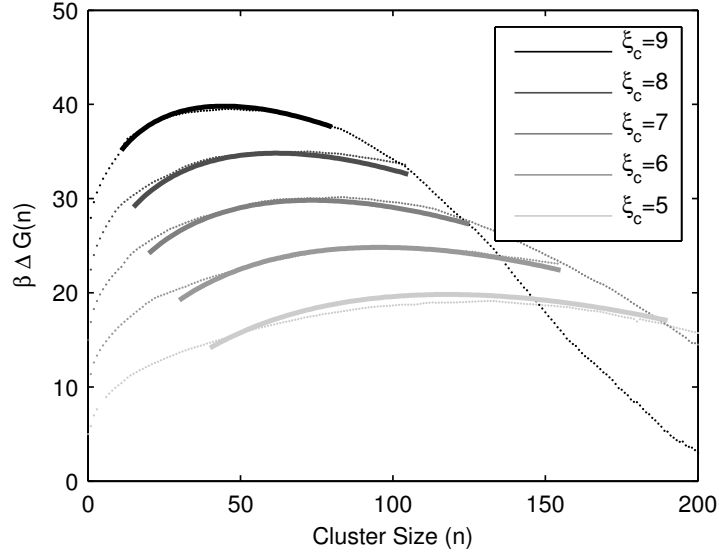

**Figure 3.3**:  Classical nucleation theory fits (thick lines) to the Gibbs free energy barriers obtained from umbrella sampling simulations at a reduced pressure of $\beta p \sigma^3 = 17$ for varying $\xi_c$ as labelled. Note that the CNT radius ($R_{CNT}$) is related to the radius ($R(\xi_c)$) measured by umbrella sampling by $R(\xi_c) = R_{\text{CNT}} + \alpha(\xi_c)$, where $\alpha(\xi_c)$ is a constant that corrects for the different ways the various order parameters identify the particles at the fluid-solid interface. The fit parameters are given in Table 3.3. We have shifted the barriers for $\xi_c = 6 - 9$ by $5, 10, 15, 20$ $k_B T$ respectively for clarity

theory (Eq. 1), there exists a unique definition of the liquid-solid interface and this a unique radius associated with CNT which we define as $R_{CNT}$. To a first approximation, for each definition of the order parameter, this radius ($R_{CNT}$) differs from that measured by our simulation ($R(\xi_c)$) by a constant which we denote as $\alpha(\xi_c)$, which is also dependent on $\xi_c$. Thus, we fit the barriers corresponding to $\xi_c = 5, 6, 7, 8$ and $9$ using CNT where we have

$$R(\xi_c) = R_{\text{CNT}} + \alpha(\xi_c). \tag{3.12}$$

Note that we have assumed that the cluster size $n$ can be related to the cluster radius

|      | $\beta \lvert \Delta \mu \rvert$ | $\beta \gamma \sigma^2$ | $R^*_{\text{CNT}}$ |
|------|------|------|------|
| CNT  | 0.54 | 0.76 | 2.49 |
| ACNT | 0.54 | 0.63 | 2.06 |

|      | $\alpha(5)$ | $\alpha(6)$ | $\alpha(7)$ | $\alpha(8)$ | $\alpha(9)$ | $c(5)$ | $c(6)$ | $c(7)$ | $c(8)$ | $c(9)$ |
|------|------|------|------|------|------|------|------|------|------|------|
| CNT  | -0.425 | -0.231 | -0.000 | 0.139 | 0.380 |  |  |  |  |  |
| ACNT | -0.879 | -0.698 | -0.464 | -0.335 | -0.076 | 7.80 | 8.56 | 8.84 | 8.87 | 8.34 |

**Table 3.3**: Numerical values for the parameters associated with the fits in Figs. 3.3 and 3.4 for classical nucleation theory and the adjusted classical nucleation theory presented in this chapter.

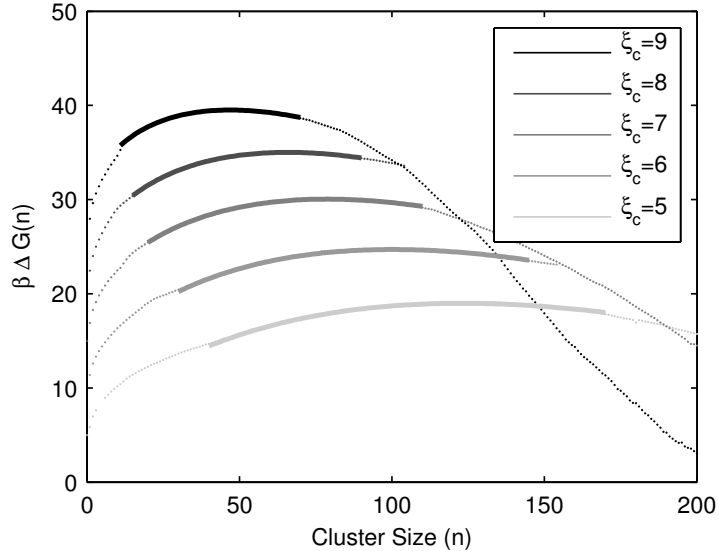

**Figure 3.4**: Fits of an adjusted classical nucleation theory (ACNT) presented in Section 3.5.1 to the Gibbs free energy barriers predicted using umbrella sampling simulations at a reduced pressure of $\beta p \sigma^3 = 17$ and for varying $\xi_c$ as labelled. Note that the CNT radius ($R_{\mathrm{CNT}}$) is related to the radius measured by umbrella sampling by $R(\xi_c) = R_{CNT} + \alpha(\xi_c)$, where $\alpha(\xi_c)$ is a constant. The fit parameters are given in Table 3.3. We have shifted the barriers for $\xi_c = 6 - 9$ by $5, 10, 15, 20\ k_B T$ respectively for clarity.

$R(\xi_c)$ by

$$n(\xi_c) = \frac{4\pi R(\xi_c)^3 \rho_s}{3}. \tag{3.13}$$

Only the top part of the free energy barriers are expected to fit to classical nucleation theory, so we take the top of the barrier corresponding to the region where the difference between $\beta\Delta G(n)$ and $\beta\Delta G(n^*)$ is approximately 5. Fitting all barriers simultaneously for the interfacial free energy density $\gamma$, the classical nucleation theory radius $R_{CNT}$, and the various $\alpha(\xi_c)$, we obtain the fits displayed in Fig. 3.3. From the various values of $\alpha$, the associated critical CNT radius ($R_{\mathrm{CNT}}^*$) can be determined. We find $R_{\mathrm{CNT}}^* = 2.49\sigma$. Additionally, we find an interfacial free energy density of $\beta\gamma\sigma^2 = 0.76$ which roughly agrees with the results of Auer and Frenkel who obtained $\beta\gamma\sigma^2 = 0.699, 0.738$ and $0.748$ for pressures $\beta p \sigma^3 = 15, 16$ and $17$ respectively [18]. However, recent calculations by David-chack *et al.* [54] of the interfacial free energy density at the fluid-solid coexistence find $\beta\gamma\sigma^2 = 0.574, 0.557$ and $0.546$ for the crystal planes (100), (110), and (111) respectively. For a spherical nucleus, the interfacial free energy density is expected to be an average over the crystal planes and was found to be $\beta\gamma\sigma^2 = 0.559$ [54]. Thus our result for the interfacial free energy density and that of Ref. [18] appear to be an overestimation.

There have been a number of papers discussing possible corrections to CNT (eg. Refs. [55, 56]). Recent work on the 2D Ising model, a system where both the interfacial free energy density and supersaturation are known analytically, demonstrated that in order to match a nucleation barrier obtained from US to CNT, two correction terms were required, specifically a term proportional to $\log(N)$ as well as a constant shift in $\Delta G$ which we define as $c$ [55]. The US barrier is only expected to match CNT near the top of the barrier where the $\log(N)$ term is almost a constant. Thus, we propose fitting the barrier to an adjusted expression for CNT (ACNT), by adding a constant $c$ to Eq. 1. Fitting the US barriers with this proposed form for the Gibbs free energy barrier, where we assume $c$ is a function of $\xi_c$, we obtain the fits displayed in Fig. 3.4. In this case we find an interfacial free energy density $\beta\gamma\sigma^2 = 0.63$, and the values for $\alpha(\xi_c)$ and $c(\xi_c)$ are given in Table 3.3. We note that this fit is much better than the fits in Fig. 3.3. The difference in the various $c(\xi_c)$ are around $1\mathrm{k_B T}$ and correspond well to the difference in heights of the barriers. More strikingly, the interfacial free energy density predicted from this proposed free energy barrier is in much better agreement with recent calculations of Davidchack $et$ $al.$ [54], than the interfacial free energy density we calculate using classical nucleation theory directly. We would also like to point out that it has been proposed that the effective interfacial free energy density will increase with pressure. However, an increase from the $\beta\gamma\sigma^2 = 0.559$ at coexistence predicted by Ref. [54] to $\beta\gamma\sigma^2 = 0.76$ predicted from CNT is larger than what would be expected (see e.g. Refs. [57, 58]). For a more thorough examination on the interfacial free energy densities of the hard sphere model, see Ref. [58]. We would like to point out here that due to the simple form of the nucleation barrier, it is difficult to be certain of any fit with more than one fitting parameter, as there are many combinations of parameters which fit almost equally well. To examine in more detail the accuracy of these fits, we have calculated the root mean square of the residual for the two fits which we denote as $\sigma_{\mathrm{RMSR}}$. In the case of the CNT fit we find $\sigma_{\mathrm{RMSR}} = 0.50$ while for the ACNT fit we find $\sigma_{\mathrm{RMSR}} = 0.11$ indicating that the ACNT fit is much better than the CNT fit. Additionally, we examined the ACNT fits for various interfacial free energy densities $\gamma$. Fixing the interfacial free energy density in the ACNT fit to the value found by CNT ($\beta\gamma\sigma^2 = 0.76$), we find $\sigma_{\mathrm{RMSR}} = 0.27$ and when we use interfacial free energy density at coexistence [54] ($\beta\gamma\sigma^2 = 0.559$) we find $\sigma_{\mathrm{RMSR}} = 0.18$.

Using either expressions for the Gibbs free energy barrier, namely CNT and ACNT, we were unable to fit the barrier corresponding to $\beta p\sigma^3 = 17$ and $\xi_c = 10$ simultaneously with the other predicted barriers for the same pressure. We speculate that our difficulty in fitting the barrier at $\xi_c = 10$ stems from an "over-biasing" of the system. Specifically, by using $\xi_c = 10$ the biasing potential could cause the system to sample more frequently more ordered clusters, and hence change slightly the region of phase space available to the US simulations. In general, the least biased systems would be expected to explore the largest region of phase space resulting in the best results. It should be noted that, in fact, this problem is simply an equilibration and measuring problem, but it does emphasize the difficulty caused by using an overly strong biasing potential.

In conclusion, with the exception of $\xi_c = 10$, the value of $\xi_c$ used in the order parameter did not appear to have an effect on the nucleation barriers once the difference in their measurements of the solid-liquid interface was taken into consideration. Finally, for use in our nucleation rate calculations (section 3.5.2) we also calculated the Gibbs free energy $\Delta G(n)$ for reduced pressures $\beta p\sigma^3 = 15$ and 16 using umbrella sampling simulations. We present the barrier heights in Table 3.4.

### 3.5.2 Umbrella Sampling Nucleation Rates

The nucleation barriers as obtained from US simulations can be used to determine the nucleation rates. The crystal nucleation rate $I$ is related to the Gibbs free energy barrier $(G(n))$ by [18]

$$I = \kappa e^{-\beta \Delta G(n^*)} \tag{3.14}$$

where

$$\kappa \approx \rho f_{n^*} \sqrt{\frac{|\beta \Delta G''(n^*)|}{2\pi}}, \tag{3.15}$$

$n^*$ is the number of particles in the critical nucleus, $\rho$ is the number density of the supersaturated fluid, $f_{n^*}$ is the rate particles are attached to the critical cluster, and $G''$ is the second derivative of the Gibbs free energy barrier. Auer and Frenkel [18] showed that the attachment rate $f_{n^*}$ could be related to the mean square deviation of the cluster size at the top of the barrier by

$$f_{n^*} = \frac{1}{2} \frac{\langle \Delta n^2(t) \rangle}{t}. \tag{3.16}$$

The mean square deviation (MSD) of the cluster size $\Delta n^2(t) = \langle n(t) - n^* \rangle$ can then be calculated by either employing a kinetic MC simulation or a MD simulation at the top of the barrier. For simplicity, in the remainder of this chapter the nucleation rate determined using this method will be referred to as umbrella sampling (US) nucleation rates, although to calculate the nucleation rates both US simulations and dynamical simulations (KMC or MD) are necessary.

The mean square deviation, or variance, in the cluster size appearing in Eq. 3.16 has both a short- and long-time behaviour. At short times, fluctuations are due to particles performing Brownian motion around their average positions while the long-time behaviour is caused by rearrangements of particles required for the barrier crossings. The slope of the variance is large at short times where only the fast rattling is sampled. However, the longer the time the further the system has diffused away from the critical cluster size at the top of the nucleation barrier. Auer [59] states that runs need to be selected that remain at the top of the barrier. However, when this is done the attachment rate is lower than when the average over all runs is taken since it excludes the runs that move off the barrier fast and have the largest attachment rate. This problem is analogous to determining the diffusion constant of a particle performing a random walk. By only including walks which remain in the vicinity of the origin, the measurement is biased and excludes trajectories which quickly move away from the origin. This results is an underestimation of the diffusion constant, and similarly, in this case, an underestimate of the attachment rate. Hence, in this chapter we do not attempt to prevent the trajectories from falling off the barrier and we include all trajectories. In Fig. 3.5 we demonstrate how, starting from a critical cluster, the size of the nucleus fluctuates as a function of time and, in fact, can completely disappear or double in size within $0.3\tau_l$ where $\tau_l$ is the time that it takes a particle on average to diffuse over a distance equal to its diameter i.e. $\tau_l = \sigma^2/(6D_l)$.

The kinetic prefactor was determined using KMC simulations with 3000 particles in an NVT ensemble in a cubic box with periodic boundary conditions. The initial configurations were taken from US simulations in one of the windows at the top of the barrier. We examined the results from both Gaussian and uniformly distributed Monte Carlo steps and found agreement within statistical errors. For all the simulations, the MC stepsize was between $0.01\sigma$ and $0.1\sigma$. The variance of the cluster size for a typical system is shown in Fig. 3.6. We observed a large variance in the attachment rates calculated for different nuclei. Specifically, some nuclei have attachment rates more than an order of magnitude higher than other nuclei of similar size. The nuclei with low attachment rates appeared to have a smoother surface than the nuclei with a high attachment rate. In calculating the attachment rates we used 10 independent configurations on the top of the barrier and followed 10 trajectories from each.

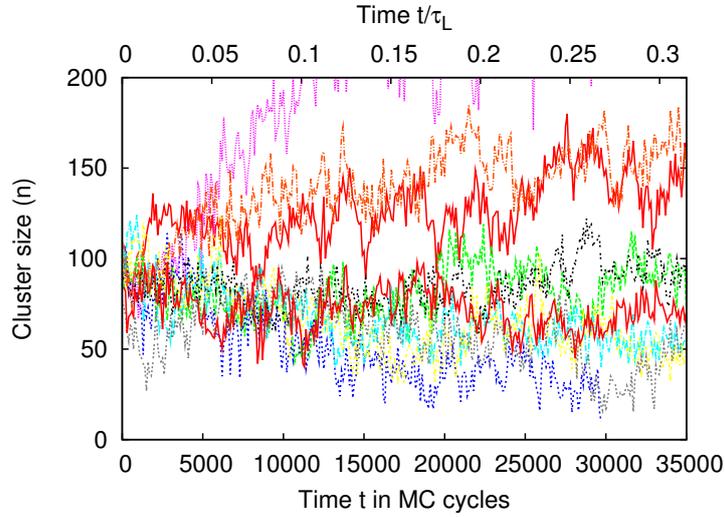

**Figure 3.5**: The cluster size $(n(t))$ as a function of time in MC cycles for a random selection of clusters that start at the top of the nucleation barrier.

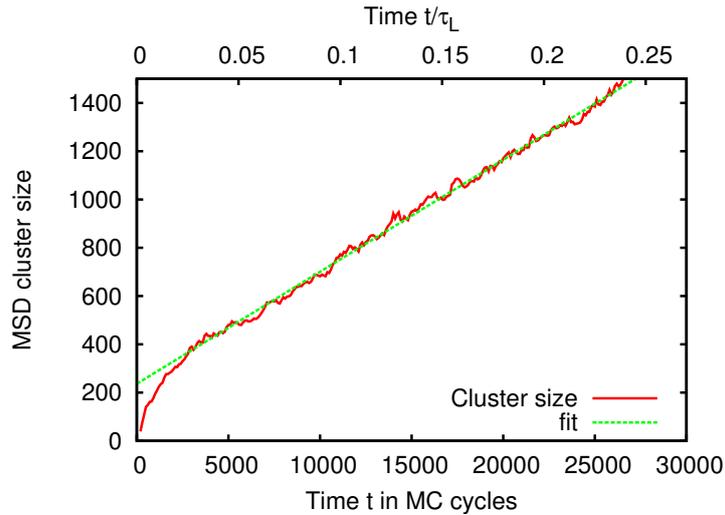

**Figure 3.6**: The mean squared deviation (MSD) of the cluster size $\langle \Delta n^2(t) \rangle$ as function of time $t$ in MC cycles. The cluster size has been measured every cycle and averaged over 100 cycles to reduce the short-time fluctuations. The slope of this graph is twice the attachment rate (Eq. 3.16).

| $\beta p\sigma^3$ | $\xi_c$ | $n^*$ | $\beta\Delta G(n^*)$ | $\beta\Delta G''(n^*)$ | $f_{n^*}/6D_L$ | $I\sigma^5/6D_L$ |
|---|---|---|---|---|---|---|
| 15 | 8 | 212 | $42.1 \pm 0.2$ | $-9.6 \cdot 10^{-4}$ | 2150 | $1.4 \cdot 10^{-17}$ |
| 16 | 8 | 112 | $27.5 \pm 0.6$ | $-1.6 \cdot 10^{-3}$ | 1950 | $3.5 \cdot 10^{-11}$ |
| 17 | 6 | 102 | $19.6 \pm 0.3$ | $-1.2 \cdot 10^{-3}$ | 3980 | $1.7 \cdot 10^{-7}$ |
| 17 | 8 | 72 | $20.0 \pm 0.4$ | $-2.0 \cdot 10^{-3}$ | 2620 | $9.9 \cdot 10^{-8}$ |
| 17 | 10 | 30 | $19.4 \pm 0.7$ | $-9.4 \cdot 10^{-3}$ | 1760 | $2.5 \cdot 10^{-7}$ |

**Table 3.4**: Nucleation rates $I$ in units of $6D_L/\sigma^5$ with $D_0$ the short time diffusion coefficient as a function of reduced pressure ($\beta p\sigma^3$) as predicted by umbrella sampling. $G''(n^*)$ is the second order derivative of the Gibbs free energy at the critical nucleus size $n^*$.

Our results for the kinetic prefactors and nucleation rates for pressures $\beta p\sigma^3 = 15, 16, 17$ are reported in Table 3.4.

## 3.6 Forward Flux Sampling

### 3.6.1 Method

The forward flux sampling method was introduced by Allen *et al.* [39] in 2005 to study rare events and has since been applied to a wide variety of systems. Two review articles (Refs. [60, 61]) on the subject have appeared recently and provide a thorough overview of the method. In the present chapter we discuss FFS as it pertains to the liquid to solid nucleation process in hard spheres. In general, FFS follows the progress of a reaction coordinate during a rare event. For hard-sphere nucleation, a reasonable reaction coordinate ($Q$) is the number of particles in the largest crystalline cluster in the system ($n$). For the remainder of this chapter, for all FFS calculations, we take the reaction coordinate to be the order parameter discussed in Sec. 3.3 with $\xi_c = 8$, $r_c = 1.3$, and $d_c = 0.7$. In general, the reaction coordinate is used to divide phase space by a sequence of interfaces ($\lambda_0$, $\lambda_1$, ... $\lambda_N$) associated with increasing values $n(\mathbf{r}^N)$ such that the nucleation process between any two interfaces can be examined. In our case the liquid is composed of all states with $n < \lambda_0$ and the solid contains all states with $n > \lambda_N$. While the complete nucleation event is rare, the interfaces are chosen such that the part of the nucleation process between consecutive interfaces is not rare, and can thus be thoroughly studied.

In the FFS methodology, the nucleation rate from the fluid phase $A$ to the solid phase $B$ is given by

$$k_{AB} = \Phi_{A\lambda_0} P(\lambda_N | \lambda_0) \tag{3.17}$$

$$= \Phi_{A\lambda_0} \prod_{i=0}^{N-1} P(\lambda_{i+1} | \lambda_i) \tag{3.18}$$

where $\Phi_{A\lambda_0}$ is the steady-state flux of trajectories leaving the $A$ state and crossing the interface $\lambda_0$ in a volume $V$, and $P(\lambda_{i+1} | \lambda_i)$ is the probability that a configuration starting at interface $\lambda_i$ will reach interface $\lambda_{i+1}$ before it returns to the fluid ($A$).

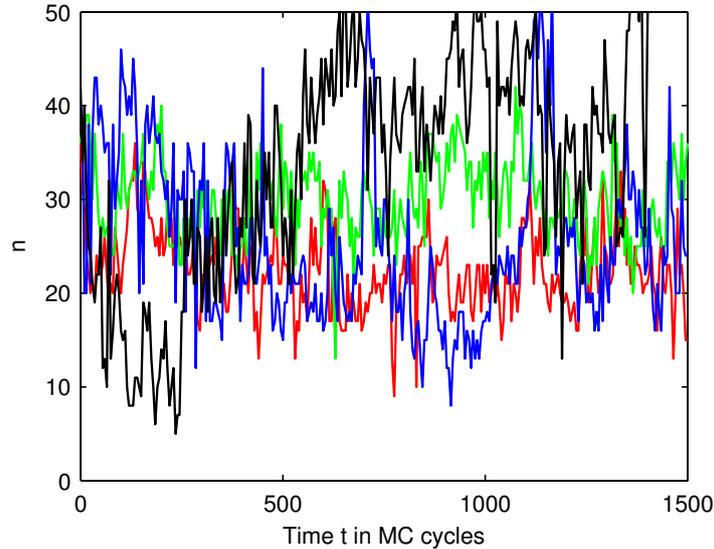

**Figure 3.7**: The cluster size as a function of time $t$ in MC cycles for 4 random trajectories at pressure $\beta p \sigma^3 = 17$ starting with a cluster size of $n = 43$ using kinetic MC simulations with stepsize $\Delta_{\text{KMC}} = 0.1\sigma$ and measuring the order parameter every $\Delta t_{\text{ord}} = 5$ MC steps.

If we apply this method directly to a hard-sphere system a number of difficulties arise. As shown in Fig. 3.5, on short times the size of a cluster measured by the order parameter fluctuates wildly. The variance in the cluster size displays two different types of behaviour, short-time fluctuations related to surface fluctuations of the cluster, and a longer time cluster growth (Fig. 3.6). Thus, if we try to measure the flux $\Phi_{A\lambda_0}$ directly, we encounter difficulties due to these short-time surface fluctuations. In theory, FFS should be able to handle these types of fluctuations, however, they increase the amount of statistics necessary to properly measure the flux and the first probability window. In the second part of FFS calculations, probabilities of the form $P(\lambda_{i+1}|\lambda_i)$ need to be determined. In calculating these probabilities it is important to be able to determine if a cluster has returned to the fluid (A). For pre-critical clusters we find large fluctuations of the order parameter, as shown in Fig. 3.7, which can lead to a cluster being misidentified as the fluid (A). Specifically, in this figure the darkest trajectory (black) shows a cluster containing 43 particles that shrinks to 5 particles before it returns to 40, and finally reaches a cluster size of 60 particles. Hence, if we had set $\lambda_0 = 5$, this trajectory would have been identified as melting back to the fluid phase (A). However, since the growth of a cluster from size 5 to 60 is a rare event in our system, we presume that this was simply a short-time fluctuation of the cluster and not a 'real' melting of the instantaneously measured cluster. For pre-critical clusters, these fluctuations result in cluster sizes that are smaller than the cluster 'really' is. We suggest that these fluctuations are largely related to the difficulty that this order parameter has in distinguishing between solid- and fluid-like particles at the fluid-solid interface. For larger clusters, where the surface to volume ratio is small, this problem is minimal. However, for elongated or rough pre-critical clusters, where the surface to volume ratio is large, these surface fluctuations and rearrangements are important, and can cause problems in measuring the order parameter.

Thus, to try and address these problems, in this chapter, we apply forward flux sampling in a slightly novel way. We regroup the elements of the rate calculation such that

$$k_{AB} = \tilde{\Phi}_{A\lambda_1} \prod_{i=1}^{N-1} P(\lambda_{i+1}|\lambda_i). \tag{3.19}$$

where

$$\tilde{\Phi}_{A\lambda_1} = \Phi_{A\lambda_0} P(\lambda_1|\lambda_0). \tag{3.20}$$

We note that if $\lambda_1$ is chosen such it is a relatively rare event for trajectories starting in $A$ to reach $\lambda_1$, then

$$\tilde{\Phi}_{A\lambda_1} \approx \frac{1}{\langle t_{A\lambda_1}\rangle V} \tag{3.21}$$

where $\langle t_{A\lambda_1}\rangle$ is the average time it takes a trajectory in $A$ to reach $\lambda_1$. The approximation made here, in contrast to normal FFS simulations, is that the time the system spends with an order parameter greater than $\lambda_1$ is negligible. Since even reaching this interface is a rare event, this approximation should have a minimal effect on the resulting rate. Additionally, in this way we are relatively free to place the first interface ($\lambda_0$) anywhere under $\lambda_1$ *. We choose to use $\lambda_0 = 1$ to minimize the effect of fluctuations, as seen in Fig. 3.7, on the probability to reach the following interface. Here we assume that any crystalline order in a system with an order parameter of 1 likely does not arise from fluctuation of a much larger cluster, but rather is very close to the fluid, and is expected to fully melt and not grow out to the next interface. In this manner we are able to start several parallel trajectories from the fluid in order to measure $\langle t_{A\lambda_1}\rangle$, stopping whenever the trajectory first hits interface $\lambda_1$.

In our implementation of FFS, we employ kinetic Monte Carlo (KMC) simulations at fixed pressure to follow the trajectories from the liquid to the solid. The KMC simulations are characterized by two parameters, the maximum stepsize ($\Delta_{KMC}$) per attempt to move each particle, and the frequency with which the order parameter (reaction coordinate) is measured $\Delta t_{ord}$.

### 3.6.2   Simulation details and results

All simulations were performed with 3000 particle in a cubic box with periodic boundary conditions. Initial configurations were produced using $NpT$ MC simulations of a liquid phase with a packing fraction of $\phi \approx 0.4$ and then simulated at a reduced pressure of $\beta p\sigma^3 = 1000$. The simulations were stopped when the packing fraction associated with the pressure of interest was reached. In this way the system volume decreased rapidly to the target density. This initial configuration was then relaxed using an $NpT$ simulation at the pressure of interest ($\beta p\sigma^3 = 15, 16, 17$). The relaxation consisted of at least 10,000 MC

---

*While it does appear that Eq. 3.19 is completely independent of $\lambda_0$, this is not strictly correct as $\lambda_0$ creates the border for state A and state A is expected to be a metastable, equilibrated state. For the purposes of this chapter, the difference is insignificant as the average time for a nucleation event is much longer than the relaxation time for the fluid.

| $\Delta_{\text{KMC}}$ | 0.1 | 0.1 | 0.1 | 0.2 | 0.2 | 0.2 |
|---|---|---|---|---|---|---|
| $\Delta t_{\text{ord}}$ | 2 | 2 | 2 | 2 | 2 | 2 |
| $P(\lambda_2\|\lambda_1)$ | 0.112 | 0.103 | 0.139 | 0.101 | 0.105 | 0.132 |
| $P(\lambda_3\|\lambda_2)$ | 0.096 | 0.117 | 0.090 | 0.104 | 0.093 | 0.112 |
| $P(\lambda_4\|\lambda_3)$ | 0.128 | 0.117 | 0.074 | 0.116 | 0.111 | 0.161 |
| $P(\lambda_5\|\lambda_4)$ | 0.180 | 0.159 | 0.082 | 0.156 | 0.115 | 0.241 |
| $P(\lambda_6\|\lambda_5)$ | 0.167 | 0.154 | 0.149 | 0.225 | 0.148 | 0.256 |
| $P(\lambda_7\|\lambda_6)$ | 0.071 | 0.074 | 0.060 | 0.128 | 0.093 | 0.118 |
| $P(\lambda_8\|\lambda_7)$ | 0.104 | 0.078 | 0.051 | 0.109 | 0.091 | 0.109 |
| $P(\lambda_9\|\lambda_8)$ | 0.100 | 0.100 | 0.105 | 0.083 | 0.075 | 0.089 |
| $P(\lambda_9\|\lambda_1)$ | $3 \cdot 10^{-8}$ | $2 \cdot 10^{-8}$ | $4 \cdot 10^{-9}$ | $5 \cdot 10^{-8}$ | $1 \cdot 10^{-8}$ | $2 \cdot 10^{-7}$ |
| $\Delta_{\text{KMC}}$ | 0.2 | 0.2 | 0.2 | 0.2 | 0.2 | 0.2 |
| $\Delta t_{\text{ord}}$ | 1 | 1 | 1 | 10 | 10 | 10 |
| $P(\lambda_2\|\lambda_1)$ | 0.112 | 0.146 | 0.138 | 0.122 | 0.127 | 0.146 |
| $P(\lambda_3\|\lambda_2)$ | 0.115 | 0.097 | 0.079 | 0.103 | 0.081 | 0.080 |
| $P(\lambda_4\|\lambda_3)$ | 0.151 | 0.110 | 0.110 | 0.121 | 0.091 | 0.116 |
| $P(\lambda_5\|\lambda_4)$ | 0.209 | 0.189 | 0.173 | 0.121 | 0.073 | 0.150 |
| $P(\lambda_6\|\lambda_5)$ | 0.274 | 0.151 | 0.189 | 0.189 | 0.121 | 0.187 |
| $P(\lambda_7\|\lambda_6)$ | 0.121 | 0.052 | 0.092 | 0.169 | 0.077 | 0.064 |
| $P(\lambda_8\|\lambda_7)$ | 0.119 | 0.077 | 0.126 | 0.132 | 0.087 | 0.064 |
| $P(\lambda_9\|\lambda_8)$ | 0.101 | 0.081 | 0.129 | 0.101 | 0.109 | 0.068 |
| $P(\lambda_9\|\lambda_1)$ | $2 \cdot 10^{-7}$ | $1 \cdot 10^{-8}$ | $6 \cdot 10^{-8}$ | $8 \cdot 10^{-8}$ | $6 \cdot 10^{-9}$ | $1 \cdot 10^{-8}$ |

**Table 3.5**: Probabilities $P(\lambda_{i+1}|\lambda_i)$ for the first 8 interfaces for a pressure of $\beta p\sigma^3 = 15$ where the KMC simulations stepsize ($\Delta_{\text{KMC}}$) and the number of MC steps between measuring the order parameter $\Delta t_{\text{ord}}$ are varied. The following interfaces were used: $\lambda_2 = 20$, $\lambda_3 = 26$, $\lambda_4 = 32$, $\lambda_5 = 38$, $\lambda_6 = 44$, $\lambda_7 = 54$, $\lambda_8 = 65$, and $\lambda_9 = 78$. In all cases, 100 configurations were started in the fluid and reached the first interface, and at each interface, $C_i = 10$ copies of each successful configuration were used.

| $\beta p\sigma^3$ | $\lambda_1$ | $\tilde{\Phi}_{A\lambda_1}/6D_l$ | $P(\lambda_B\|\lambda_1)$ | $I\sigma^5/6D_l$ |
|---|---|---|---|---|
| 17 | 27 | $2.66 \cdot 10^{-5}$ | $7.6 \cdot 10^{-3}$ | $2.0 \cdot 10^{-7}$ |
| 17 | 27 | $2.68 \cdot 10^{-5}$ | $1.4 \cdot 10^{-2}$ | $3.7 \cdot 10^{-7}$ |
| 16 | 20 | $8.57 \cdot 10^{-6}$ | $3.1 \cdot 10^{-7}$ | $2.6 \cdot 10^{-12}$ |
| 16 | 20 | $8.57 \cdot 10^{-6}$ | $2.1 \cdot 10^{-7}$ | $1.8 \cdot 10^{-12}$ |
| 15 | 15 | $8.72 \cdot 10^{-6}$ | $1.9 \cdot 10^{-15}$ | $1.6 \cdot 10^{-20}$ |

**Table 3.6**: Nucleation rates predicted using forward flux sampling in long-time self diffusion coefficient units ($D_l$). The probabilities $P(\lambda_B|\lambda_1)$, number of steps between the order parameter measurements $\Delta_{\text{ord}}$, and kinetic MC stepsize are as in Tables 3.7, 3.8, and 3.9. At each interface, $C_i$ copies of each successful configuration were used.

| | | trial 1 | | trial 2 | |
|---|---|---|---|---|---|
| i | $\lambda_i$ | $C_{i-1}$ | $P(\lambda_i\|\lambda_{i-1})$ | $C_{i-1}$ | $P(\lambda_i\|\lambda_{i-1})$ |
| 2 | 43 | 10 | 0.137 | 10 | 0.157 |
| 3 | 60 | 10 | 0.272 | 10 | 0.312 |
| 4 | 90 | 10 | 0.350 | 10 | 0.414 |
| 5 | 150 | 2 | 0.594 | 2 | 0.691 |
| 6 | 250 | 2 | 0.988 | 2 | 0.988 |

**Table 3.7**: Probabilities $P(\lambda_{i+1}|\lambda_i)$ for the interfaces used in calculating the nucleation rate for pressure $\beta p\sigma^3 = 17$ with step size $\Delta_{\mathrm{KMC}} = 0.1\sigma$ and measuring the order parameter every $\Delta t_{\mathrm{ord}} = 5$ MC cycles.

| | | trial 1 | | trial 2 | |
|---|---|---|---|---|---|
| i | $\lambda_i$ | $C_{i-1}$ | $P(\lambda_i\|\lambda_{i-1})$ | $C_{i-1}$ | $P(\lambda_i\|\lambda_{i-1})$ |
| 2 | 28 | 10 | 0.105 | 10 | 0.110 |
| 3 | 38 | 10 | 0.075 | 10 | 0.077 |
| 4 | 50 | 10 | 0.070 | 10 | 0.089 |
| 5 | 70 | 10 | 0.114 | 10 | 0.089 |
| 6 | 90 | 10 | 0.095 | 10 | 0.101 |
| 7 | 110 | 10 | 0.339 | 10 | 0.278 |
| 8 | 250 | 10 | 0.152 | 10 | 0.112 |
| 9 | 350 | 1 | 1.000 | 1 | 1.000 |

**Table 3.8**: Same as Table 3.7 but for $\beta p\sigma^3 = 16$.

| i | $\lambda_i$ | $C_{i-1}$ | $P(\lambda_i\|\lambda_{i-1})$ | i | $\lambda_i$ | $C_{i-1}$ | $P(\lambda_i\|\lambda_{i-1})$ |
|---|---|---|---|---|---|---|---|
| 2 | 20 | 10 | 0.101 | 10 | 92 | 10 | 0.101 |
| 3 | 26 | 10 | 0.104 | 11 | 110 | 10 | 0.085 |
| 4 | 32 | 10 | 0.116 | 12 | 135 | 10 | 0.062 |
| 5 | 38 | 10 | 0.156 | 13 | 160 | 10 | 0.131 |
| 6 | 44 | 10 | 0.225 | 14 | 190 | 10 | 0.131 |
| 7 | 54 | 10 | 0.128 | 15 | 230 | 10 | 0.134 |
| 8 | 65 | 10 | 0.109 | 16 | 400 | 10 | 0.058 |
| 9 | 78 | 10 | 0.083 | | | | |

**Table 3.9**: Same as Table 3.7 but for $\beta p\sigma^3 = 15$ and with $\Delta t_{\mathrm{ord}} = 2$.

cycles, after which the simulation continued until a measurement of the order parameter found no crystalline particles in the system.

In order to determine the flux and the probabilities, 100 trajectories were started in the liquid and terminated when $n(\mathbf{r}^N) = \lambda_1$. These trajectories were produced using KMC simulations. The probability $P(\lambda_2|\lambda_1)$ was then found by making $C_1$ copies of the configurations that reached $\lambda_1$, and following these configurations until they either reached $\lambda_2$ or returned to the fluid. By taking different random number seeds, the various copies of the same configurations follow different trajectories. The fraction of successful trajectories corresponds to the required probability. The successful trajectories were then copied $C_2$ times to determine $P(\lambda_3|\lambda_2)$. The remaining $P(\lambda_{i+1}|\lambda_i)$'s are calculated similarly.

To study the effect of the two KMC parameters, namely $\Delta_{\mathrm{KMC}}$ and $\Delta t_{\mathrm{ord}}$, on the nucleation rates, we have examined the first 8 FFS windows for $\beta p\sigma^3 = 15$ for various values of the number of MC steps between the order parameter measurements $\Delta t_{\mathrm{ord}}$ and the maximum displacement $\Delta_{\mathrm{KMC}}$ for the KMC simulations. The results are shown in Table 3.5. As shown in this table we do not find a significant effect on the rate from either parameter. Thus for numerical efficiency, unless otherwise indicated, the rates in this section come from $\Delta t_{\mathrm{ord}} = 5$ MC cycles and $\Delta_{\mathrm{KMC}} = 0.2\sigma$.

For pressures $\beta p\sigma^3 = 16$ and $17$ we have performed two separate FFS calculations to determine the nucleation rates, and for pressure $\beta p\sigma^3 = 15$ we have the result from a single FFS simulation. A summary of the results are given in Table 3.6. A complete summary of the results for $P(\lambda_i + 1|\lambda_i)$ for each simulation is given in Tables 3.7, 3.8, and 3.9.

## 3.7 Summary and Discussion

### 3.7.1 Nucleation Rates

In this section we examine hard-sphere nucleation rates predicted using US simulations, MD simulations and FFS simulations together with the experimental results of Harland and Van Megen, [33] Sinn *et al.* [35] and Schätzel and Ackerson [36] and the US simulations of monodisperse and 5% polydisperse hard-spheres mixtures examined by Auer and Frenkel [18]. The experimental volume fractions have been scaled to yield the coexistence densities of monodisperse hard spheres[42]. Similarly, we scale the polydisperse results of Auer and Frenkel with the coexistence densities determined in Ref. [62]. Inspired by the recent work of Pusey *et al.* [42], we plot the nucleation rates in units of the long-time self diffusion coefficient. In experiments with colloidal particles, the influence of the solvent on the dynamics cannot be ignored. Specifically, the system slows down due to hydrodynamic interactions when the density is increased. However, by presenting the nucleation rates in terms of the long-time self diffusion coefficient, we expect our simulated nucleation rates from the hard sphere model without an explicit solvent to be in agreement with the experimental rates with a solvent. The time in experiments is typically measured in units of $D_0$, the free diffusion at low density. We convert the short-time self diffusion coefficient

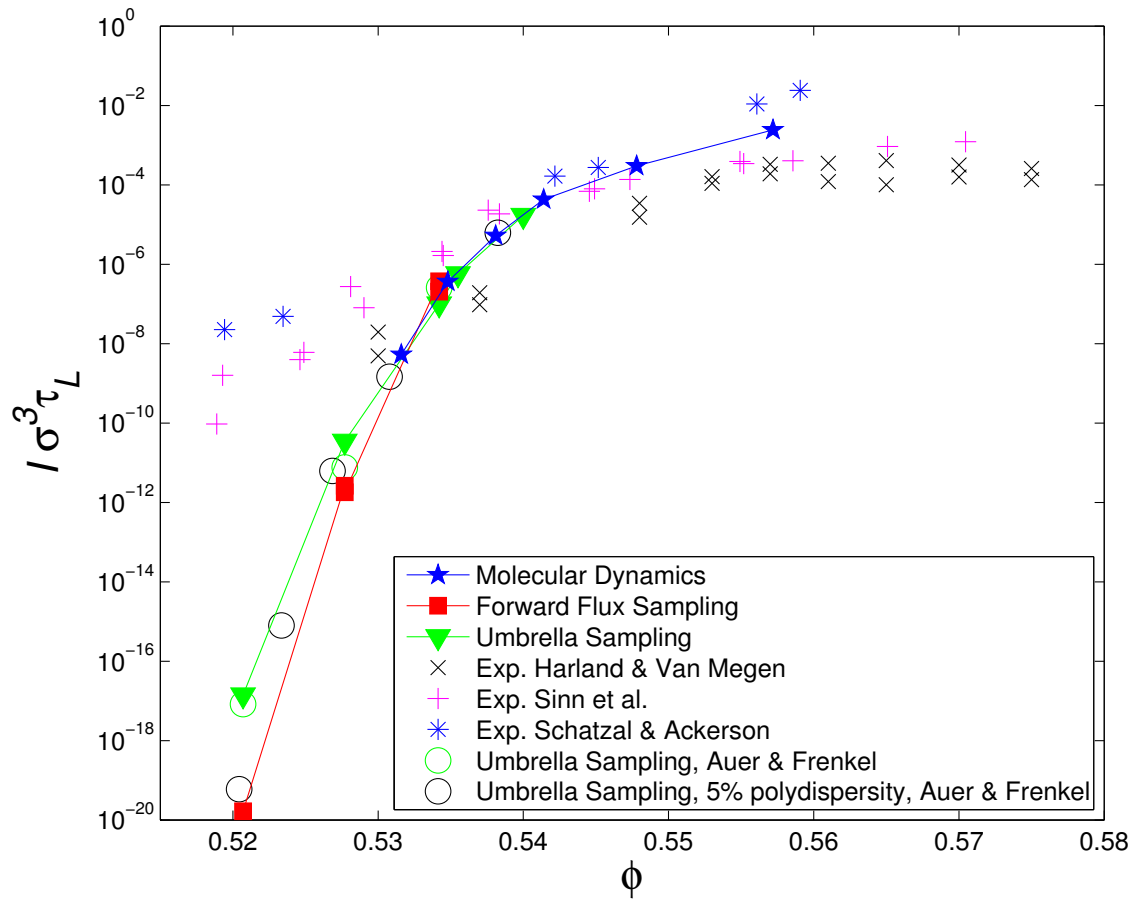

**Figure 3.8**: A comparison of the crystal nucleation rates of hard spheres as determined by the three methods described in this chapter FFS, US, and MD with the experimental results from Refs. [33, 35, 36] and previous theoretical results from Ref. [18]. Note that error bars have not been included in this plot but are discussed in the main text. Within these estimated error bars, the simulated nucleation rates are all in agreement, while the experimentally obtained rates show a markedly different behaviour, particularly for low supersaturations where the difference between the simulations and experiments can be as large as 12 orders of magnitude.

$D_0$ to the long-time self diffusion coefficient $D_l$ using

$$\frac{D_l(\phi)}{D_0} = \left(1 - \frac{\phi}{0.58}\right)^{\delta}. \tag{3.22}$$

Harland and Van Megen [33] claim that $\delta = 2.6$ gives a good fit to their system and Sinn *et al.* [35] use $\delta = 2.58$. Since the system Schätzel and Ackerson [36] examine is very similar to the other two, we use $\delta = 2.6$ to convert their nucleation rates to long-time units. We note that both $\delta = 2.58$ and $\delta = 2.6$ give very similar results. The results for both the theoretical and experimental rates in long time units are shown in Fig. 3.8. Note that for clarity reasons the error bars have not been included in this plot. In general, the error bars of the simulated nucleation rates are largest for lower supersaturations (i.e. lower volume fractions), as the barrier height is higher. For the FFS and US simulations, the error for $\beta p\sigma^3 = 15$ ($\phi = 0.5214$) is between 2 and 3 orders of magnitude, and for $\beta p\sigma^3 = 17$ ($\phi = 0.5352$) is approximately one to two orders of magnitude. The MD results are quite accurate around $\beta p\sigma^3 = 17$, however the error bars are larger for the higher pressure MD results.

In Ref. [42], Pusey *et al.* showed that the nucleation rates for various polydispersities (0 to 6%) of hard-sphere mixtures collapsed onto the same curve when the rates were plotted in units of the long-time self diffusion coefficient. We find similar results here. Both the monodisperse and polydisperse US results of Auer and Frenkel [18], in addition to our own US predictions of the nucleation rate, agree well within the expected measurement error. Additionally, we find that the simulation results of the US, FFS, and MD all agree. Whereas the simulation results agree well with the experimental results for the nucleation rate at high supersaturation there is still a significant difference at low supersaturations. Unfortunately, the origin of this discrepancy remains unsolved.

However, on the experimental side, the nucleation rates of Harland and Van Megen [33] are approximately one to two orders of magnitude below the experiments of Sinn *et al.* [35] and Schätzel and Ackerson [36]. This is unexpected due to the similarity between the experimental systems. The main difference between these experiments is the polydispersity of the particle mixtures: 5% in the case of Harland and Van Megen [33], 2.5% in the case of Sinn *et al.* [35], and < 5% for Schätzel and Ackerson [36]. However, as demonstrated by Pusey *et al.* [42], and now also in Fig. 3.8, the nucleation rate when measured in long-time self diffusion coefficient units should not be affected by the polydispersity. Thus, this seems unlikely as an explanation.

### 3.7.2   Nuclei

To examine whether the structure and shape of the critical clusters from US simulations depended on the precise threshold values used for the crystalline order parameters, we compared and analysed the critical clusters obtained when three different crystalline order parameters were used to bias the US simulations, namely, $\xi_c = 5, 7$ and 9. Subsequently we analyzed these critical clusters using the three different order parameters. In Fig. 3.9, two typical critical clusters from different biasing order parameters are shown on the top and bottom rows. The nucleus of the cluster, shown in blue, was identified by all three cluster criteria ($\xi_c = 5, 7$ and 9). The main difference between the criteria is the location of the fluid-solid interface as shown by the green and red particles. The strictest order parameter finds only the more ordered center whereas the loosest version detects the more disordered particles at the interface as well. In Fig. 3.10 we show some of the nuclei obtained from MD simulations. These snapshots were taken just before the nuclei grew out so they are not necessarily precisely at the top of the nucleation barrier. They appear very similar in roughness and aspect ratio to those obtained from US simulations. We note here that this is not meant to be a thorough study of the critical clusters, but rather just a rough comparison to demonstrate that to a first approximation the clusters formed by the three simulation techniques are the same. A more thorough examination of the structure of the nuclei for high supersaturations can be found in Ref. [63]

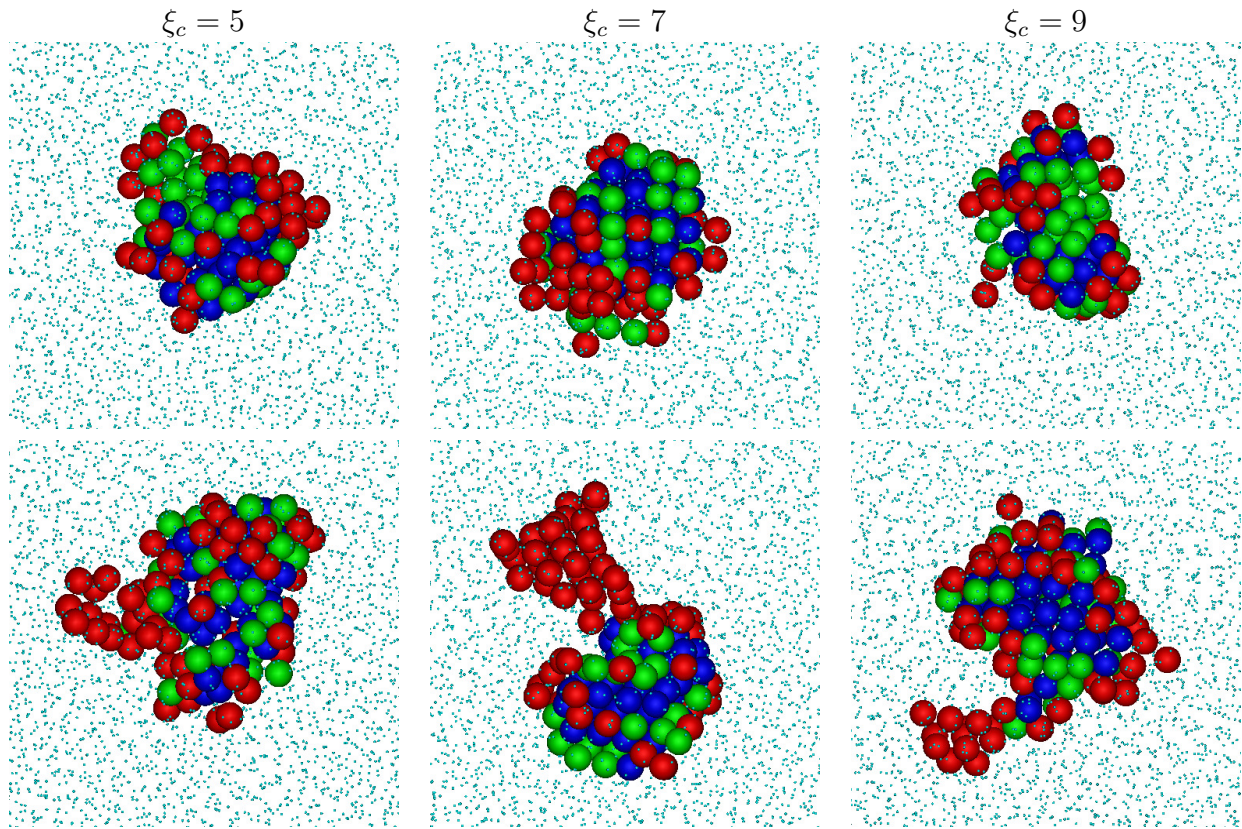

**Figure 3.9**: Two typical snapshots (top and bottom) of the critical nuclei as obtained with US at a volume fraction $\phi = 0.5355$ using different values of the critical number of crystalline bonds $\xi_c = 5$ (left), 7 (middle) and 9 (right) in the biasing potential. The clusters are analyzed with three different crystalline order parameters. The blue particles are found by all three cluster criteria, the green particles have $\xi = 7$ or 8 crystalline bonds and the red particles have only $\xi = 5$ or 6 crystalline bonds.

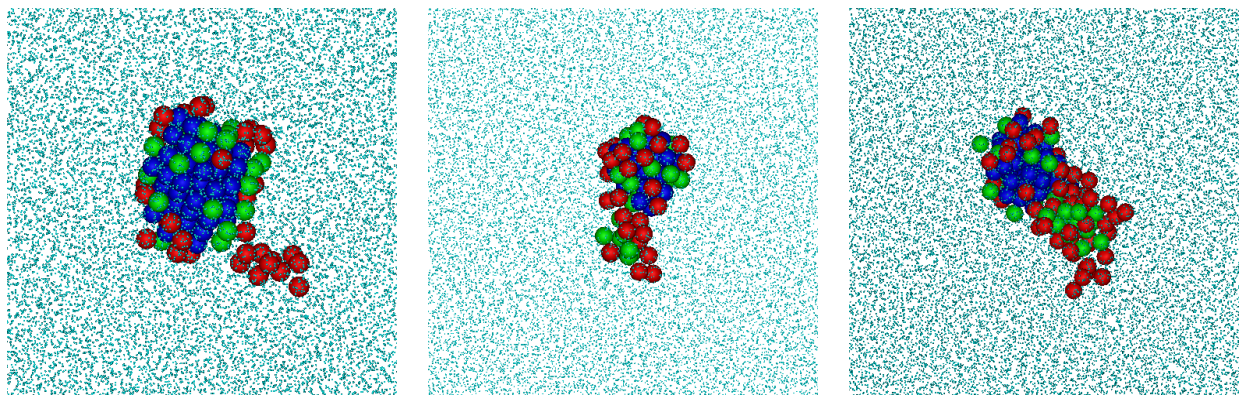

**Figure 3.10**: Snapshots of spontaneously formed nuclei during an MD simulation at a volume fraction of $\phi = 0.537$. The snapshots were taken just before the nuclei grew. The color coding of the particles is the same as in Fig. 3.9.

To further examine whether the choice of method influenced the resulting clusters, particularly the presence of the biasing potential in the US simulations and the choice of reaction coordinate and interfaces in FFS, we calculated the radius of gyration tensor for each of the methods for pressure $\beta p \sigma^3 = 17$ as a function of cluster size (see Figure 3.11). There is no indication that the clusters in any of the simulation methods differed substantially.

Additionally, we examined whether the simulation technique influenced the type of pre-critical nuclei that formed in the simulations, i.e. face-centered-cubic (FCC), and hexagonal-close-packed (HCP). To do this we used the order parameter introduced by Ref. [64] which allows us to identify each particle in the cluster as either FCC-like or HCP-like. The results for a wide range in nucleus size is shown in Fig. 3.12. We find complete agreement between the three simulation techniques. Specifically, in all cases we find that the nucleus is composed of approximately 80% FCC-like particles. This was unexpected as the free energy difference between the bulk FCC and HCP phases is about $0.001 k_B \mathrm{T}$ per particle at melting [65] and hence random-hexagonal-close-packing order in the nuclei would be expected [66]. Note that using our order parameter this would appear as an approximately 50% occurrence of FCC- and HCP-like particles in the nucleus. We speculate that this predominance of FCC-like particles in the nuclei arises from surface effects.

## 3.8 Conclusions

In this chapter we have examined in detail three independent simulation techniques for studying nucleation processes and predicting nucleation rates, namely forward flux sampling, umbrella sampling and molecular dynamics. We have shown that the three simulation techniques are completely consistent in their prediction of the nucleation rates for hard spheres over the large range of volume fractions studied, despite the fact that they treat the dynamics differently. Additionally, in agreement with the recent work of Pusey *et al.* [42], we find that by measuring the nucleation rates in terms of the long-time self diffusion constant and scaling to the coexistence density of monodisperse hard spheres, the 5% polydisperse results of Auer and Frenkel [18] also agree. On examining the critical clusters, we also do not find a difference in the nuclei formed using the three simulation techniques. Hence we conclude that the origional prediction of Auer and Frenkel [18] for the nucleation rates in hard sphere systems was indeed robust.

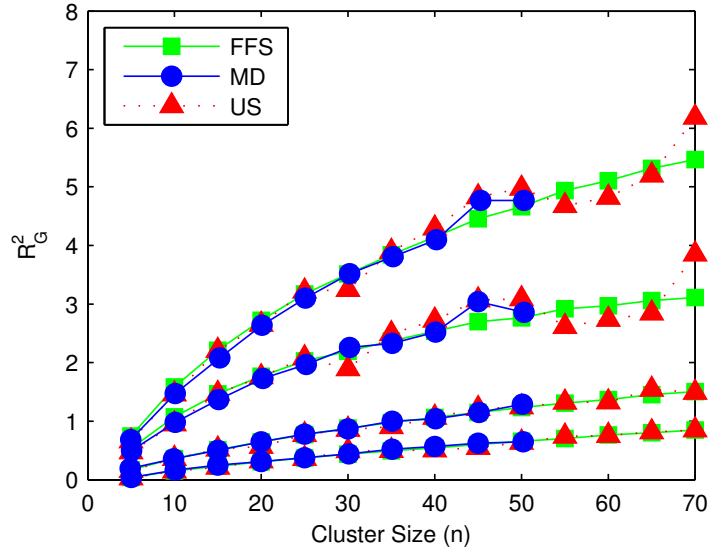

**Figure 3.11**: A comparison of the three components of the radius of gyration tensor as a function of cluster size $n$, as well as the sum of the three components, for clusters produced using FFS, MD, and US simulations.

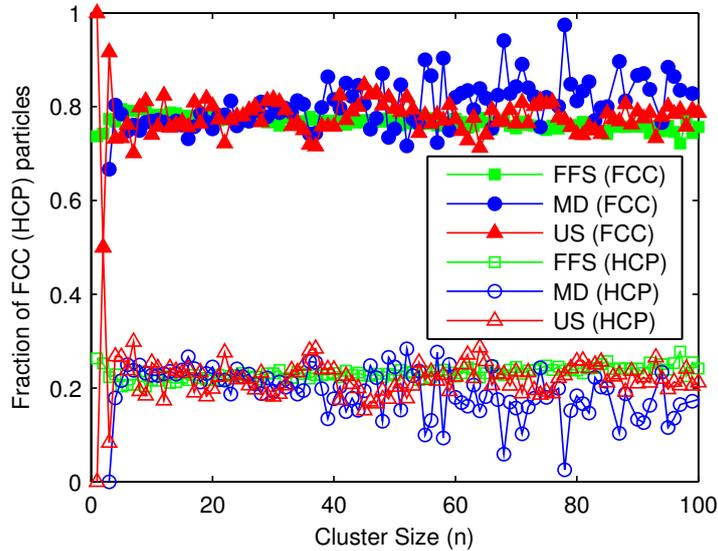

**Figure 3.12**: Fraction of particles identified as either FCC or HCP respectively in the clusters produced via molecular dynamics (MD), forward flux sampling (FFS), and umbrella sampling (US) simulations as a function of cluster size $n$. All three methods agree and find the pre-critial clusters predominately FCC.

We have also compared our nucleation rates with previous experimental data, specifically, the nucleation rates predicted by Harland and Van Megen [33], Sinn *et al.* [35] and Schätzel and Ackerson [36]. As was found first by Auer and Frenkel [18], while the simulation results agree well with the experimental results for high supersaturations, there is a significant difference between the simulations and experiments for smaller volume fractions. The agreement between the three theoretical methods examined in this chapter, namely molecular dynamics, umbrella sampling, and forward flux sampling, seems to indicate that either there is a fundamental difference between the simulations and theory which we are not taking into account, such as some form of collective hydrodynamics which are included in the experiments but not considered in the thoery or some difficulty in interpreting the experimental data. In either case, the origin of the huge discrepancy in the theoretical and experimental nucleation rates remains a mystery.

## 3.9 Acknowledgements

I would like to thank Dr. L. Filion and Dr. M. Hermes for simulations on forward flux sampling and molecular dynamics, respectively.

# 4

---

# Simulation of nucleation in almost hard-sphere colloids: the discrepancy between experiment and simulation persists

---

In this chapter we examine the phase behaviour of the Weeks-Chandler-Andersen (WCA) potential with $\beta\epsilon = 40$. Crystal nucleation in this model system was recently studied by Kawasaki and Tanaka [Proc. Natl. Acad. Sci. U.S.A. **107**, 14036 (2010)], who argued that the computed nucleation rates agree well with experiment, a finding that contradicted earlier simulation results. Here we report an extensive numerical study of crystallization in the WCA model, using three totally different techniques (Brownian Dynamics, Umbrella Sampling and Forward Flux Sampling). We find that all simulations yield essentially the same nucleation rates. However, these rates differ significantly from the values reported by Kawasaki and Tanaka and hence we argue that the huge discrepancy in nucleation rates between simulation and experiment persists. When we map the WCA model onto a hard-sphere system, we find good agreement between the present simulation results and those that had been obtained for hard spheres [S. Auer and D. Frenkel, Nature **409**, 1020 (2001), L. Filion *et al.*, J. Chem. Phys. **113**, 244115 (2010)].

## 4.1   Introduction

In a recent article, Kawasaki and Tanaka [20] examined the crystal nucleation in an almost hard-sphere system using Brownian dynamics simulations. The nucleation rates reported in Ref. [20] appear to be in good agreement with those that were found in earlier light-scattering experiments [33, 35, 36]. This is in sharp contrast with previous simulation studies of hard spheres, which show a large discrepancy between the experimental and simulated rates for low volume fractions [18, 19]. In this chapter we revisit the system examined by Kawasaki and Tanaka in order to determine the origin of the difference between the simulated rates and, in particular, to clarify if there is indeed a discrepancy between the experimental and simulated nucleation rates. We study the system using a variety of simulation techniques, including brute force Brownian dynamics, umbrella sampling and forward flux sampling.

Colloidal solutions consist of small particles suspended in another medium and are typically characterized by the dynamics of these suspended particles, i.e., colloidal particles exhibit Brownian motion. As a result, Brownian dynamics simulations (BD) are the natural choice to use when examining dynamical properties of colloidal systems, such as crystal nucleation. Brownian dynamics are based on a simplified version of Langevin dynamics and correspond to the "overdamped" limit. Specifically, in BD it is assumed that the particles' inertial motion is completely damped out by frictional forces. As a result, the motion of the particles is determined by the instantaneous forces acting on the colloid plus a stochastic, diffusive displacement. However, unlike molecular dynamics simulations (MD) where an event driven formalism exists which allows one to apply MD to systems with hard-core interactions (see e.g. Ref. [50]), no such formalism exists for BD of hard particles. Hence, when Brownian dynamics are applied to hard-core interactions, the hard core is typically approximated. One such approximation is the Weeks-Chandler-Andersen potential.

The Weeks-Chandler-Andersen (WCA) potential was introduced in 1971 in order to address the short-range repulsive part of the Lennard-Jones liquid separately from the longer range attractive tail. In contrast to the Lennard-Jones system, the phase diagram for the WCA potential consists simply of liquid and solid phases; i.e., the liquid-gas phase coexistence is not present in this model. The WCA potential [67] is given by

$$\beta U_{\text{WCA}}\left(r\right) = \begin{cases} 4\beta\epsilon\left[\left(\frac{\sigma}{r}\right)^{12} - \left(\frac{\sigma}{r}\right)^{6} + \frac{1}{4}\right] & r/\sigma \leq 2^{1/6} \\ 0 & r/\sigma > 2^{1/6} \end{cases} \tag{4.1}$$

where $\sigma$ is a length scale, $\epsilon$ is the energy scale, and $\beta = 1/k_B T$ where $k_B$ is Boltzmann's constant and $T$ is the temperature. Note that the WCA potential is simply the Lennard-Jones potential where the cutoff is chosen such that only the repulsive part remains and the potential is shifted upwards so that the minimum occurs at zero. A plot of this potential is shown in Fig. 4.1. The "hardness" of the interaction can be set by tuning the interaction strength, $\beta\epsilon$. In Ref. [20], Kawasaki and Tanaka studied a WCA model at an interaction strength $\beta\epsilon = 40$, which corresponds to a low temperature.

This chapter is organized as follows: in section 4.2 we use free-energy calculations to determine the phase diagram for this model, in section 4.3 we describe the nucleation rates, in section 4.4 we compare our results to the previous work of Kawasaki and Tanaka [20] and to hard-sphere crystal nucleation rates found both in simulations as well as light scattering experiments. Our conclusions are found in section 4.5.

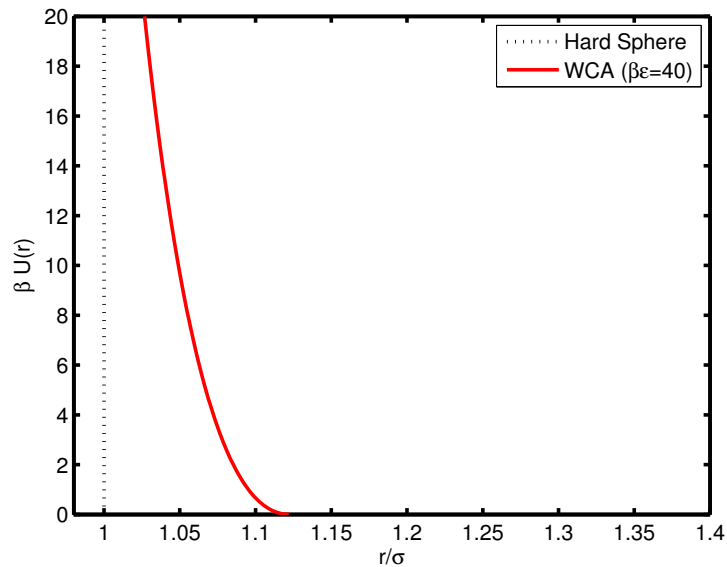

**Figure 4.1**: The WCA potential and hard sphere potential $\beta U(r)$ as a function of center-to-center distance $r$.

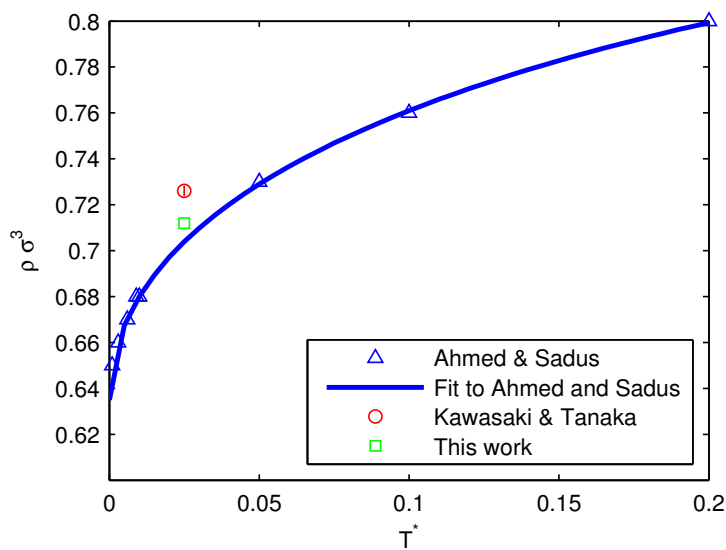

**Figure 4.2**: The *triangles* correspond to the freezing number density ($\rho_F^A$) from Ahmed and Sadus [68] as a function of $T^*$ where $T^* = k_B T/\epsilon$. The fit corresponds to $\sigma \rho_F = 0.635 + 0.473(T^*)^{1/2} - 0.236 T^*$. The *square* corresponds to the freezing number density ($\rho_F^*$) determined using full free-energy calculations as described in this chapter. The *circle* corresponds to the freezing number density determined by Kawasaki and Tanaka. [20]

## 4.2   Phase diagram

To calculate the coexistence densities for the WCA potential, we used full free-energy calculations in combination with common tangent constructions. For the crystal phase, the excess free energy $F_{ex}$ was calculated using Einstein integration [10, 69, 70] at a density of $\rho\sigma^3 = 0.8$ for systems of $N = 500, 864, 1372$ and $2048$ particles. Note that the excess free energy is defined by $F_{ex} = F_{tot} - F_{id}$ where $F_{tot}$ is the total free energy and $F_{id}$ is the ideal gas free energy. Following Ref. [70], we plotted $\beta F_{ex}/N + \log N/N$ as a function of $1/N$ and extrapolated to an infinite system yielding a free energy of $\beta F_{tot}/N = 4.8975$. The free energy at other densities was determined using thermodynamic integration of the equation of state [10]. The equation of state was determined using Monte Carlo $NpT$ simulations with $N = 4000$ particles. We note that no significant difference was found in the coexistence densities for equations of state determined using $N = 1372$ and $N = 4000$. To test our Einstein integration and integration over the equation of state, we determined the free energy at $\rho\sigma^3 = 0.9$ for $N = 1372$ and integrated over the equation of state calculated for $N = 1372$. The free energies agreed within $0.00046$ $k_BT$ per particle. The fluid chemical potential was determined using the Widom insertion technique [10] at $\rho\sigma^3 = 0.4$ with $N = 4000$ and was found to be $\beta\mu = 3.3173$; for $N = 1372$ we find $\beta\mu = 3.3194$. Again integration over the equation of state was used to determine the free energy as a function of density. To test the Widom insertion calculations, and our integration over the equation of state, we also calculated the chemical potential at $\rho\sigma^3 = 0.3$ for $N = 1372$. The difference in the free energy at $\rho\sigma^3 = 0.3$ associated with the Widom insertions and integration over the equation of state results in a free energy difference of $0.00075$ $k_BT$ per particle, and hence we concluded that the Widom insertions and integration over the equation of state were correct. Using these free energies and common tangent constructions we find freezing and melting coexistence densities $\rho_F^*\sigma^3 = 0.712$ and $\rho_M^*\sigma^3 = 0.785$ respectively.

The phase diagram for the WCA potential has been examined previously by Ahmed and Sadus [68] for a range of $T^* = 1/\beta\epsilon$ using a phenomenological method based on non-equilibrium MD simulations. The results of Ref. [68] for the freezing density are plotted in Fig. 4.2. We find that their results for the freezing number density $\rho_F$ as a function of $\beta\epsilon$ fit well to $\rho_F\sigma^3 = 0.635 + 0.473(T^*)^{1/2} - 0.236T^*$. From this fit we approximate a freezing number density of $\rho_F^A\sigma^3 = 0.704$ at $\beta\epsilon = 40$. We note that this is in good agreement with our predictions. Hence, our free-energy calculations support the phenomenological procedure of Ref. [68]. However, we find that the non-equilibrium MD estimate of the freezing density is slightly lower than the true equilibrium coexistence density reported here. Additionally, Kawasaki and Tanaka [20] found the freezing number density for $\beta\epsilon = 40$ to be $\rho_F^K\sigma^3 = 0.725$. To locate the freezing point, these authors performed BD simulations of a face-centered cubic (FCC) crystal and identified the density at which the crystal becomes mechanically unstable as the freezing density [71]. Such calculations cannot be used to accurately determine the coexistence densities, but rather give an approximate lower bound for the *melting* density. As can be seen in Fig. 2, the freezing density estimated in Ref. [20] is some 2% higher than the value that we find using free-energy calculations.

| $\beta p \sigma^3$ | $\beta \lvert \Delta\mu \rvert$ | $\rho_{\text{liq}}\sigma^3$ | $\rho_{\text{sol}}\sigma^3$ | $\phi_{\text{eff}}$ |
|---|---|---|---|---|
| 12.0 | 0.41 | 0.762 | 0.844 | 0.526 |
| 13.0 | 0.54 | 0.775 | 0.858 | 0.535 |
| 13.3 | 0.58 | 0.778 | 0.862 | 0.538 |
| 13.4 | 0.59 | 0.780 | 0.863 | 0.539 |
| 13.6 | 0.61 | 0.782 | 0.865 | 0.540 |
| 13.9 | 0.65 | 0.785 | 0.868 | 0.542 |
| 14.0 | 0.66 | 0.787 | 0.870 | 0.544 |
| 14.4 | 0.71 | 0.791 | 0.874 | 0.547 |
| 14.6 | 0.73 | 0.793 | 0.876 | 0.548 |

**Table 4.1:** Reduced pressure ($\beta p \sigma^3$), reduced chemical potential difference between the fluid and solid phases ($\beta \lvert \Delta\mu \rvert$), reduced number density of the metastable liquid $\rho_{\text{liq}}\sigma^3$, reduced number density of the solid phase $\rho_{\text{sol}}\sigma^3$, and the effective hard-sphere packing fraction $\phi_{\text{eff}}$ for the state points studied in this chapter.

## 4.3 Nucleation rates

In this section we apply Brownian dynamics, umbrella sampling (US) and forward flux sampling (FFS) to study the crystal nucleation of the WCA model. The methods for predicting nucleation rates have been discussed in detail in Chapter 3 and Ref. [19] and so only a short overview will be presented here. An overview of the state points discussed in this chapter is found in Table 4.1 where we list for various pressures $\beta p \sigma^3$ the corresponding chemical potential difference between the fluid and solid phases $\lvert \beta\Delta\mu \rvert$, the reduced number density of the metastable liquid phase $\rho_{liq}\sigma^3$ and the stable solid phase $\rho_{sol}\sigma^3$, and the effective packing fraction $\phi_{\text{eff}}$ (as defined below).

In all of the simulation methods examined in this chapter, an order parameter is needed to differentiate between liquid-like and solid-like particles and a cluster algorithm is used to identify the solid clusters. The order parameter we use is the local bond-order parameter introduced by ten Wolde *et al.* [48, 49]. Please see Chapter 3.3 for details, and in this chapter we used $d_c = 0.7$, $\xi_c = 6$ or 8 as identified and $r_c$ is always either $1.5\sigma$ or $1.6\sigma$ and is explicitly indicated in each section.

### 4.3.1 Brownian dynamics

Brownian dynamics is a simplified Langevin dynamics which can be used to describe the motion of Brownian particles. In Brownian dynamics simulations, the motion of each particle $i$ is described by [72]

$$\frac{\mathrm{d}\mathbf{r}_i}{\mathrm{d}t} = \frac{1}{m\gamma}\left[-\nabla_i U + \mathbf{W}_i(t)\right], \tag{4.2}$$

where $\gamma$ and $\mathbf{W}_i(t)$ are the friction coefficient and the stochastic force of the solvent, $m$ is the mass of the particles and $U$ is the potential energy of the system. They are linked through the dissipation-fluctuation theorem $\langle \mathbf{W}_i(t) \cdot \mathbf{W}_j(t') \rangle = 6m\gamma k_B T \delta_{ij}\delta(t-t')$ where $\delta$ is the Kronecker delta function. In our simulations, $\gamma$ and $m$ are both set to 1 and we use the time step $\Delta t = 10^{-5}\tau_B$ to integrate Eq. 9.12. Note that $\tau_B$ is the Brownian time which is defined as $\tau_B = \sigma^2/D_0$ where $D_0$ is the diffusion coefficient of the particle in the infinitely dilute system.

| $\rho\sigma^3$ | $n_{tr}$ | $n_e$ | $\langle t\rangle/\tau_B$ | $I\sigma^5/D_0$ |
|---------|------|------|--------|-------------------|
| 0.79228 | 5    | 5    | 13.8   | $1.4 \times 10^{-5}$ |
| 0.78507 | 5    | 5    | 159    | $1.2 \times 10^{-6}$ |
| 0.78153 | 10   | 10   | 260    | $7.3 \times 10^{-7}$ |
| 0.77700 | 20   | 10   | 3282   | $5.8 \times 10^{-8}$ |
| 0.77468 | 50   | 5    | 23340  | $8.1 \times 10^{-9}$ |

**Table 4.2**: Nucleation rates, $I\sigma^5/D_0$, obtained from $(NVT)$ Brownian dynamics simulations for various densities $\rho\sigma^3$ with $n_{tr}$ and $n_e$ the number of simulations and the number of observed nucleation events, respectively and $\langle t\rangle$ is the average waiting time for a nucleation event.

To calculate the nucleation rates from Brownian dynamics simulations, we perform multiple independent simulations of systems with $N = 4096$ particles and with the volume $V$ chosen such that the density of interest is acquired. Each simulation stops when a nucleation event happens, and the nucleation rate is determined by

$$I = \frac{1}{\langle t\rangle V},\qquad(4.3)$$

where $\langle t\rangle$ is the average waiting time for a single nucleation event. Thus $\langle t\rangle = \sum_i t_i/n_e$ where $t_i$ is the simulation time of the independent simulation $i$ and where $n_e$ is the number of nucleation events observed. The results from our BD simulations for varying densities are shown in Table 4.2.

Additionally, for $\rho\sigma^3 = 0.77000$, we performed 50 independent Brownian dynamics simulations. After a total simulation time $\sum_i t_i = 116700\tau_B$ we have not observed a single nucleation event in a system of $N = 4096$ particles. Since nucleation is a rare event, the probability distribution of a nucleation event happening at time $t$ is an exponential distribution given by

$$p(t) = \frac{1}{\langle t\rangle} \exp\left(-\frac{t}{\langle t\rangle}\right),\qquad(4.4)$$

where $\langle t\rangle$ is the average waiting time for a nucleation event. The probability of a nucleation event happening before time $t$ is $\int_0^t p(t)\mathrm{d}t = 1 - \exp(-t/\langle t\rangle)$. Thus for $\rho\sigma^3 = 0.77$, we can estimate the upper boundary for the nucleation rate. We find that if the nucleation rate is $4.85503 \times 10^{-9}D_0/\sigma^5$, the probability to observe a nucleation event before $116700\tau_B$ in a system of $N = 4096$ is 95%. Additionally, if the nucleation rate is $1.48499 \times 10^{-9}D_0/\sigma^5$, this probability is 60%.

### 4.3.2   Umbrella sampling

In this section, we use umbrella sampling to determine the Gibbs free-energy barriers, and then calculate the crystal nucleation rates from these barriers. The method is described in details in Chapter 3.5. For pressures $\beta p\sigma^3 = 12, 13$ and $13.3$, the free-energy barriers are shown in Fig. 4.3 and the attachment rates $f_{n^*}$ and nucleation rates $I$ are listed in Table 4.3. Note that in these simulations we used a neighbour cutoff of $r_c = 1.5\sigma$ and coupling parameter $\lambda = 0.2$.

| $\beta P\sigma^3$ | $n^*$ | $f_{n^*}/D_0$ | $\beta\Delta G(n^*)$ | $\beta\Delta G''(n^*)$ | $\rho\sigma^3$ | $I\sigma^5/D_0$ |
|---|---|---|---|---|---|---|
| 12 | 130 | 586.17 | 32.5 | 0.0015 | 0.762 | $5.23 \times 10^{-14}$ |
| 13 | 60 | 319.05 | 18.5 | 0.0030 | 0.774 | $4.98 \times 10^{-8}$ |
| 13.3 | 50 | 361.86 | 17.200 | 0.0030 | 0.777 | $2.08 \times 10^{-7}$ |

**Table 4.3**: Nucleation rates, $I\sigma^5/D_0$, as obtained from ($NpT$) umbrella sampling MC simulations at various pressures, $\beta P\sigma^3$, with $\rho\sigma^3$ the corresponding density of the supersaturated fluid. $\beta\Delta G(n^*)$ is the height of the free-energy barriers with $n^*$ the size of the critical cluster, and $\beta\Delta G''(n^*)$ and $f_{n^*}/D_0$ are the second order derivative and attachment rate at the top of the free-energy barrier, respectively.

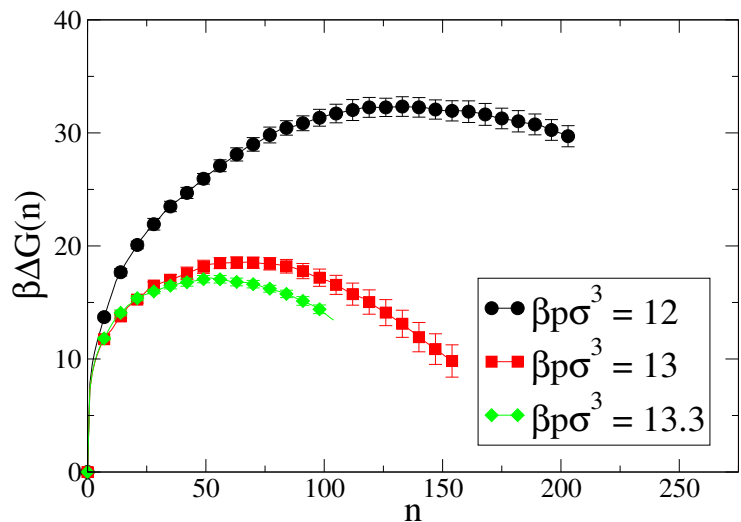

**Figure 4.3**: Gibbs free-energy barriers $\beta\Delta G(n)$ as a function of cluster size $n$ as obtained from umbrella sampling MC simulations at reduced pressures $\beta p\sigma^3 = 12, 13$ and $13.3$ as labelled.

### 4.3.3   Forward flux sampling

We use forward flux sampling to determine the nucleation rates, and the detailed description of the method is in Chapter 3.6. The dynamics in the forward flux sampling simulations were approximated using kinetic Monte Carlo simulations with a step size of $\Delta_{\text{KMC}} = 0.05\sigma$ and measuring the order parameter every $\Delta t_{\text{ord}} = 2$ MC cycles. The nearest neighbour cutoff for the order parameter was taken to be $r_c = 1.5\sigma$. The probabilities $P(\lambda_i|\lambda_{i-1})$ of going from interface $\lambda_{i-1}$ to $\lambda_i$ required in the forward flux sampling rate calculation for pressures $\beta p\sigma^3 = 12, 13$ and $14$ are given in Tables 4.4, respectively. The resulting rates in terms of the short-time diffusion coefficient $D_0$ are given in Table 4.5 .

| $\beta p\sigma^3$ | $i$ | $\lambda_i$ | $P(\lambda_i|\lambda_{i-1})$ | $\beta p\sigma^3$ | $i$ | $\lambda_i$ | $P(\lambda_i|\lambda_{i-1})$ | $\beta p\sigma^3$ | $i$ | $\lambda_i$ | $P(\lambda_i|\lambda_{i-1})$ |
|---|---|---|---|---|---|---|---|---|---|---|---|
| 12 | 2 | 20 | 0.133 | 12 | 9 | 150 | 0.130 | 13 | 6 | 100 | 0.166 |
| 12 | 3 | 26 | 0.132 | 12 | 10 | 200 | 0.317 | 13 | 7 | 150 | 0.633 |
| 12 | 4 | 34 | 0.107 | 12 | 11 | 250 | 0.842 | 14 | 2 | 40 | 0.164 |
| 12 | 5 | 45 | 0.068 | 13 | 2 | 20 | 0.132 | 14 | 3 | 70 | 0.453 |
| 12 | 6 | 60 | 0.066 | 13 | 3 | 30 | 0.124 | 14 | 4 | 100 | 0.847 |
| 12 | 7 | 80 | 0.041 | 13 | 4 | 40 | 0.193 | | | | |
| 12 | 8 | 110 | 0.036 | 13 | 5 | 60 | 0.132 | | | | |

**Table 4.4**: Probabilities $P(\lambda_{i+1}|\lambda_i)$ for the interfaces used in calculating the nucleation rate for pressure $\beta p\sigma^3 = 12, 13$ and 14.

| $\beta p\sigma^3$ | $\Phi_{A\lambda_1}\sigma^5/D_0$ | $P(\lambda_B|\lambda_1)$ | $I\sigma^5/D_0$ |
|---|---|---|---|
| 12 | $2.96 \times 10^{-6}$ | $4.32 \times 10^{-10}$ | $1.27 \times 10^{-15}$ |
| 13 | $1.10 \times 10^{-5}$ | $4.38 \times 10^{-5}$ | $4.80 \times 10^{-10}$ |
| 14 | $1.06 \times 10^{-5}$ | $6.29 \times 10^{-2}$ | $6.69 \times 10^{-7}$ |

**Table 4.5**: Nucleation rate $I\sigma^5/D_0$, flux $\Phi_{A\lambda_1}$, and $P(\lambda_B|\lambda_1)$ at various pressures $\beta p\sigma^3$ as obtained by $(NpT)$ forward flux sampling.

## 4.4 Discussion

In this section we compare our predicted nucleation rates to previous theoretical and experimental studies. In Fig. 4.4 we show our predicted WCA crystal nucleation rates and compare them with those found in Ref. [20]. Note that the nucleation rates shown in Fig. 4.4 (and Fig. 4.6, see below) cannot be obtained directly from Ref. [20] as there is a mistake in that paper regarding the mapping from effective packing fraction units to number densities [71].

We first note that our BD results match well with previous BD nucleation rates [20][*]. We also note that the uncertainty in the BD results is approximately one order of magnitude and the uncertainty in the US and FFS results is approximately two orders of magnitude. Within this uncertainty, the BD, US, and FFS nucleation rates all agree. This is consistent with Chapter 3 and Ref. [19] which found that molecular dynamics and FFS rates agreed well with the US rates of Auer and Frenkel [18].

We note that the US and FFS simulations were performed at constant pressure, i.e. in an $NpT$ ensemble, while the BD simulations were at constant volume $(NVT)$. While we have not examined in detail the nuclei appearing in these simulations, no significant difference was found between the nuclei forming in the BD simulations and the nuclei forming in the FFS and US simulations. This question was addressed in more detail our nucleation study on hard spheres as described in Chapter 3 and Ref. [19]. In that case, the radius of gyration tensor of the resulting clusters was measured as a function of cluster

---

[*]The number densities appearing in Figs. 4.4 and 4.6 were obtained directly via communications with the authors of Ref. [20].

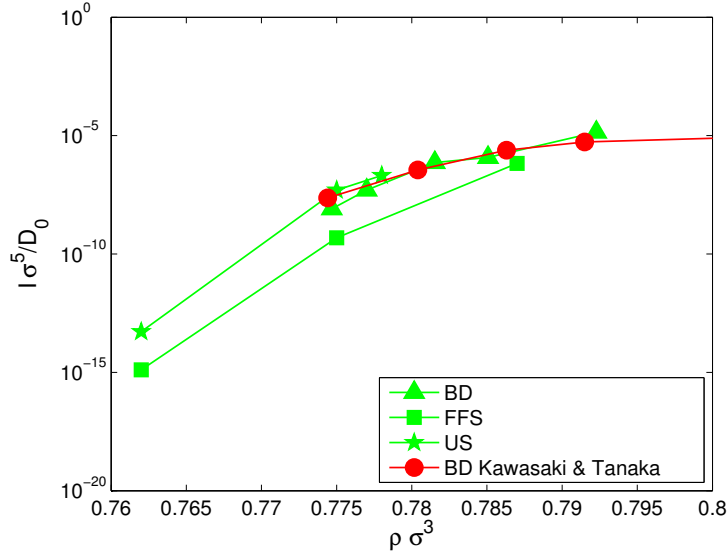

**Figure 4.4**: Crystal nucleation rates $I\sigma^5/D_0$ as a function of number density $\rho\sigma^3$ where $D_0$ is the short-time diffusion coefficient. While we have not included error bars in this plot, note that the uncertainty in the US and FFS nucleation rates is approximately 2 orders of magnitude while the uncertainty in the BD results is approximately 1 order of magnitude. Note also that the US and FFS simulations were performed at constant pressure ($NpT$) while the BD simulations were at constant volume ($NVT$).

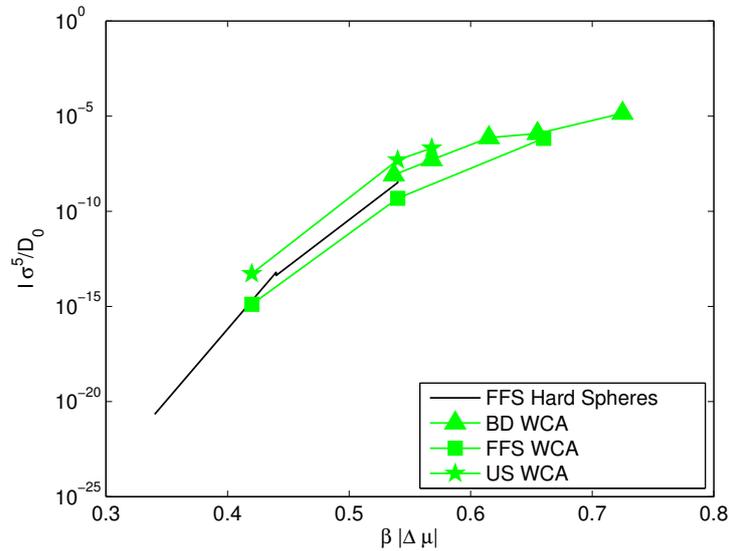

**Figure 4.5**: Crystal nucleation rates $I\sigma^5/D_0$ as a function of supersaturation $\beta|\Delta\mu|$ where $D_0$ is the short-time diffusion coefficient. The hard-sphere (HS) nucleation rates are taken from Ref. [19].

size for constant volume molecular dynamics simulations, and constant pressure FFS, and US simulations. No difference between the resulting nuclei was found. Additionally, in an $NVT$ ensemble, the formation of a nucleus depletes the number of particles in the fluid

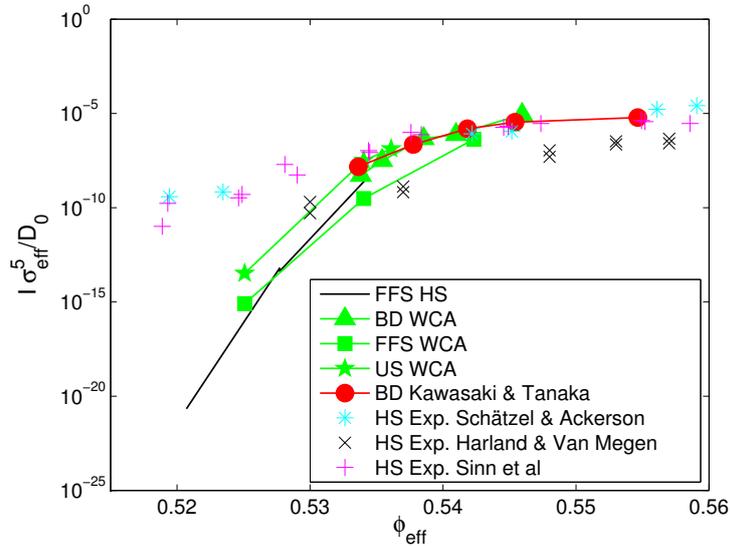

**Figure 4.6**: Crystal nucleation rates $I\sigma_{\text{eff}}^5/D_0$ as a function of effective packing fraction $\phi_{\text{eff}} = \frac{\pi}{6}\rho\sigma_{\text{eff}}^3$ where $D_0$ is the short-time diffusion coefficient and $\rho$ is the number density of the fluid. Note that $\sigma_{\text{eff}}$ is the size of a hard-sphere particle which has the same freezing number density as the WCA model. The hard-sphere (HS) nucleation rates are taken from Ref. [19].

and lowers slightly the number density of the fluid. However, when the system size is sufficiently large, this effect is negligible. While this effect was not studied in this article, it was examined by Kawasaki and Tanaka [20] who found that the nucleation rates for this model (i.e. the WCA potential) at high supersaturation converged for approximately 4000 particles. For lower supersaturations, we can approximate the effect of the system size by determining the number density of the fluid when a critical nucleus is present at fixed volume. For a system containing $N = 4096$ particles, at the lowest density we studied using BD simulations, namely $\rho\sigma^3 = 0.775$, we find the fluid density to be approximately $\rho\sigma^3 = 0.774$ when a critical nucleus containing 60 particles is present. As a result, we expect at the very most a horizontal error bar of 0.001 in the number density of the BD rates. Hence, we do not expect a significant effect from the system size in our BD simulations

In Fig. 4.5 we compare our predicted WCA rates with the crystal nucleation rates of hard spheres as a function of supersaturation, i.e. the chemical potential difference between the bulk crystal and the supersaturated fluid ($\Delta\mu$). We find good agreement between the nucleation rates in these two systems.

We further compare our WCA results with those of the hard-sphere system examined in Ref. [19] and the experimental light scattering results from Refs. [33, 35, 36]. To do this, we scale our WCA results in terms of an effective packing fraction in the same manner as is done experimentally. Specifically, we scale the freezing number density of the WCA model ($\rho_F\sigma^3 = 0.712$) to the freezing packing fraction of hard spheres. Note that in literature there is a range of freezing packing fractions for hard spheres, namely $0.491 \leq \phi_F^{HS} \leq 0.494$ (see, e.g. Refs. [10, 34, 73]). Here we follow Frenkel and Smit [10] which we believe to be the most accurate. In their work, finite size effects are taken into consideration when calculating the free energy of the face-centered-cubic (FCC) crystal, i.e. they use the result from Ref. [70]. In addition, the Speedy equations of state for the solid and fluid phase were employed [43, 44]. The resulting freezing packing fraction is found to be $\phi_F^{HS} = 0.492$ [10]. The WCA nucleation rates $I\sigma_{\text{eff}}^5/D_0$ scaled to $\phi_F^{HS} = 0.492$ are compared to the hard-sphere results in Fig. 4.6 where $\sigma_{\text{eff}}$ is the size of a hard-sphere particle which has the same freezing number density as the WCA model. We stress here that any error in the freezing coexistence results in a horizontal shift in the nucleation rates. Hence, in addition to an uncertainty of approximately 2 orders of magnitude in the nucleation rates, there is an additional uncertainty of approximately $\Delta\phi_{\text{error}} = \pm 0.005$ in the effective packing fractions. Thus, within these error bars, we find good agreement between our predicted hard-sphere and WCA crystal nucleation rates.

Previous studies [46, 47] have shown that softness in the potential increases the nucleation rate, however, this can not be confirmed from our predictions as the uncertainty in the nucleation rates is too large. We stress that the experimental hard-sphere nucleation rates differ significantly from our predicted rates for low supersaturations. This is in contrast to the results presented in Ref. [20] where good agreement was found between the WCA rates and the light scattering experimental nucleation rates. This difference originates from the freezing number density which was used to map the WCA number densities to effective packing fractions. As described in Section 4.2, in this chapter we have determined the freezing densities using full free-energy calculations which are known to be very accurate. In contrast, the method used in Ref. [20] appears to yield results that differ significantly from the "exact" coexistence densities.

The large difference between the nucleation rates when plotted in terms of effective packing fractions emphasizes one possible problem in the comparison between the experimental and simulated nucleation rates: the determination of the effective packing fractions. A difference of 1-2% in the freezing density has a significant effect on the position of the drop-off of the nucleation rates. Whereas it is straightforward to evaluate the correct effective volume fractions in simulations, the procedure required to deduce the same information from experiments is more subtle. Hence, part of the discrepancy between the computed and measured crystal-nucleation rates of "hard-sphere" colloids may be due to a small difference in the definition of the effective packing fraction. Yet, this is certainly not the whole story: the very large discrepancy between experimental and numerical nucleation rates at lower densities cannot be accounted for by a simple rescaling of the density axis. Hence, unlike Kawasaki and Tanaka, we conclude that the discrepancy between simulation and experiment is as large as ever, and still unexplained.

## 4.5   Conclusions

In conclusion, we have examined the crystal nucleation of particles interacting with the WCA potential with $\beta\epsilon = 40$ using Brownian dynamics, umbrella sampling and forward flux sampling. As in Ref. [19], we find good agreement between the nucleation rates predicted using these different methods. Additionally, we find that the nucleation rates predicted for the WCA model agree well with those of hard spheres as a function of the effective packing fraction $\phi_{\text{eff}}$ defined such that $\phi_{\text{eff}}$ at freezing matches that of hard spheres.

## 4.6 Acknowledgements

I would like to thank Dr. L. Filion for the calculation of phase diagram and forward flux sampling simulations.

# 5

---

# Crystal nucleation of colloidal hard dumbbells

---

Using computer simulations we investigate homogeneous crystal nucleation in suspensions of colloidal hard dumbbells. The free energy barriers are determined by Monte Carlo simulations using the umbrella sampling technique. We calculate the nucleation rates for the plastic crystal and the aperiodic crystal phase using the kinetic prefactor as determined from event driven molecular dynamics simulations. We find good agreement with the nucleation rates determined from spontaneous nucleation events observed in event driven molecular dynamics simulations within error bars of one order of magnitude. We study the effect of aspect ratio of the dumbbells on the nucleation of plastic and aperiodic crystal phases and we also determine the structure of the critical nuclei. Moreover, we find that the nucleation of the aligned CP1 crystal phase is strongly suppressed by a high free energy barrier at low supersaturations and slow dynamics at high supersaturations.

## 5.1 Introduction

Recent breakthroughs in particle synthesis produced a spectacular variety of anisotropic building blocks [3]. Colloidal particles with the shape of a dumbbell are one of the simplest anisotropic building blocks [74]. Their unique morphologies lead to novel self-organized structures. For instance, it was found that magnetic colloidal dumbbells can form chain-like clusters with tunable chirality [75], while novel crystal structures have been predicted recently for asymmetric dumbbell particles consisting of a tangent large and small hard sphere, which are atomic analogs of NaCl, CsCl, $\gamma$CuTi, CrB, and $\alpha$IrV, when we regard the two individual spheres of each dumbbell independently [76]. Moreover, colloidal dumbbells also gain increasing scientific attention in recent years due to its potential use in photonic applications. It has been shown that dumbbells on a face-centered-cubic lattice where the spheres of the dumbbells form a diamond structure exhibit a complete band gap [77, 78] while it is impossible to obtain a complete band gap in systems consisting of spherical particles. A very recent calculation showed that for midrange aspect ratios, both asymmetric and symmetric dumbbells have 2 - 3 large band gaps in the inverted lattice [79]. Although these structures are not thermodynamically stable for hard dumbbells [80–86], it does show the promising potential of anisotropic particles in photonic applications.

New routes of synthesizing colloidal dumbbells make it easy to control the aspect ratio [74]. In addition, by adding salt to the solvent, the interactions between dumbbells can be tuned from long-ranged repulsive to hard interactions. Although hard dumbbells were originally modeled for simple non-spherical diatomic molecules, such as nitrogen, they are also a natural model system for studying the self-assembly of colloidal dumbbells [87–90]. The phase behavior of hard dumbbells has been extensively studied by density functional theory [80, 81] and computer simulations [82–86]. The bulk phase diagram of hard dumbbells displays three types of stable crystal structures [80–86]. For small aspect ratio, the dumbbells form a plastic crystal phase at low densities. The freezing into a cubic plastic crystal phase in which the dumbbells are positioned on a face-centered-cubic lattice but are free to rotate, has been determined using Monte Carlo simulations [82]. These results have been refined by Vega, Paras, and Monson, who showed that at higher densities the cubic plastic crystal phase transforms into an orientationally ordered crystal CP1 phase. Additionally, these authors showed that the fluid-plastic crystal coexistence region terminates at $L/\sigma \simeq 0.38$, where $L$ is the distance between the centers of spheres and $\sigma$ is the diameter of the dumbbells. For longer dumbbells a fluid-CP1 coexistence region was found, whereas the relative stability of the close-packed crystal structures CP1, CP2, and CP3, which only differ in the way the hexagonally packed dumbbell layers are stacked remained undetermined as the free energies are very similar [83–85]. Moreover, these authors showed by making an estimate for the degeneracy contribution to the free energy that dumbbells with $L/\sigma = 1$ may form an aperiodic crystal phase [83]. The stability of such an aperiodic crystal structure in which both the orientations and positions of the particles are disordered, while the spheres of each dumbbell are located on the lattice positions of a random-hexagonal-close-packed (rhcp) lattice, has been verified recently for $L/\sigma > 0.88$ [86]. In addition, it has been shown that the plastic crystal phase with the hexagonal-close-packed structure is more stable than the cubic plastic crystal for a large part of the stable plastic crystal region [86]. Although, the bulk phase diagram is well-studied, the kinetic pathways of the fluid-solid phase transitions are still unknown, and only a few studies have been devoted to the crystal nucleation of anisotropic particles [91, 92]. In the present work, we investigate the nucleation of the plastic crystal phase of hard dumbbells using computer simulations and study the effect of aspect ratio of the dumbbells on the resulting nucleation rates and the structure and size of the critical nuclei. Moreover, for longer dumbbells we investigate crystal nucleation of the aperiodic crystal phase. First, we calculate the Gibbs free energy barriers for nucleation using Monte Carlo (MC) simulations with the umbrella sampling technique, which are then combined with event driven molecular dynamics (EDMD) simulations to determine the kinetic prefactor and the nucleation rates. Additionally, we determine the nucleation rates from spontaneous nucleation events observed in EDMD simulations. We compare the nucleation rates and critical nuclei obtained from the umbrella sampling MC simulations with those from EDMD simulations.

The remainder of this chapter is organized as follows. In Sec. 5.2, we describe the methodology including the model and simulation methods used. We present the results and discussions on the nucleation of three types of crystal phases in suspensions of hard dumbbells in Sec. 5.3. We end with some discussions and conclude in Sec. 5.4.

## 5.2   Methodology

We consider a system of hard dumbbells consisting of two overlapping hard spheres with diameter $\sigma$ with the centers separated by a distance $L$. We define the aspect ratio as $L^* \equiv L/\sigma$, such that the model reduces to hard spheres for $L^* = 0$ and to tangent spheres for $L^* = 1$. We study crystal nucleation of hard dumbbells for $0 \leq L^* \leq 1$. We focus on the nucleation of the plastic crystal phase ($0 \leq L^* < 0.4$) and the aperiodic crystal phase ($0.88 < L^* \leq 1$) [81, 83–86].

### 5.2.1   Order parameter

In order to study the nucleation of the crystal phase, we require a cluster criterion that identifies the crystalline clusters in a metastable fluid. In this work, we employ the order parameter based on the local bond order parameter analysis of Steinhardt *et al.* [93]. We define for every particle $i$, a $2l + 1$-dimensional complex vector $\mathbf{q}_l(i)$ given by

$$q_{lm}(i) = \frac{1}{N_b(i)} \sum_{j=1}^{N_b(i)} \Upsilon_{lm}(\hat{\mathbf{r}}_{ij}),\tag{5.1}$$

where $N_b(i)$ is the total number of neighboring particles of particle $i$, and $\Upsilon_{lm}(\hat{\mathbf{r}}_{ij})$ is the spherical harmonics for the normalized direction vector $\hat{\mathbf{r}}_{ij}$ between particle $i$ and $j$, $l$ is a free integer parameter, and $m$ is an integer that runs from $m = -l$ to $m = +l$. Neighbors of particle $i$ are defined as those particles which lie within a given cutoff radius $r_c$ from particle $i$. In order to determine the correlation between the local environments of particle

$i$ and $j$, we define the rotationally invariant function $d_l(i, j)$

$$d_l(i, j) = \sum_{m=-l}^{l} \tilde{q}_{lm}(i) \cdot \tilde{q}_{lm}^*(j), \tag{5.2}$$

where $\tilde{q}_{lm}(i) = q_{lm}(i)/\sqrt{\sum_{m=-l}^{l} |q_{lm}(i)|^2}$ and the asterisk is the complex conjugate [94]. If $d_l(i, j) > d_c$, the bond between particle (sphere) $i$ and $j$ is regarded to be solid-like or connected, where $d_c$ is the dot-product cutoff. We identify a particle (sphere) as solid-like when it has at least $\xi_c$ solid-like bonds. We have chosen the symmetry index $l = 6$ as the particles (spheres) display hexagonal order in the plastic crystal and the aperiodic crystal phase. We have chosen $r_c = 1.3\sigma$, $d_c = 0.7$, and $\xi_c = 6$ in our simulations. It has been shown recently that the choice of order parameter ($r_c$, $d_c$, and $\xi_c$) does not affect the resulting nucleation rate if it is not too restrictive [19, 92].

To analyze the structure of the critical nuclei, we use the averaged local bond order parameter $\bar{q}_l$ and $\bar{w}_l$ proposed by Lechner and Dellago [64], which allows us to identify each particle as fcc-like or hcp-like, provided the number of neighboring particles $N_b(i) \geq 10$:

$$\bar{q}_l(i) = \sqrt{\frac{4\pi}{2l+1} \sum_{m=-l}^{l} |\bar{q}_{lm}(i)|^2}, \tag{5.3}$$

$$\bar{w}_l(i) = \frac{\sum_{m_1+m_2+m_3=0} \begin{pmatrix} l & l & l \\ m_1 & m_2 & m_3 \end{pmatrix} \bar{q}_{lm_1}(i) \bar{q}_{lm_2}(i) \bar{q}_{lm_3}(i)}{\left( \sum_{m=-l}^{l} |\bar{q}_{lm}(i)|^2 \right)^{3/2}}, \tag{5.4}$$

where

$$\bar{q}_{lm}(i) = \frac{1}{N_b(i)+1} \sum_{k=0}^{N_b(i)} q_{lm}(i). \tag{5.5}$$

The sum from $k = 0$ to $N_b(i)$ runs over all neighbors of particle (sphere) $i$ plus the particle (sphere) $i$ itself. While $q_{lm}(i)$ takes into account the structure of the first shell around particle $i$, the averaged $\bar{q}_{lm}(i)$, contains also the information of the structure of the second shell, which increases the accuracy of the crystal structure determination. In order to distinguish fcc-like and hcp-like particles, we employ $\bar{q}_4$ and $\bar{w}_4$, as the order parameter distributions of pure fcc and hcp phases of Lennard-Jones and Gaussian core systems are well separated in the $\bar{q}_4 - \bar{w}_4$ plane [64].

## 5.2.2   Umbrella sampling

The Gibbs free energy $\Delta G(n)$ for the formation of a crystalline cluster of size $n$ is given by $\Delta G(n)/k_B T = \text{const} - \ln[P(n)]$, where $P(n)$ is the probability distribution function of finding a cluster of size $n$, $k_B$ is Boltzmann's constant, and $T$ the temperature. As nucleation is a rare event and the probability to find a spontaneous nucleation event is very small in a brute force simulation within a reasonable time, one has to resort to specialized simulation techniques such as forward flux sampling, umbrella sampling or transition path sampling. Here, we employ the method developed by Frenkel and coworkers [18] to calculate the free energy of the largest cluster. In this method, the sampling is biased towards configurations that contain clusters with a certain size. To this end, we introduce a biasing potential $\omega(\mathbf{r}^N)$, which is a harmonic function of the cluster size $n$:

$$\beta\omega(\mathbf{r}^N) = \frac{1}{2}k\left[n(\mathbf{r}^N) - n_0\right]^2,\tag{5.6}$$

where $n(\mathbf{r}^N)$ is the size of largest cluster and $n_0$ is the center of the umbrella sampling window whose width depends on $k$. In this work we set $k = 0.2$. By increasing the value of $n_0$, we increase the size of the largest crystalline cluster in our system, which enables us to cross the nucleation barrier. If we define the average number of crystalline clusters with $n$ particles by $\langle N_n \rangle$, one can calculate the probability distribution $P(n) = \langle N_n \rangle / N$ from which we can determine the Gibbs free energy $\Delta G(n)$.

### 5.2.3   Event driven molecular dynamics simulations

Since the potential between particles in systems of hard dumbbells is discontinuous, the pair interactions only change when particles collide. The particles perform elastic collisions when they encounter each other. We numerically identify and handle these collisions by using an EDMD simulation [13, 95].

Using MD simulations to determine the nucleation rate is straightforward. Starting with an equilibrated fluid configuration, an MD simulation is used to evolve the system until the largest cluster in the system exceeds the critical nucleus size. Then the nucleation rate is given by

$$I = \frac{1}{\langle t \rangle V},\tag{5.7}$$

where $\langle t \rangle$ is the averaged waiting time of forming a critical nucleus in a system of volume $V$.

## 5.3   Results and discussions

In this section, we present the results on the nucleation of the plastic crystal, the aperiodic crystal and the CP1 crystal phase in suspensions of hard dumbbells.

### 5.3.1   Nucleation of the plastic crystal phase

We first investigate the nucleation of the plastic crystal phase of hard dumbbells. Monte Carlo simulations with the umbrella sampling technique are performed on hard-dumbbell fluids with $L^* = 0, 0.15$ and $0.3$ at supersaturation $\beta|\Delta\mu| = 0.34$ and with $L^* = 0, 0.15$ and $0.2$ for $\beta|\Delta\mu| = 0.54$ with $\beta = 1/k_BT$. We have chosen a shorter aspect ratio for the highest supersaturation as the plastic crystal phase for dumbbells with $L^* = 0.3$ becomes metastable with respect to the aligned CP1 phase for $P^* = P\sigma^3/k_BT > 30$, i.e., $\beta|\Delta\mu| > 0.47$. The Gibbs free energy $\beta\Delta G(n)$ as a function of cluster size $n$ is shown in Fig. 5.1. We clearly observe that at low supersaturation, i.e. $\beta|\Delta\mu| = 0.34$, the heights of the free energy barriers increase slightly ($\sim 8\%$) with aspect ratio. More specifically, $\beta\Delta G^* = 42.9 \pm 0.3, 44.5 \pm 1.1$, and $45.2 \pm 2$ for $L^* = 0, 0.15$ and $0.3$, respectively. According to classical nucleation theory (CNT), the nucleation barrier for a spherical nucleus with radius $R$ is given by $\Delta G(R) = 4\pi\gamma R^2 - 4\pi|\Delta\mu|\rho_s R^3/3$ with $\gamma$ the interfacial tension, $|\Delta\mu|$ the chemical potential difference between the solid and fluid phase, and $\rho_s$ the bulk density of the solid phase. CNT predicts a nucleation barrier height $\Delta G^* = (16\pi/3)\gamma^3/(\rho_s|\Delta\mu|)^2$ and a critical radius $R^* = 2\gamma/\rho_s|\Delta\mu|$. The small increase in barrier height with aspect ratio can be explained by the small increase in the crystal-melt interfacial tensions that have been determined recently for the crystal planes (100), (110), (111) using nonequilibrium work measurements with a cleaving procedure in MC simulations [96]. For a spherical cluster, the surface tension is expected to be an average over the crystal planes, i.e., $\beta\gamma d^2 = 0.58, 0.57$, and $0.60$, for $L^* = 0, 0.15$ and $0.3$, respectively, where $d^3 = \sigma^3(1 + 3/2L^* - 1/5L^{*3})$. Another paper by Davidchack *et al.* found a slightly lower value for the averaged interfacial tension of hard spheres, i.e, $\beta\gamma d^2 = 0.559$ [54]. Using $\beta\Delta G^* = 42.9$ and the more precise value for the surface tension $\beta\gamma d^2 = 0.559$, and the values for $\beta\gamma d^2$ and the bulk density $\rho_s$ for varying $L^*$ presented in Table 5.1, CNT predicts a slightly larger increase in barrier height upon increasing $L^*$, i.e., $\beta\Delta G^* = 45.6$ and 50.49, for $L^* = 0.15$ and 0.3, respectively. However, when the supersaturation is increased to $\beta|\Delta\mu| = 0.54$, we find a decrease in barrier height upon increasing the aspect ratio as shown in Fig. 5.1 and Table 5.1, which cannot be explained by CNT. Apparently, the pressure dependence of the surface tension is different for dumbbells with various aspect ratios.

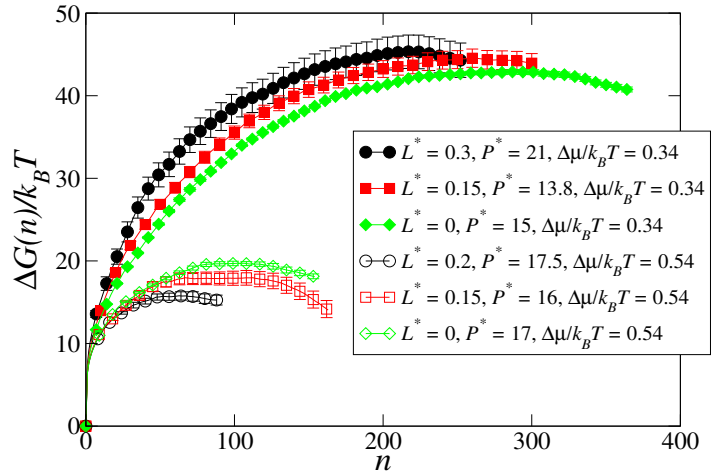

**Figure 5.1**: Gibbs free energy $\Delta G(n)/k_B T$ as a function of cluster size $n$ for the nucleation of the plastic crystal phase of hard dumbbells with various aspect ratios $L^* = L/\sigma$ as displayed and supersaturation $\beta|\Delta\mu| = 0.34$ (filled symbols) and 0.54 (open symbols).

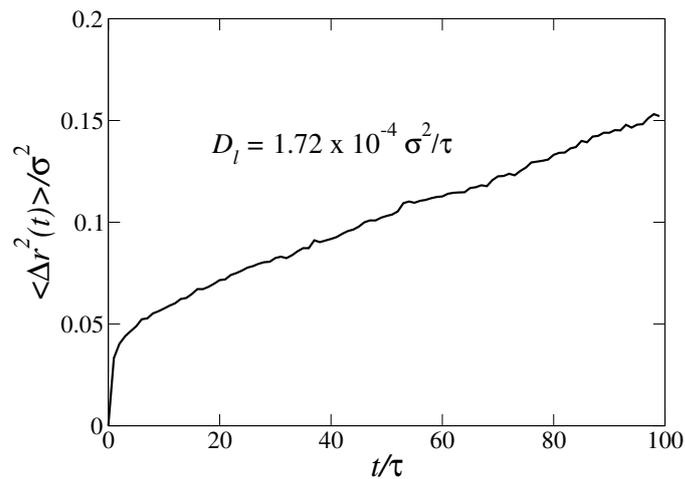

**Figure 5.2**: Mean square displacement $\langle \Delta r^2(t) \rangle$ as a function of time $t/\tau$ in a fluid of hard dumbbells with $L^* = 0.3$ at $P^* = 30$ (for $\beta|\Delta\mu| = 0.47$).

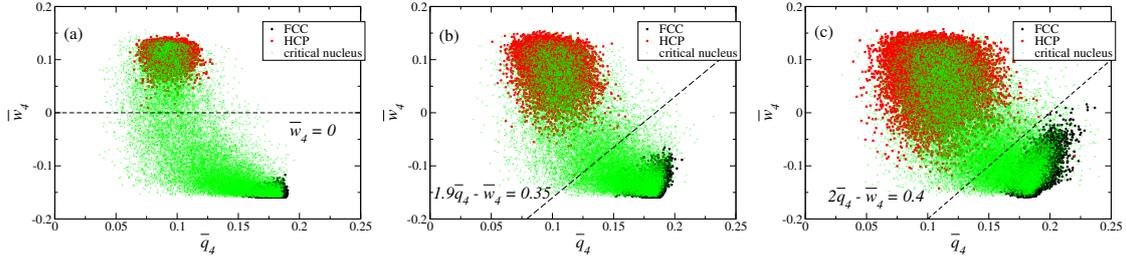

**Figure 5.3**: Distribution of particles in the critical nuclei for the plastic crystal nucleation in hard dumbbell systems with $L^* = 0$ (a), 0.15 (b), and 0.3 (c) as obtained from umbrella sampling MC simulations at a supersaturation $\beta|\Delta\mu| = 0.34$ in the $\bar{q}_4 - \bar{w}_4$ plane compared with those for pure fcc and hcp plastic crystal phases with corresponding pressures. The dashed lines are used to distinguish the fcc-like and hcp-like particles, and the formulas are next to them.

The nucleation barriers obtained from umbrella sampling MC simulations can also be used to determine the nucleation rates as given by [18]:

$$I = \kappa \exp\left(-\beta\Delta G^*\right) \tag{5.8}$$

where $\kappa$ is the kinetic prefactor given by $\kappa = \rho_l f_{n^*} \sqrt{|\Delta G''(n^*)|/2\pi k_B T}$, $\rho_l$ is the number density of particles in the fluid phase, $f_{n^*}$ the rate at which particles are attached to the critical nucleus, $\Delta G''(n^*)$ is the second derivative on the top of the Gibbs free energy barrier. The attachment rate can be calculated from the mean square deviation of the cluster size at the top of the free energy barrier by

$$f_{n^*} = \frac{1}{2} \frac{\langle [n(t) - n(0)]^2 \rangle}{t} \tag{5.9}$$

where $n(t)$ is the cluster size at time $t$. The mean square deviation of the cluster size can be determined from EDMD simulations starting from configurations at the top of the free energy barriers. Using the results for the attachment rates and the nucleation barriers obtained from umbrella sampling MC simulations, we can determine the nucleation rates, which we compare with those obtained directly from spontaneous nucleation events in EDMD simulations. We observed a large variance in the attachment rates calculated for different nuclei. We used 10 independent configurations on the top of the barrier and followed 10 trajectories for each of them to determine the attachment rates. Taking into account the statistical errors in the free energy barriers and attachment rates, we estimate that the error in the resulting nucleation rates is one order of magnitude. In order to exclude the effect of dynamics, we compare the nucleation rates for the plastic crystal phase in long-time self diffusion times, i.e. $\tau_L = \sigma^2/6D_l$ with $D_l$ the long-time self diffusion coefficient. We calculate $D_l$ by measuring the mean square displacement at supersaturation $\beta|\Delta\mu| = 0.34$ and 0.54 as shown in Table 5.1 for various aspect ratios. We clearly observe that the dynamics becomes slower for increasing aspect ratio $L^*$, resulting in long-time self diffusion coefficients $D_l\tau/\sigma^2 = 0.012, 0.01, 0.0023$ for $L^* = 0, 0.15$ and 0.3 at $\beta|\Delta\mu| = 0.34$ with $\tau = \sigma\sqrt{m/k_B T}$. At higher supersaturation $\beta|\Delta\mu| = 0.54$, we find even smaller values for $D_l$, i.e., $D_l\tau/\sigma^2 = 0.0078, 0.006, 0.003$ for $L^* = 0, 0.15$ and 0.2, respectively.

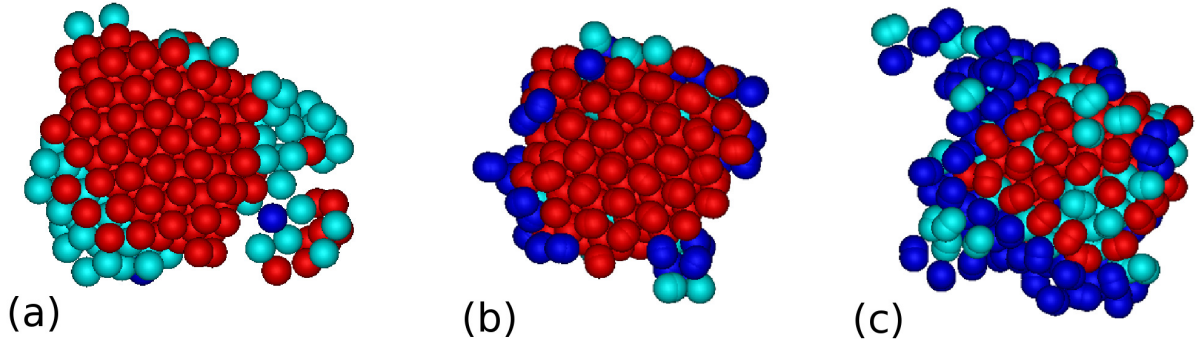

**Figure 5.4**: Typical configurations of critical nuclei for the plastic crystal nucleation of hard dumbbells with aspect ratios $L^* = 0$ (a), 0.15 (b), and 0.3 (c) at supersaturation $|\Delta\mu| = 0.34 k_B T$. The red (dark grey) particles are fcc-like, the blue particles are hcp-like particles, while the light blue (light grey) particles are undetermined.

The resulting nucleation rates in units of the long-time self diffusion coefficient are shown in Table 5.1. We wish to make a few remarks here. First, the nucleation rates obtained from spontaneous nucleation events observed in EDMD simulations agree well with the ones obtained from umbrella sampling MC simulations within error bars of one order of magnitude, which means that the nucleation results obtained from the umbrella sampling MC simulations are reliable. Secondly, we clearly observe that the nucleation rates for the different aspect ratios ranging from $L^* = 0$ to 0.3 are remarkably similar as the differences are within the errorbars for both supersaturations.

Finally, we made an attempt to study spontaneous nucleation of dumbbells with $L^* = 0.3$ at supersaturation $\beta|\Delta\mu| = 0.54$ using event-driven MD simulations. As already mentioned above, the plastic crystal phase for dumbbells with $L^* = 0.3$ becomes metastable with respect to an aligned CP1 phase for $P^* = P\sigma^3/k_BT > 30$, i.e., $\beta|\Delta\mu| > 0.47$. Hence, we would expect to find nucleation of the CP1 phase here. However, we find that the nucleation is severely hampered due to slow dynamics, which can be appreciated from Fig. 5.2, where we plot the mean square displacement for $\beta|\Delta\mu| = 0.47$. The resulting long-time self diffusion coefficient $D_l = 1.72 \times 10^{-4}\sigma^2/\tau$ is at least one order of magnitude smaller than the long-time self diffusion coefficients at $\beta|\Delta\mu| = 0.54$, where we observed spontaneous nucleation for $L^* = 0, 0.5$, and 0.2.

| $L^*$ | $P\sigma^3/k_BT$ | $\beta|\Delta\mu|$ | $\rho_s d^{3*}$ | $n^*$ | $\beta\Delta G(n^*)$ | $|\beta\Delta G''(n^*)|$ |
|---|---|---|---|---|---|---|
| 0 | 15 | 0.34 | 1.107 | 300 | $42.9 \pm 0.3$ | $5.1 \times 10^{-4}$ |
| 0.15 | 13.8 | 0.34 | 1.104 | 265 | $44.5 \pm 1.1$ | $6.0 \times 10^{-4}$ |
| 0.3 | 21 | 0.34 | 1.163 | 220 | $45.2 \pm 2$ | $1.0 \times 10^{-3}$ |
| 0 | 17 | 0.54 | 1.136 | 102 | $19.6 \pm 0.3$ | $1.2 \times 10^{-3}$ |
| 0.15 | 16 | 0.54 | 1.131 | 70 | $18.0 \pm 0.7$ | $9.7 \times 10^{-4}$ |
| 0.2 | 17.5 | 0.54 | 1.143 | 65 | $15.8 \pm 0.5$ | $2.0 \times 10^{-3}$ |

| $L^*$ | $P\sigma^3/k_BT$ | $f_{n^*}/6D_l$ | $D_l\tau/\sigma^2$ | $I\sigma^5/6D_l$ (US) | $I\sigma^5/6D_l$ (MD) |
|---|---|---|---|---|---|
| 0 | 15 | 4550 | 0.012 | $9.6 \times 10^{-18\pm1}$ | - |
| 0.15 | 13.8 | 3700 | 0.01 | $1.4 \times 10^{-18\pm1}$ | - |
| 0.3 | 21 | 7464 | 0.0023 | $1.7 \times 10^{-18\pm1}$ | - |
| 0 | 17 | 3980 | 0.0078 | $1.7 \times 10^{-7\pm1}$ | $1.6 \times 10^{-7\dagger}$ |
| 0.15 | 16 | 3779 | 0.006 | $6.1 \times 10^{-7\pm1}$ | $3.5 \times 10^{-7\pm1}$ |
| 0.2 | 17.5 | 2682 | 0.003 | $5.5 \times 10^{-6\pm1}$ | $4.4 \times 10^{-6\pm1}$ |

$^*d^3 = \sigma^3(1 + 3/2L^* - 1/2L^{*3})$ [96]
$^\dagger$Extrapolated from Ref. [19]

**Table 5.1**: Nucleation rates $I\sigma^5/6D_l$ for the nucleation of the plastic crystal phase in systems of hard dumbbells with elongation $L^*$, at pressure $P\sigma^3/k_BT$, and supersaturation $\beta|\Delta\mu|$. $\rho_s d^3$ is the number density of dumbbells in the solid phase, $\beta\Delta G(n^*)$ is the barrier height, and $|\beta\Delta G''(n^*)|$ is the second derivative of the Gibbs free energy at the critical nucleus size $n^*$, i.e., the number of dumbbells in the critical cluster. $f_{n^*}/6D_l$ is the attachment rate in units of the long-time self diffusion coefficient $D_l$.

In umbrella sampling MC simulations, we can "fix" the simulations at the top of the nucleation barrier which allows us to study the properties of the critical nuclei. We investigate the effect of the particle anisotropy on the structure of the critical nuclei using the order parameters $\overline{q}_4$ and $\overline{w}_4$ as defined above. At supersaturation $\beta|\Delta\mu| = 0.34$, the size of the critical nuclei is $n \simeq 250$ which is sufficiently large to determine the crystal structure of the nuclei. For each dumbbell, we calculate the averaged local bond order parameter $\overline{q}_4$ and $\overline{w}_4$, provided the particle has $N_b(i) \geq 10$ neighbors. The distribution of particles in the critical nuclei are presented as scatter plots in the $\overline{q}_4 - \overline{w}_4$ plane along with those for pure fcc and hcp plastic crystal phases of dumbbells with $L^* = 0, 0.15$, and $0.3$, at corresponding pressures. From Fig. 5.3, we clearly observe that the critical nuclei for $L^* = 0$ and $0.15$, contains predominantly fcc-like rather than hcp-like particles. In order to distinguish the fcc-like and hcp-like particles more quantitatively, we divide the $\overline{q}_4 - \overline{w}_4$ plane by a straight line in such a way that the particle distributions for the pure fcc and hcp plastic crystal phases are maximally separated. We plot the criteria to distinguish fcc-like and hcp-like particles as dashed straight lines in Fig. 5.3 with the corresponding formula. We note, however, that the criteria seem to be arbitrarily chosen, but the identification of fcc-like and hcp-like particles for typical nuclei seems to be less sensitive on the precise details of these criteria. Typical snapshots of the critical nuclei for $L^* = 0, 0.15$, and $0.3$ are shown in Fig. 5.4, where the color-coding denotes the identity (fcc-like, hcp-like or undetermined) of the particle using these criteria. As we did not calculate the averaged local bond order parameter $\overline{q}_4$ and $\overline{w}_4$ for particles with $N_b(i) < 10$ neighbors, the identity of these particles remains undetermined. We clearly observe that the critical nuclei for $L^* = 0, 0.15$ contains mainly fcc-like particles. The particle distributions becomes broader for the pure fcc and hcp plastic crystal phases upon increasing $L^*$ and consequently it becomes more difficult to distinguish fcc-like and hcp-like particles. However, the fraction of hcp-like particles seems to increase with increasing particle elongation. This agrees with the results from free energy calculations of hard dumbbell systems, where it has been shown that the hcp plastic crystal phase is more stable than the one with an fcc structure at $L^* \geq 0.15$ [86]. It is worth noting here that recent nucleation studies of hard spheres showed that the critical nuclei contain approximately 80% fcc-like particles, see Chapter 3 and Ref. [19] As the free energy difference per particle between bulk fcc and hcp phases is only about $0.001\ k_BT$ at melting, the predominance for fcc-like particles is attributed to surface effects.

### 5.3.2   Nucleation of the aperiodic crystal phase

For more elongated dumbbells, i.e., $L^* > 0.88$, the orientationally disordered aperiodic crystal phase becomes stable [83–86], in which the individual spheres of the dumbbells are on a random hcp lattice, and in addition the orientations of the dumbbells are random. In this section, we investigate the nucleation of the aperiodic crystal phase of hard dumbbells with different aspect ratios. We perform Monte Carlo simulations using the umbrella sampling technique to determine the Gibbs free energy as a function of cluster size for hard dumbbells with $L^* = 1.0$ and supersaturation $P^* = 16$ and 17. The order parameter that is employed here in the umbrella sampling technique is equal to the number of spheres $n$ (and thus not the number of dumbbells) in the largest crystalline cluster in the system. Thus, we check for each individual sphere whether or not it belongs to the largest crystalline cluster, and as a consequence, the whole dumbbell can be part of the largest cluster or only one sphere of the dumbbell can belong to the cluster, or the whole dumbbell is regarded to be fluid-like. Consequently, it is convenient to introduce a bulk chemical potential per sphere, which equals 0.5 times the bulk chemical potential per dumbbell $\mu^{sph} = \mu/2$. We compare the results with those for hard spheres at the same pressure in Fig. 5.5. Since the bulk pressure for the solid-fluid transition of hard dumbbells with $L^* = 1$ is remarkably close to that of hard spheres $\beta P_{coex}\sigma^3 = 11.8$ [83–86], one might naively expect that the nucleation barriers should be compared at the same dimensionless pressure. However, we observe that at the same pressure, the nucleation barrier for the aperiodic crystal phase of hard dumbbells is slightly higher than that of hard spheres. CNT predicts that the barrier height is given by $\Delta G^* = (16\pi/3)\gamma^3/(\rho_s^{sph}|\Delta\mu^{sph}|)^2$, and hence a difference in barrier height should be due to a difference in the interfacial tension $\gamma$, the density of spheres in the solid phase $\rho_s^{sph}$, or in $|\Delta\mu^{sph}|$. As the reduced density of spheres $\rho_s^{sph}\sigma^3$ in the aperiodic crystal phase is very close to that of a solid phase of hard spheres at $P^* = 16$ and $17$, and the interfacial tensions $\beta\gamma\sigma^2$ are also expected to be very similar, the difference in barrier height can only be caused by a difference in $|\Delta\mu^{sph}|$. We therefore calculated more accurately the bulk chemical potential difference per sphere between the solid and the fluid phase using

$$|\Delta\mu^{sph}| = \int_{P_{coex}}^{P} \left(\frac{1}{\rho_l^{sph}} - \frac{1}{\rho_s^{sph}}\right)\mathrm{d}P \tag{5.10}$$

where $\rho_l^{sph}$ and $\rho_s^{sph}$ are the density of spheres in the liquid and solid phase. In Fig. 5.6, we plot the equation of state for the fluid and solid phase of hard spheres from Ref. [44, 97] along with the equation of state for the fluid phase of hard dumbbells for $L^* = 1$ from Ref. [98]. In addition, we determined the equation of state for the solid phase using EDMD simulations. Using these results and Eq. 5.10, we indeed find that the supersaturation $\beta|\Delta\mu^{sph}|$ per sphere is $\sim 2.3\%$ smaller for hard dumbbells than for hard spheres, resulting in an increase in barrier height of $\sim 5\%$, which perfectly matches our results. We conclude that the difference in the height of the nucleation barrier between the aperiodic crystal phase of dumbbells with $L^* = 1.0$ and the hard-sphere crystal is mostly due to the difference in $|\Delta\mu^{sph}|$.

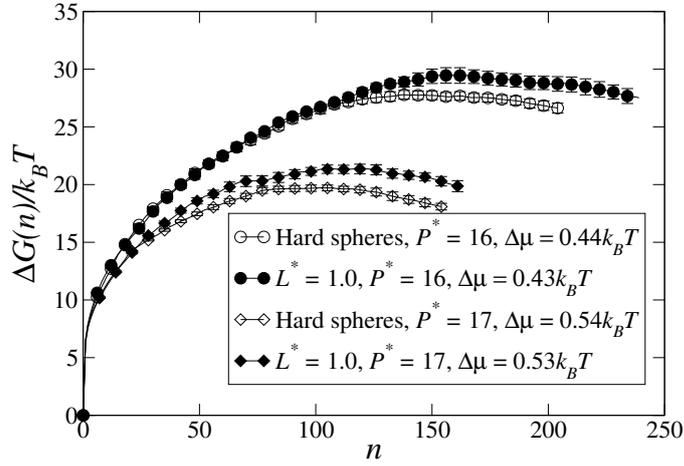

**Figure 5.5**: Gibbs free energy $\Delta G(n)$ as a function of number of spheres $n$ in the largest cluster for the nucleation of the (aperiodic) crystal phase of hard dumbbells and of hard spheres.

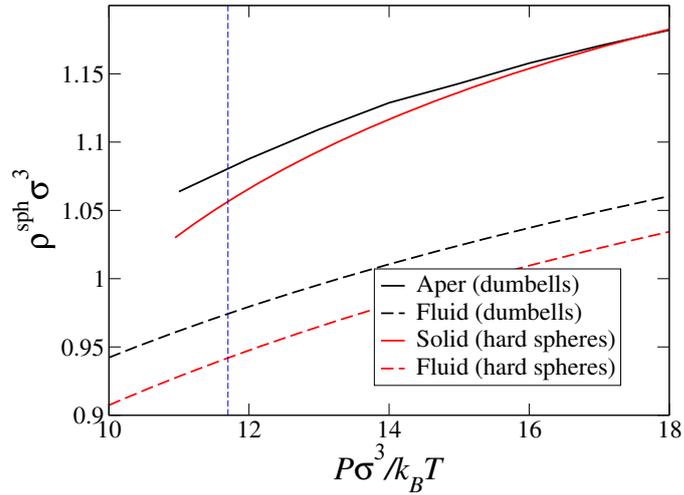

**Figure 5.6**: Equation of state (EOS, i.e., $\beta P \sigma^3$ vs the number density of spheres $\rho^{sph}\sigma^3$ for a system of hard spheres and hard dumbbells with $L^* = 1.0$. For the fluid and solid phase of hard spheres, the Carnahan-Starling [97] and Speedy [44] EOS are plotted. The EOS of hard dumbbells for the fluid phase is obtained from Ref. [98]. The dashed vertical line denotes the bulk coexistence pressure of hard dumbbells with $L^* = 1.0$.

Moreover, we also performed EDMD simulations for the spontaneous nucleation of the aperiodic crystal phase of dumbbells at $P^* = 17$ in system of $N = 16000$ hard dumbbells. The number of spheres in the biggest cluster as a function of time from a typical MD simulation is shown in Fig. 5.7. We find that the size of critical nuclei in spontaneous nucleation is around 100 spheres which agrees well with the result obtained from umbrella sampling MC simulations shown in Fig. 5.5. The nucleation rate obtained from spontaneous nucleation events observed in MD simulations is $I\sigma^5/6D_l = 7.3 \times 10^{-8\pm1}$ which agrees very well with the rate obtained from umbrella sampling MC simulations, $I\sigma^5/6D_l = 2.8 \times 10^{-8\pm1}$, within the error bars of one order of magnitude.

Furthermore, we study the effect of aspect ratio on the nucleation of the aperiodic crystal phase, and the free energy barriers for hard dumbbells with aspect ratios $L^* = 0.95, 0.97$, and $1.0$ at supersaturation $\beta|\Delta\mu^{sph}| = 0.43$. We plot $\Delta G(n)$ as a function of cluster size $n$, i.e., the number of spheres in the cluster, in Fig. 5.8. We observe that at the same supersaturation the barrier height decreases upon decreasing the elongation of the dumbbells. According to classical nucleation theory, $\Delta G^* \propto \gamma^3/(\rho_s^{sph}|\Delta\mu^{sph}|)^2$, where $\Delta\mu^{sph}$ is the supersaturation per sphere with $\rho_s^{sph}$ the bulk density of spheres in the solid phase. As shown in Table 5.2, $\rho_s^{sph}\sigma^3$ is very similar for $L^* = 0.95, 0.97$, and $1.0$, and we argue that the interfacial tension of the aperiodic crystal decreases upon decreasing the elongation of the dumbbells. In order to calculate the nucleation rates, we perform EDMD simulations starting from configurations on the top of the free energy barriers. We plot the mean square deviation of the cluster size as a function of time in Fig. 5.9. We find that the attachment rate decreases significantly as the anisotropy of the dumbbells decreases. The resulting nucleation rates in units of the long time diffusion coefficient are shown in Fig. 5.10. We clearly observe that at fixed supersaturation the nucleation rate increases with decreasing dumbbell elongation. However, in the phase diagram of hard dumbbells [83–86], the pressure range where the aperiodic crystal phase is thermodynamically stable shrinks significantly when the aspect ratio decreases. As a result, it is not possible to increase the supersaturation further for shorter dumbbells, although the nucleation rates are already much higher for shorter ones than for longer ones at the same supersaturation.

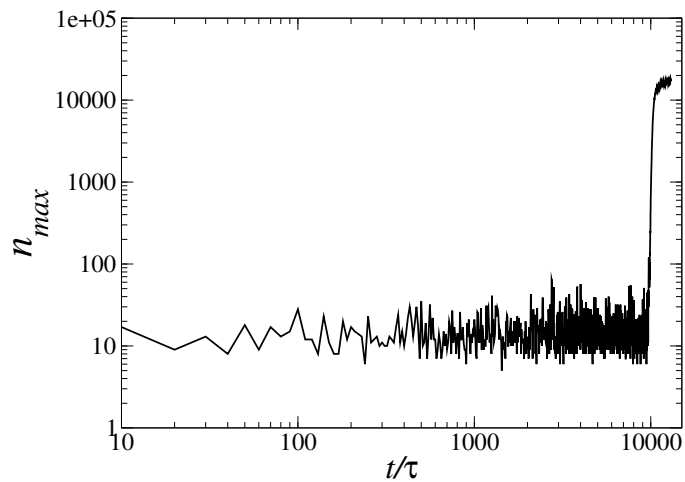

**Figure 5.7**: Number of spheres in the largest cluster $n_{max}$ as a function of time $t/\tau$ for a typical trajectory obtained from EDMD simulations for the aperiodic crystal nucleation of hard dumbbells with $L^* = 1.0$, $N = 16000$ and $P^* = 17$.

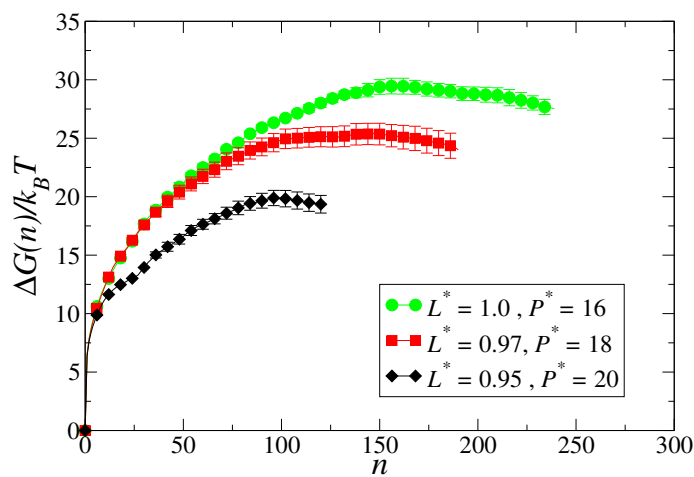

**Figure 5.8**: Gibbs free energy $\Delta G(n)$ as a function of the number of spheres $n$ in the largest crystalline cluster for the aperiodic crystal nucleation of hard dumbbells with $L^* = 0.95, 0.97$, and 1 at supersaturation $\beta|\Delta\mu^{sph}| = 0.43$.

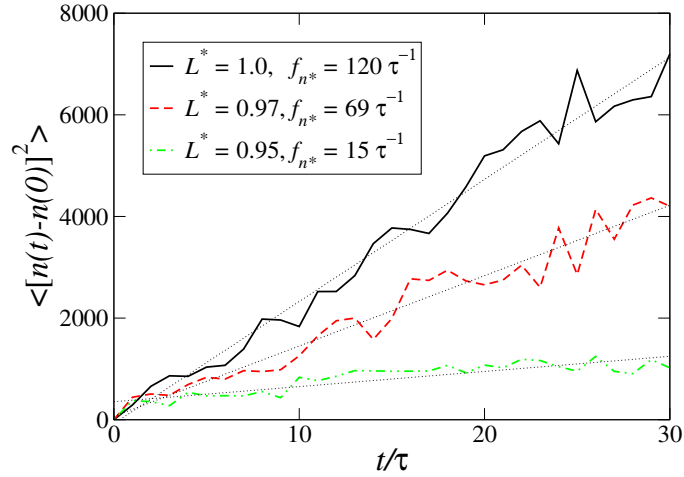

**Figure 5.9**: Mean square deviation of the cluster size $< [n(t) - n(0)]^2 >$ as a function of time $t/\tau$ for hard dumbbells with $L^* = 0.95, 0.97$, and $1.0$ at supersaturation $\beta |\Delta \mu^{sph}| = 0.43$. The resulting attachment rates $f_{n^*}$ are listed in units of $\tau^{-1}$.

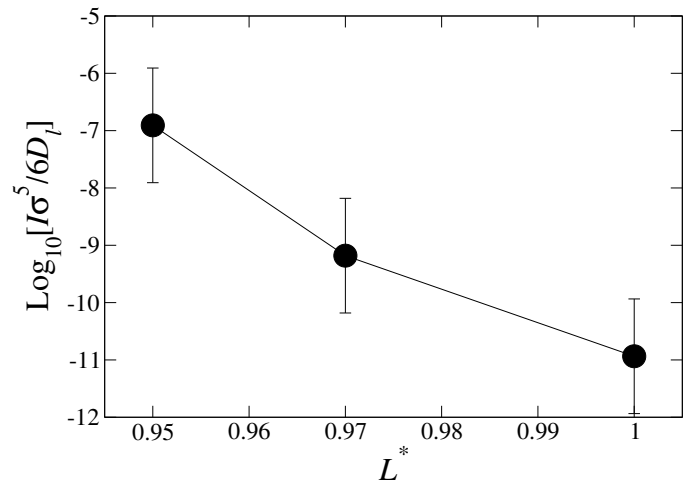

**Figure 5.10**: Nucleation rate $I \sigma^5 / 6 D_l$ for the aperiodic crystal phase as a function of the aspect ratio $L^*$ of hard dumbbells at supersaturation $\beta |\Delta \mu^{sph}| = 0.43$.

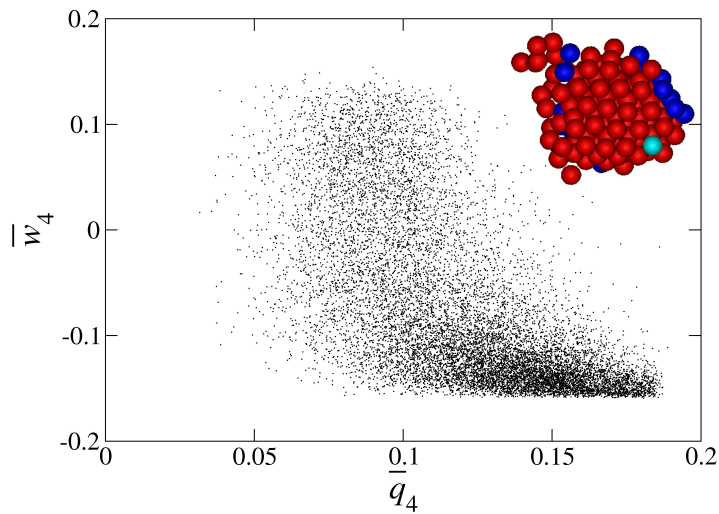

**Figure 5.11**: Distribution of the spheres in the critical nuclei as obtained from umbrella sampling MC simulations in $\bar{q}_4 - \bar{w}_4$ plane in systems of hard dumbbells with $L^* = 1.0$ at supersaturation $\beta|\Delta\mu^{sph}| = 0.43$. Inset: Typical configuration of a critical nucleus. Red denotes fcc-like spheres and blue denotes hcp-like spheres while the light blue are the undetermined ones.

Additionally, we also study the structure of the critical nuclei by calculating the averaged local bond order parameter $\bar{q}_4$ and $\bar{w}_4$, provided the sphere has $N_b(i) \geq 10$ neighbors. The distribution of spheres in the critical nuclei are presented as scatter plots in the $\bar{q}_4 - \bar{w}_4$ plane in Fig. 5.11 for $L^* = 1.0$. We observe only a few spheres with $\bar{w}_4 > 0$ and $\bar{q}_4 < 0.1$, as most of the spheres are in the area of $\bar{w}_4 < 0$ and $\bar{q}_4 > 0.1$, which is very similar to the scatter plots for hard spheres shown in Fig. 5.3a. Consequently, the critical nucleus of the aperiodic crystal phase of hard dumbbells contains also more fcc-like than hcp-like particles, similar to the critical nuclei observed in hard-sphere nucleation [19]. A typical configuration of a critical nucleus is shown in the inset of Fig. 5.11, where the spheres are considered to be fcc-like if $\bar{w}_4 < 0$.

| $L^*$ | $P\sigma^3/k_BT$ | $\beta\|\Delta\mu^{sph}\|$ | $\rho_s^{sph}\sigma^3$ | $n^*$ | $\beta\Delta G(n^*)$ | $\|\beta\Delta G''(n^*)\|$ |
|---|---|---|---|---|---|---|
| 1 | 17 | 0.53 | 1.170 | 115 | $21.4 \pm 0.4$ | $1.2 \times 10^{-3}$ |
| 1 | 16 | 0.43 | 1.158 | 170 | $29.5 \pm 0.6$ | $9.4 \times 10^{-4}$ |
| 0.97 | 18 | 0.43 | 1.171 | 140 | $25.3 \pm 0.9$ | $8.4 \times 10^{-4}$ |
| 0.95 | 20 | 0.43 | 1.182 | 100 | $19.9 \pm 0.7$ | $3.0 \times 10^{-3}$ |

| $L^*$ | $P\sigma^3/k_BT$ | $f_{n^*}/6D_l$ | $D_l\tau/\sigma^2$ | $I\sigma^5/6D_l$ (US) | $I\sigma^5/6D_l$ (MD) |
|---|---|---|---|---|---|
| 1 | 17 | 2813 | 0.0026 | $2.0 \times 10^{-8\pm1}$ | $7.3 \times 10^{-8\pm1}$ |
| 1 | 16 | 5556 | 0.0036 | $1.1 \times 10^{-11\pm1}$ | - |
| 0.97 | 18 | 5228 | 0.0022 | $6.6 \times 10^{-10\pm1}$ | - |
| 0.95 | 20 | 2273 | 0.0011 | $1.2 \times 10^{-7\pm1}$ | - |

**Table 5.2**: Nucleation rates $I\sigma^5/6D_l$ for the nucleation of the aperiodic crystal phase in systems of hard dumbbells with elongation $L^*$, at pressure $P\sigma^3/k_BT$, and supersaturation per sphere $\beta|\Delta\mu^{sph}|$. $\rho_s^{sph}\sigma^3$ is the number density of spheres in the solid phase, $\beta\Delta G(n^*)$ is the barrier height, and $|\beta\Delta G''(n^*)|$ is the second derivative of the Gibbs free energy at the critical nucleus size $n^*$, i.e., the number of spheres in the critical cluster. $f_{n^*}/6D_l$ is the attachment rate in units of the long-time self diffusion coefficient $D_l$.

| $L^*$ | $P^*$ | $\phi$ | $D_l\tau/\sigma^2$ |
|---|---|---|---|
| 0.4 | 34.5 | 0.64 | $1.02 \times 10^{-4}$ |
| 0.5 | 31.2 | 0.63 | $2.47 \times 10^{-4}$ |
| 0.8 | 24.8 | 0.61 | $2.78 \times 10^{-4}$ |

**Table 5.3**: Long-time diffusion coefficients $D_l$ in units of $\sigma^2/\tau$ with $\tau = \sigma\sqrt{m/k_BT}$ for hard dumbbells with elongation $L^*$ at pressure $P^*$, packing fraction $\phi$, and supersaturation $\beta|\Delta\mu| = 1.0$.

## 5.3.3   Slow dynamics of hard dumbbells

The phase diagram of hard dumbbells shows a stable aligned CP1 crystal phase at infinite pressure for all aspect ratios of the dumbbells, and a fluid-CP1 coexistence region for $0.4 \leq L^* \leq 0.8$ [83–86]. The surface tension for the fluid-CP1 interface of hard dumbbells with $L^* = 0.4$ is $\beta\gamma\sigma^2 \simeq 1.8$ [96]. The height of the free energy barrier is given by $\Delta G^* = 16\pi\gamma^3/3(\rho_s|\Delta\mu|)^2$ in CNT. If we assume that the interfacial tension does not change significantly with increasing pressure, we can estimate the free energy barrier height as a function of pressure by integrating the Gibbs-Duhem equation to obtain $|\Delta\mu|$. The barrier height $\Delta G^*$ and the packing fraction $\phi$ for the fluid phase are shown in Fig. 5.12 as a function of the pressure $P^*$. We find that the barrier height $\Delta G^*$ is extremely high, and only becomes less than $50k_BT$ for $P^* > 45$, corresponding to a packing fraction of the fluid phase $\phi > 0.67$. However, if the interfacial tension increases with increasing pressure as shown in Ref. [99], the "actual" height of free energy barrier can become even higher. As a consequence nucleation of the CP1 crystal phase is an extremely rare event.

Additionally, we calculate mean square displacements $\langle\Delta r^2(t)\rangle$ and the second-order orientational correlator $L_2(t) = \langle P_2[\cos(\theta(t))]\rangle$ for a metastable fluid of hard dumbbells with $L^* = 0.4, 0.5$, and 0.8 at supersaturation $\beta|\Delta\mu| = 1.0$ as shown in Fig. 5.13. We find that at a supersaturation $\beta|\Delta\mu| = 1.0$, where the barrier height is still very high, $\Delta G^*/k_B T \sim 170$ for $L^* = 0.4$, the long-time self diffusion coefficients $D_l \simeq 10^{-4}\sigma^2/\tau$ obtained from $\langle\Delta r^2(t)\rangle$ is extremely small, see Table 5.3), whereas $L_2(t)$ exhibits slow relaxation. Our findings are consistent with predictions obtained from mode-coupling theory for a liquid-glass transition, in which the structural arrest is due to steric hindrance for both translational and reorientational motion [100–104]. Moreover, mode-coupling theory predicts that the steric hindrance for reorientations becomes stronger with increasing elongation, which is consistent with our results for $L_2(t)$ in Fig. 5.13 [100–104]. Increasing the supersaturation will lower $\Delta G^*$, but $D_l$ will decrease as well, while at lower supersaturation the barrier height will only increase. As a result, the nucleation of CP1 phase of hard dumbbells is severely hindered by a high free energy barrier at low supersaturations and slow dynamics at high supersaturations, which explains why the CP1 phase of colloidal hard dumbbells has never been observed in experiments [90] or in direct simulations. It is worth noting that the phase diagram might also display (meta)stable CP2 and CP3 close-packed crystal structures [83], which only differ in the way the hexagonally packed dumbbell layers are stacked. As the free energy difference for the three close-packed structures is extremely small, we expect the surface tensions and the nucleation barrier height to be very similar. Hence, we expect that also the nucleation of the CP2 and CP3 phases are hindered by either a high free energy barrier or slow dynamics.

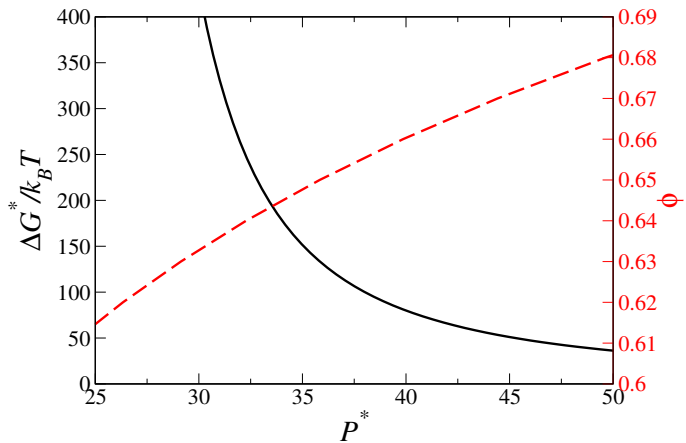

**Figure 5.12**: Estimated height of the Gibbs free energy barrier $\Delta G^*/k_B T$ obtained from classical nucleation theory (solid line) and the packing fraction $\phi$ in the supersaturated fluid phase [98] (dashed line) as a function of pressure $P^*$ for the nucleation of the CP1 phase of hard dumbbells with $L^* = 0.4$.

## 5.4 Conclusions

In conclusion, we investigated the homogeneous nucleation of the plastic crystal, aperiodic crystal and CP1 crystal phase of hard dumbbells using computer simulations. Hard dumbbells serve as a model system for colloidal dumbbells for which the self-assembly is mainly determined by excluded volume interactions. For charged colloidal dumbbells or diatomic molecules, screened Coulombic interactions and Van der Waals interactions may significantly change the kinetic pathways for nucleation. For instance, crystal nucleation of hard rods proceeds via multi-layered crystalline nuclei whereas attractive depletion interactions between the rods in a polymer solutions favor the nucleation of single-layered nuclei [92, 105]. For the nucleation of the plastic crystal phase of hard dumbbells, we found that at low supersaturations the free energy barriers increases slightly with increasing dumbbell anisotropy, which can be explained by a small increase in surface tension for more anisotropic dumbbells [96]. When the supersaturation increases, the barrier height decreases with increasing dumbbell aspect ratio, which can only be explained by a different pressure-dependence of the interfacial tension for hard dumbbells with different aspect ratios. Although the nucleation rate for the plastic crystal phase does not vary much with aspect ratio, the dynamics do decrease significantly. We also carried out EDMD simulations and compared the nucleation rates obtained from spontaneous nucleation events with those obtained from the umbrella sampling Monte Carlo simulations, and found good agreement within the error bars of one order of magnitude. Additionally, we investigated the structure of the critical nuclei of the plastic crystal phase of hard dumbbells with various aspect ratios. We found that the nuclei of the plastic crystal tend to include more fcc-like particles rather than hcp-like ones, which is similar to the critical nuclei of hard spheres [19]. However, the amount of hcp-like particles increases with increasing dumbbell aspect ratio, which agrees with the free energy calculations [86] where it has been shown that the hcp structure is more stable than fcc structure for $L^* \geq 0.15$.

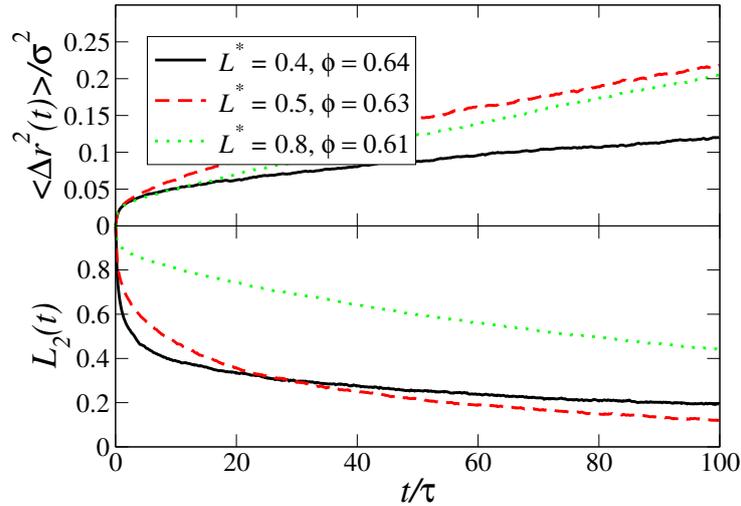

**Figure 5.13**: Mean square displacements $\langle \Delta r^2(t) \rangle$ and $L_2(t) = \langle P_2[\cos(\theta(t))] \rangle$ as a function of time for a fluid of hard dumbbells with an aspect ratio $L^* = 0.4, 0.5$, and $0.8$ at supersaturation $\beta|\Delta\mu| = 1.0$.

Moreover, we also studied the nucleation of the aperiodic crystal phase of hard dumb-bells, and our results showed that at the same pressure, the nucleation barrier of the aperiodic crystal phase of hard dumbbells with $L^* = 1.0$ is slightly higher than that of hard spheres which is mostly due to a small difference in supersaturation $\beta|\Delta\mu^{sph}|$. We also performed EDMD simulations for the spontaneous nucleation of the aperiodic crystal from hard-dumbbell fluid phase, and we found that the nucleation rate obtained from spontaneous nucleation agrees very well with the one obtained from umbrella sampling MC simulations. Furthermore, we studied the effect of aspect ratio on the nucleation of the aperiodic crystal phase, and found that at the same supersaturation, the nucleation rate in units of long-time self diffusion coefficients increases for shorter hard dumbbells. However, when the aspect ratio of dumbbells decreases, the pressure range where the aperiodic crystal phase is stable becomes smaller. Additionally, we also found that the structure of the critical nuclei of the aperiodic crystal phase formed by hard dumbbells with $L^* = 1.0$ is very similar to that of hard spheres which tend to have more fcc-like particles rather than hcp-like ones.

Finally, we estimated the height of the free energy barrier for the nucleation of the CP1 crystal phase of hard dumbbells according to classical nucleation theory, which turns out to be extremely high in the normal pressure range due to a high interfacial tension. Furthermore, we calculated the long-time self diffusion coefficients for hard dumbbells at a moderate supersaturation, i.e., $\beta|\Delta\mu| = 1.0$, which appears to be very small. As a result, we conclude that the high free energy barrier as well as the slow dynamics suppress significantly the nucleation of CP1 phase.

## 5.5 Acknowledgments

We thank Dr. M. Marechal for offering the equation of state for the crystal structures of hard dumbbell particles and fruitful discussions.

# 6

## Glassy dynamics, spinodal fluctuations, and the kinetic limit of nucleation in suspensions of colloidal hard rods

Using simulations we identify three dynamic regimes in supersaturated isotropic fluid states of short hard rods: (i) for moderate supersaturations we observe nucleation of multi-layered crystalline clusters; (ii) at higher supersaturation, we find nucleation of small crystallites which arrange into long-lived locally favored structures that get kinetically arrested, while (iii) at even higher supersaturation the dynamic arrest is due to the conventional cage-trapping glass transition. For longer rods we find that the formation of the (stable) smectic phase out of a supersaturated isotropic state is strongly suppressed by an isotropic-nematic spinodal instability that causes huge spinodal-like orientation fluctuations with nematic clusters diverging in size. Our results show that glassy dynamics and spinodal instabilities set kinetic limits to nucleation in a highly supersaturated fluid.

## 6.1   Introduction

Nucleation is the process whereby a thermodynamically metastable state evolves into a stable one, via the spontaneous formation of a droplet of the stable phase. According to classical nucleation theory (CNT), the Gibbs free energy associated with the formation of a spherical cluster of the stable phase with radius $R$ in the metastable phase is given by a volume term, which represents the driving force to form the new phase, and a surface free energy cost to create an interface, i.e., $\Delta G = -4\pi R^3 \rho |\Delta\mu|/3 + 4\pi R^2 \gamma$ with $\gamma$ the surface tension between the coexisting phases, $\rho$ the density of the cluster, and $|\Delta\mu| > 0$ the chemical potential difference between the metastable and stable phase. For a given $|\Delta\mu|$ and $\rho$, CNT predicts a nucleation barrier $\Delta G_{crit} = (16\pi/3)\gamma^3/(\rho|\Delta\mu|)^2$ and a critical nucleus radius $R_{crit} = 2\gamma/\rho|\Delta\mu|$. CNT predicts an infinite barrier at bulk coexistence ($\Delta\mu = 0$), which decreases with increasing supersaturation. However, CNT incorrectly predicts a *finite* barrier at the spinodal, whereas a non-classical approach yields a vanishing barrier at the spinodal, with a diffuse critical nucleus that becomes of infinite size [106, 107]. Both approaches explain why liquids must be supercooled substantially before nucleation occurs, and one might expect that nucleation should always occur for sufficiently high supersaturation. For deep quenches close to the spinodal, but not beyond it, simulation studies show either nucleating anisotropic and diffuse clusters [108], or precritical clusters that grow further [109] or that coalesce in ramified structures [110]. These results contrast the mean-field predictions that the critical size should diverge at the spinodal [106, 107]. On the other hand, Wedekind *et al.* showed that the system can become unstable by a so-called kinetic spinodal, where the largest cluster in the system has a vanishing barrier, i.e. $\Delta G_{crit}^{large} = 0$, implying the immediate formation of a critical cluster in the system [111]. Beyond this kinetic limit, which is system-size dependent as $\Delta G_{crit}^{large} = \Delta G_{crit} - k_B T \ln N$, the system is kinetically unstable, and the phase transformation proceeds immediately via growth of the largest cluster. Here $N$ is the number of particles, $k_B$ the Boltzmann constant, and $T$ the temperature. This scenario also explains why it is hard to reach the thermodynamic spinodal and why a divergence of the critical cluster size was never observed in simulations, as the system already becomes kinetically unstable at much lower supersaturations. Interestingly, recent simulations of silica also showed a kinetic limit of the homogeneous nucleation regime that is strongly influenced by glassy dynamics, without any spinodal effects [112]. Clearly, the nucleation kinetics at high supersaturation is still poorly understood.

In this chapter, we investigate not only the nucleation pathways of the isotropic-crystal (IX) transition of rod-like particles as a function of supersaturation, but also those of the isotropic-smectic (ISm) transition. The nucleation pathways of structures with both orientational and positional order are still unknown, as nucleating smectic or crystalline clusters have never been observed in experiments or simulations [91, 105] *. We show for the first time that crystal nucleation proceeds via nucleation of multi-layer crystalline clusters, while previous studies found that nucleation is hampered by self-poisoning [91]. Additionally, we identify two mechanisms of dynamic arrest that sets a kinetic limit of the crystal nucleation regime, one based on dynamic arrest of small crystalline nuclei

---

*Very recently, Kuijk *et al.* observed the crystal nucleation of hard rods in experiments [113].

that form locally favored structures, and one based on a conventional cage-trapping glass
transition. Moreover, for longer rods we show that the (metastable) isotropic-nematic
(IN) spinodal severely hinders and even prevents ISm nucleation.

## 6.2 Crystal nucleation of colloidal short rods ($L/\sigma = 2.0$)

We consider a suspension of $N$ hard spherocylinders with a diameter $\sigma$ and a cylindrical
segment of length $L = 2\sigma$ in a volume $V$ or at pressure $P$. The equilibrium bulk phase
diagram of these rods with a length-to-diameter ratio $L^* = L/\sigma = 2$ is well known as
shown in Fig. 6.1 [114]; it features the IX phase transition at pressure $P^* = \beta P \sigma^3 = 5.64$
with $\beta = 1/k_B T$.

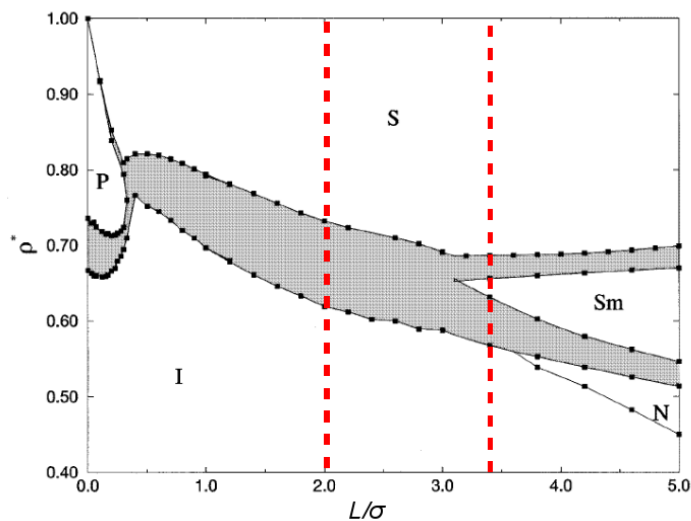

**Figure 6.1**: Phase diagram in the $\rho^*$ (density) versus $L/\sigma$ (aspect ratio) representation from
Ref. [114], where I, P, N, Sm, S denote isotropic, plastic crystal, nematic, smectic and crystal
phase respectively.

We first use *NPT*-Monte Carlo (MC) simulations to compress an isotropic fluid of
10,000 rods at the moderate pressure $P^* = 7.6$ corresponding to a chemical potential
difference $\beta|\Delta\mu| = 1.11$ between the (metastable) fluid and the crystal phase. We then
take random MC configurations as initial configurations for molecular dynamics (MD)
simulations in the *NVT* ensemble to study spontaneous crystal nucleation, employing the
cluster criterion as described in Ref. [115, 116]. We find spontaneous nucleation of a multi-
layered crystalline cluster in the isotropic fluid. Fig. 6.2 shows the time evolution from a
typical MD trajectory. In the initial stage of the MD simulation the system remains in the
metastable isotropic fluid for a long time, with small multi-layered crystalline clusters
appearing and disappearing along the simulation. After time $t = 1000\tau$, with time unit
$\tau = \sigma\sqrt{m/k_B T}$ and $m$ the mass of the particle, a nucleus consisting of multiple crystalline

layers starts to grow gradually until the whole system has been transformed into the bulk crystal phase. We note that the cluster prefers to grow laterally as was also found for attractive rods [105]. We observed similar spontaneous nucleation at $P^* = 7.4$. The long waiting time $t_w$ before a postcritical cluster starts to appear by a spontaneous fluctuation is typical for nucleation and growth. We calculate the nucleation rate $I = 1/\langle t_w \rangle V$, and find from our MD simulations that $I = 5 \times 10^{-9\pm2} \tau^{-1} \sigma^{-3}$ and $1.7 \times 10^{-8\pm1} \tau^{-1} \sigma^{-3}$, for $P^* = 7.4$ and 7.6, respectively.

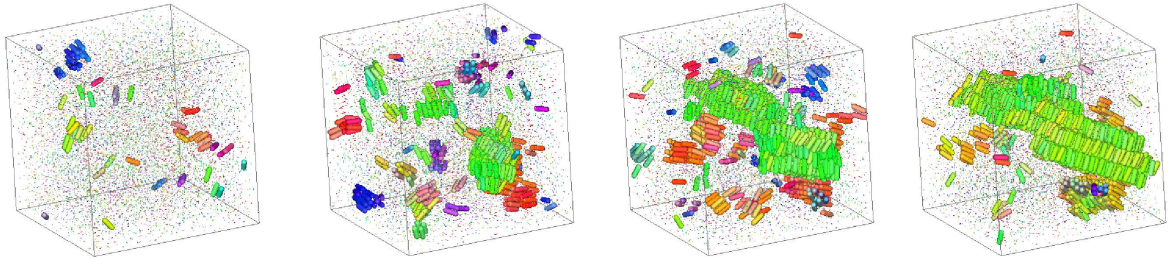

**Figure 6.2**: Configurations for spontaneous crystal nucleation from a typical molecular dynamics trajectory at $P^* = 7.6$ and $t/\tau = 0, 1000, 2000$ and $3000$ (from left to right) with $\tau = \sigma \sqrt{m/k_B T}$ and $m$ the mass of the particle. Isotropic-like particles are drawn 10 times smaller than their actual size. A movie can be found in [116].

As our MD simulations provide evidence that the IX transformation can occur via nucleation of multilayer crystalline clusters, we determine the nucleation barrier using umbrella sampling (US) in MC simulations. We bias the system to configurations with a certain cluster size and we sample the equilibrium probability $P(n)$ to find a cluster of $n$ rods. The Gibbs free energy of a cluster of size $n$ is then given by $\beta \Delta G(n) = -\ln P(n)$. We perform MC simulations of 2000 particles at $P^* = 7.0, 7.2$, and 7.4 corresponding to $\beta |\Delta \mu| = 0.78, 0.89$ and $1.0$, respectively. Fig. 6.3 shows $\Delta G(n)$, which for $P^* = 7.2$ and 7.4 display a maximum of $\beta \Delta G_{crit} \approx 27 \pm 1.5$ and $20 \pm 1.5$ at critical cluster sizes $n_{crit} \approx 140$ and $80$, respectively. A typical configuration of the critical cluster, consisting of three crystalline layers at $P^* = 7.4$, is shown in the inset of Fig. 6.3; its structure agrees with those observed in our MD simulations of spontaneous nucleation of multilayer crystallites. For $P^* = 7.0$ the free-energy barrier is too high to be calculated in our simulations as the cluster starts to percolate the simulation box before the top is reached. For even lower pressures, i.e. $P^* = 6.0$ (not shown), this problem is even more severe. For clusters up to $n \simeq 100$, however, the barrier can be calculated with the US scheme, revealing multilayered structures very similar to the one shown for $P^* = 7.4$. Our MC simulation results for $P^* = 7.2$ and 7.4 can also be used to calculate the nucleation rate from $I = \kappa \exp(-\beta \Delta G_{crit})$ with kinetic prefactor $\kappa = |\beta \Delta G''_{crit}/(2\pi)|^{1/2} \rho_I f_{n_{crit}}$, where $\rho_I$ is the number density of the isotropic fluid and $f_{n_{crit}}$ is the attachment rate of particles to the critical cluster (which we compute using MD simulations starting with independent configurations at the top of the nucleation barrier [18]). For $P^* = 7.2$ and 7.4 we find $I = 1 \times 10^{-13\pm1}$ and $2 \times 10^{-10\pm1} \tau^{-1} \sigma^{-3}$, respectively, in agreement within the errorbars with the MD simulations.

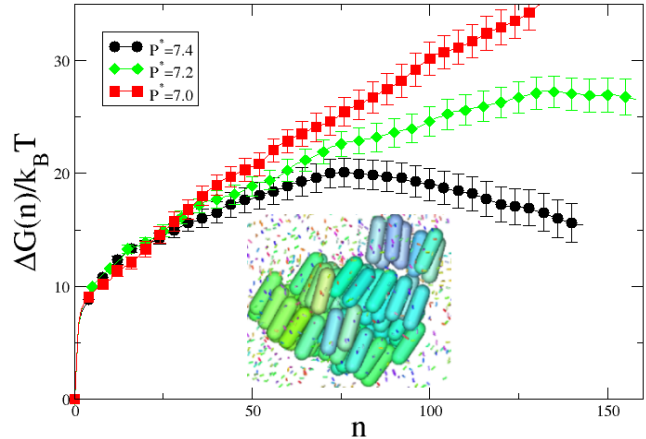

**Figure 6.3**: Gibbs free energy $\Delta G(n)$ as a function of the number of rods $n$ in the crystalline cluster at pressure $P^* = 7.0, 7.2,$ and $7.4$. Inset: A typical configuration of a critical cluster ($n = 81$) at $P^* = 7.4$.

Our observation of spontaneous nucleation of bulk crystals of short rods is in marked contrast with an earlier study, which showed that the free energy never crosses a nucleation barrier [91]. These simulations showed the formation of a single crystalline layer, while subsequent crystal growth is hampered. The authors attribute the stunted growth of this monolayer to self-poisoning by rods that lie flat on the cluster surface. If we use the same cluster criterion as in Ref. [91] for the biasing potential, we indeed also find crystalline monolayers at $P^* = 7.4$, which cannot grow further as $\Delta G(n)$ increases monotonically with $n$. These results for the nucleation barrier agree with theoretical predictions that for sufficiently low supersaturations $\Delta G(n)$ for a single layer is always positive, while multilayer crystalline clusters can grow spontaneously when the nucleus exceeds the critical size [117]. However, our detailed check [116] of the order parameter in Ref. [91] actually reveals a strong (unwanted) bias to form single-layered clusters in US simulations.

We also study the IX transformation at higher oversaturation. To this end, we compress 1000 rods ($L^* = 2$) in $NPT$-MC simulations at $P^* = 8$ ($\beta|\Delta\mu| = 1.33$). Using $\beta\gamma\sigma^2 \simeq 0.44$, which follows from fitting the two barriers of Fig. 6.3 to CNT, we estimate barriers as low as $\beta\Delta G_{crit} \sim 9$ and $\beta\Delta G_{crit}^{large} \sim 2$ for $P^* = 8$. Indeed, many small crystallites nucleate immediately after the compression quench, indicative of the proximity of a kinetic spinodal. These crystallites are oriented in different directions, and have a large tendency to align perpendicular to each other. The subsequent equilibration is extremely slow, since the growth of a single crystal evolves via collective re-arrangements of smaller clusters that subsequently coalesce. In fact, after $3 \times 10^7$ MC cycles, our system is dynamically arrested. Interestingly, Frank proposed more than 50 years ago that dynamic arrest may be attributed to the formation of *locally favored structures* in which the system gets kinetically trapped in local potential-energy minima [118], while direct observation of such a mechanism for dynamic arrest was only recently reported in the gel phase of a colloid-polymer mixture [119]. In our simulations, we clearly observe the formation of long-lived locally favored structures consisting of perpendicularly oriented crystallites. Only via cooperative rearrangements (rotation of the whole cluster) the system can escape from the kinetic traps, but these events are rare in MC simulations. So despite the large supersaturation and the low barrier as predicted by CNT, the actual formation of a single crystal is impeded dramatically by slow dynamics. In fact, our observations agree with experiments on soft-repulsive selenium rods, where transient structures of 5-10 aligned particles tend to form locally favored structures with perpendicularly aligned clusters, which gradually merge into larger clusters [120]. Only attractive $\beta$-FeOOH rods form crystalline monolayers in agreement with [105].

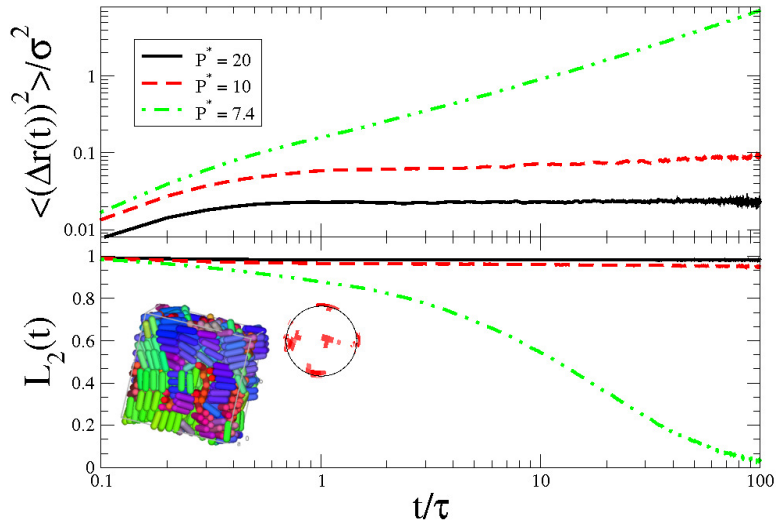

**Figure 6.4**: Mean square displacement $\langle(\Delta \mathbf{r}(t))^2\rangle$ and second-order orientational correlator $L_2(t)$ for hard rods with $L^* = 2$ and pressures as labeled. The inset shows a typical configuration of a glassy state with cubatic order at $P^* = 10$.

In order to investigate whether the system can be quenched *beyond* a thermodynamic spinodal (such that the transformation should proceed via spinodal decomposition), we also perform simulations at $P^* = 10$. We find again the immediate nucleation of many small crystallites, which is expected beyond the kinetic spinodal. As the phase transformation sets in right away, we cannot determine whether the nucleation barrier is finite or zero; it is therefore unclear whether or not we have crossed a thermodynamic spinodal (if there is one for freezing). We note, however, that we did not find any characteristics of spinodal decomposition in the early stages. The small crystallites tend to align perpendicular, and in fact the system displays clear orientational ordering along three perpendicular directions (cubatic order), as shown by the orientation distribution on the surface of a unit sphere in the inset of Fig. 6.4. In order to check for finite size effects, we studied a system of $N = 4000$ rods, which again show long-range cubatic order. However, we cannot make any definite conclusions on the range of the cubatic order due to slow dynamics of larger systems. The mean-square displacement $\langle(\Delta \mathbf{r}(t))^2\rangle$ and the second-order orientational correlator $L_2(t) = \langle(3\cos^2\theta(t) - 1)/2\rangle$ are also displayed in Fig. 6.4, which show the characteristic plateau of structural arrest. For comparison, we also present data
for $P^* = 7.4$, which show relatively fast relaxation of the translational and orientational
degrees of freedom. At an even larger supersaturation, $P^* = 20$, we find that the system
is kinetically arrested immediately after the quench. We find hardly any crystalline or-
der, while the orientation distribution remains isotropic (not shown). Clearly, the system
crossed the conventional cage trapping glass transition [121] that prevents the formation
of any ordering. The dynamic arrest can be appreciated by the plateau in $\langle (\Delta \mathbf{r}(t))^2 \rangle$ and
$L_2(t)$ in Fig. 6.4. Our results thus show that nucleation at high supersaturation is strongly
affected by vitrification, either due to locally favored structures or by the conventional
glass transition, yielding glasses with and without small crystallites, respectively.

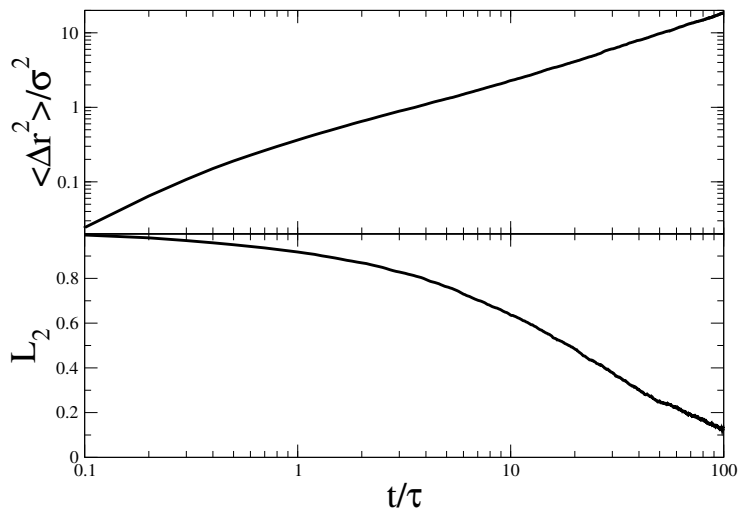

**Figure 6.5**: Mean square displacement $\langle (\Delta \mathbf{r}(t))^2 \rangle$ and second-order orientational correlator
$L_2(t)$ for hard rods with $L^* = 3.4$ at $P* = 3.0$.

## 6.3 Nucleation of smectic phase of colloidal short rods ($L/\sigma = 3.4$)

We also study longer hard rods with $L^* = 3.4$, which show ISm coexistence at $P^* =
2.828$ as shown in Fig. 6.1. A previous MC simulation study [122] indeed showed the
formation of the smectic phase out of the highly supersaturated I phase at $P^* = 3.1$
via spinodal decomposition. However, nucleation and growth of the smectic phase out
of weakly supersaturated I phases at $P^* = 2.85 - 3.0$ was *not* observed [122]. As strong
pre-smectic ordering and huge nematic-like clusters were observed in the isotropic fluid
phase, the hampered nucleation was attributed to slow dynamics. Here we reinvestigate
the regime $P^* = 2.828 - 3.0$ at much longer time scales by MD simulations. We confirm
the earlier findings as regards the structure, but did not find any evidence for structural
arrest in $\langle (\Delta \mathbf{r}(t))^2 \rangle$ and $L_2(t)$ as shown in Fig. 6.5. Instead we find huge and strongly
fluctuating nematic-like clusters [116]. The nematic character of the clusters is evident from the structure factor $S(k)$ and orientational structure factor $S_{or}(k)$, shown in Fig. 6.6a, revealing a small-$k$ divergence for $S_{or}(k)$ but not for $S(k)$ [121]. The correlation length $\xi$ of the orientational fluctuations obtained from fitting the orientational correlation function $g_{or}(r) \sim \exp[-r/\xi]/r$ is shown in Fig. 6.6b to satisfy a power law $\xi \sim |P - P_c|^{-\nu}$ with $P_c^* = 3.01$ the alleged IN spinodal pressure and $\nu = 0.47$ as shown in Fig. 6.7a, which is close to the expected mean-field exponent $\nu = 1/2$ of the IN-spinodal [123]. Apparently, the ISm nucleation is prevented, or severely slowed down, by an intervening metastable IN spinodal. Our observation that the metastable isotropic fluid is more susceptible to nematic than to smectic fluctuations is corroborated by second-virial calculations of the Zwanzig model of block-like $H \times D \times D$ rods with three orthogonal orientations [124].

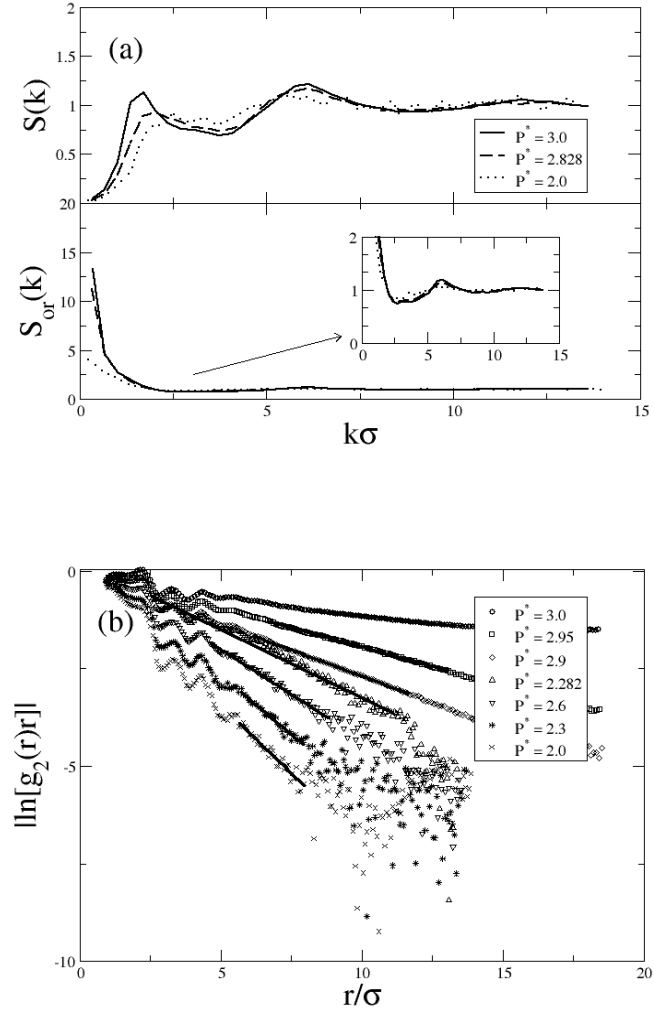

**Figure 6.6**: (a) Positional (top) and orientational (bottom) structure factor of hard rods with $L/\sigma = 3.4$ at $P^* = 2, 2.828$ and 3. (b) Orientational correlation $g_2(r) = \langle P_2[\cos(\theta(r))]\rangle$, where $P_2$ is the second order Legendre polynomial, as a function of distance $r$ for the systems of hard rods with $L/\sigma = 3.4$ at various pressures. The solid line represents the fit $\sim \exp(-r/\xi)/r$ with $\xi$ the correlation length.

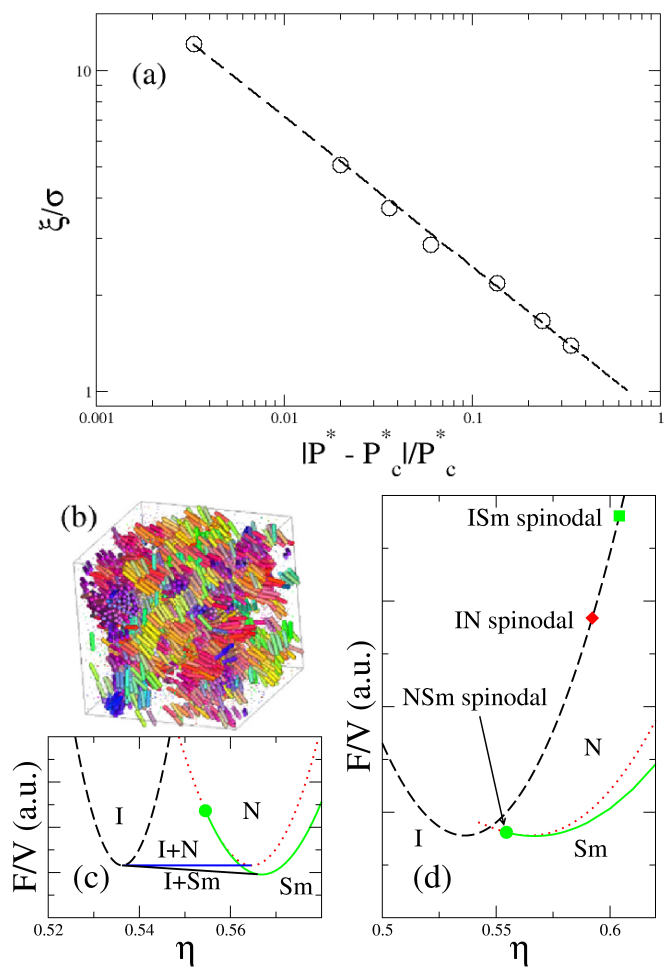

**Figure 6.7**: (a) Correlation length as a function of $|P^* - P_c^*|/P_c^*$ where $P_c^* = 3.01$ is the nematic spinodal pressure and the line is a power-law fit with critical exponent $\nu = 0.47$ which agrees well with the prediction obtained from Landau theory ($\nu = 1/2$) [123]. (b) Typical snapshot of rods with $L/\sigma = 3.4$ at $P^* = 3.0$. (c,d) Helmholtz free energy density of Zwanzig rods in the I, N, and Sm phases from a numerical minimization of the second-virial free-energy functional. Common-tangent constructions (black and blue lines) and bifurcation analyses (symbols) reveal equilibrium ISm coexistence and metastable IN coexistence and IN and ISm spinodal instabilities on the supersaturated isotropic free-energy branch. A movie is shown in Ref. [116]. Isotropic-like particles are drawn 10 times smaller than their actual size.

The dimensionless Helmholtz free-energy density $f(\eta)$ of the I, N, and Sm phase for $H/D = 4.3$, shown in Fig. 6.7, reveal equilibrium ISm coexistence, and a metastable N branch. Moreover, the IN spinodal on the metastable isotropic branch occurs at a lower packing fraction $\eta$ than the ISm spinodal. In other words, the isotropic fluid is predicted to exhibit spinodal nematic fluctuations upon increasing the supersaturation, consistent with the diverging $\xi$ as observed in our simulations. One might have expected that the presence of these orientationally ordered nematic clusters facilitate the formation of the smectic phase. However, although we do find some layering of the rods, the density within these nematic clusters is too low and the orientational fluctuations change too rapidly to form the smectic layers.

## 6.4 Conclusions

In conclusion, our results show that nucleation of short hard rods from a supersaturated isotropic fluid phase to orientationally and positionally ordered crystal and smectic phases is much more rare than perhaps naively anticipated. Only for very short rods and moderate supersaturations, we find, for the first time, nucleation of multi-layered crystals; at higher supersaturations we identified two mechanism for dynamic arrest. The first one occurs close to the kinetic spinodal, and is such that (locally favored) crystalline clusters appear immediately after the quench, followed by extremely slow dynamics due to geometric constraints of these tightly packed clusters. The second type of dynamic arrest occurs at very high supersaturation and is due to the conventional cage-trapping glass transition. For these very short rods ($L^* = 2$) we have thus identified a competition between the nucleation of multi-layered crystals, the vitrification of small crystallites, and the formation of a cage-trapped amorphous glass state. In the supersaturated isotropic state of slightly longer rods ($L^* = 3.4$), the nucleation of the (equilibrium) smectic phase is found to be hampered by huge nematic fluctuations due to the existence of a *metastable* IN spinodal instability. In fact, we showed for the first time that for quenches close to a spinodal the clusters diverge in size.

Our findings are of fundamental and practical interest. They provide strong evidence for a local structural mechanism for dynamic arrest in a system with orientational and positional degrees of freedom. They also explain why the self-organization of ordered assemblies of nanorods is difficult and why most of the nanorod self-assembly techniques require additional alignment of the rods by applied electric fields, fluid flow, or substrates in order to facilitate the formation of the desired self-assembled structures [125]. Our simulations show that this additional "steering" is required since the spontaneous nucleation of the rods is strongly affected by glassy dynamics and spinodal instabilities.

## 6.5 Acknowledgments

I would like to thank S. Belli for doing the theoretical calculations on Zwanzig model and Dr. F. Smallenburg for fruitful discussions.

# 7

## Effect of bond connectivity on crystal nucleation of hard polymeric chains

We study the spontaneous nucleation and crystallization of linear and cyclic chains of flexibly connected hard spheres using extensive molecular dynamics simulations. To this end, we present a novel event-driven molecular dynamics simulation method, which is easy to implement and very efficient. We find that the nucleation rates are predominately determined by the number of bonds per sphere in the system, rather than the precise details of the chain topology, chain length, and polymer composition. Our results thus show that the crystal nucleation rate of bead chains can be enhanced by adding monomers to the system. In addition, we find that the resulting crystal nuclei contain significantly more face-centered-cubic than hexagonal-close-packed ordered particles. More surprisingly, the resulting crystal nuclei possess a range of crystal morphologies including structures with a five-fold symmetry.

## 7.1  Introduction

Although crystal nucleation from a supersaturated fluid is one of the most fundamental processes during solidification, the mechanism is still far from being well understood. Even in a relatively simple model system of pure hard spheres, the nucleation rates obtained from Monte Carlo (MC) simulations using the umbrella sampling technique differ by more than 6 orders of magnitude from those measured in experiments [18]. This discrepancy in the nucleation rates has led to intense ongoing debates in the past decade on the reliability of various techniques as employed in simulations and experiments to obtain the nucleation rates [19, 20]. Recently, it was shown that the theoretical prediction of the nucleation rates for hard spheres is consistent for three widely used rare-event techniques, e.g., forward flux sampling, umbrella sampling and brute force molecular dynamics simulations, despite the fact that the methods treat the dynamics very differently [19]. Moreover, the structure of the resulting crystal nuclei as obtained from the different simulation techniques all agreed and showed that the nuclei consist of approximately 80% face-centered-cubic-like particles. The predominance of face-centered-cubic-like particles in the critical nuclei is unexpected, as the free energy difference between the bulk face-centered-cubic (fcc) and hexagonal-close-packed (hcp) phases is about 0.001 $k_B T$ per particle, and one would thus expect to find a random-hexagonal-close-packed (rhcp) crystal phase [65]. More surprisingly, simulation studies showed that the subsequent growth of these critical nuclei resulted in a range of crystal morphologies with a predominance of multiply twinned structures exhibiting in some cases structures with a five-fold symmetry [126, 127]. Such structures are intriguing as the fivefold symmetry is incompatible with space-filling periodic crystals. Moreover, the formation mechanism of these fivefold structures is still unknown. Bagley speculated that the fivefold structures are due to the growth of fivefold local structures (a decahedral or pentagonal dipyramid cluster of spheres) [128] that are already present in the supersaturated fluid phase [129]. Another mechanism that has been proposed is that these multiple twinned structures with a fivefold symmetry originates from nucleated fcc domains that are bound together by stacking faults [128]. For instance, five tetrahedral fcc domains can form a cyclic multiple twinned structure with a pentagonal pyramid shape. A recent event-driven Molecular Dynamics simulation study on hard spheres showed, however, no correlation between the fivefold local clusters that are already present in the supersaturated fluid and the multiple twinned structures in the final crystal phase [127]. Hence, it was concluded that crystalline phases with multiple stacking directions may possess fivefold structures, whereas crystals with a unique stacking direction do not show any five-fold symmetry patterns. These authors also showed using Monte Carlo simulations that hard-sphere chains never formed crystalline structures with a fivefold symmetry, and hence, they argued that chain connectivity prohibits the formation of twinned structures and forces the crystals to grow in a single stacking direction [127, 130]. However, several unphysical MC moves had to be introduced to study polymer crystallization, which may affect the chain dynamics and the resulting crystal morphologies.

In this chapter, we present an event driven molecular dynamics (EDMD) scheme that is easy to implement and mimics closer the actual dynamics in these polymer systems. We study the effect of bond connectivity on the nucleation rates and crystal morphology of flexibly connected hard spheres. These hard-sphere chains can serve as a simple model
for polymeric systems and a better understanding of the behavior of these bead chains
may shed light on the glass transition and crystallization of polymers. In fact, it has been
shown recently that random packings of granular ball chains show striking similarities
with the glass transition in polymers [131]. We also mention that recently, a colloidal
model system of bead chains has been realized consisting of colloidal spheres that are
bound together with "flexible linkers" [132].

## 7.2 Model

We consider a system of $M$ polymer chains consisting of $N$ identical hard spheres with
diameter $\sigma$ in a volume $V$. In addition, the hard-sphere beads are connected by flexible
bonds with a bond energy $U_{bond}(r_{ij})$ given by

$$\frac{U_{bond}(r_{ij})}{k_B T} = \begin{cases} 0 & \sigma < r_{ij} < \sigma + \delta \\ \infty & \text{otherwise} \end{cases} \tag{7.1}$$

where $r_{ij}$ is the center-to-center distance between two connected spherical beads $i$ and
$j$, $\delta$ is the maximum bond length, $k_B$ the Boltzmann constant, and $T$ the temperature.
The maximum bond length $\delta$ varies from 0 to $0.05\sigma$ in our simulations, such that $\delta = 0$
corresponds to a freely jointed chain of tangent hard spheres. Since the pair potentials
between all beads (spheres) are discontinuous, the pair interactions only change when the
beads collide or when the maximum bond length is reached. Hence, the particles move in
straight lines (ballistically) until they encounter another particle or reach the maximum
bond length. The particles then perform an elastic collision. These collisions are identified
and handled in order of occurrence using an EDMD simulation [133].

Using EDMD simulations, it is straightforward to determine the nucleation rate. Start-
ing from an equilibrated fluid phase, an EDMD simulation is employed to evolve the sys-
tem until a spontaneous nucleation event occurs. The nucleation rate is then given by
$I = 1/\langle t \rangle V$, where $\langle t \rangle$ is the average waiting time before an nucleation event occurs in a
system of volume $V$. In order to identify the crystalline clusters in the fluid phase, we em-
ploy the local bond-order parameter analysis as introduced by Steinhardt *et al.* [93, 134].
We define for every spherical bead $i$, a $2l + 1$-dimensional complex vector $\mathbf{q}_l(i)$ given by

$$q_{lm}(i) = \frac{1}{N_b(i)} \sum_{j=1}^{N_b(i)} \Upsilon_{lm}(\hat{\mathbf{r}}_{ij}), \tag{7.2}$$

where $N_b(i)$ is the total number of neighboring particles of particle $i$, $\Upsilon_{lm}(\hat{\mathbf{r}}_{ij})$ are the
spherical harmonics for the normalized distance vector $\hat{\mathbf{r}}_{ij}$ between bead $i$ and $j$, and
$m \in [-l, l]$. Neighbors of particle $i$ are defined as those particles that are within a
given cutoff radius $r_c$ from particle $i$. To determine the correlation between the local
environments of particle $i$ and $j$, we define the rotationally invariant function $d_l(i, j)$
given by

$$d_l(i, j) = \sum_{m=-l}^{l} \tilde{q}_{lm}(i) \cdot \tilde{q}_{lm}^*(j), \tag{7.3}$$

where $\tilde{q}_{lm}(i) = q_{lm}(i)/\sqrt{\sum_{m=-l}^{l}|q_{lm}(i)|^2}$ and the asterisk is the complex conjugate [94]. If $d_l(i,j) > d_c$, the "bond" between sphere $i$ and $j$ is regarded to be solid-like or connected, where $d_c$ is the dot-product cutoff. We identify a sphere as solid-like when it has at least $\xi_c$ solid-like bonds. Previous studies on the crystallization of hard-sphere chains have shown that the beads are located on the lattice positions of a random-hexagonal-close-packed crystal lattice whereas the bonds are randomly oriented [130, 135]. We therefore choose the symmetry index $l = 6$, and we employ $r_c = 1.3\sigma$, $d_c = 0.7$, and $\xi_c = 6$ in our simulations.

## 7.3 Crystal Nucleation of linear and ring polymers

We first perform EDMD simulations to study the spontaneous nucleation of linear hard-sphere chains at a packing fraction $\phi = 0.55$. In order to obtain the initial configuration for our EDMD simulations, we use the Lubachevsky-Stillinger algorithm [136] to grow the particles in the simulation box to the packing fraction of interest with a very fast speed, i.e., $d\sigma(t)/dt = 0.01\sigma/\tau$ where $\sigma(t)$ is the size of spheres at time $t$ with $\sigma$ the target sphere size and $\tau = \sigma\sqrt{m/k_BT}$ the MD time scale. In order to exclude the effect of dynamics on the nucleation rates [19, 134], we first calculate the long-time diffusion coefficient $D_L$, and use the long-time diffusion time, $\tau_L = 1/6D_L$, as the unit of time in the nucleation rate. We found that $D_L$ decreases with increasing $N$, and is rather independent of the bond length $\delta$, at least for the range of values that we studied. Subsequently, we calculate the crystal nucleation rates from EDMD simulations for various linear and ring-like hard-sphere polymers. In Fig. 7.1, we show the nucleation rate for linear hard-sphere polymers with a chain length $N = 1$ (hard spheres), 2 (dumbbells), 3, 5, 10, and 20 and bond length $\delta = 0.05\sigma$. We find that the nucleation rate decreases monotonically with increasing chain length $N$. However, it is remarkable that hard-sphere chains with a maximum bond length $\delta = 0.05\sigma$ can crystallize into a crystal phase of $\phi = 0.55$ which corresponds to an average surface-to-surface distance of about $0.10\sigma$ between the spheres in the fcc phase. We also note that the nucleation rate is very similar for long polymers, i.e., $N \geq 10$. To check this surprising result, we also determined the nucleation rate of a single polymer of length $N = 10^4$, where all the beads are doubly connected except the two end beads. We observe that the system remains in the fluid phase for $\simeq 7000\tau$, before a critical nucleus of $\sim 100$ beads forms in the middle of the chain, which subsequently grows further until the whole system is crystalline. In Fig. 7.2, we show the size of the largest crystalline cluster as a function of simulation time from a typical MD trajectory. Additionally, we find that the nucleation rate does not decrease significantly for $N = 10^4$, which is highly unexpected as the beads can only move collectively.

We also determined the nucleation rate for linear chains with a smaller bond length $\delta = 0.04\sigma$ and chain length $N = 1, 2, 3, 4$, and 5. For longer chains, we did not observe spontaneous nucleation within the simulation times that we considered. We find that the nucleation rate decreases with bond length $\delta$, which is to be expected as bead chains with shorter bond lengths are even more frustrated in an fcc crystal at $\phi = 0.55$. For comparison, we also plot the nucleation rates for dumbbells with a bond length of $\delta = 0.02\sigma$ and $\delta = 0$ [134]. Our results clearly show that the nucleation rate decreases by several orders of magnitude upon decreasing the bond length. Finally, we also determine the nucleation rates for cyclic bead-chains (ring polymers) with bond length $\delta = 0.05\sigma$. Figure 7.1 shows that the nucleation rate of ring polymers is always lower than for linear polymers with the same length. However, the difference in nucleation rate is small for $N \geq 10$. Furthermore, the nucleation rate does not decrease monotonically with chain length $N$ for small ring polymers. For instance, the nucleation rate of rings of 4 beads is an order of magnitude higher than for rings consisting of 5 beads.

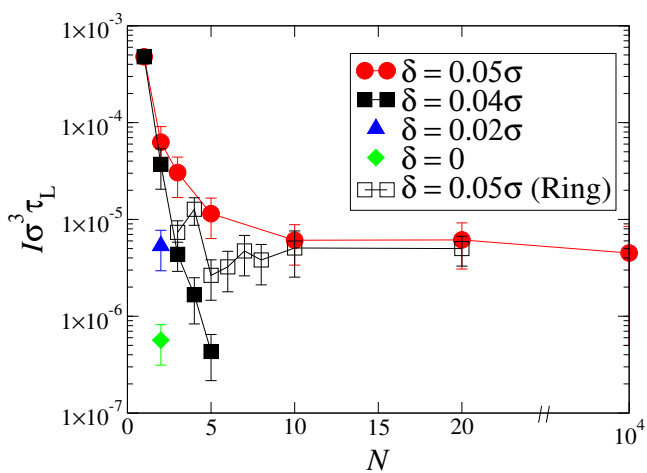

**Figure 7.1**: Nucleation rates $I\sigma^3\tau_L$ of linear hard-sphere polymers with maximum bond length $\delta = 0.05\sigma$ and chain length $N = 1, 2, 3, 5, 10,$ and 20 (solid circles), with $\delta = 0.04\sigma$ and chain length $N = 1, 2, 3$ and 5 (solid squares), and of ring-like hard-sphere polymers with maximum bond length $\delta = 0.05\sigma$ and chain length $N = 3, 4, 5, 6, 7, 8, 10,$ and 20 (open squares). For comparison, we also plot the nucleation rate for hard dumbbells with a maximum bond length $\delta = 0.02\sigma$ (blue triangles) and $\delta = 0$ (diamonds).

Our simulations on linear polymers show that the nucleation rate decreases with chain length $N$. One may argue that the nucleation rate is largely determined by the chain connectivity or the average number of bonds per sphere in the system. In Fig. 3, we plot the nucleation rate for linear bead chains with maximum bond length $\delta = 0.04\sigma$ and chain length $N = 1, 2, 3, 4,$ and 5, which correspond to an average number of bonds per bead of $n_b = (N-1)/N = 0, 1/2, 2/3, 3/4$ and $4/5$. In order to investigate the effect of average number of bonds per bead in the system on the nucleation rate, we also perform simulations for binary mixtures of linear polymers with different chain lengths and the same maximum bond length $\delta = 0.04\sigma$. We consider mixtures of chain length $N = 2$ and 5, $N = 1$ and 6, and $N = 1$ and 10. The composition of the mixture is chosen such that the value of $n_b$ matches with one of the values for the pure systems. We compare the nucleation rates for the pure and binary systems in Fig. 7.3. We indeed observe a nice data collapse, suggesting that the nucleation rate is mainly determined by the number of bonds per sphere in the system, and the frustration imposed by the chain connectivity in these systems.

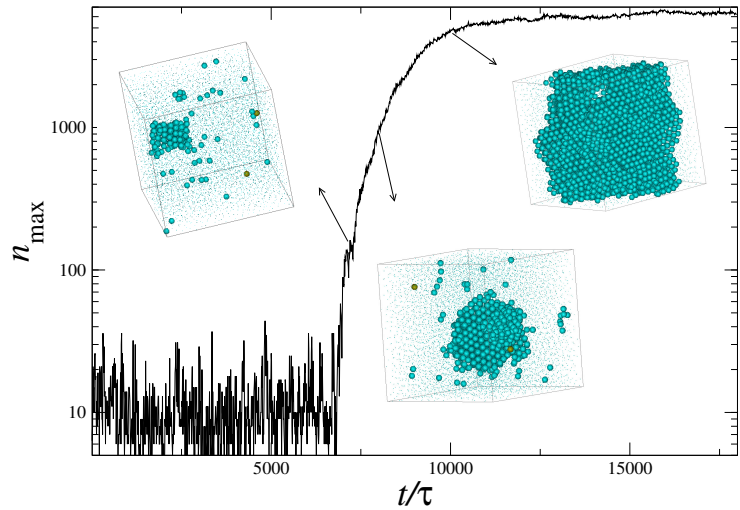

**Figure 7.2:** Size of the largest crystalline cluster $n_{\max}$ as a function of simulation time for a single linear hard-sphere polymer with chain length $N = 10^4$. The insets are snapshots at $t = 7060\tau$, $8000\tau$ and $10000\tau$, respectively, where only the solid-like beads are shown, and the two dark yellow spheres are the two ends of the chain.

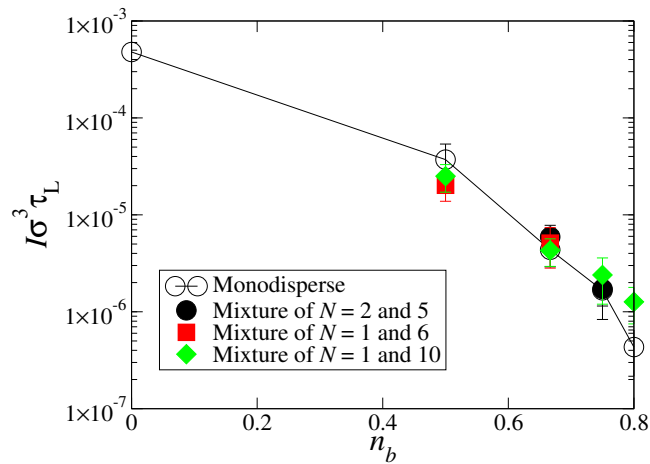

**Figure 7.3:** Nucleation rates $I\sigma^3\tau_L$ of linear hard-sphere polymers with chain length $N = 1, 2, 3,$ and $5$ and of binary mixtures of linear hard-sphere polymers with chain length $N = 2$ and $5$, $N = 1$ and $6$, and $N = 1$ and $10$, as a function of the average number of bonds per bead $n_b$. The composition of the mixture was chosen such that the value for $n_b$ matches with one of the values for the pure systems. The maximum bond length equals $\delta = 0.04\sigma$ for all bead chains.

Additionally, we also investigate the structure of the resulting crystals by calculating the averaged local bond order parameters $\overline{q}_4$ and $\overline{w}_4$ for each sphere $i$ that has $N_b(i) \geq 10$ neighbours. This analysis allows us to check whether a bead is fcc-like or hcp-like [64, 134]. We find that the resulting crystal structures contain more fcc-like than hcp-like particles, which is very similar to the critical nuclei observed in hard-sphere [19] and hard-dumbbell nucleation [134]. In addition, we find that the resulting crystal structures display a range of crystal morphologies including structures with a five-fold symmetry pattern for all polymer systems that we considered, even for ring-like polymers with chain length as small as $N = 3$, which contrasts previous simulations where no five-fold structures were observed [127]. Exemplarily, Fig. 7.4 shows typical configurations of these five-fold symmetry patterns formed by linear hard-sphere chains of length $N = 20$, and cyclic bead chains of length $N = 3$ and 5. We note that the crystal structures resemble closely those observed in MD simulations of hard spheres [126, 127]. As the crystal morphology is mainly determined by the crystallization kinetics rather than the bulk and surface contributions to the free energy of the nucleus, it is tempting to speculate that the crystallization dynamics of hard polymeric chains is similar to that of hard spheres, and is thus not strongly affected by the chain connectivity.

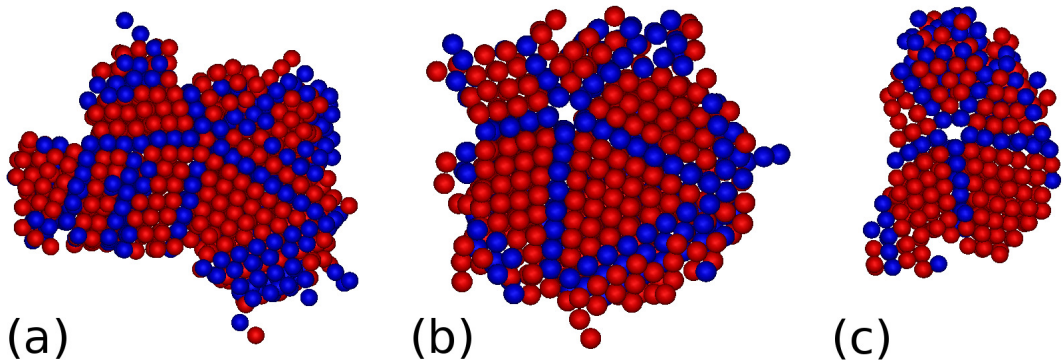

**Figure 7.4**: Typical configurations of the crystal structures for linear hard-sphere chains with chain length $N = 20$ (a) and for ring-like polymers with $N = 3$(b) and $N = 5$(c). Only crystalline spheres are shown here. The blue and red spheres are hcp-like and fcc-like particles, respectively.

Furthermore, we determine the bond angle distribution function $p(\cos\theta)$ in order to quantify the distribution of bond angles between two neighboring polymer bonds in the supersaturated fluid and crystal nuclei. Fig. 7.5 shows $p(\cos\theta)$ for systems consisting of linear and cyclic bead chains with $N = 10$. For the fluid phase, $p(\cos\theta)$ displays two peaks at $\cos\theta = -0.5$ and 0.5, i.e., $\theta = 120$ and 60 degrees, which corresponds with the most frequent three particle structures observed in random packings of spheres [135]. However, $p(\cos\theta)$ of the crystal structures exhibits four pronounced peaks located around $\cos\theta = -0.5, 0, 0.5$ and 1.0, which corresponds with $\theta = 120, 90, 60$ and 0 degrees, respectively. In order to compare $p(\cos\theta)$ with that for a self-avoiding random walk on an fcc crystal lattice, we integrate the peaks of $p(\cos\theta)$ between the two neighboring local minima. The results are shown in Fig. 7.5 together with the bond angle distribution function for a self-avoiding random walk on an fcc crystal lattice. We find that $p(\cos\theta)$ for crystals of linear and cyclic polymers agree well with that of a self-avoiding random walk [130]. Hence, $p(\cos\theta)$ seems not to be affected by the chain connectivity of the polymer chains. Free energy calculations show indeed that the stable solid of freely jointed hard-sphere chains is an aperiodic crystal phase, where the spheres are positioned on an fcc lattice with the bonds randomly oriented [137].

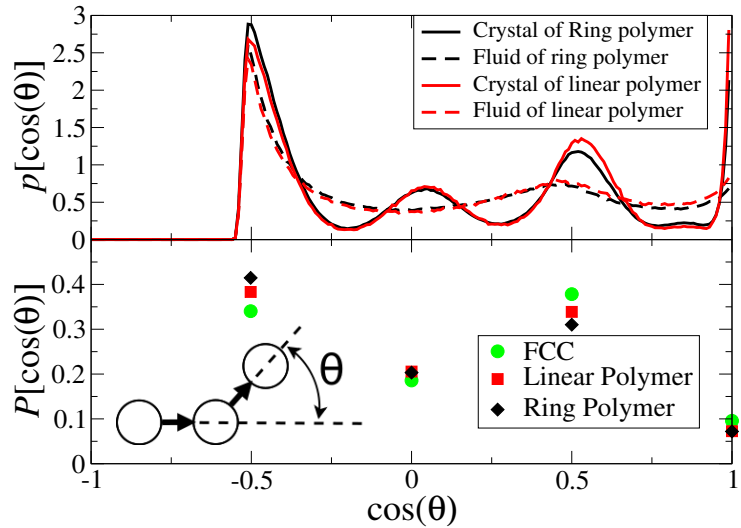

**Figure 7.5**: Bond angle distribution function $p(\cos\theta)$ for a supersaturated fluid and crystal of linear and cyclic hard-sphere polymers with chain length $N = 10$ (top). Integration of the peaks of $p(\cos\theta)$ between the two neighboring local minima (bottom). For comparison, we also plot the bond angle distribution for self-avoiding random walks consisting of 10 steps on an fcc crystal lattice (bottom). The inset shows the definition of the angle $\theta$.

## 7.4   Conclusions

In conclusion, we presented an efficient event-driven Molecular Dynamics simulation method for linear and cyclic hard-sphere chains, which can easily be extended to semi-flexible polymers by taking into account a bending energy [138, 139]. We performed extensive EDMD simulations to study the spontaneous nucleation and crystallization in systems of linear and cyclic chains of flexibly connected hard spheres. We found that the nucleation rate decreases significantly upon decreasing the maximum bond length, as the bond connectivity frustrates the crystal formation of hard-sphere chains. Surprisingly, we find that the nucleation rate is determined by the average number of bonds per bead in the system rather than the chain length, chain topology, and polymer composition. We thus find that the crystal nucleation rate can be enhanced by the addition of monomers. Furthermore, we find that the final crystal structures show a wide range of crystal morphologies including structures with a five-fold symmetry pattern, which are remarkably similar to those observed in MD simulations of hard spheres [126, 127]. We also find that the bond angle distribution function in the hard-sphere chain crystals resembles closely that of self-avoiding random walks on an fcc crystal lattice. Hence, our observations suggest that the nucleation and crystallization of hard-sphere polymers is remarkably similar to that observed in hard-sphere systems, but frustrated by the bond connectivity.

## 7.5 Acknowledgments

I would like to thank Dr. F. Smallenburg for fruitful discussions.

# 8

# Crystal nucleation in binary hard-sphere mixtures: The effect of the order parameter on the cluster composition

We study crystal nucleation in a binary mixture of hard spheres and investigate the composition and size of the (non)critical clusters using Monte Carlo simulations. In order to study nucleation of a crystal phase in computer simulations, a one-dimensional order parameter is usually defined to identify the solid phase from the supersaturated fluid phase. We show that the choice of order parameter can strongly influence the composition of noncritical clusters due to the projection of the Gibbs free-energy landscape in the two-dimensional composition plane onto a one-dimensional order parameter. On the other hand, the critical cluster is independent of the choice of the order parameter, due to the geometrical properties of the saddle point in the free-energy landscape, which is invariant under coordinate transformation. We investigate the effect of the order parameter on the cluster composition for nucleation of a substitutional solid solution in a simple toy model of identical hard spheres but tagged with different colors and for nucleation of an interstitial solid solution in a binary hard-sphere mixture with a diameter ratio $q = 0.3$. In both cases, we find that the composition of noncritical clusters depends on the order parameter choice, but are well explained by the predictions from classical nucleation theory. More importantly, we find that the properties of the critical cluster do not depend on the order parameter choice.

## 8.1  Introduction

The process of nucleation in colloidal systems has attracted significant attention in recent years, both in experimental and simulation studies. The framework with which phenomena like these have been described traditionally is classical nucleation theory (CNT), which is based on the notion that a thermal fluctuation spontaneously generates a small droplet of the thermodynamically stable phase into the bulk of the metastable phase. In CNT as developed by Volmer [22], Becker [25], and Zeldovich [140], the free energy of formation of small nuclei of the new phase in the parent phase is described by using the "capillary approximation", i.e., the free energy to form a cluster of the new phase relative to the homogeneous metastable phase is described by their difference in bulk free energy and a surface free-energy term that is given by that of a planar interface between the two coexisting phases at the same temperature. Thus the droplet is assumed to be separated from the metastable bulk by a sharp step-like interface in CNT. The bulk free-energy term is proportional to the volume of the droplet and represents the driving force to form the new phase, while the surface free-energy cost to create an interface is proportional to the surface area of the cluster. Hence, small droplets with a large surface-to-volume ratio have a large probability to dissolve, while droplets that exceed a critical size and cross the free-energy barrier, can grow further to form the new stable bulk phase.

CNT has successfully explained simulation results for the nucleation of spherical particles, such as the fluid-solid and gas-liquid nucleation in Lennard-Jones systems [52, 141, 142] and crystal nucleation of hard spheres [18, 19]. A modified CNT has been used to explain the nucleation of anisotropic clusters of the nematic or solid phase (also called tactoids) from a supersaturated isotropic phase of colloidal hard rods [92, 115, 143] and the nucleation of 2D assemblies of attractive rods [105, 117]. This state of affairs should be contrasted with the case of binary nucleation for which various nucleation theories have been developed that differ substantially in the way they describe the composition of the cluster [106, 144, 145]. For instance, Reiss assumed the surface tension to be independent of composition [144], while Doyle extended CNT by taking into account a surface tension that depends on the cluster composition [146]. However, more than 20 years later, it was shown by Renninger [147], Wilemski [148, 149], and Reiss [150] that Doyle's derivation leads to thermodynamic inconsistencies. A revised thermodynamically consistent classical binary nucleation theory was developed by Wilemski in which the composition of the surface layer and the interior of the cluster could vary independently [148, 149]. However, in the case of strong surface enrichment effects, this approach can lead to unphysical negative particle numbers in the critical clusters [151, 152]. In addition, it was shown in Ref. [153] that the derivation by Wilemski starts off with the wrong equations, but the resulting equations are correct. Moreover, binary nucleation can be accompanied with huge fractionation effects, i.e., the compositions of the metastable phase and of the phase to be nucleated can differ enormously from the compositions of the two coexisting bulk phases. It is therefore unclear i) how to determine the surface free-energy term for a cluster, which is in quasi-equilibrium with a metastable parent phase with a composition that is very different from those of the two coexisting bulk phases, ii) whether the interfacial tension depends on composition, curvature, and surface enrichment effects, and finally iii) whether or not one can use the capillary approximation in the first place to describe binary nucleation in systems where fractionation and surface activity of the species are important. To summarize, there is no straightforward generalization to multicomponent systems of classical nucleation theory that is thermodynamically consistent, does not lead to unphysical effects, and can be applied to small nuclei [145, 154].

Numerical studies may shed light on this issue, as the nucleation barrier can be determined directly in computer simulations using the umbrella sampling technique [37, 38]. In this method, an order parameter is chosen and configuration averages for sequential values of the order parameter are taken. While this makes it possible to measure properties of clusters with specific values for the order parameter, it should be noted that the results can depend on the choice of order parameter. In the present chapter, we investigate whether the size and composition of (non)critical clusters can be affected by the order parameter choice employed in simulation studies of multicomponent nucleation. For simplicity, we focus here on crystal nucleation in binary hard-sphere mixtures, where surface activity of the species can be neglected, and we assume the surface tension to be composition independent. This chapter is organized as follows. In Sec. 8.2, we describe the general nucleation theorem as derived by Oxtoby and Kashchiev [145], which does not rely on the "capillary approximation" and can even be employed to describe small clusters. Starting from the multicomponent nucleation theorem, it is straightforward to reproduce the usual CNT for binary nucleation, which is the focus in the remainder of this chapter. In Sec. 8.3 and 8.4, we define the (Landau) free energy as a function of an order parameter, and we describe the order parameter that is employed to study crystal nucleation. Additionally, we discuss the effect of order parameter choice on the nucleation barrier in more detail. We present results for binary nucleation for a simple toy model of hard spheres in Sec. 8.5, and subsequently, we study the nucleation of an interstitial solid solution in an asymmetric binary hard-sphere mixture in Sec. 8.6.

## 8.2   Classical Nucleation Theory for multi-component systems

We study the formation of a multicomponent spherical cluster of the new phase in a supersaturated homogeneous bulk phase $\alpha$ consisting of species $i = 1, 2, \ldots$. We note that the thermodynamic variables corresponding to the metastable phase $\alpha$ are denoted by the subscript $\alpha$, whereas those corresponding to the new phase do not carry an extra subscript to lighten the notation. We first consider a homogeneous bulk phase $\alpha$ characterized by an entropy $S_\alpha^o$, volume $V_\alpha^o$, and particle numbers $N_{i,\alpha}^o$. Note that the superscripts denote the original bulk phase. The internal energy $U_\alpha^o$ of the original bulk phase reads

$$U_\alpha^o = T^o S_\alpha^o - P_\alpha^o V_\alpha^o + \sum \mu_{i,\alpha}^o N_{i,\alpha}^o \tag{8.1}$$

with $T^o$ the temperature, $P_\alpha^o$ the bulk pressure, $\mu_{i,\alpha}^o$ the bulk chemical potential of species $i$, and the summation runs over all species.

Following the derivation in Refs. [145, 154], we now consider a spherical cluster of the new phase with a volume $V$ separated from the original phase by an arbitrarily chosen Gibbs dividing surface. The volume of the interface is set to zero, and the particle number of species $i$ in the cluster is given by $N_i + N_{i,s}$, where $N_i$ is the number of particles of species $i$ in a volume $V$ which is homogeneous in the new bulk phase, and $N_{i,s}$ is the surface excess number of particles of species $i$ that corrects for the difference between a step-like interfacial density profile and the actual one. The surface excess number $N_{i,s}$ depends on the choice of dividing surface. The internal energy $U$ of the resulting system is then given by

$$U = TS_\alpha + TS - P_\alpha V_\alpha - PV + \Psi + \sum \mu_{i,\alpha} N_{i,\alpha} + \sum \mu_i N_i + \sum \mu_{i,s} N_{i,s}, \qquad (8.2)$$

where $P$ and $S$ denote the bulk pressure and entropy of the nucleated phase, and $\mu_i$ and $\mu_{i,s}$ are the chemical potentials of species $i$ in the new phase and the surface phase, $T$ is the temperature of the system with the cluster, and $\Psi = \Psi(\{N_i\}, \{N_{i,s}\}, V)$ is the total surface energy of the spherical cluster. As the volume of the surface layer is zero, the corresponding pressure is not defined.

The difference in the appropriate thermodynamic potential as a function of cluster size depends on the quantities that are kept fixed during the nucleation process. If the nucleus is formed at constant temperature and constant total number of particles of each species $i$, and if we keep the pressure of the original phase fixed, then $T = T^o$, $N_{i,\alpha} + N_i + N_{i,s} = N^o_{i,\alpha}$, and $P^o_\alpha = P_\alpha$. The corresponding Gibbs free energy of the initial system $G^o_\alpha$ and that of the final system $G$ are then given by the Legendre transformation

$$\begin{aligned}
G^o_\alpha &= U^o_\alpha - T^o S^o_\alpha + P^o_\alpha V^o_\alpha = \sum \mu^o_{i,\alpha} N^o_{i,\alpha} \\
G &= U - TS + P^o_\alpha (V_\alpha + V) \\
&= (P^o_\alpha - P)V + \Psi + \sum \mu_{i,\alpha} N_{i,\alpha} + \sum \mu_i N_i \\
&\quad + \sum \mu_{i,s} N_{i,s}.
\end{aligned} \qquad (8.3)$$

If we now assume that the composition of the metastable phase $\alpha$ remains unchanged and we consider the Maxwell relation

$$\left( \frac{\partial V_\alpha}{\partial N_{i,\alpha}} \right)_{T, P_\alpha, \{N_{j \neq i, \alpha}\}} = v_{i,\alpha} = \left( \frac{\partial \mu_{i,\alpha}}{\partial P_\alpha} \right)_{T, \{N_{i,\alpha}\}} \qquad (8.4)$$

with $v_{i,\alpha}$ the partial particle volumes of species $i$ in phase $\alpha$, we find that at constant pressure, the chemical potential for each species $i$ remains constant $\mu^o_{i,\alpha} = \mu_{i,\alpha}$. Subsequently, we obtain for the change in Gibbs free energy $\Delta G = G - G^o_\alpha$ when a nucleus is formed in the bulk of the original phase:

$$\Delta G = (P^o_\alpha - P)V + \Psi + \sum (\mu_i(P) - \mu^o_{i,\alpha}(P^o_\alpha))N_i + \sum (\mu_{i,s} - \mu^o_{i,\alpha}(P^o_\alpha))N_{i,s}. \qquad (8.5)$$

Consequently, the Gibbs free energy $\Delta G$ of a growing cluster depends on the number of particles $N_i$ and $N_{i,s}$ in the cluster and the surface energy of the cluster. Hence, one can define a free-energy surface in the multi-dimensional composition plane with a saddle point that corresponds to the critical nucleus [144]. The conditions for the critical cluster

read

$$\left(\frac{\partial \Delta G}{\partial N_i}\right)_{V,\{N_{j\neq i}\},\{N_{i,s}\}} = 0,$$

$$\left(\frac{\partial \Delta G}{\partial N_{i,s}}\right)_{V,\{N_i\},\{N_{j\neq i,s}\}} = 0, \qquad (8.6)$$

$$\left(\frac{\partial \Delta G}{\partial V}\right)_{\{N_i\},\{N_{i,s}\}} = 0.$$

To recover the chemical and mechanical equilibrium conditions, we use the above conditions as well as the Gibbs-Duhem equation and the Gibbs adsorption equation. The Gibbs-Duhem equation at constant temperature for the nucleated bulk phase is

$$-V\mathrm{d}P + \sum N_i \mathrm{d}\mu_i = 0, \qquad (8.7)$$

and the Gibbs adsorption equation for the surface at constant temperature is

$$A\mathrm{d}\gamma + \sum N_{i,s}\mathrm{d}\mu_{i,s} = 0, \qquad (8.8)$$

where we have employed $\Psi = \gamma A$. Note that $\gamma$ denotes the surface free energy per unit area and $A$ is the surface area of the cluster. The resulting equilibrium conditions for all particle species $i$ in the critical cluster, the surface, and the metastable parent phase are then given by

$$\mu_i^*(P^*) = \mu_{i,s}^* = \mu_{i,\alpha}^o(P_\alpha^o), \qquad (8.9)$$

and for the pressure difference inside and outside the droplet we find

$$P^* - P_\alpha^o = \frac{\partial \gamma^* A^*}{\partial V^*}, \qquad (8.10)$$

where $^*$ denotes quantities associated with a system where a critical cluster is present. Hence, the composition of the critical cluster can be determined from these saddle point conditions.

In order to obtain the usual classical nucleation theory for multicomponent systems, we assume a spherical droplet with radius $R$. Note that the surface area is then $A = 4\pi R^2$. In addition, we use the fact that the volume of a spherical droplet can be expressed in terms of the partial particle volumes $v_i$ of species $i$:

$$V = \frac{4}{3}\pi R^3 = \sum N_i v_i. \qquad (8.11)$$

Combining this with Eq. 8.10, we arrive at the generalised Laplace equation:

$$P^* - P_\alpha^o = \frac{2\gamma^*}{R^*} + \left[\frac{\partial \gamma^*}{\partial R^*}\right], \qquad (8.12)$$

where the square brackets denote a derivative associated with the displacement of the dividing surface. One can now choose the dividing surface so that

$$\left[\frac{\partial \gamma^*}{\partial R^*}\right] = 0, \tag{8.13}$$

and hence one recovers the usual Laplace equation. This choice for the dividing surface, corresponding to a specific value for $R^*$ and $\gamma^*$, is called the surface of tension. In addition, if we use the Gibbs adsorption isotherm (8.8) and the Maxwell relation (Eq. 8.4) for the bulk phase of the nucleated cluster, we find for the critical cluster

$$\mathrm{d}\mu_{i,s}^* = \mathrm{d}\mu_i^* = v_i \mathrm{d}P \tag{8.14}$$

and

$$\left[A\frac{\partial \gamma^*}{\partial R^*}\right] = -\sum N_{i,s}v_i\left[\frac{\partial P^*}{\partial R^*}\right] = 0, \tag{8.15}$$

which is the condition for a curvature independent surface tension. Since $\partial P^*/\partial R^* \neq 0$, Eq. 8.15 implies that the dividing surface has to be chosen such that

$$\sum N_{i,s}v_i = 0, \tag{8.16}$$

which is called the equimolar surface, as for one-component systems $N_{i,s} = 0$, i.e. the number of particles in the cluster equals the number of particles in a uniform bulk phase with the same volume. It is generally not possible in a multicomponent system to choose the dividing surface such that $N_{i,s} = 0$ for all species. Thus, as $v_i$ is usually positive, $N_{i,s} < 0$ for at least one of the species. This may lead to (unphysical) negative particle numbers when $N_i + N_{i,s} < 0$ as noted in Refs. [151, 152]. However, as will be discussed in sections 8.5 and 8.6, there are cases in which the assumption $N_{i,s} = 0$ for all $i$ is valid.

If the nucleated phase is assumed to be incompressible, one can integrate the Gibbs-Duhem equation (8.7) at constant temperature to arrive at

$$V(P_\alpha^o - P) = \sum(\mu_i(P_\alpha^o) - \mu_i(P))N_i, \tag{8.17}$$

and using Eq. 8.5, we find

$$\Delta G = \gamma A + \sum(\mu_i(P_\alpha^o) - \mu_{i,\alpha}^o(P_\alpha^o))N_i + \sum(\mu_{i,s} - \mu_{i,\alpha}^o(P_\alpha^o))N_{i,s}. \tag{8.18}$$

Again using the Gibbs-Duhem equation at constant temperature and pressure and the Gibbs adsorption isotherm, and minimizing the free energy with respect to $N_i$ at fixed $\{N_{i,s}\}$, we recover the Gibbs-Thomson (also called Kelvin) equations for multi-component spherical critical clusters

$$\Delta\mu_i^* = -\frac{2\gamma^* v_i}{R^*}, \tag{8.19}$$

where $\Delta\mu_i^* = \mu_i^*(P_\alpha^o) - \mu_{i,\alpha}^o(P_\alpha^o)$. The radius of the critical cluster $R^*$ and the barrier height $\Delta G^*$ read

$$R^* = \frac{2\gamma^* v_i}{|\Delta\mu_i^*|} \tag{8.20}$$

$$\Delta G^* = \frac{4\pi R^{*2}\gamma^*}{3} = \frac{16\pi\gamma^{*3}}{3(\Delta\mu_i^*/v_i)^2}. \tag{8.21}$$

Using Eq. 8.20 or the Maxwell relation (8.4), one can show:

$$v_i \Delta \mu_i = v_j \Delta \mu_j, \qquad (8.22)$$

and the radius of the critical cluster $R^*$ can be expressed in terms of the bulk composition $x_i = N_i / \sum N_i$ of the critical cluster and $v = V / \sum N_i$:

$$R^* = \frac{2\gamma^* v}{\sum x_i |\Delta \mu_i^*|}. \qquad (8.23)$$

In order to study multi-component nucleation, MC simulations are often performed in the isobaric-isothermal ensemble, in which the number of particles $N_{1,\alpha}^o$ and $N_{2,\alpha}^o$, the pressure of the original bulk phase $P_\alpha^o$, and the temperature $T$ are kept fixed. One of the assumptions of classical nucleation theory is that the composition of the metastable bulk phase remains constant, while nucleating the new phase, see Eq. 8.4. In simulations this can only be achieved if the system is sufficiently large, i.e., the volume of the metastable bulk phase is much larger than that of the nucleating cluster. Especially, for binary (multicomponent) nucleation, where the composition of the stable phase is very different from that of the metastable phase, this can lead to a huge depletion of one of the components in the metastable fluid phase, and therefore a change in composition. In order to circumvent this problem, simulation studies on binary nucleation are often carried out in the semi-grand canonical ensemble [155, 156], i.e. the total number of particles $N_\alpha^o = \sum N_{i,\alpha}^o$, the chemical potential difference $\Delta \mu_{12,\alpha}^o = \mu_{2,\alpha}^o - \mu_{1,\alpha}^o$ between the two species, the pressure $P_\alpha^o$, and the temperature $T$ are kept fixed of the original bulk phase. The corresponding thermodynamic potential is obtained by a Legendre transformation

$$Y(N, \Delta \mu_{12}, P, T) = G(N, N_2, P, T) - N_2 \Delta \mu_{12} \qquad (8.24)$$

Combining Eq. 8.3 with the conditions that the total number of particles are fixed $N_{1,\alpha}^o + N_{2,\alpha}^o = N_1 + N_2 + N_{1,\alpha} + N_{2,\alpha}$, the chemical potential difference in the metastable phase is kept fixed $\Delta \mu_{12,\alpha}^o = \Delta \mu_{12,\alpha}$, constant pressure of the metastable phase $P_\alpha^o = P_\alpha$ and constant temperature $T = T^o$, we find for the corresponding thermodynamic potentials

$$
\begin{aligned}
Y_\alpha^o &= G_\alpha^o - N_{2,\alpha}^o \Delta \mu_{12,\alpha}^o = \mu_{1,\alpha}^o (N_{1,\alpha}^o + N_{2,\alpha}^o) \\
Y &= G - (N_{2,\alpha} + N_2) \Delta \mu_{12,\alpha} \\
&= (P_\alpha^o - P)V + \Phi + \mu_{1,\alpha}(N_{1,\alpha} + N_{2,\alpha}) + \\
&\quad \mu_1(N_1 + N_2) - \Delta \mu_{12} N_2 - \Delta \mu_{12,\alpha} N_2,
\end{aligned} \qquad (8.25)
$$

where we have set the surface excess numbers $N_{i,s}$ to zero. Using the Maxwell equation

$$\left( \frac{\partial \mu_1}{\partial P} \right)_{N, \Delta \mu_{12}, T} = \left( \frac{\partial V}{\partial N} \right)_{\Delta \mu_{12}, P, T} = v, \qquad (8.26)$$

we find that due to constant pressure, the chemical potential of species 1 remains unchanged $\mu_{1,\alpha}^o = \mu_{1,\alpha}$. Hence, we find that the change in free energy due to the formation of a nucleus $\Delta Y = Y - Y_\alpha^o$ equals $\Delta G$ as given in Eq. 8.5 and the nucleation barrier can be calculated in the semi-grand canonical ensemble. Similarly, one can show that in any statistical ensemble (grand canonical, canonical, etc. ), the change in the corresponding thermodynamic potential as a function of cluster size is always the same, provided that the metastable parent phase is sufficiently large. A similar result was also obtained by Oxtoby and Bob Evans, who showed that the nucleation free-energy barriers in the isobaric-isothermal and grand canonical ensemble are identical, i.e., $\Delta G = \Delta \Omega$ for a one-component system [157].

## 8.3 Free-energy Barrier

While nucleation is an inherently non-equilibrium process, the assumption of local equilibrium is often made to describe the behavior of the system during the nucleation process. In essence, this assumption states that the nucleus is in quasi-equilibrium with the parent phase for every cluster size. This is approximately true if the time required to reach an equilibrium distribution of clusters is short compared to the time needed to nucleate. After the system crosses the free-energy barrier, the cluster of the new phase grows too rapidly for this assumption to be accurate, but during the nucleation process itself, local equilibrium has proven to be a useful assumption.

In order to compute the free-energy barrier that separates the metastable phase from the stable phase, an order parameter $\Phi$ (or reaction coordinate) should be defined that quantifies how much the system has transformed to the new phase. A common order parameter that is employed in nucleation studies is the size of the largest cluster in the system as defined by a certain cluster criterion. In the present chapter, we restrict ourselves to binary nucleation. From Eq. 8.18, we find that the Gibbs free energy $\Delta G$ of a growing binary cluster depends on the number of particles of species 1 and 2 in the cluster, and hence, one can define a free-energy surface in the $(N_1, N_2)$-plane with a saddle point that corresponds to the critical nucleus [144]. By projecting the phase space of the system onto the (usually) one-dimensional order parameter, one can define the (Landau) Gibbs free energy $\Delta G(\Phi)$ as a function of this order parameter $\Phi$

$$\beta \Delta G(\Phi) = G_c - \ln P(\Phi), \tag{8.27}$$

where $\beta = 1/k_B T$, $k_B$ Boltzmann's constant, $T$ the temperature, $G_c$ is a normalization constant generally taken to correspond to the free energy of the homogeneous metastable phase, and $P(\Phi)$ is the probability of observing an order parameter of value $\Phi$. In a system of $N$ particles, at fixed pressure $P$, and constant temperature $T$, the probability $P(\Phi)$ is given by:

$$P(\Phi) = \frac{\int dV \int d\mathbf{r}^N \exp[-\beta(U(\mathbf{r}^N) + PV)]\delta(\Phi - \Phi(\mathbf{r}^N))}{\int dV \int d\mathbf{r}^N \exp[-\beta(U(\mathbf{r}^N) + PV)]} \tag{8.28}$$

with $V$ the volume of the system, $U$ the potential energy, and $\delta$ the Kronecker delta function. The order parameter function $\Phi(\mathbf{r}^N)$ is a function that assigns to each configuration $\mathbf{r}^N$ of the system a value for the order parameter. The probability distribution $P(\Phi)$ can be obtained from Monte Carlo (MC) simulations via the umbrella sampling technique [37, 38]. In this method, an additional external potential $U_{\text{bias}}$ is added to the system to bias the sampling towards configurations corresponding to a certain window of order parameter values centered around $\Phi_o$. By increasing $\Phi_o$ sequentially, the entire free-energy barrier as a function of $\Phi$ can be sampled. The typical biasing potential used in umbrella sampling simulations is given by:

$$\beta U_{\mathrm{bias}}(\mathbf{r}^N) = k(\Phi(\mathbf{r}^N) - \Phi_o)^2, \qquad (8.29)$$

where the constants $k$ and $\Phi_o$ determine the width and location of the window, and $\mathbf{r}^N$ are the positions of all $N$ particles in the simulation.

## 8.4   Order Parameter

In order to follow a phase transformation, a cluster criterion is required that is able to identify the new phase from the supersaturated phase. In this chapter, we focus on the formation of a solid cluster in a supersaturated fluid phase. In order to study crystal nucleation, the local bond-order parameter is used to differentiate between liquid-like and solid-like particles and a cluster algorithm is employed to identify the solid clusters [52]. In the calculation of the local bond order parameter a list of "neighbours" is determined for each particle. The neighbours of particle $i$ include all particles within a radial distance $r_c$ of particle $i$, and the total number of neighbours is denoted $N_b(i)$. A bond orientational order parameter $q_{l,m}(i)$ for each particle is then defined as

$$q_{l,m}(i) = \frac{1}{N_b(i)} \sum_{j=1}^{N_b(i)} \Upsilon_{l,m}(\theta_{i,j}, \phi_{i,j}), \qquad (8.30)$$

where $\Upsilon_{l,m}(\theta, \phi)$ are the spherical harmonics, $m \in [-l, l]$ and $\theta_{i,j}$ and $\phi_{i,j}$ are the polar and azimuthal angles of the center-of-mass distance vector $\mathbf{r}_{ij} = \mathbf{r}_j - \mathbf{r}_i$ with $\mathbf{r}_i$ the position vector of particle $i$. Solid-like particles are identified as particles for which the number of connections per particle $\xi(i)$ is at least $\xi_c$ and where

$$\xi(i) = \sum_{j=1}^{N_b(i)} H(d_l(i,j) - d_c), \qquad (8.31)$$

$H$ is the Heaviside step function, $d_c$ is the dot-product cutoff, and

$$d_l(i,j) = \frac{\displaystyle\sum_{m=-l}^{l} q_{l,m}(i) q_{l,m}^*(j)}{\left(\displaystyle\sum_{m=-l}^{l} |q_{l,m}(i)|^2\right)^{1/2} \left(\displaystyle\sum_{m=-l}^{l} |q_{l,m}(j)|^2\right)^{1/2}}. \qquad (8.32)$$

A cluster contains all solid-like particles which have a solid-like neighbour in the same cluster. Thus each particle can be a member of only one cluster.

The parameters contained in this algorithm include the neighbour cutoff $r_c$, the dot-product cutoff $d_c$, the critical value for the number of solid-like neighbours $\xi_c$, and the symmetry index for the bond orientational order parameter $l$. The hard-sphere crystals considered in this chapter are expected to have random hexagonal order, thus the symmetry index is chosen to be 6 in the present study.

This choice of order parameter $\Phi$, defined as the number of solid-like particles in the largest crystalline cluster, has been used to study crystal nucleation in various one-component systems, e.g., Lennard-Jones systems [52], hard-sphere systems [18], and Yukawa systems [47].

On the other hand, for binary systems, a variety of crystal structures can appear in the bulk phase diagram, e.g., substitutionally ordered (superlattice) crystal structures with varying stoichiometries, substitutionally disordered solid solutions, interstitial solid solutions, crystalline phases of species 1 with a dispersed fluid of species 2, etc. Nucleation of a substitutionally disordered solid solution and a crystal with the CsCl structure has been studied in a binary mixture of hard spheres using the total number of particles in the largest crystalline cluster as an order parameter, i.e. $\Phi = N_1 + N_2$ [155]. This order parameter has also been employed in a crystal nucleation study of a substitutionally disordered face-centered cubic crystal and a crystal with the CsCl structure of oppositely charged colloids [28], and nucleation of the NaCl salt crystal from its melt using the symmetry index $l = 4$ instead of $l = 6$ for the bond orientational order parameter [158]. However, one can also define other linear combinations of $N_1$ and $N_2$ as an order parameter. When the partial particle volumes of the two species are very different, one can employ the volume of the largest crystalline cluster $\Phi = V = N_1 v_1 + N_2 v_2$ as an order parameter. While, if the crystal structure consists of only one species, say species 1, with the other species randomly dispersed, the number of particles of species 1 in the largest crystalline cluster would be more appropriate to use as an order parameter $\Phi = N_1$. On the other hand, one can also use the stoichiometry $n$ of the $AB_n$ superlattice structure to define the order parameter $\Phi = N_1 + N_2/n$ in order to prevent a strong bias towards one of the species. More generally, if the cluster size is measured by the order parameter $\Phi = N_1 + \lambda N_2$, the sensitivity of the order parameter to particles of species 2 can be tuned via the parameter $\lambda$. For $\lambda = 1$, this corresponds to the total number of particles in the cluster, while for $\lambda = 0$, this corresponds to the number of particles of type 1.

As already mentioned above, the umbrella sampling technique is often employed to determine the probability distribution $P(\Phi)$ and the Gibbs free energy $\Delta G(\Phi)$. To this end, a biasing potential is introduced to sample configurations with certain values for this order parameter $\Phi$. In this chapter, we investigate the effect of the choice of order parameter on the properties of the clusters during nucleation in a binary mixture of hard spheres, where we assume that the surface excess numbers of species $i$ are negligible. Using Eq. 8.18, we now write down explicitly the change in Gibbs free energy for binary nucleation

$$\Delta G \;\; = \;\; \gamma A + \Delta \mu_1 N_1 + \Delta \mu_2 N_2, \tag{8.33}$$

where $\Delta \mu_i = \mu_i(P^o_\alpha) - \mu^o_{i,\alpha}(P^o_\alpha)$. The Gibbs free energy $\Delta G$ depends on the particle numbers $N_1$ and $N_2$ and the composition of the critical cluster can be determined from the saddle point conditions for $\Delta G$. The free-energy surface in the two-dimensional composition plane $(N_1, N_2)$ is projected in umbrella sampling MC simulations onto a one-dimensional order parameter, e.g. $\Phi = N_1 + \lambda N_2$. Hence, the projected $\Delta G(\Phi)$ and the averaged (or projected) cluster composition of noncritical clusters both depend on the order parameter. We note that this is not an artifact of the umbrella sampling MC simulations, but merely the projection of a correctly measured equilibrium distribution. To determine the averaged composition of noncritical spherical clusters with radius $R$ as a function of $\Phi$, we can minimize $\Delta G$ with respect to $N_2$ while keeping the order parameter $\Phi$ fixed:

$$\left(\frac{\partial \Delta G}{\partial N_2}\right)_\Phi \;=\; \Delta\mu_2 - \lambda\Delta\mu_1 + \frac{2\gamma v_1}{R}(\omega - \lambda) = 0, \tag{8.34}$$

where $\omega = v_2/v_1$. If we use the umbrella sampling technique in MC simulations to determine the Gibbs free energy $\Delta G(\Phi)$ as a function of $\Phi$, one can easily determine the slope of the barrier from the simulations, which is equal to

$$\frac{\mathrm{d}\Delta G}{\mathrm{d}\Phi} = \left(\Delta\mu_2 + \frac{2\gamma v_1}{R}\omega\right)\left(\frac{\partial N_2}{\partial \Phi}\right) + \left(\Delta\mu_1 + \frac{2\gamma v_1}{R}\right)\left(\frac{\partial N_1}{\partial \Phi}\right) \tag{8.35}$$

with

$$\frac{\partial N_1}{\partial \Phi} = \frac{1 - x - \lambda N \frac{\partial x}{\partial \Phi}}{1 - x + \lambda x} \tag{8.36}$$

$$\frac{\partial N_2}{\partial \Phi} = \frac{x + N \frac{\partial x}{\partial \Phi}}{1 - x + \lambda x}, \tag{8.37}$$

where we define the composition $x = N_2/N$ and $N = N_1 + N_2$. Combining Eqs. (8.34) and (8.35) yields

$$\omega\Delta\mu_1 - \Delta\mu_2 = (\omega - \lambda)\frac{\mathrm{d}\Delta G}{\mathrm{d}\Phi}. \tag{8.38}$$

We wish to make a few remarks here. First, we recover the Gibbs-Thomson equations for the critical cluster (8.19) when we set $\mathrm{d}\Delta G/\mathrm{d}\Phi$ in Eq. 8.35 to zero, and we recover Eq. 8.22 from Eq. 8.38 for critical clusters. Consequently, the size and composition of the critical cluster are independent of the choice of $\lambda$. This can also be understood from the fact that the saddlepoint in the free-energy landscape is invariant under coordinate transformations. As long as the top of the nucleation barrier corresponds to this saddle point, the average properties of the cluster will be dominated by the configurations around this saddlepoint, regardless of the chosen order parameter. While most reasonable choices of the order parameter fulfill this requirement, it is possible to design order parameters that shift the top of the barrier away from the saddle point. In this case, the clusters at the top of the barrier are non-critical clusters, and rates calculated from the resulting free energy barrier are unreliable. It is important to note that a different choice of order parameter can change the height of the nucleation barrier, since the barrier height is determined by the fraction of phase space mapped to the same order parameter value at the top of the barrier. However, this effect should be small, as the probability of finding a cluster at the top of the nucleation barrier is dominated by the probability of being in the saddle point of the free-energy landscape. For noncritical clusters, we clearly find that the slope of the barrier, and hence the composition of the cluster, depends on the choice of order parameter via $\lambda$. Below, we study the effect of the choice of order parameter for a simple toy model of hard spheres and for the nucleation of an interstitial solid solution in an asymmetric binary hard-sphere mixture. It is interesting to compare this to past studies investigating one-component systems with higher-dimensional order parameters [159, 160]. For the Lennard-Jones system, Moroni *et al.*, have shown that the number of particles in the cluster alone is insufficient to provide a good prediction for the probability a cluster will grow out to a large crystal [159]. Using a two-dimensional order parameter, they observed a strong correlation between the crystallinity and the size of clusters with a 50% probability of growing out. Specifically, clusters with a large amount of face-centered-cubic (fcc) ordering require much smaller sizes to grow out than those with more body-centered-cubic (bcc) ordering. They found that this correlation was not visible in the two-dimensional free-energy landscape, and argued that the shape and structure of a nucleus could determine whether it will grow out. However, we note that the two-dimensional order parameter is still a projection from a higher-dimensional phase space. Thus, the properties of non-critical clusters likely depend on the choice of order parameter as well.

## 8.5 A substitutional solid solution

In order to obtain more insight in the effect of order parameter choice on the cluster composition of noncritical clusters, we first investigate binary crystal nucleation in a toy model of hard spheres. Here, we consider a system consisting of two species of hard spheres with identical sizes, but tagged with different colors, say species 1 is red and species 2 is blue. Obviously, the stable solid phase to be nucleated is a substitutional disordered face-centered-cubic (fcc) crystal phase with the red and blue particles randomly distributed on an fcc lattice. Refs. [18, 19] and Chapter 3 showed that the nucleation barriers for pure hard spheres are well-described by the predictions from classical nucleation theory, where because of the condition of the equimolar surface, the surface excess number $N_s = 0$. It is therefore safe to neglect the surface excess numbers for the present model as well. In addition, it is clear that the partial particle volumes $v_i$ and volume per particle $v$ are identical, and $\omega = v_2/v_1 = 1$. Using the Gibbs-Thomson equations for a binary critical cluster (8.19), we find that the supersaturation $\Delta\mu_1^* = \Delta\mu_2^* = -2\gamma^* v/R^*$, and hence the composition of the critical cluster follows straightforwardly from the bulk chemical potentials $\mu_1^*(P_\alpha^o)$ and $\mu_2^*(P_\alpha^o)$, which depends on the bulk chemical potentials of the original bulk phase and the supersaturation.

As already mentioned above, the composition of noncritical clusters depends on the choice of order parameter, i.e., the projection of the two-dimensional composition plane onto a one-dimensional order parameter $\Phi$. Using Eq. 8.38, we find that for $\lambda = 1$, the composition of the noncritical cluster is determined by the supersaturation $\Delta\mu_1 = \Delta\mu_2$ and the bulk chemical potentials of the original bulk phase. For $\lambda = 0$, we only measure the number of particles of one color, say red, in the cluster. However, a thermodynamic average of all clusters with $N_1$ red particles also includes all post-critical clusters with many blue particles, and as a result, the order parameter fails to work for $\lambda = 0$. For non-zero values of $\lambda$, the ensemble of clusters of each size is well-defined, and we can perform umbrella sampling MC simulations to measure the average cluster composition.

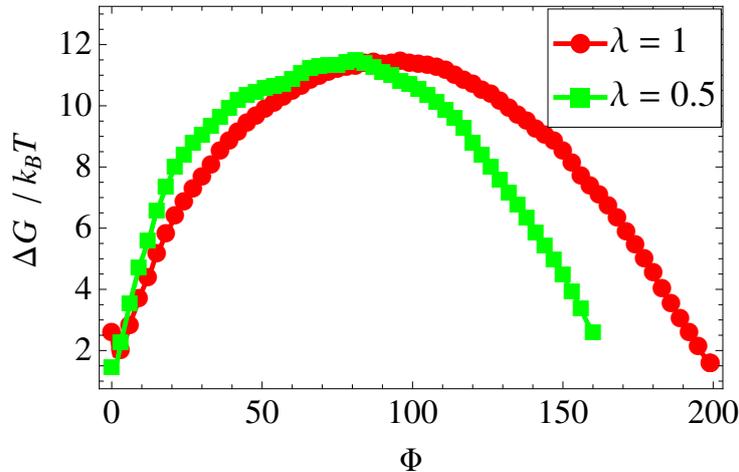

**Figure 8.1**: Gibbs free energy $\Delta G(\Phi)/k_BT$ as a function of order parameter $\Phi = N_1 + \lambda N_2$ for a binary mixture of red (species 1) and blue (species 2) hard spheres with equal diameter $\sigma$ as obtained from umbrella sampling MC simulations at a reduced pressure of $P^* = P_\alpha^o \sigma^3/k_BT = 17$ with $\lambda = 1$ and $\lambda = 0.5$.

In order to keep the composition of the metastable fluid fixed, we perform Monte Carlo simulations on a binary mixture with $N = 1000$ hard spheres in the semi-grand canonical ensemble. Both species of hard spheres are identical in size with diameter $\sigma$, and are either tagged red (species 1) or blue (species 2). The simulations were carried out in a cubic box with periodic boundary conditions and the Metropolis sampling consists of particle displacements and volume changes, and attempts to switch the identity (color) of the particles. The acceptance rule for the identity swap moves is determined by the chemical potential difference $\Delta\mu_{12,\alpha}^o$ [155, 156]. We use the umbrella sampling technique to determine the nucleation barrier $\Delta Y = \Delta G$ as a function of an order parameter $\Phi = N_1 + \lambda N_2$, where $N_1(N_2)$ denotes the number of red (blue) solid-like particles in the largest crystalline cluster in the system as determined by the local bond-order parameter and cluster criterion described in Sec. 8.4 with cutoff radius $r_c = 1.3\sigma$, dot-product cutoff $d_c = 0.7$, and number of solid bonds $\xi_c \geq 6$. We first calculate the nucleation barrier for $\lambda = 1$, for which the order parameter $\Phi$ is simply the total number of solid-like particles in the largest cluster. We set the reduced pressure $P^* = P_\alpha^o \sigma^3/k_BT = 17$, and $\Delta\mu_{12,\alpha}^o = 0$, which corresponds on average to an equimolar mixture of red and blue hard spheres for the metastable fluid phase. We plot the resulting nucleation barriers $\Delta G$ as a function of $\Phi$ in Fig. 8.1. We note that the nucleation barrier for $\lambda = 1$ is equivalent to the nucleation barrier for a pure system of hard spheres [18, 19]. In addition, we show the composition of the largest cluster as a function of $\Phi$ in Fig. 8.2. We find that the averaged composition $x = N_2/N = 0.5$ as it should be since $\Delta\mu_1 = \Delta\mu_2$ and the bulk chemical potentials of the metastable fluid are equal $\mu_{1,\alpha}^o = \mu_{2,\alpha}^o$. Using the binomial coefficients and the measured one-dimensional free-energy barrier, we determine the two-dimensional free-energy landscape $\Delta G(N_1, N_2)/k_BT = -\ln P(N_1, N_2)$ from the probability distribution

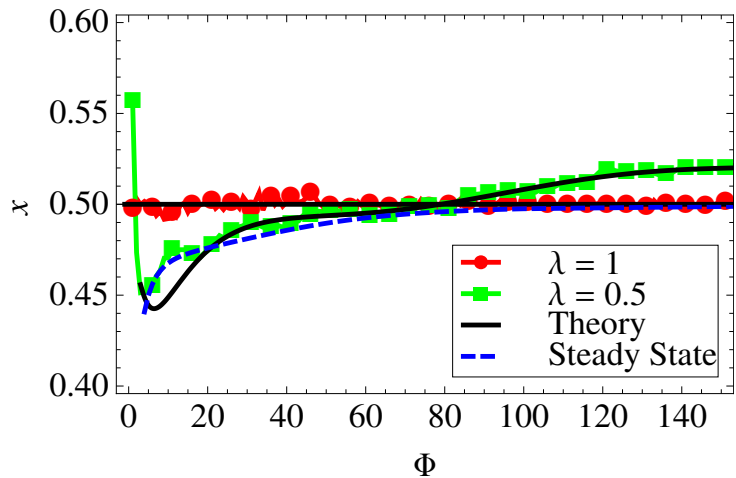

**Figure 8.2**: Composition $x = N_2/N$ of the largest crystalline cluster as a function of order parameter $\Phi = N_1 + \lambda N_2$ for a binary mixture of red (species 1) and blue (species 2) hard spheres with equal diameter $\sigma$ as obtained from umbrella sampling simulations at pressure $P^* = P^o_\alpha \sigma^3/k_B T = 17$ with $\lambda = 1$ (red circles) and $\lambda = 0.5$ (green squares). For comparison, we plot the theoretical prediction (8.38) using the measured nucleation barrier of Fig. 8.1 (black solid line) and the composition determined from a steady-state cluster size distribution for $\lambda = 0.5$ (blue dashed line). The critical cluster size is $\Phi \simeq 79$ and 96 for $\lambda = 0.5$ and 1, respectively.

function

$$P(N_1, N_2) = \exp[-\Delta G(N_1 + N_2)/k_B T]\ 2^N \left( \begin{array}{c} N \\ N_1 \end{array} \right). \tag{8.39}$$

Fig. 8.3 presents a contour plot of the two-dimensional free-energy landscape $\beta\Delta G(N_1, N_2)$ as a function of $N_1$ and $N_2$. Exemplarily, we also plot isolines for the order parameter $\Phi = N_1 + \lambda N_2$ for $\lambda = 1$ and 0.5 to show the projection of the two-dimensional composition plane onto a one-dimensional order parameter.

In order to check the effect of order parameter choice in the biasing potential (8.29) on the nucleation barrier and the composition of the clusters, we also calculate the nucleation barrier for $\lambda = 0.5$ at the same reduced pressure. We plot the nucleation barrier in Fig. 8.1 and the averaged composition of the cluster as a function of $\Phi$ in Fig. 8.2. While the barrier height is not significantly affected by the choice of order parameter in the biasing potential, in agreement with our predictions in Sec. 8.4, the critical cluster "size" as measured by $\Phi$, i.e. $\simeq 79$ and 96 for $\lambda = 0.5$ and 1, respectively, depends on the order parameter choice as expected. In addition, we determine the theoretical prediction for the cluster composition using Eq. 8.38. Using the measured slope of the nucleation barrier from Fig. 8.1 , we obtain the chemical potential difference $\Delta\mu_{12}(\Phi)$ of species 1 and 2 in the cluster from Eq. 8.38. Using Eq. 8.39, we find

$$P(N_1, N_2) \propto 2^N \frac{N!}{N_2!(N - N_2)!} \exp[-\beta N_2 \Delta\mu_{12}(\Phi)] \tag{8.40}$$

from which we determine the most probable composition $x = 1 - \exp[-\beta\Delta\mu_{12}(\Phi)]$. The theoretical prediction for the composition is plotted in Fig. 8.2. We find good agreement with the measured composition, except for very small cluster sizes, where we do not expect CNT to match our nucleation barriers. For comparison, we also plot the same predictions for the nucleation paths in Fig. 8.3. We clearly observe that the two nucleation paths cross at the saddle point yielding the same size and composition of the critical cluster for both order parameters, as expected.

Finally, we also determine the composition of the clusters from the steady-state distribution. In systems where the nucleation of the new phase is measured directly, either in experiments or simulations, the measured cluster size distribution corresponds to a steady-state distribution rather than an equilibrium distribution. The steady-state distribution observed during the nucleation process is different from the equilibrium distribution, as clusters that exceed the critical cluster size during the steady-state process will continue to grow further. The steady-state distribution depends both on the free-energy landscape and the dynamics of the system, and includes a flux across the free-energy barrier, whereas the equilibrium distribution can only be determined by preventing the system from nucleating, i.e, constraining the maximum cluster size by e.g. umbrella sampling MC simulations. While the equilibrium and steady-state distributions are in good agreement for small cluster sizes, they disagree strongly for postcritical cluster sizes, i.e., when the system crosses the free-energy barrier. In particular, the equilibrium cluster size distribution shows a minimum corresponding with the maximum in the free-energy barrier, and the steady-state distribution generally decreases continuously (even) beyond the critical cluster size.

We calculate the cluster composition from the steady-state distribution for our binary mixture of hard spheres. To this end, we determine the free energy as a function of cluster size $N_1$ and $N_2$ from Eq. 8.39 using a fit to the free-energy barrier obtained from umbrella sampling MC simulations with $\lambda = 1$. The dynamics of the cluster are described by the following rates:

$$
\begin{aligned}
k_{N_1,N_2}^{+,1} &= 1 \\
k_{N_1,N_2}^{+,2} &= 1 \\
k_{N_1,N_2}^{-,1} &= \exp[-\beta(G(N_1 - 1, N_2) - G(N_1, N_2))] \\
k_{N_1,N_2}^{-,2} &= \exp[-\beta(G(N_1, N_2 - 1) - G(N_1, N_2))].
\end{aligned}
$$

Here, $k_{N_1,N_2}^{+(-),i}$ is the rate associated with adding (removing) a particle of species $i$ to (from) the nucleus consisting of $N_1$ and $N_2$ particles. Hence, clusters can only grow or shrink by one particle at a time with a rate determined by the corresponding free-energy difference. In order to determine the steady-state cluster size distribution, we set a limit to the steady-state distribution by defining a maximum cluster size, which exceeds the critical cluster size. As a barrier crossing can be considered as a one-way event, subsequent nucleation events should start again from the metastable fluid phase. To this end, we impose that the addition of an extra particle to a nucleus with this maximum cluster size falls back to size zero. We note that this step is not reversible, and results in slightly modified rates for nuclei with the maximum cluster size and for clusters of zero size. With the exception of these steps, the dynamics obey detailed balance.

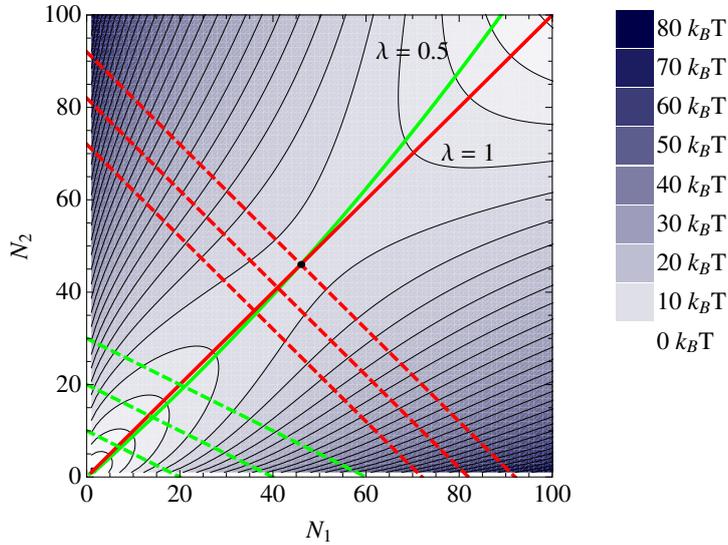

**Figure 8.3**: Contour plot of the two-dimensional free-energy landscape $\Delta G(N_1, N_2)/k_B T$ as a function of $N_1$ and $N_2$. We also plot a few isolines for the order parameter $\Phi = N_1 + \lambda N_2$ for $\lambda = 0.5$ and 1 (dashed lines), and we plot the nucleation path (solid lines labeled with $\lambda = 1$ and $\lambda = 0.5$) for the two order parameters that we considered as predicted by (8.38). The two nucleation paths cross at the saddle point corresponding to the critical cluster size.

In order to determine the steady-state distribution, we set the rate at which clusters of size $(N_1, N_2)$ are created to zero. Hence, the flux with which clusters of size $(N_1, N_2)$ are created should balance the flux with which clusters of this size disappear:

$$P_{ss}(N_1, N_2) \sum_i (k_{N_1, N_2}^{+, i} + k_{N_1, N_2}^{-, i}) =$$
$$P_{ss}(N_1 + 1, N_2)k_{N_1+1, N_2}^{-, 1} + P_{ss}(N_1 - 1, N_2)k_{N_1-1, N_2}^{+, 1} +$$
$$P_{ss}(N_1, N_2 + 1)k_{N_1, N_2+1}^{-, 2} + P_{ss}(N_1, N_2 - 1)k_{N_1, N_2-1}^{+, 2}.$$

Here, $P_{ss}(N_1, N_2)$ denotes the steady-state cluster size distribution. The equations for cluster size zero and the maximum cluster size are slightly different due to a flux of clusters from maximum to zero cluster size. By solving this set of linear equations numerically, we obtain the steady-state distribution. Subsequently, the average cluster composition can be obtained from the steady-state distribution by averaging over clusters with equal $\Phi = N_1 + \lambda N_2$. The resulting cluster composition is shown in Fig. 8.2 for $\lambda = 0.5$. Since the two-dimensional steady-state cluster size distribution, which is symmetric in $N_1$ and $N_2$ decreases monotonically with cluster size, the resulting projected composition is always lower than 0.5 and matches well with the cluster compositions obtained from umbrella sampling MC simulations and the theoretical prediction, except at small cluster sizes as expected. Moreover, in the limit of large (postcritical) clusters, the cluster growth rate approaches a constant for the current choice of dynamics, resulting in a nearly flat steady-state cluster size distribution and a cluster composition of 0.5.

In conclusion, we have shown using a simple model for a binary mixture of hard spheres that the composition of the critical cluster does not depend on the choice of order parameter, while the composition of noncritical clusters is affected by the order parameter. This is a direct consequence of the projection of the two-dimensional free-energy landscape onto a one-dimensional order parameter, say $\Phi = N_1 + \lambda N_2$, which influence directly the projected (Landau) $\Delta G(\Phi)$ and the averaged (or projected) cluster composition. Moreover, as the umbrella sampling method allows us to equilibrate the system for various values of the order parameter, the system can be regarded to be in local equilibrium for each value of the order parameter. The nucleation paths that the system then follows remain close to the minimum free-energy path (see Fig. 8.3), and thus the height of the nucleation barrier is largely unaffected by the choice of order parameter.

## 8.6   An interstitial solid solution

We consider crystal nucleation of an interstitial solid solution in a highly asymmetric binary mixture of large and small hard spheres with size ratio $q = \sigma_2/\sigma_1 = 0.3$, where $\sigma_{1(2)}$ denotes the diameter of species 1 (large spheres) and 2 (small spheres). The interstitial solid solution consists of a face-centered-cubic crystal phase of large spheres with a random occupancy of the octahedral holes by small spheres, and hence the composition of the interstitial solid solution can vary from $x = N_2/N \in [0, 1]$ [161]. As the volume of this solid phase is not largely affected by the density of small spheres, we set the partial particle volume $v_2$ and $\omega = v_2/v_1$ to zero. Using Eq. 8.38, we find the following relation if the system is in local equilibrium at fixed order parameter $\Phi = N_1 + \lambda N_2$

$$\Delta\mu_2 = \lambda \frac{\mathrm{d}\Delta G}{\mathrm{d}\Phi}. \tag{8.41}$$

For $\lambda = 0$, the order parameter $\Phi = N_1$ measures only the large spheres in the cluster, and the cluster composition of both critical and noncritical clusters is determined by the chemical equilibrium condition for the small spheres in the cluster and the metastable fluid phase, i.e., $\Delta\mu_2 = 0$. For $\lambda = 1$, when all particles in the clusters are counted by the order parameter $\Phi = N_1 + N_2$, the composition of precritical clusters will have a higher density of small particles compared to the chemical equilibrium condition for the small particles in the cluster and the metastable fluid phase, as the slope of the nucleation barrier is positive, and similarly postcritical clusters will have a lower density of small particles. For both order parameters, we find that the critical cluster satisfies the Gibbs-Thomson equation (8.19), and thus for a partial particle volume $v_2 = 0$, we obtain chemical equilibrium for the small particles in the critical cluster and the fluid phase independent of the order parameter choice.

As the composition and size of the critical cluster are not affected by the choice of order parameter, we set $\lambda = 0$ in order to investigate whether or not we observe diffusive equilibrium for species 2 for all noncritical clusters. To keep the composition of the fluid fixed, it would be convenient to use again Monte Carlo simulations in the semi-grand canonical ($NPT - \Delta\mu_{1,2,\alpha}$) ensemble. However, the acceptance probability of changing small spheres into large spheres is extremely small, which makes the equilibration time of the simulation prohibitively long, even when we use the augmented semigrand ensemble presented in Ref. [155], where the diameter of the particles is changed gradually in different stages. In order to solve this problem, we determine the free-energy barrier using the umbrella sampling technique in isothermal-isobaric MC simulations, in which the pressure $P_\alpha^o$, the temperature $T$, and the particle numbers $N_{1,\alpha}^o$ and $N_{2,\alpha}^o$ are kept fixed of the original metastable bulk phase. We perform successive simulations for each window, but in such a way that the composition $x_\alpha^o = N_{2,\alpha}/(N_{1,\alpha}^o + N_{2,\alpha}^o)$ of the metastable fluid phase is on average kept fixed during the growth of the nucleus. To this end, we first measure the instantaneous composition $x_\alpha$ of the fluid phase in the initial configuration for the successive umbrella sampling windows centered around a new order parameter value $\Phi$. If the composition of the fluid has changed more than 0.1%, we resize random particles in the fluid phase during an equilibration run until the fluid phase reaches its original composition $x_\alpha^o$. We then start the production run to measure the probability distribution function $P(\Phi)$ and the corresponding part of the free-energy barrier in a normal isobaric-isothermal MC simulation. We assume that the composition of the fluid phase during MC simulations of a single umbrella sampling window does not change significantly, since the cluster size is approximately constant. In order to determine the composition of the fluid phase, we first determine the largest crystalline cluster in the system by using the local bond-order parameter and cluster criterion as described in Sec. 8.4 with cutoff radius $r_c = 1.1\sigma_1$, dot-product cutoff $d_c = 0.7$, and number of solid bonds $\xi_c \geq 6$. The composition of the fluid is defined as $x_\alpha = (N_{2,\alpha}^o - N_2)/(N_{2,\alpha}^o + N_{1,\alpha}^o - N_2 - N_1)$ where $N_1$ is the number of large spheres in the cluster and $N_2$ is the number of small spheres which have at least 6 neighbors of large spheres in the cluster within cut-off distance $r_c = 1.1\sigma_1$. $N_{1,\alpha}^o$ and $N_{2,\alpha}^o$ denote the total number of large and small spheres in the MC simulation.

In addition, we determine the composition of the solid nucleus $x = N_2/N$. In order to avoid surface effects and defects in the crystal structure of the solid nucleus, we determine the fraction of octahedral holes that is occupied by a small sphere in the fcc lattice of the large spheres in the solid cluster. An octahedral hole is defined as a set of 6 large particles, where each particle is a neighbour of 4 other particles in the same set, and the octahedral hole is occupied by a small particle if all 6 large particles are within a cutoff radius of $0.22\sigma_1$ of the center-of-mass of this small sphere.

We first determine the nucleation barrier in a normal $N_{1,\alpha}^o N_{2,\alpha}^o P_\alpha^o T$ MC simulation using the umbrella sampling technique for system sizes $N_\alpha^o = N_{1,\alpha}^o + N_{2,\alpha}^o = 3000, 6000,$ and 9000 particles. The initial fluid composition is set to $x_\alpha^o = 0.5$ and reduced pressure $P^* = \beta P_\alpha^o \sigma_1^3 = 25$. We plot the Gibbs free energy $\Delta G/k_B T$ as a function of the number of large spheres $N_1$ in the largest crystalline cluster in Fig. 8.4. We observe that the nucleation barrier height and critical cluster size decreases upon increasing system size. This can be explained by a change in the composition of the metastable fluid phase during the growth of a crystalline cluster. In Fig. 8.5, we plot the composition of the metastable fluid phase as a function of the cluster size $N_1$ for the various system sizes. We clearly find that the fluid composition changes significantly during the growth of a solid nucleus for smaller system sizes. In order to corroborate this result, we perform umbrella sampling MC simulations in which the composition of the metastable fluid phase is kept fixed in each successive umbrella sampling window using the method as described above. The composition of the fluid phase is indeed kept fixed by this method as shown in Fig. 8.5. The nucleation barrier as obtained by fixing the composition of the metastable fluid phase is presented in Fig. 8.4. As the nucleation barrier calculated at fixed fluid composition should correspond to an infinitely large system size, we plot the barrier heights $\Delta G^*/k_B T$ as a function of $1/N_\alpha^o$ with $N_\alpha^o = N_{1,\alpha}^o + N_{2,\alpha}^o$. We find that the barrier height depends linearly on $1/N_\alpha^o$ within errorbars. Moreover, extrapolating the barrier heights obtained from $N_{1,\alpha}^o N_{2,\alpha}^o P_\alpha^o T$ MC simulations to the thermodynamic limit, we find that the finite-size corrected barrier height agrees well within errorbars with the barrier height determined from umbrella sampling MC simulations with fixed fluid composition corresponding to an infinitely large system size. In addition, we plot the composition of the solid cluster as a function of cluster size $N_1$ in Fig. 8.5, and we find no strong dependence of the cluster composition on system size.

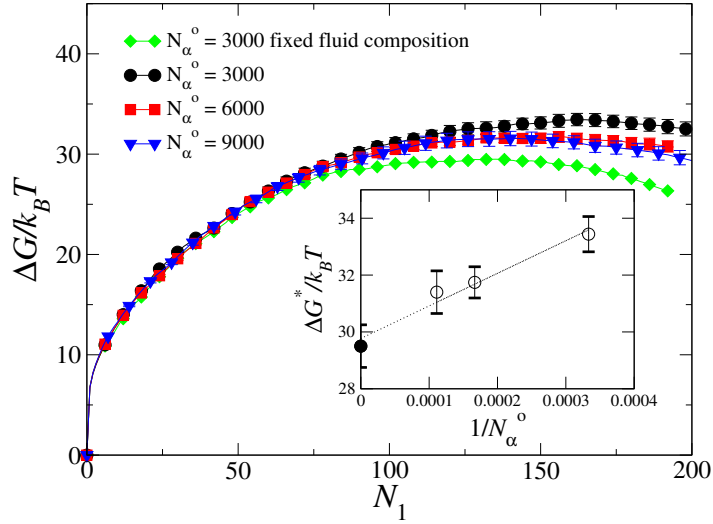

**Figure 8.4**:  a) Gibbs free energy $\Delta G/k_B T$ as a function of number of large particles $N_1$ in the largest crystalline cluster using normal $N_1^o N_2^o PT$ MC simulations and isobaric-isothermal MC simulations with fixed fluid composition $x_\alpha = 0.5$, and pressure $P^* = \beta P_\alpha^o \sigma_1^3 = 25$.  b) Free-energy barrier height $\Delta G^*/k_B T$ as a function of $1/N_\alpha^o$.

Finally, we determine the composition of (non)critical clusters for the nucleation of the interstitial solid solution for four different fluid compositions $x_\alpha^o = 0.2, 0.5, 0.7$ and $0.8$ at statepoints well-inside the fluid-solid coexistence region using umbrella sampling MC simulations with fixed fluid composition and system size $N_\alpha^o = 3000$. Following Ref. [155], the "supercooling" was kept fixed, i.e., $P^*/P_{coex}^* = 1.2$, where $P_{coex}^*$ is the pressure at the bulk fluid-solid coexistence at the corresponding fluid composition. We note however that these statepoints correspond to different values for the supersaturation, and can therefore lead to significantly different barrier heights. We determine the Gibbs free-energy barrier and the cluster composition as a function of cluster size $N_1$ using umbrella sampling MC simulations, and plot the results in Fig. 8.6 and 8.7 for the four different fluid compositions. In Fig. 8.7, the dashed lines indicate the compositions predicted by Eq. 8.41 with $\lambda = 0$, i.e., chemical equilibrium for species 2 in the clusters and the metastable fluid phase. For comparison, we also plot the composition of the coexisting solid phase at $P^*$. We clearly observe that the measured cluster compositions obtained from umbrella sampling MC simulations are in good agreement with the predictions from CNT for cluster sizes larger than 30, which predicts chemical equilibrium for the small spheres in the cluster and the metastable parent phase. If we now take a closer look at the statepoint defined by $x^o_\alpha = 0.2$ and $P^*/P^*_{coex} = 1.2$ for the metastable fluid phase, we find from Ref. [161] that the composition of the coexisting fluid and solid phase after full phase separation should be $x \simeq 0.47$ and 0.15, respectively. Interestingly, we find that the composition of the nucleating clusters is much lower ($x \simeq 0.07$) than that of the coexisting bulk crystal phase. Hence, the phase separation is mainly driven by the nucleation of large spheres while maintaining chemical equilibrium for the smaller species throughout the whole system. Only when the chemical potential of the large spheres in the metastable fluid is sufficiently low due to a depletion of large spheres as a result of crystal nucleation and crystal growth, small spheres will diffuse into the crystal phase in order to increase the composition of the solid phase. However, we note that the chemical equilibrium condition for the smaller species only holds for the present order parameter choice $\Phi = N_1$, whereas any other choice of order parameter would certainly yield different results for the cluster composition.

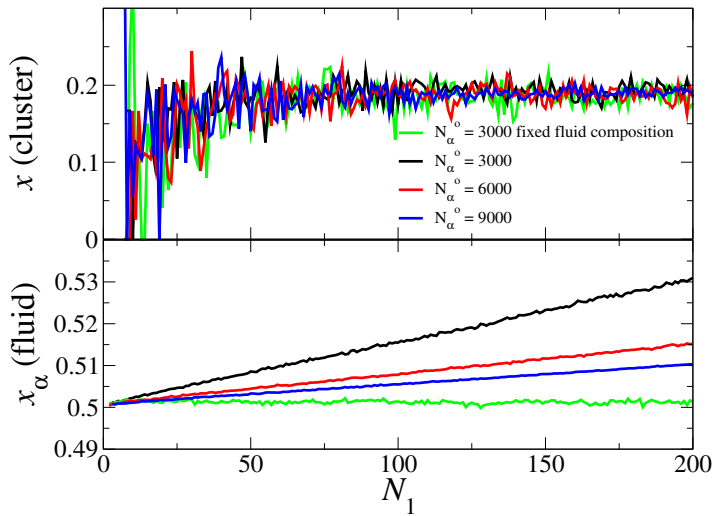

**Figure 8.5**: The composition of the largest crystalline cluster $x = N_2/N$ (top) and the metastable fluid phase $x_\alpha$(bottom) as a function of cluster size $N_1$. The green lines denotes the simulation in which the composition of the fluid was reset to the original value at the start of each US window. The other lines correspond to normal $N^o_{1,\alpha}N^o_{2,\alpha}P^o_\alpha T$ MC simulations, where the overall composition of the system is kept fixed for various system sizes.

For highly asymmetric binary hard-sphere mixtures, where the stable solid phase corresponds to an fcc of large spheres with a dispersed fluid of small particles, one would naively expect that the small particles are always in chemical equilibrium during the nucleation process. Hence, in order to study crystal nucleation in highly asymmetric mixtures, one can employ an effective pairwise depletion potential description as described by Bob Evans and coworkers in Ref. [162–164] provided that three- and higher-body interactions are negligible and the depletion potentials are determined at fixed chemical potential of the small spheres. Such an effective pair potential approach was employed in a nucleation study in the vicinity of a critical isostructural solid-solid transition in a binary mixture of hard spheres with size ratio $q = \sigma_2/\sigma_1 = 0.1$, but this study showed according to the authors a breakdown of classical nucleation theory [165]. It would be interesting to investigate whether or not the breakdown is caused by the (false) assumption of chemical equilibrium of small spheres during the nucleation process. For less asymmetric binary hard-sphere mixtures, where the small spheres cannot diffuse freely in the solid cluster, chemical equilibrium of the smaller species is harder to maintain, especially when the nucleated crystal phase has long-range crystalline order for both species as in the case of a superlattice structure where the chemical potentials of the two species are not independent as it is determined by the stoichiometry of the crystal structure. It would be interesting to investigate at which size ratio and pressures this crossover occurs.

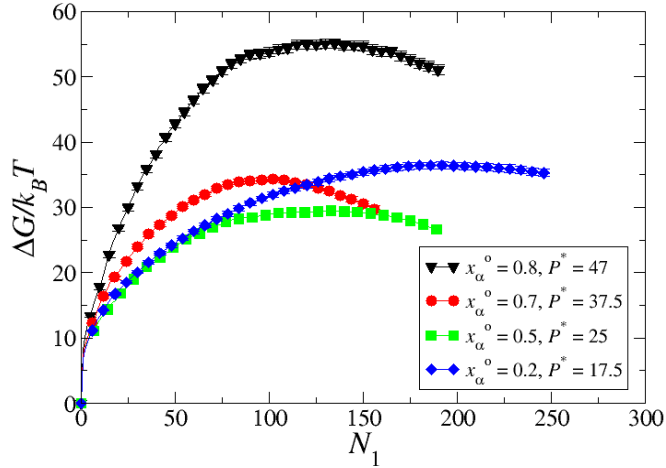

**Figure 8.6**: Gibbs free energy $\Delta G/k_B T$ as a function of cluster size $N_1$ for four different fluid compositions $x_\alpha^o = 0.2, 0.5, 0.7$, and $0.8$ for a binary mixture of hard spheres with size ratio 0.3 at 20% supercooling, i.e., $P^*/P_{coex}^* = 1.2$ with $P_{coex}^*$ the bulk coexistence pressure.

## 8.7 Conclusions

In this chapter, we have studied crystal nucleation in a binary mixture of hard spheres and we have investigated what the effect is of the choice of order parameter on the composition and size of both critical and noncritical clusters. We have studied nucleation of a substitutional solid solution in a simple toy model of identical hard spheres but tagged with different colors and we investigate the nucleation of an interstitial solid solution in a binary hard-sphere mixture with a diameter ratio $q = 0.3$. In order to study nucleation of a crystal phase in computer simulations, a one-dimensional order parameter is usually defined to identify the solid phase from the supersaturated fluid phase. We have shown that the choice of order parameter can strongly influence the composition of noncritical clusters, as the free-energy landscape in the two-dimensional composition plane $(N_1, N_2)$ is projected onto a one-dimensional order parameter, say $\Phi = N_1 + \lambda N_2$, in umbrella sampling MC simulations. This is supported by the good agreement that we found between our results on the composition of noncritical clusters obtained from umbrella sampling MC simulations and the predictions from CNT for the nucleation of a substitutional solid solution in a toy model. While the effect is clearly visible in the case of a binary system, it should occur more generally whenever a higher-dimensional free-energy landscape is projected onto a single order parameter. For the nucleation of an interstitial solid solution in a highly asymmetric hard-sphere system, we found that the composition of noncritical clusters is determined by the chemical equilibrium condition of the small spheres in the cluster and the fluid phase, as the partial particle volume of the small spheres in the solid phase can be neglected. We compared the composition of the noncritical clusters obtained from umbrella sampling MC simulations and the theoretical prediction from CNT, and found again good agreement. More importantly, we find that the barrier height and the composition of the critical cluster are not significantly affected by the choice of order parameter. As a result, critical clusters and the barrier height should be comparable even with different order parameters.

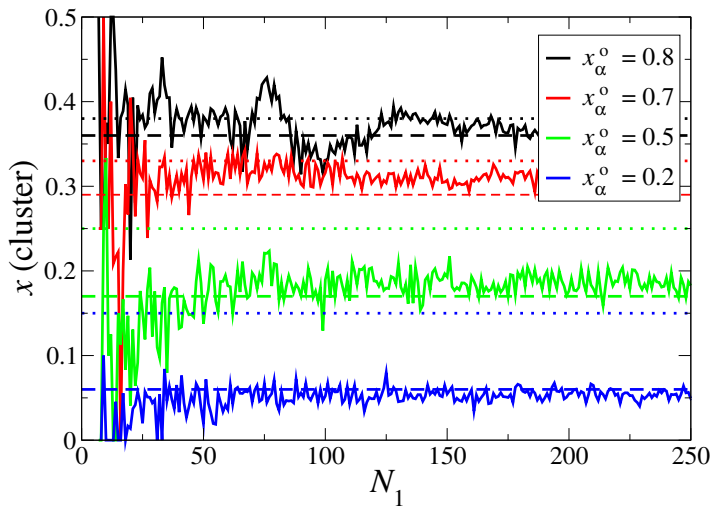

**Figure 8.7**:  Cluster compositions $x = N_2/N$ as a function of cluster size $N_1$ for four different fluid compositions $x_\alpha^o = 0.2, 0.5, 0.7$, and $0.8$ for a binary mixture of hard spheres with size ratio 0.3 at supercooling $P^*/P_{coex}^* = 1.2$.  The long dashed lines denote the composition predicted by CNT, which corresponds to chemical equilibrium of species 2 in the solid clusters and the metastable fluid phase, while the dotted lines denote the composition of the coexisting bulk crystal phase.

## 8.8   Acknowledgements

I would like to thank Dr. F. Smallenburg for the simulations on the nucleation of substitutional solid solution and Dr. L. Filion for the calculation of phase diagram.

# Part II

# Phase behavior in colloidal systems

# 9

## Phase diagram of colloidal hard superballs: from cubes via spheres to octahedra

The phase diagram of colloidal hard superballs, of which the shape interpolates between cubes and octahedra via spheres, is determined by free-energy calculations in Monte Carlo simulations. We discover not only a stable face-centered cubic (fcc) plastic crystal phase for near-spherical particles, but also a stable body-centered cubic (bcc) plastic crystal close to the octahedron shape, and in fact even coexistence of these two plastic crystals with a substantial density gap. The plastic fcc and bcc crystals are, however, unstable in the cube and octahedron limit, suggesting that the rounded corners of superballs play an important role in stablizing the rotator phases. In addition, we observe a two-step melting phenomenon for hard octahedra, in which the Minkowski crystal melts into a metastable bcc plastic crystal before melting into the fluid phase.

## 9.1   Introduction

Recent breakthroughs in particle synthesis have resulted in a spectacular variety of anisotropic nanoparticles such as cubes, octapods, tetrapods, octahedra, icecones, etc. [3]. A natural starting point to study the self-assembled structures of these colloidal building blocks is to view them as hard particles [1]. Not only can these hard-particle models be used to predict properties of suitable experimental systems, but such models also provide a stepping stone towards systems where soft interactions play a role [31, 166]. Moreover, the analysis of hard particles is of fundamental relevance and raises problems that influence fields as diverse as (soft) condensed matter [3, 167–169], mathematics [168, 170], and computer science [171]. In this light the concurrent boom in simulation studies of hard anisotropic particles is not surprising [85, 86, 114, 168–170, 172–176].

The best-known hard-particle system consists of hard spheres, which freeze into close-packed hexagonal (cph) crystal structures [171], of which the ABC-stacked cph crystal, better known as the face-centered cubic (fcc) crystal phase, is thermodynamically stable [65]. Hard anisotropic particles can form liquid-crystalline equilibrium states if they are sufficiently rod- or disclike [114, 175], but particles with shapes that are close-to-spherical tend to order into plastic crystal phases, also known as rotator phases [85, 86, 114]. In fact, simple guidelines were recently proposed to predict the plastic- and liquid-crystal formation only on the basis of rotational symmetry and shape anisotropy of hard polyhedra [169]. In this chapter we will take a different approach, based on free-energy calculations, and address the question whether and to what extent rounding the corners and faces of polyhedral particles affects the phase behavior. Such curvature effects are of direct relevance to experimental systems, in which sterically and charged stabilised particles can often not be considered as perfectly flat-faced and sharp-edged [177]. For instance, recent experiments on nanocube assemblies show a continuous phase transformation between simple cubic and rhombohedral phases by increasing the ligand thickness and hence the particle sphericity [166].

## 9.2   Methodology

### 9.2.1   Model

In this chapter, we study a system of colloidal hard superballs in order to address these problems. A superball is defined by the inequality

$$\left|\frac{x}{a}\right|^{2q} + \left|\frac{y}{a}\right|^{2q} + \left|\frac{z}{a}\right|^{2q} \leq 1, \tag{9.1}$$

where $x$, $y$ and $z$ are Cartesian coordinates with $q$ the deformation parameter and with $a$ the radius of the particle. The shape of the superball interpolates smoothly between two Platonic solids, namely the octahedron ($q = 0.5$) and the cube ($q = \infty$) via the sphere ($q = 1$) as shown in Fig. 9.1. The volume of a superball with the shape parameter $q$ is

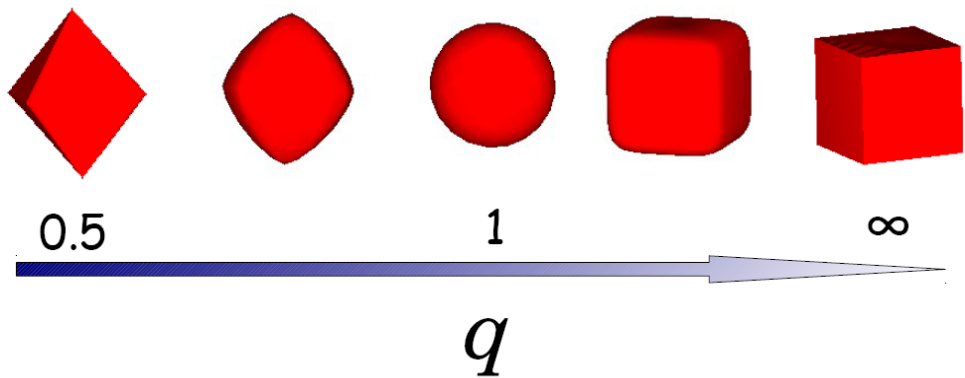

0.5      1      ∞

$q$

**Figure 9.1**: The shape of superballs interpolates between octahedra ($q = 0.5$) and cubes ($q = \infty$) via spheres ($q = 1$).

given by

$$
\begin{aligned}
V(q) &= 8a^3 \int_0^1 \int_0^{(1-x^{2q})^{1/2q}} (1 - x^{2q} - y^{2q})^{1/2q} \mathrm{d}\, y \, \mathrm{d}\, x \\
&= \frac{8a^3 \left[ \Gamma \left( 1 + 1/2q \right) \right]^3}{\Gamma \left( 1 + 3/2q \right)}.
\end{aligned}
\tag{9.2}
$$

By determining the phase diagram of these superballs as a function of $q$, we discovered a thermodynamically stable body-centered (bcc) plastic crystal phase for octahedron-like superballs. To the best of our knowledge *no* plastic crystals other than cph structures have so far been observed for hard particles. Moreover, we find that bcc and fcc plastic crystal phases are unstable for hard octahedra and hard cubes, respectively. Therefore, rounded faces and edges play an important role in stabilizing rotator phases, while flat faces tend to stabilize crystals.

### 9.2.2 Fluid phase

We employ standard $NPT$ Monte Carlo simulations to obtain the equation of state (EOS) for the fluid phase, and determine the free energy by measuring the free energy at density $\rho$ by integrating the EOS from reference density $\rho_0$ to $\rho$:

$$
\frac{F(\rho)}{N} = \frac{F(\rho_0)}{N} + \int_{\rho_0}^{\rho} \frac{P(\rho')}{\rho'^2} \, \mathrm{d}\, \rho'
\tag{9.3}
$$

where $F(\rho_0)/N = \mu(\rho_0) - P(\rho_0)/\rho_0$ is the Helmholtz free energy per particle at density $\rho_0$ with $N$ the number of particles and $\mu(\rho_0)$ the chemical potential which is calculated by the Widom's particle insertion method [10]. The calculated chemical potentials at the reference density $\rho_0$ for various superballs are listed in Table 9.1.

| $q$ | $\rho_0 a^3$ | $N$ | $\mu(\rho_0)/k_B T$ | $F(\rho_0)/Nk_B T$ |
|------|--------|-----|-------------|-------------|
| 0.7 | 0.1 | 500 | 1.85523 | -1.63213 |
| 0.79248 | 0.1 | 500 | 3.32035 | -1.14744 |
| 0.85 | 0.05 | 500 | -0.945694 | -3.07702 |
| 1.75 | 0.02 | 500 | -2.56824 | -4.29418 |
| 2.5 | 0.036 | 500 | 0.718079 | -2.6957 |
| 3.0 | 0.028 | 500 | -0.609041 | -3.30777 |

**Table 9.1**: Helmholtz free energy per particle, $F(\rho_0)/Nk_B T$, and the chemical potential, $\mu(\rho_0)$, in the fluid phase of hard superballs with various shape parameter $q$ at density $\rho_0$ calculated from the Widom's particle insertion method.

## 9.2.3 Crystal phases

For the free energy of a crystal we use the Einstein integration method, and the Helmholtz free energy $F$ of a crystal is

$$F(N, V, T) = F_{\text{Einst}}(N, V, T) - \int_0^{\lambda_{\max}} \mathrm{d}\lambda \left\langle \frac{\partial U_{\text{Einst}}(\lambda)}{\partial \lambda} \right\rangle \tag{9.4}$$

where $V$ and $T$ are the volume and temperature of the system, respectively, with $k_B$ the Boltzmann constant, and $F_{\text{Einst}}$ is the free energy of the ideal Einstein crystal given by

$$\begin{aligned}
\frac{F_{\text{Einst}}(N, V, T)}{k_B T} &= -\frac{3(N-1)}{2} \ln\left(\frac{\pi k_B T}{\lambda_{\max}}\right) + N \ln\left(\frac{\Lambda_t^3}{a^3}\right) + N \ln\left(\frac{\Lambda_r}{a}\right) \\
&+ \ln\left(\frac{a^3}{V N^{1/2}}\right) - \ln\left\{\frac{1}{8\pi^2} \int \mathrm{d}\theta \sin(\theta) \mathrm{d}\phi \mathrm{d}\chi \right. \\
&\left. \times \exp\left[-\frac{\lambda_{\max}}{k_B T}(\sin^2 \psi_{ia} + \sin^2 \psi_{ib})\right]\right\}
\end{aligned} \tag{9.5}$$

and

$$U_{\text{Einst}}(\lambda) = \lambda \sum_{i=1}^{N} [(\mathbf{r}_i - \mathbf{r}_{i,0})^2 + (\sin^2 \psi_{ia} + \sin^2 \psi_{ib})] \tag{9.6}$$

is the aligning potential for fixing the particles onto a crystal lattice where $(\mathbf{r}_i - \mathbf{r}_{i,0})$ is the displacement of the particle $i$ from its rest position in the ideal Einstein crystal and the angles $\psi_{ia}$ and $\psi_{ib}$ are the minimum angles formed by the two field vectors, i.e. a and b, in the ideal Einstein crystal and the vectors defining the orientation of the particle in the crystal. $\Lambda_t$ and $\Lambda_r$ in Eq. 9.5 are the translational and orientational thermal wavelengths of the particles, respectively, and both are set to $a$. The last term on the right hand side of Eq. 9.5 is calculated by Monte Carlo integrations with $\lambda_{\max}/k_B T = 1000$ as

$$\ln\left\{\frac{1}{8\pi^2} \int \mathrm{d}\theta \sin(\theta) \mathrm{d}\phi \mathrm{d}\chi \exp\left[-\frac{\lambda_{\max}}{k_B T}(\sin^2 \psi_{ia} + \sin^2 \psi_{ib})\right]\right\}\bigg|_{\lambda_{\max}=1000} = -10.180034$$

### 9.2.4 Plastic crystal phases

For the free energy calculations of a plastic crystal phase, we use a soft potential between particles given by

$$\frac{\varphi(i,j)}{k_B T} = \begin{cases} \gamma[1 - A(1 + \zeta(i,j))] & \zeta(i,j) < 0 \\ 0 & \text{otherwise} \end{cases} \tag{9.7}$$

where $\zeta(i,j)$ is the overlapping potential as defined on Eq. 9.12 which is negative when two particle $i$ and $j$ are overlapping and positive otherwise [178], and $\gamma$ is the integration parameter with the constant $A = 0.9$ [179]. This method was introduced in Ref. [179], and allow us to change gradually from a non-interacting system, i.e., $\gamma = 0$, to a plastic crystal phase of hard superballs where $\gamma = \gamma_{max}$. The Helmholtz free energy of the plastic crystal is then given by

$$\begin{aligned} F(N,V,T) \;=\; & F_{\text{Einst}}(N,V,T) - \int_0^{\lambda_{max}} d\lambda \left\langle \frac{\partial U_{\text{Einst}}(\lambda)}{\partial \lambda} \right\rangle_{\gamma_{max}} \\ & + \int_0^{\gamma_{max}} d\gamma \left\langle \frac{\partial \sum_{i \neq j} \varphi(i,j)}{\partial \gamma} \right\rangle_{\lambda_{max}} \end{aligned} \tag{9.8}$$

The calculated free energies $F(\rho_0)$ for the crystals and plastic crystal structures at reference densities $\rho_0 a^3$ are listed in Table 9.2.

| $q$ | crystal type | $N$ | $\rho_0 a^3$ | $F(\rho_0)/Nk_B T$ |
|------|-------------|------|-----------|-----------------|
| 0.7 | plastic bcc | 512 | 0.212 | 4.22893 |
| 0.7 | plastic fcc | 500 | 0.212 | 4.27468 |
| 0.79248 | plastic bcc | 512 | 0.175892 | 3.80535 |
| 0.79248 | plastic fcc | 500 | 0.175892 | 3.67894 |
| 0.85 | plastic fcc | 500 | 0.171887 | 4.45765 |
| 0.85 | plastic bcc | 432 | 0.171887 | 4.84055 |
| 0.85 | bct | 512 | 0.207163 | 11.9065 |
| 1.75 | deformed $C_1$ | 512 | 0.11654 | 9.42461 |
| 1.75 | plastic fcc | 500 | 0.0976 | 4.54847 |
| 2.5 | deformed $C_1$ | 512 | 0.10675 | 9.58457 |
| 2.5 | plastic fcc | 500 | 0.077 | 3.13303 |
| 3.0 | deformed $C_1$ | 512 | 0.10522 | 10.0684 |
| 3.0 | plastic fcc | 500 | 0.076 | 3.77605 |

**Table 9.2**: Helmholtz free energy per particle, $F(\rho_0)/Nk_B T$, for the crystal phases of hard superballs with various shape parameter values $q$ at density $\rho_0$ calculated from Einstein integration method, where fcc, bcc and bct mean face-centered cubic, body-centered cubic and body-centered tetragonal crystal phases, respectively, and the $C_1$ crystal is defined in Ref. [173].

## 9.3  Results and Discussions

### 9.3.1  Cube-like superballs ($1 < q < \infty$)

Following Refs. [76, 172], we first calculate the close-packed structures for systems of hard superballs. For cube-like particles, it is found that at close packing there are so-called $C_0$ and $C_1$ crystal phases in accordance with Ref. [173]. When we perform *NPT* Monte Carlo simulations with variable box shape to determine the EOS of the crystal phase as shown in Fig. 9.2, our simulation results show that both the $C_0$ and the $C_1$ crystals deform with decreasing density. The lattice vectors for $C_1$ crystals are given by $\mathbf{e}_1 = 2^{1-1/2q}\,\mathbf{i} + 2^{1-1/2q}\,\mathbf{j}$, $\mathbf{e}_2 = 2^{1-1/2q}\,\mathbf{i} + 2^{1-1/2q}\,\mathbf{k}$, $\mathbf{e}_3 = 2(s+2^{-1/2q})\,\mathbf{i} - 2s\,\mathbf{j} - 2s\,\mathbf{k}$, where $\mathbf{i}$, $\mathbf{j}$ and $\mathbf{k}$ are the unit vectors along the axes of the particle, $s$ is the smallest positive root of the equation $(s+2^{-1/2q})^{2q} + 2s^{2q} - 1 = 0$, and there is one particle in the unit cell [173]. For instance, when $q = 2.5$, one finds that $\langle \mathbf{e}_1, \mathbf{e}_2 \rangle = 0.5$, $\langle \mathbf{e}_1, \mathbf{e}_3 \rangle = \langle \mathbf{e}_3, \mathbf{e}_2 \rangle = 0.60552$, $|\mathbf{e}_2|\,/\,|\mathbf{e}_1| = 1$ and $|\mathbf{e}_3|\,/\,|\mathbf{e}_1| = 0.825737$, where $\langle \mathbf{e}_i, \mathbf{e}_j \rangle$ is the cosine of the angle between $\mathbf{e}_i$ and $\mathbf{e}_j$. The calculated angles and the length ratios between lattice vectors as a function of packing fraction $\phi$ for the cube-like particles with $q = 2.5$ are shown in Fig. 9.3. We find that at packing fractions approaching close packing, the crystal remains in the $C_1$ phase. With decreasing packing fraction, the crystal lattice deforms towards an fcc structure: $\langle \mathbf{e}_1, \mathbf{e}_2 \rangle = \langle \mathbf{e}_1, \mathbf{e}_3 \rangle = \langle \mathbf{e}_2, \mathbf{e}_3 \rangle = 0.5$ and $|\mathbf{e}_2|\,/\,|\mathbf{e}_1| = |\mathbf{e}_3|\,/\,|\mathbf{e}_1| = 1$.

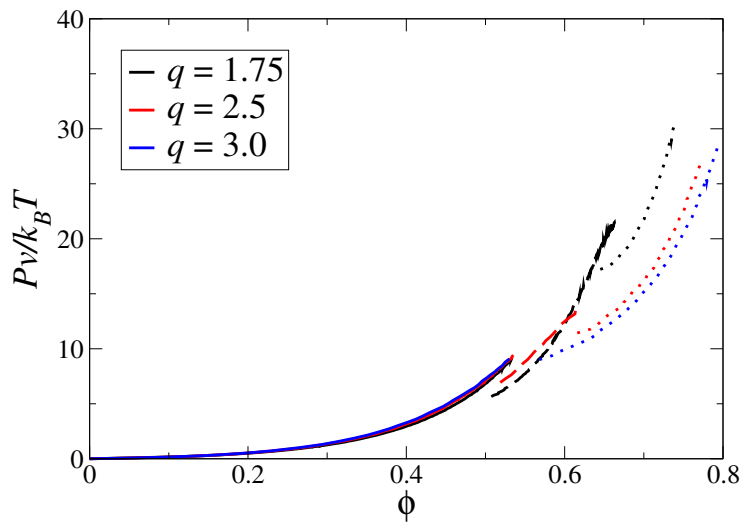

**Figure 9.2**:  Equation of state for cube-like hard superballs with various shape parameters $q$. The solid, dashed and dotted lines are fluid, plastic fcc and deformed $C_1$ phases, respectively. $P$ and $\phi$ are the pressure and packing fraction of the system, and $v$ is the volume of a particle.

Moreover, when $1 < q < 3$, it is found that the deformed $C_0$ and deformed $C_1$ crystal melt into an fcc plastic crystal phase. Using the Einstein integration method, we calculated the Helmholtz free energy as a function of packing fraction for both the fcc plastic crystal and the deformed $C_1/C_0$ crystal phases [10]. Combined with the free-energy calculations for the fluid phase done by the Widom's particle insertion method, we obtain the phase boundaries in the phase diagram shown in Fig. 9.4. The part of the phase diagram for hard cube-like superballs roughly agrees with the empirical phase diagram by Batten *et al.* [174]. At high packing fractions, there are stable deformed $C_0$ and $C_1$ phases. When $q > 1.1509$, the close-packed structure is the $C_1$ crystal, whereas it is the $C_0$ crystal whenever $1 < q < 1.1509$ [173]. To determine the location of the transition from the deformed $C_0$ crystal to the deformed $C_1$ crystal, we performed two series of $NPT$ MC simulations with increasing value of $q$ for the first series and decreasing $q$ for the second series of simulations at pressure $P^* = Pv/k_BT \simeq 250$, with $k_B$ the Boltzmann constant, $T$ the temperature, and $v$ the volume of the particle [175]. The first series started from a $C_0$ crystal phase, while the second series of simulations started from a $C_1$ crystal phase. Our simulations show that the phase transition occurred around $q = 1.09$ at packing fraction $\phi = 0.736$ as shown by the asterix in Fig. 9.4. Moreover, for hard cubes ($q = \infty$) the $C_1$ crystal is a simple cubic (sc) crystal. In contrast to the result of Ref. [169], the free-energy calculations show that there is no stable cubatic phase between the sc crystal and the fluid phase for systems of hard cubes [176].

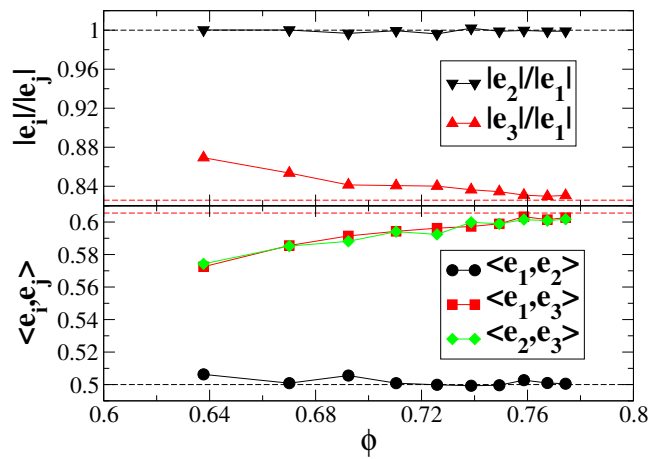

**Figure 9.3**: The deformation of the crystal unit cell with lattice vectors $\mathbf{e}_i$ as a function of packing fraction $\phi$ in a system of hard superballs with $q = 2.5$. The dashed lines in the figures indicate the values for the $C_1$ crystal.

### 9.3.2 Octahedron-like superballs ($0.5 \leq q < 1$)

The other part of the phase diagram concerns the octahedron-like superballs. For $0.79248 < q < 1$, we obtained a denser structure than the predicted $O_0$ lattice of Ref. [173]. For instance, after compressing the system to pressures around $P^* = 10^7$ at $q = 0.85$, we obtained a bct crystal with $\phi = 0.7661$. This is denser than the $O_0$ crystal, which achieves $\phi = 0.7656$ at $q = 0.85$. Note however that these two crystals are very similar to each other, since $O_0$ is also a form of a bct lattice. The only difference is that the orientation of the particles in the $O_0$ crystal is the same as the symmetry of the axes in the crystal lattice, while in our bct crystal there is a small angle between these two orientations in the square plane of the crystal. Furthermore, for $q < 0.79248$, we also found a crystal with a denser packing than the predicted $O_1$ crystal in Refs. [173, 180]. For $q = 0.7$, we performed floppy-box MC simulations with several particles to compress the system to a high pressure state, i.e., $P^* = 10^7$. We found a deformed bcc (dbcc) crystal shown in Fig. 9.4, which is an intermediate form between the bcc lattice and the Minkowski crystal. The lattice vectors are $\mathbf{e}_1 = 0.912909\mathbf{i} + 0.912403\mathbf{j} - 0.912165\mathbf{k}$, $\mathbf{e}_2 = -0.271668\mathbf{i} + 1.80916\mathbf{j} - 0.288051\mathbf{k}$ and $\mathbf{e}_3 = 0.28834\mathbf{i} - 0.272001\mathbf{j} - 1.80882\mathbf{k}$, where $\mathbf{i}$, $\mathbf{j}$ and $\mathbf{k}$ are the unit vectors along the axes of the particle. One can find that our dbcc crystal are very close the predicted $O_1$ crystal, whose lattice vectors are $\mathbf{e}_1 = 0.912492\mathbf{i} + 0.912492\mathbf{j} - 0.912492\mathbf{k}$, $\mathbf{e}_2 = -0.2884\mathbf{i} + 1.80629\mathbf{j} - 0.2884\mathbf{k}$, and $\mathbf{e}_3 = 0.2884\mathbf{i} - 0.2884\mathbf{k} - 1.80629\mathbf{k}$. However, it has a packing fraction of $\phi = 0.832839$ which is denser than the predicted $O_1$ crystal with $\phi = 0.824976$ in Refs. [173, 180] by roughly 1%. In Ref. [173], the $O_0$ and $O_1$ phases are found to switch at $q = 0.79248$. We also observed that the bct and dbcc crystals both transform into the bcc phase at $q = 0.79248$. Moreover, the equation of state for hard octahedron-like superballs is shown in Fig. 9.5.

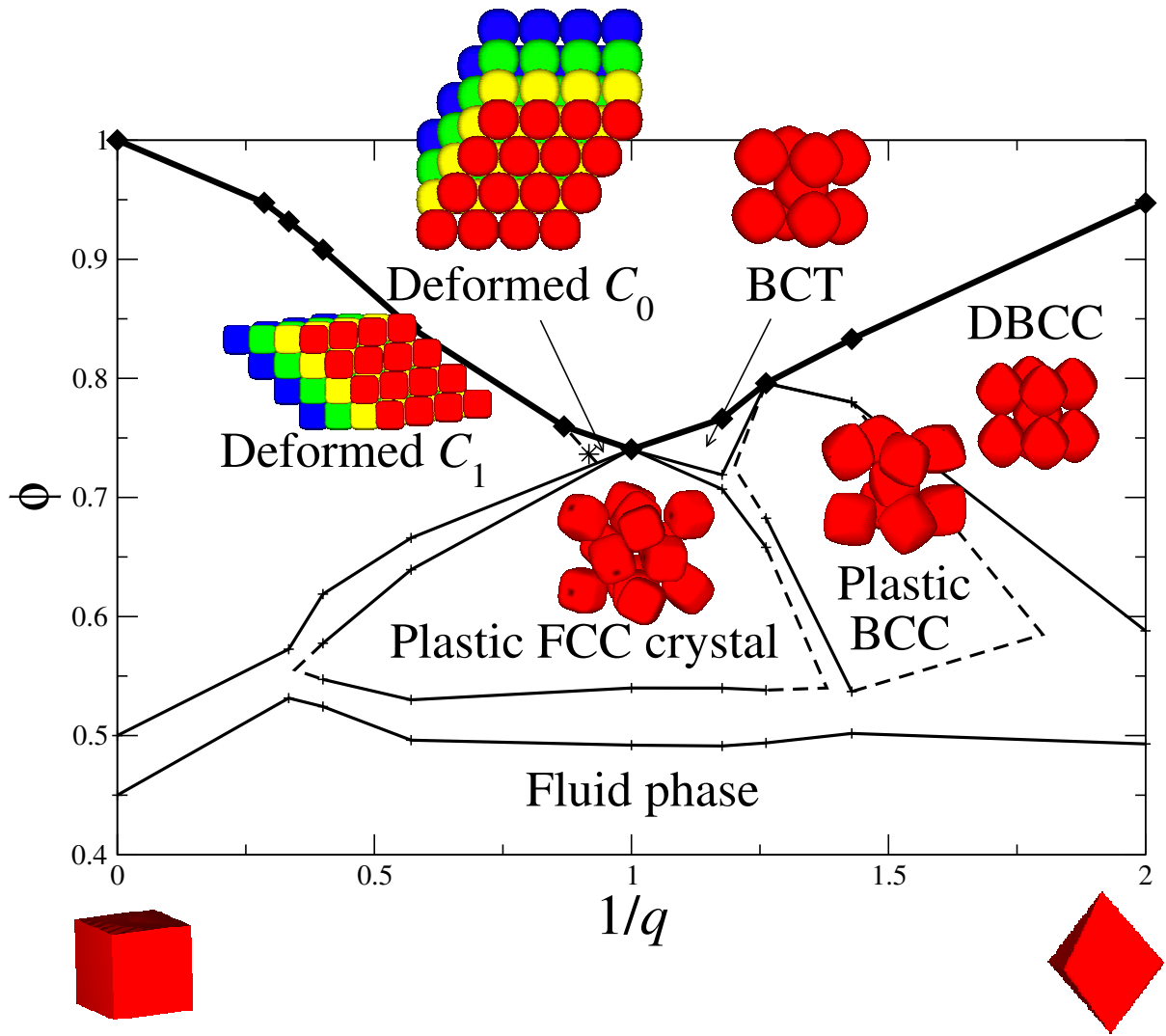

**Figure 9.4**: Phase diagram for hard superballs in the $\phi$ (packing fraction) versus $1/q$ representation where $q$ is the deformation parameter. Here the $C_1$ and $C_0$ crystals are defined in Ref. [173], where the particles of the same color are in the same layer of stacking. The solid diamonds indicate the close packing, and the locations of triple points are determined by extrapolation as shown by the dashed lines. The phase boundaries for hard cubes are taken from Ref. [176].

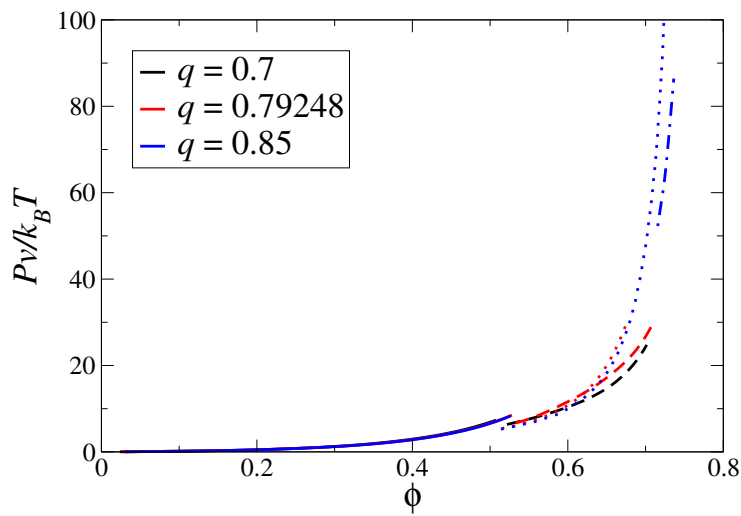

**Figure 9.5**: Equation of state for octahedron-like hard superballs with various shape parameters $q$. The solid, dashed, dotted and dash-dotted lines are fluid, plastic bcc, plastic fcc and bct phases, respectively. $P$ and $\phi$ are the pressure and packing fraction of the system, and $v$ is the volume of a particle.

As shown in Fig. 9.4, when the shape of the superballs is close to spherical, i.e., $0.7 < q < 3$, there is always a stable fcc plastic crystal phase. Surprisingly, when the shape of superballs is octahedron-like, we find a stable bcc plastic crystal phase. Moreover, around $q = 0.8$ we even find a fairly broad two-phase regime where a low-density fcc plastic crystal coexists with a high-density bcc plastic crystal phase. In order to quantify the orientational order in the bcc plastic crystal, we calculate the cubatic order parameter $S_4$ given by [174]

$$S_4 = \max_{\mathbf{n}} \left\{ \frac{1}{14N} \sum_{i,j} \left( 35|\mathbf{u}_{ij} \cdot \mathbf{n}|^4 - 30|\mathbf{u}_{ij} \cdot \mathbf{n}|^2 + 3 \right) \right\}, \qquad (9.9)$$

where $\mathbf{u}_{ij}$ is the unit vector of the $j$-th axis of particle $i$, $N$ the number of particles, and $\mathbf{n}$ is the unit vector for which $S_4$ is maximized. The cubatic order parameter $S_4$ is shown in Fig. 9.6 as a function of packing fraction for a system of superballs with $q = 0.7$. We observe that at low packing fractions $\langle S_4 \rangle \simeq 0.2$, which means that there is a very weak orientational order in the system. With increasing packing fraction, the cubatic order parameter increases monotonically to around 0.65 at a packing fraction of 0.7, which is indicative of a medium-ranged orientationally ordered system. This suggests that the entropic repulsion due to the rotation of the octahedron-like superballs stablizes the bcc lattice.

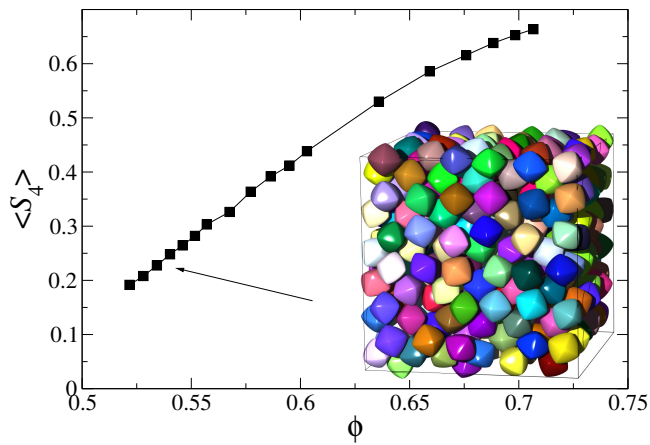

**Figure 9.6**: Cubatic order parameter $S_4$ as a function of packing fraction $\phi$ for a bcc plastic crystal phase of hard superballs with $q = 0.7$. The inset shows a typical configuration of a bcc plastic crystal of hard superballs with $q = 0.7$ and $\phi = 0.54$.

Due to the numerical instability in the overlap algorithm, we are not able to investigate systems of superballs with $q < 0.7$ [178]. However, we can use the separating axis theorem [169] to simulate hard superballs with $q = 0.5$, i.e., perfect octahedra. When we compressed the system from a fluid phase, we did not observe the spontaneous formation of a crystal phase in our simulation box within our simulation time. When we expand the Minkowski crystal, which is the close-packed structure of octahedra, in $NPT$ MC simulations by decreasing the pressure, the system melts into a bcc plastic crystal phase as shown in Fig. 9.7. We also calculated the free energy for these three phases to determine the phase boundaries. To exclude finite-size effects in the free-energy calculation of crystal phases, we performed Einstein integration for systems of $N = 1024$, 1458, and 2000 particles, and applied a finite-size correction [10]. We confirmed the errors in the free-energy calculations to be on the order of $10^{-4}k_BT$ per particle. The calculated free-energy densities for the three phases are shown in Fig. 9.7. Employing a common tangent construction, we found that there is only phase coexistence between a fluid phase and a Minkowski crystal phase, while the bcc plastic crystal phase is metastable. However, the free-energy differences between the fluid and the plastic crystal phase at the bulk coexistence pressure is very small , i.e., $\sim 10^{-2}k_BT$ per particle, and the Minkowski crystal does melt into a bcc plastic crystal before melting into the fluid phase. This explains why in a recent report the bcc (plastic) crystal was misidentified as a stable phase in an empirical phase diagram of octahedra [181]. Our results thus show that the rounded corners of octahedra play an important role in stablizing the bcc plastic crystal phase which is a new plastic crystal phase for systems of hard particles.

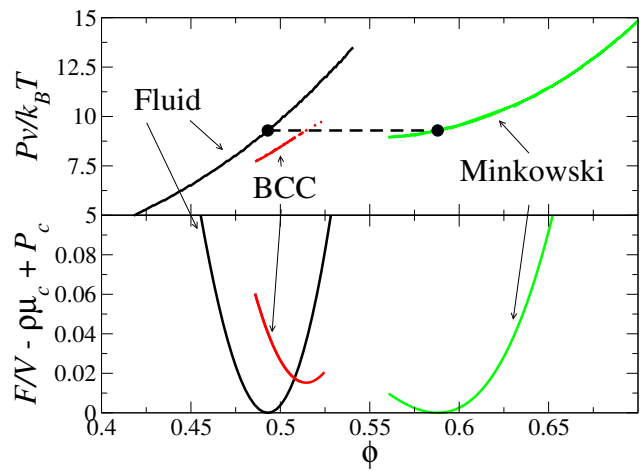

**Figure 9.7:** A part of the equation of state for hard octahedra. The pressure $Pv/k_BT$ and free-energy density $F/V - \rho\mu_c + P_c$ as a function of packing fraction $\phi$. Here $v$ is the volume of the particle; $F$ and $V$ are the Helmholtz free energy and the volume of the system (in units of particle volume) respectively; $\mu_c$ and $P_c$ are the chemical potential and pressure at bulk coexistence respectively with $\rho$ the number density of the particles. The solid lines in the EOS for the Minkowski and the bcc plastic crystal phases are obtained by melting the close packed Minkowski crystal in floppy box $NPT$ MC simulations, and the dotted line for the bcc plastic crystal is obtained by compressing the crystal in cubic box $NPT$ MC simulations. The black points and dashed line show the coexistence between the fluid phase and the Minkowski crystal phase.

## 9.4 Conclusions

In conclusion, using free-energy calculations we determined the full phase diagram of hard superballs with shapes interpolating between cubes and octahedra, i.e., $0.5 \le q < \infty$. In systems of cube-like superballs ($q > 1$), we find a stable deformed $C_1$ phase at high packing fraction, except close to the sphere-limit ($q = 1$) where a deformed $C_0$ crystal is stable. For $q < 3$ the crystal phase melts into an fcc plastic crystal before melting into a fluid phase of cube-like superballs. In systems of octahedron-like superballs ($0.5 < q < 1$), we find a stable bct or a deformed bcc crystal phase upon approaching close packing, with a crossover at $q = 0.79248$. Moreover, a stable fcc plastic crystal appears at intermediate densities for $0.7 < q \le 1$. Interestingly, for $q < 0.85$, we find a novel stable bcc plastic crystal phase, which can even coexist with the fcc plastic crystal phase at around $q = 0.8$. More surprisingly, the bcc and fcc rotator phases are unstable for the flat-faced and sharp-edged hard octahedra and hard cubes, respectively, which suggests that the rounded corners play an important role in stabilizing rotator phases. It is tempting to argue that entropic directional forces [181] that tend to align the flat faces of the polyhedral-shaped particles destabilize rotator phases in favor of crystals. Finally, we also observed a two-step melting phenomenon in the system of hard octahedra, such that the Minkowski crystal melts into a metastable bcc plastic crystal before melting into the fluid phase. Nanoparticle self-assembly is surprisingly sensitive to particle curvature.

## 9.5   Acknowledgments

We thank A. P. Gantapara and J. de Graaf for the simulations on hard octahedra and Dr. F. Smallenburg for fruitful discussions and Dr. D. Ashton for making the snapshot in Fig. 9.6.

## 9.6   Appendices

### 9.6.1   Overlap algorithm for superballs

The algorithm that we used to check for overlaps between superballs is based on the Perram and Wertheim (PW) potential [182]. The details of the application of this general method to the specific case of superballs can be found in Ref. [178]. A superball with shape parameter $q$ and size $a$, located at $\mathbf{r}_0$, and orientation matrix $\mathbf{O} = (\mathbf{o}_1, \mathbf{o}_2, \mathbf{o}_3)$ is given by the set of points $\{\mathbf{r} \,|\, \zeta(\mathbf{r}) \leq 0, \mathbf{r} \in \mathbb{R}^3\}$ with $\zeta$ an appropriate shape function. The shape function is strictly convex and defined by

$$\zeta(\mathbf{r}) = g\left[\tilde{\zeta}(\tilde{\mathbf{r}})\right] - 1 \qquad (9.10)$$

with

$$
\begin{aligned}
g(x) &= x^{1/q} \\
\tilde{\zeta}(\tilde{\mathbf{r}}) &= \left(\frac{\tilde{r}_1}{a}\right)^{2q} + \left(\frac{\tilde{r}_2}{a}\right)^{2q} + \left(\frac{\tilde{r}_3}{a}\right)^{2q}
\end{aligned}
$$

where $\tilde{\mathbf{r}} = (\tilde{r}_1, \tilde{r}_2, \tilde{r}_3)^T = \mathbf{O}^{-1}(\mathbf{r} - \mathbf{r}_0)$ gives the relative coordinates of $\mathbf{r}$ with respect to the particle centered at $\mathbf{r}_0$ with the reference orientation $\mathbf{O}$.

The condition for overlap between a pair of particles $A$ and $B$ can be thought of as an inequality between the position and orientation of the particles. For this purpose, we measure the distance between the two superballs using the overlap potential $\zeta(A, B)$, where $A$ and $B$ contain the information for the location and orientation of the two superballs. The sign of $\zeta(A, B)$ gives us an overlap criterion through

$$
\begin{cases}
\zeta(A, B) > 0 & \text{if } A \text{ and } B \text{ are disjoint} \\
\zeta(A, B) = 0 & \text{if } A \text{ and } B \text{ are externally tangent} \qquad (9.11)\\
\zeta(A, B) < 0 & \text{if } A \text{ and } B \text{ are overlapping}
\end{cases}
$$

$\zeta(A, B)$ is also at least twice continuously differentiable in the position and orientation of $A$ and $B$, respectively.

In the following we describe the procedure by which $\zeta(A, B)$ can be determined for two superballs with given position and orientation. We define and compute the overlap conditions using a procedure originally developed for ellipsoids by Perram and Wertheim [182].

The PW overlap potential is defined by

$$\zeta(A, B) = \max_{0 \leq \lambda \leq 1} \min_{\mathbf{r}_C} \left[ \lambda \zeta_A(\mathbf{r}_C) + (1 - \lambda) \zeta_B(\mathbf{r}_C) \right],  \tag{9.12}$$

where $\zeta_A(\mathbf{r}_C)$ and $\zeta_B(\mathbf{r}_C)$ are the shape functions that define the two superballs $A$ and $B$, respectively. Here $\mathbf{r}_C$ can be thought of as the first point of contact between $A$ and $B$, when these particles are uniformly expanded/scaled, whilst keeping their orientation and position fixed. This is illustrated in Fig. S1.

For every $\lambda$, the solution of the inner optimization over $\mathbf{r}_C$ is unique due to the strict convexity of $A$ and $B$, and it satisfies the gradient condition

$$\nabla \zeta(A, B) = \lambda \nabla \zeta_A(\mathbf{r}_C) + (1 - \lambda) \nabla \zeta_B(\mathbf{r}_C),  \tag{9.13}$$

which shows that the normal vectors are anti-parallel as shown in Fig. S1. The solution of the outer optimization problem over $\lambda$ is specified by the condition

$$\zeta(A, B) = \zeta_A(\mathbf{r}_C) = \zeta_B(\mathbf{r}_C).  \tag{9.14}$$

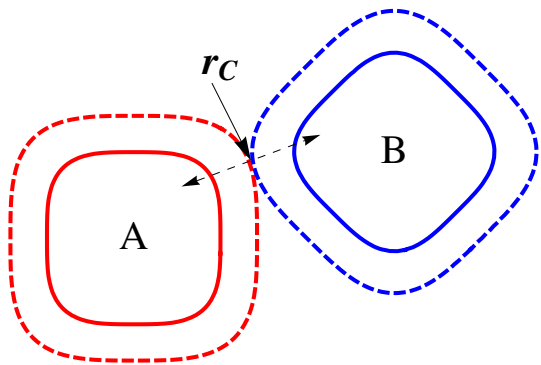

FIG. S1: An illustration of the scaling procedure applied to the two superballs A and B, which results in the contact point at $\mathbf{r}_C$, and the two anti-parallel vectors that are normal to the scaled surfaces of the particles at $\mathbf{r}_C$.

Calculating the PW overlap potential can be done by solving for $\mathbf{r}_C(\lambda)$ in Eq. 9.13, and then determining the $\lambda$ that satisfies Eq. 9.14. The solution to Eq. 9.12 by solving a set of ordinary differential equations (ODEs) and by making use of the ODE event location method [183] to achieve $\zeta_A(\mathbf{r}_C) = \zeta_B(\mathbf{r}_C)$. This method is rigorous in the sense that the optimal $\lambda$ can be determined within an arbitrary accuracy, however, it is inefficient since it requires solving ODEs.

If a good-enough initial guess can be provided for $\lambda$, one can directly use Newton-Raphson (NR) method for the two equations. The method has the advantage that it is more efficient than the one used in Ref. [183]. The Newton steps are determined as follows

$$\Delta\lambda = \frac{-1}{\zeta_{\lambda\lambda}}\left[(\zeta_B - \zeta_A) - \Delta\mathbf{g}^T\mathbf{M}^{-1}\nabla\zeta_{AB}\right], \qquad (9.15)$$

$$\Delta\mathbf{r}_C = \mathbf{M}^{-1}\left(\Delta\mathbf{g}\Delta\lambda - \nabla\zeta_{AB}\right), \qquad (9.16)$$

where

$$\mathbf{M} = \lambda\nabla^2\zeta_A + (1-\lambda)\nabla^2\zeta_B,$$

$$\Delta\mathbf{g} = \nabla\zeta_A - \nabla\zeta_B,$$

$$\zeta_{\lambda\lambda} = \Delta\mathbf{g}^T\mathbf{M}^{-1}\Delta\mathbf{g},$$

$$\nabla\zeta(\mathbf{r}_C) = g'\left[\tilde{\zeta}(\tilde{\mathbf{r}}_C)\right]\nabla\tilde{\zeta}(\tilde{\mathbf{r}}_C),$$

$$\nabla^2\zeta(\mathbf{r}_C) = g'\left[\tilde{\zeta}(\tilde{\mathbf{r}}_C)\right]\left(\nabla^2\tilde{\zeta}(\tilde{\mathbf{r}}_C)\right) + g''\left[\tilde{\zeta}(\tilde{\mathbf{r}}_C)\right]\left(\nabla\tilde{\zeta}(\tilde{\mathbf{r}}_C)\right)\left(\nabla\tilde{\zeta}(\tilde{\mathbf{r}}_C)\right)^T,$$

$$\nabla\tilde{\zeta}(\tilde{\mathbf{r}}_C) = \mathbf{O}\,\nabla_{\tilde{\mathbf{r}}_C}\tilde{\zeta}(\tilde{\mathbf{r}}_C),$$

$$\nabla^2\tilde{\zeta}(\tilde{\mathbf{r}}_C) = \mathbf{O}\,\nabla^2_{\tilde{\mathbf{r}}_C}\tilde{\zeta}(\tilde{\mathbf{r}}_C)\,\mathbf{O}^T,$$

with $\mathbf{O}$ an orthogonal matrix, and $\nabla_{\tilde{\mathbf{r}}_C}$ and $\nabla^2_{\tilde{\mathbf{r}}_C}$ the gradient and Hessian matrix with respect to $\tilde{\mathbf{r}}_C$, respectively. Here we also corrected the typographical errors in Ref. [178].

We have found that this NR method is only sufficiently numerically stable for simulations of superballs with $0.85 \leq q \leq 1.7$. Therefore, in order to improve the range of stability, we make the following modifications to the Newton steps:

$$\Delta\lambda^* = \frac{\Delta\lambda \cdot \alpha}{\max\left(|\Delta\lambda|, |\Delta\mathbf{r}_C|\right)} \qquad (9.17)$$

$$\Delta\mathbf{r}_C^* = \frac{\Delta\mathbf{r}_C \cdot \alpha}{\max\left(|\Delta\lambda|, |\Delta\mathbf{r}_C|\right)} \qquad (9.18)$$

where $\alpha$ is a uniform random number in the interval $[0, 1)$. Essentially the modification makes the length of the Newton step randomly smaller than unity. This helps to avoid the divergence of the iterations in the NR procedure around singularities. With this modification, we have shown that we are able to study systems of superballs with $0.7 \leq q \leq 3.5$.

### 9.6.2   Lattice vectors of Minksowski crystal

The lattice vectors of a Minkowski crystal are given by

$$\mathbf{e}_1 = \frac{2}{3}\mathbf{i} + \frac{2}{3}\mathbf{j} - \frac{2}{3}\mathbf{k}$$

$$\mathbf{e}_2 = -\frac{1}{3}\mathbf{i} + \frac{4}{3}\mathbf{j} - \frac{1}{3}\mathbf{k}$$

$$\mathbf{e}_3 = \frac{1}{3}\mathbf{i} - \frac{1}{3}\mathbf{j} - \frac{4}{3}\mathbf{k}$$

where $\mathbf{i}$, $\mathbf{j}$ and $\mathbf{k}$ are the unit vectors along the axes of the octahedra, i.e., superball with $q = 1/2$, given by

$$|x| + |y| + |z| \leq 1 \tag{9.19}$$

# 10

## Surface roughness directed self-assembly of asymmetric dumbbells into colloidal micelles

Colloidal particles with site-specific directional interactions, so called "patchy particles", have gained an increasing scientific attention in the past decades, because of their promising properties for bottom-up assembly routes towards complex structures. In this chapter, we create a type of patchy particle by controlling the surface roughness of the specific site on the particle. In particular, we study a system of particles with only one attractive patch both in experiments and simulations. We found that when the interaction range is relatively large, it can be well described with a Wertheim type theory. However, in experiments, the interaction range is usually very small, which makes it very difficult to reach equilibrium. Direct Monte Carlo simulations give rise to cluster size distributions that are in good agreement with those found in experiments, although they both disagree with results obtained from free energy calculations.

## 10.1  Introduction

Recent breakthroughs in particle synthesis have resulted in a spectacular variety of building blocks with anisotropic interactions [3]. In particular, colloidal particles with site-specific directional interactions, so called "patchy particles", have gained a significant amount of scientific attention in past decades, since they are promising candidates for bottom-up assembly routes towards complex structures with rationally designed properties. [184–186] The size and geometry of the patches together with the shape of the inter-particle potential are expected to determine the formed structures and phases, which may range from empty liquids [187] and crystals [188–190] to finite-sized clusters [185, 191], and lead to novel collective behavior. [192]

Recent approaches to assemble colloidal particles at specific sites include hydrophobic-hydrophilic interactions, [189, 190] and lock-and-key recognition mechanisms. [193] With a wide variability of colloidal shapes available today, the ultimate challenge is to identify general methods to render specific areas of the colloids attractive or repulsive, while not depending on a specific choice of material or surface chemistry. [3] Ideally, the attraction strength and range is tunable and interactions are reversible to allow the formation of equilibrium structures.

In this chapter, we investigate a system of "patchy" asymmetric dumbbells, of which one sphere is smooth while the other is rough [194–197]. By adding non-adsorbing polymers into the solution, the smooth spheres are shown to be exclusively attractive due to their different overlap volume of depletion zones. We study the formation of colloidal micelles and cluster size distribution. Our simulation results on the cluster size distributions agree well with theoretical predictions as obtained from free energy calculations in the case of relatively long interaction range. However, in experiments the depletion attraction is very short ranged which makes it extremely difficult to reach equilibrium. Our results obtained from direct simulations agree well with experiments but disagree with the free energy calculations.

## 10.2  Methodology

### 10.2.1  Model

Our approach to achieve patchy particles employs depletion interactions between particles with locally different surface roughness as shown in Fig. 10.1. Depletion attractions arise in dispersions of colloidal particles when a second, smaller type of non-adsorbing colloid or macromolecule, also termed depletant, is introduced in the suspension [198–201]. The center of mass of the depletant cannot approach the surface of the larger colloidal particles closer than its radius $r_p$, restricting the volume available to the depletant (see Fig. 10.1B). The volume around the colloidal particles unavailable to the depletant is called the exclusion volume. When the surfaces of two large colloids come closer together than the diameter of the depletant, $2r_p$, their exclusion volumes overlap and the volume accessible to the depletant increases by the amount of this overlap volume $\Delta V$. The volume available to the depletant, and hence its entropy, increases, and an effective

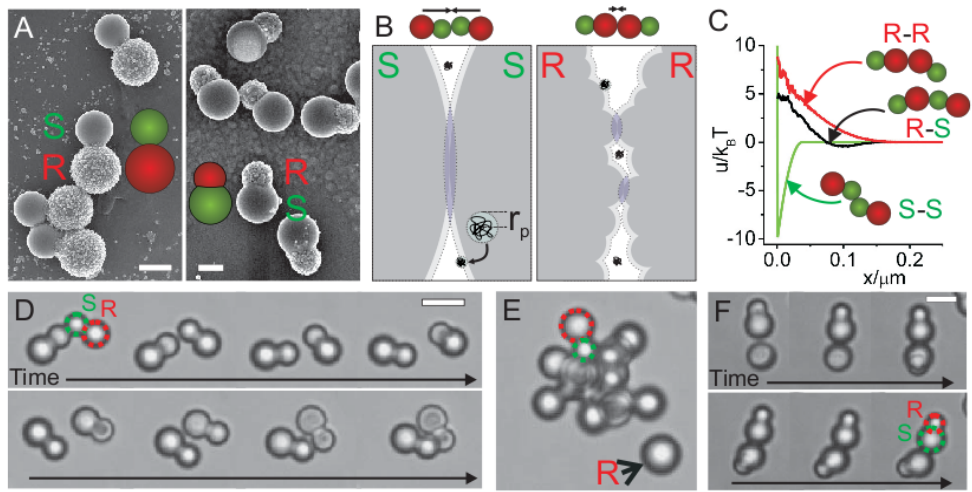

**Figure 10.1: Patchy particles by roughness specific depletion interactions** A) Colloidal
model systems consisting of one sphere with a smooth and one sphere with a rough surface. Scale
bars are $2\mu m$. B) In the presence of small depletants (here depicted as polymers with radius
$r_p$) the colloidal particles are surrounded by a layer inaccessible to the depletant (dotted line).
If colloidal particles approach such that their excluded volumes overlap, the depletant gains
entropy, which results in a net attraction between the colloids. The arising attraction, called the
depletion potential, is proportional to the overlapping excluded volume (blue regions). For two
rough spheres the overlap volume is significantly reduced compared to that for smooth particles.
Small arrows represent the effective forces on both colloids. C) Depletion potentials obtained
from simulations between two smooth, two rough and one smooth and one rough side of our
colloids, and polymer of size $r_p = 19$nm ($\rho_p = 0.038\rho_{\text{overlap}}$) as a function of the distance between
the surfaces of the colloids. D) Snapshots from a movie showing the breaking of a bond between
the smooth sides of two particles and later reformation of the bond. Dextran polymer with
radius $r_p = 19$nm was used at a concentration of $\rho_p = 0.4\rho_{\text{overlap}}$. Scale bar is $5\mu m$. E) Rough
spheres as indicated by the black arrow are left out of the colloidal micelles formed from the
particles with one attractive patch. F) Bond formation between the larger smooth sides of two
particles and subsequent rearrangement due to the flexible bond (Dextran polymer, $r_p = 8.9$nm,
$\rho_p = 0.20\rho_{\text{overlap}}$). Scale bar is $5\mu m$.

attractive depletion potential is induced between the two larger colloids. [198–201] The
depletion potential is roughly proportional to the number density of the depletant $\rho_p$ and
the overlap volume $\Delta V$ as $u_{\text{AO}}/k_B T = -\rho_p \Delta V$. Here, $k_B$ is Boltzmann's constant and $T$
is the temperature. Smooth surfaces have larger overlap volumes than incommensurate
rough surfaces, and hence are more strongly attracted towards each other by depletion
interactions as shown in Fig. 10.1B. [194–197] By employing Monte Carlo simulations,
we calculated the effective interactions between rough-rough and rough-smooth particles
in polymer solutions as shown in Fig. 10.1C. It can be observed that the depletion at-
tractions between rough-rough and rough-smooth spheres are significantly screened by
the surface roughness compared to those between smooth-smooth particles. Therefore,
we model the interactions between the rough spheres as hard-core interactions, while the depletion attraction between the smooth spheres is given by

$$\frac{u_{\mathrm{AO}}(r)}{k_B T} = \begin{cases} \infty, & r \leq \sigma \\ \epsilon \frac{\frac{r^3}{2q^3} - \frac{3r}{2q} + 1}{\frac{\sigma^3}{2q^3} - \frac{3\sigma}{2q} + 1}, & \sigma < r \leq q \\ 0, & r > q \end{cases} , \qquad (10.1)$$

where $r$ is the center-to-center distance between the smooth spheres of two dumbbells with $\sigma$ the diameter of smooth spheres. The interaction is described by the interaction strength $\epsilon$ and the interaction range $q$, where $q = \sigma + \sigma_p$ with $\sigma_p$ the diameter of depletants. In the solution of ideal polymer, the interaction strength is given by

$$\epsilon = \phi_p \frac{\frac{\sigma^3}{2q^3} - \frac{3\sigma}{2q} + 1}{(q-\sigma)^3} q^3 \qquad (10.2)$$

where $\phi_p = \pi \sigma_p^3 \rho_p / 6$ is the packing fraction of ideal polymers in a reservoir in chemical equilibrium with the system.

## 10.2.2   Simulation details

We studied this model system by simulating a system of $N = 1000$ particles at constant density $\rho$ and temperature $T$ in Monte Carlo (MC) simulations. We directly measure the cluster size distribution during the simulation. The initial configurations consist of random located particles with random orientations. The simulation is equilibrated for at least $10^7$ MC cycles. In order to speed up equilibration of the system and increase the mobility of the clusters with more than one particle, we employ cluster moves [10] to collectively move the particles belonging to the same cluster. Two particles are clustered when the center-to-center distance between their smooth spheres is less than the attraction range $q$.

# 10.3   Results and Discussions

## 10.3.1   Cluster size distributions

We compare the cluster size distributions as obtained from direct MC simulations with free energy calculations. To this end, we consider a system of $N$ particles in a volume $V$ at temperature $T$. These particles form micelles under the constraint that the total number of particles satisfies

$$N = \sum_{n=1}^{\infty} n N_n \qquad (10.3)$$

where $N_n$ is the number of micelles consisting of $n$ particles. For sufficiently dilute micelle solutions the system can be modeled as an ideal gas of clusters, non-interacting but capable of exchanging particles. This allows us to write the canonical partition function

$Z(N, V, T)$ as

$$Z(N, V, T) = \prod_{n=1}^{\infty} \frac{Q_n^{N_n}}{N_n!}, \tag{10.4}$$

$$Q_n = \frac{1}{(4\pi)^n \Lambda^{3n} n!} \int_V \mathrm{d}\mathbf{r}^n \int \mathrm{d}\omega^n \exp(-\beta U(\mathbf{r}^n, \omega^n)) c(\mathbf{r}^n), \tag{10.5}$$

where $Q_n$ is the internal configuration integral of a cluster of $n$ particles, $\beta = 1/k_B T$ with $k_B$ Boltzmann's constant, $\Lambda^3$ is the thermal volume of a particle, $\mathbf{r}^n$ and $\omega^n$ denote the position and orientation of the particles in the cluster, respectively. $c(\mathbf{r}^n)$ is the function for clustering, which equals 1 when the particles at $\mathbf{r}^n$ form a cluster, and zero otherwise.

The ratio $Q_n/Q_1$ can be measured using a grand-canonical ($\mu VT$) MC simulation (GCMC) [202]. By imposing the constraint of having only a single cluster in a GCMC simulation, the probability of observing a cluster of size $n$ is

$$\frac{P(n)}{P(1)} = \frac{Q_n}{Q_1} \exp[\beta\mu(n-1)] \tag{10.6}$$

Hence, the ratio $Q_n/Q_1$ can be directly obtained for all $n$ from GCMC simulations.

## 10.3.2 Comparison with experiments

Colloidal particles consisting of one smooth and one rough sphere were synthesized as depicted in Fig. 10.1B. The final particles had a protrusion radius of $1.11 \pm 0.06\mu m$ (smooth side) and a seed radius of $1.46 \pm 0.06\mu m$ (rough side),and the total length is $4.9 \pm 0.12\mu m$, which means that the size ratio of the two spheres of the dumbbell is $\sigma/\sigma_R = 0.76$ with center-to-center distance $d = 0.79\sigma$. Furthermore, large rough spheres of radius $1.6 \pm 0.1\mu m$ were employed in the experiments. Although NaCl was added to the system to screen the electrostatic interactions, some of the charge repulsion remains between the particles. As a result, it is difficult to calculate the effective interaction strength from the concentration of polymers. However, the size of the polymer ($\sigma_p = 38$ nm) and the packing fraction of the system ($\phi = 0.003$) are known, and we expect that the electrostatic repulsion mainly affects the effective strength of the interaction, without strongly influencing the behavior of the system. Thus, the interaction strength $\epsilon$ will be our only fitting parameter when comparing distributions to the experimental results. The direct observations from light microscopy shows that it typically takes 10 minutes for a particle to escape from a dimer. By using Kramers' approach [203, 204], we estimate that the interaction strength in Eq. 10.1 is around $\epsilon = -10k_B T$.

According to experimental settings, we perform canonical MC simulations ($NVT$) to study the self-assembly of this asymmetric dumbbell system of $\sigma/\sigma_R = 0.76$ and $d = 0.79\sigma$ at the packing fraction $\phi = 0.003$ with the attraction strength between smooth spheres $\epsilon = -10k_B T$. Due to the strong attraction between particles, they form clusters in the solution, and the representative images of the colloidal clusters containing $n = 1$ to $n = 12$ particles are presented in Fig. 10.2. Note that the colloids are free to move within the limitations of the bonds and particularly the rough parts are free to sample the accessible volume around the center of the clusters. These clusters are reminiscent of

Experiments                                         Simulations

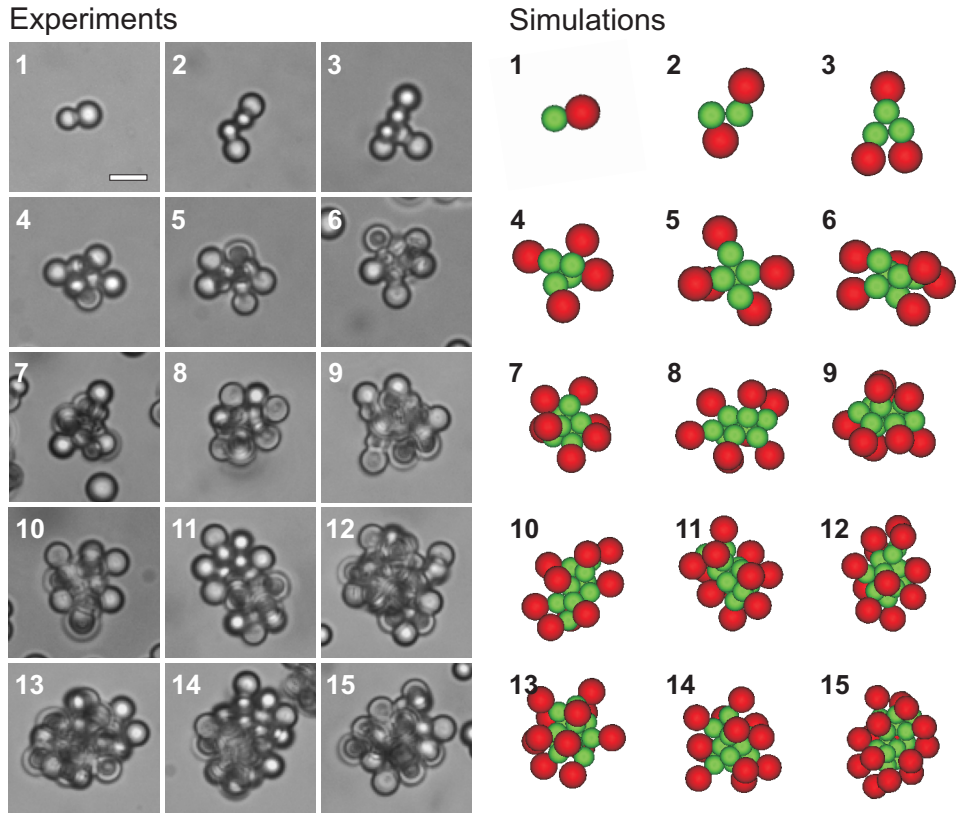

**Figure 10.2**: Typical cluster shapes obtained from colloids with attractive small smooth, and large rough (non-attractive) side containing $n = 1$ to $n = 12$ patchy particles. On the left side, experimentally observed clusters of colloids with small, smooth side are presented. Scale bar is 5$\mu$m. The right side shows clusters obtained from Monte Carlo simulations on dimers consisting of a rough and a smooth sphere. The smooth spheres interact by an attractive depletion potential (green) and the rough spheres interact with a hard-sphere potential (red). The interactions between rough and smooth spheres are also assumed to be hard-sphere-like. In experiments and simulations, the smaller attractive sides are located at the core of the clusters, reminiscent of micelles. Snapshots for experiments and MC simulations taken after the cluster size distribution stopped evolving significantly.

surfactant micelles, where the colloids specifically bind at their smaller smooth sides inside the clusters just like the hydrophobic parts of surfactants attract each other. The larger, rough sides of the particles are located outside of the clusters similar to the hydrophilic head group of surfactant micelles. These interactions together with their overall cone-like shape make our colloids the simplest realization of "colloidal surfactants" [205], which in analogy to molecular surfactants form "colloidal micelles".

The strength of the attraction between the smooth spheres and the size of the clusters can be tuned by the polymer concentration as shown in Fig. 10.3. When the polymer concentration is low, i.e., $\rho_p = 0.32\rho_{overlap}$, there are no clusters formed in the system, where $\rho_{overlap} = (\pi\sigma_p^3/6)^{-1}$ is the polymer overlap concentration. When the polymer concentration increases to $\rho_p = 0.35\rho_{overlap}$, there are small clusters consisting of two or three particles formed in the system. Colloidal micelles are obtained at a slightly higher polymer concentration, i.e., $\rho_p = 0.38\rho_{overlap}$, and the cluster size distribution shows a significant second peak located at $n = 10$, which is the most probably cluster size. The MC simulation results agree very well with the experiments, but disagree with the results from free energy calculation in equilibrium. This suggests that the MC simulations and experiments are out of equilibrium.

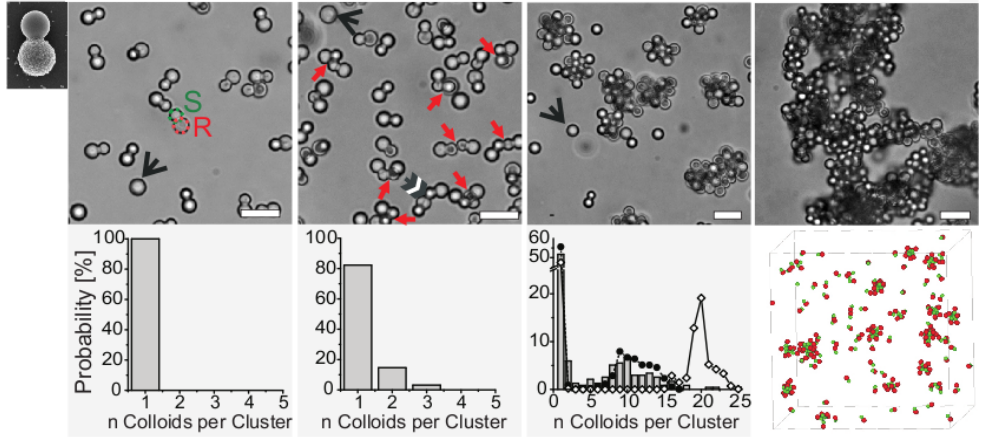

**Figure 10.3**: Transmission light microscopy images of colloidal clusters from colloids with small attractive patches at increasing polymer concentrations and corresponding cluster size distributions for experiments (bars, $\sigma_p = 38$nm), and direct MC simulations ($\epsilon = -10k_BT$, $\sigma_p = 38$nm, after $10^8$ MC cycles, shown as solid dots) and free energy calculations (shown as open diamonds). Single particles are present in solution at $\rho_p = 0.32\rho_{overlap}$ (left of the top row). Small clusters with an exponentially decaying size distribution for $\rho_p = 0.35\rho_{overlap}$ (middle of the top row). Bonds between smooth patches are indicated by red arrows, and black/white arrow indicates binding between smooth and rough sides of the particles. For $\rho_p = 0.40\rho_{overlap}$ a clear peak in the cluster size distribution appears around $n = 10$ (right of the top row). Black arrows point out rough spheres. Cluster distributions shown below the microscopy images corroborate that experiments and direct MC simulations are in agreement. However, the distributions are not in equilibrium yet as free energy calculations yield a significantly different cluster distribution. Above a critical aggregation concentration site-specificity is lost. Scale bars are $10\mu$m.

## 10.4   Conclusions

In this chapter, we performed kinetic Monte Carlo simulations to study the self-assembly of colloidal asymmetric dumbbells, of which one sphere is smooth and the other is rough. In the solution of non-adsorbing polymers with certain sizes, the smooth spheres are attractive to each other due to the depletion interaction while the depletion attraction between the rough sphere is screened out. This makes the particle behave like a patchy particle with only one attractive patch. We model such particles as asymmetric dumbbells by employing the Asakura-Oosawa (AO) potential between smooth spheres and hard-core interactions between smooth-rough and rough-rough spheres. Our simulations show that such asymmetric patchy dumbbells can form micelles in the presence of non-adsorbing polymer, and when the interaction range of AO potential is relatively large, which means that the system can easily reach equilibrium. However, in experiments, the interaction range of AO potential is very short, which makes it very difficult to reach equilibrium. Our direct simulation results agree very well with the experiments, but they both disagree with the equilibrium free energy calculations. We demonstrate that this experimental method is promising in making well controlled patchy particles while it still needs to increase the interaction range to improve the chance of reaching equilibrium.

## 10.5   Acknowledgments

I would like to thank Dr. D. J. Kraft for performing the experiments on this system, Dr. M. Hermes for calculating the depletion potential for rough-rough and smooth-rough spheres, and Dr. F. Smallenburg for the free energy calculations discussed in this chapter.

# Summary

In this thesis, we study the phase behavior and the kinetic pathway of phase transitions in colloidal systems driven by entropy. In Chapter 1, we gave a general introduction on colloid and the simulation methods we employed throught the thesis, i.e., Monte Carlo and molecular dynamics simulation methods. Then the thesis is divided into two parts: nucleation and phase behavior in colloidal systems.

In the first part of this thesis, we study the nucleation in systems consisting of different colloidal particles. In Chapter 2, we introduce the physical background of nucleation by taking an example of the gas-liquid nucleation in a system of Van der Waals fluid below the critical temperature. Then we briefly derive the classical nucleation theory, which is a widely used phenomenological theory to describe the free energy barrier and the kinetics of nucleation. Furthermore, in the rest of the first part of this thesis, we studied the nucleation in a variety of colloidal systems.

Hard sphere is almost the simplest model system for studying the phase behavior of colloidal systems, and the calculated equilibrium phase diagram of hard spheres is in good agreement with experiments [8]. However, the calculated crystal nucleation rates of hard sphere systems from Monte Carlo simulations differ from those measured in experiments by more than six orders of magnitude [18], which has induced an ongoing hot debate in the past decade. In Chapter 3, we studied the crystal nucleation in a system of hard spheres by using three different methods of rare events, i.e., umbrella sampling, forward flux sampling and molecular dynamics simulations. We found that the nucleation rates calculated from those three methods all agree with each other very well in long-time diffusion units. Moreover, the nucleation rates calculated from simulations agree with experimental results at high supersaturations, while there is still a markedly large discrepancy at low supersaturations. Furthermore, we found that the structure of nuclei is independent of simulation methods, and they contain on average significantly more face-centered-cubic (fcc) ordered particles than hexagonal-close-packed (hcp) ordered particles while the free energy difference between fcc and hcp is on the order of $10^{-3}k_BT$ per particle.

In experiments, the synthesized particles can not be perfect hard spheres, and there is always some soft repulsion between particles, thus it has be speculated that the discrepancy in the nucleation rates obtained in experiments and computer simulations may arise from such soft repulsions between particles in experiments. Therefore, in Chapter 4, we examine the phase behavior of the Weeks-Chandler-Andersen (WCA) potential with $\beta\epsilon = 40$, i.e, a hard-sphere like interaction. Crystal nucleation in this model system was recently studied by Kawasaki and Tanaka [20], who argued that the computed nucleation rates agree well with experiment, a finding that contradicted earlier simulation results. Here we report an extensive numerical study of crystallization in the WCA model, using three totally different techniques (Brownian Dynamics, Umbrella Sampling and Forward Flux Sampling). We find that all simulations yield essentially the same nucleation rates. However, these rates differ significantly from the values reported by Kawasaki and Tanaka and hence we argue that the huge discrepancy in nucleation rates between simulation and experiment persists. When we map the WCA model onto a hard-sphere system, we find good agreement between the present simulation results and those that had been obtained for hard spheres.

In addition, the particles synthesized in experiments can not be perfect spherical, and possibly the small anisotropy of the particles can influence the resulting nucleation rates. Thus, in Chapter 5, we study the homogeneous crystal nucleation in suspensions of colloidal hard dumbbells. When the shape of hard dumbbells is close to spherical, we found that the system nucleate a plastic crystal phase which agrees with the equilibrium phase diagram from free energy calculations [86]. In addition, at low supersaturations the free energy barriers increases slightly with increasing dumbbell anisotropy, which can be explained by a small increase in surface tension for more anisotropic dumbbells [96]. When the supersaturation increases, the barrier height decreases with increasing dumbbell aspect ratio, which can only be explained by a different pressure-dependence of the interfacial tension for hard dumbbells with different aspect ratios. Although the nucleation rate for the plastic crystal phase does not vary much with aspect ratio, the dynamics do decrease significantly. We also carried out molecular dynamics simulations and compared the nucleation rates obtained from spontaneous nucleation events with those obtained from the umbrella sampling Monte Carlo simulations, and found good agreement within the error bars of one order of magnitude. Moreover, we also studied the nucleation of the aperiodic crystal phase of hard dumbbells, when the shape the dumbbells is close to that of two touching spheres. Our results showed that at the same pressure, the nucleation barrier of the aperiodic crystal phase of hard dumbbells with $L^* = 1.0$ is slightly higher than that of hard spheres which is mostly due to a small difference in supersaturation.

The interest in positionally and orientationally ordered assemblies of anisotropic particles is driven by their great technological potential as they exhibit anisotropic optical properties, but arises from a more fundamental point of view as well. However, the kinetic pathways of the self-assembly of anisotropic particles are not well understood. For instance,the phase diagram of hard rods has been known for around fifteen years, and shows that there are stable isotropic, nematic, smectic and crystal phases depending on the aspect ratio [114]. Only very recently, the kinetic pathway of isotropic-nematic (IN) phase transition for long rods was reported, but the isotropic-smectic (ISm) and isotropic-crystal (IX) phase transitions of short rods still remain unknown. In Chapter 6, we study the kinetic pathways of IX and ISm in systems of hard rods, and we identify three dynamic regimes in supersaturated isotropic fluids of short hard rods: for moderate supersaturations we observe nucleation of multi-layered crystalline clusters; at higher supersaturation, we find nucleation of small crystallites which arrange into long-lived locally favored cubatic-like structures that get kinetically arrested, while at even higher supersaturation the dynamic arrest is due to the conventional cage-trapping glass transition. For longer rods we find that the formation of the (stable) smectic phase out of a supersaturated isotropic state is strongly suppressed by an isotropic-nematic spinodal instability, and we showed for the first time that for quenches close to a spinodal the clusters diverge in size.

Furthermore, in Chapter 7, by performing extensive molecular dynamics simulations, we study nucleation in a system of particles with internal degrees of freedom, i.e., colloidal polymers consisting flexibly connected hard spheres. It has been shown recently that random packings of granular ball chains show striking similarities with the glass transition in polymers [131]. Therefore, these hard-sphere chains can serve as a simple model for polymeric systems and a better understanding of the behavior of these bead chains may shed light on the glass transition and crystallization of polymers. We present a novel event-driven molecular dynamics simulation method, which is easy to implement and very efficient. We find that the nucleation rates are predominately determined by the number of bonds per sphere in the system, rather than the precise details of the chain topology, chain length, and polymer composition. Our results thus show that the crystal nucleation rate of bead chains can be enhanced by adding monomers to the system. In addition, we find that the resulting crystal nuclei contain significantly more fcc than hcp ordered particles. More surprisingly, the resulting crystal nuclei possess a range of crystal morphologies including structures with a five-fold symmetry.

Compared to the nucleation in systems consisting of monodisperse particles, crystal nucleation is much more complicated in the systems containing different types of particles. For instance, in systems of binary mixtures, dynamical heterogeneities may make the kinetic pathways of nucleation out of equilibrium [28], and they may be influenced by the order parameter used in the umbrella sampling simulations. In the last chapter of the first part of this thesis, i.e., Chapter 8, we study crystal nucleation in a binary mixture of hard spheres and investigate the composition and size of (non)critical clusters using Monte Carlo simulations. We show that the choice of order parameter can strongly influence the composition of noncritical clusters due to the projection of the Gibbs free-energy landscape in the two-dimensional composition plane onto a one-dimensional order parameter. On the other hand, the critical cluster is independent of the choice of the order parameter, due to the geometrical properties of the saddle point in the free-energy landscape, which is invariant under coordinate transformation. We investigate the effect of the order parameter on the cluster composition for nucleation of a substitutional solid solution in a simple toy model of identical hard spheres but tagged with different colors and for nucleation of an interstitial solid solution in a binary hard-sphere mixture with a diameter ratio $q = 0.3$. In both cases, we find that the composition of noncritical clusters depends on the order parameter choice, but are well explained by the predictions from classical nucleation theory. More importantly, we find that the properties of the critical cluster do not depend on the order parameter choice.

In the second part of this thesis, we study the entropy-driven phase behavior of colloidal systems. In Chapter 9, we determined the full phase diagram of colloidal hard superballs, of which the shape interpolates between cubes and octahedra via spheres, by free-energy calculations in Monte Carlo simulations. We discover not only a stable fcc plastic crystal phase for near-spherical particles, but also a stable body-centered cubic (bcc) plastic crystal close to the octahedron shape, and in fact even coexistence of these two plastic crystals with a substantial density gap. The plastic fcc and bcc crystals are, however, unstable in the cube and octahedron limit, suggesting that the rounded corners of superballs play an important role in stablizing the rotator phases. In addition, we observe a two-step melting phenomenon for hard octahedra, in which the Minkowski crystal melts into a metastable bcc plastic crystal before melting into the fluid phase.

Finally, in the last chapter of this thesis, i.e., Chapter 10, we realize a type of patchy particle by controlling the surface roughness of specific sites on the particle. In particular, we study a system of particles with only one attractive patch. We found that when the interaction range is relatively large, it can be well described with a Wertheim type theory. However, in experiments, the interaction range is usually very small, which makes it very difficult to reach equilibrium. Direct Monte Carlo simulations give rise to cluster size distributions that are in good agreement with those found in experiments, although they both disagree with results obtained from free energy calculations.

# Samenvatting

In deze scriptie bestuderen we het fasegedrag en het kinetische reactiepad van faseovergangen in colloïdale systemen gedreven door entropie. Hoofdstuk 1 bevat een algemene introductie in colloïden en de simulatiemethoden die gebruikt worden in deze scriptie, namelijk Monte-Carlo- en moleculairedynamicamethoden. De scriptie is opgedeeld in twee delen; een deel over de nucleatie en één over het fasegedrag van colloïdale systemen.

In het eerste deel van deze scriptie bestuderen we de nucleatie in systemen die bestaan uit verschillende colloïdale deeltjes. In Hoofdstuk 2 introduceren we de fysische achtergrond van nucleatie door te kijken naar een voorbeeld van de gas-vloeistof-nucleatie in een systeem van Van der Waals fludum beneden de kritische temperatuur. Vervolgens leiden we in het kort de klassieke nucleatietheorie af, hetgeen een veelvuldig toegepaste fenomenologische theorie is voor het beschrijven de vrije-energiebarrières en de kinetika van nucleatie. In de rest van het eerste deel van deze scriptie bestuderen we vervolgens nucleatie in verscheidene colloïdale systemen.

Het hardebollenmodel is zo ongeveer het simpelste model voor het bestuderen van fasegedrag in colloïdale systemen, en het berekende fasediagram van harde bollen is in goede overeenstemming met experimenten [8]. De uit Monte-Carlosimulaties berekende kristalnucleatiesnelheden van hardebollensystemen verschillen echter meer dan zes ordegroottes van de experimenteel gemeten waardes, hetgeen in het afgelopen decennium een felle, nog voortdurende discussie teweeg heeft gebracht. In Hoofdstuk 3 bestudeerden we kristalnucleatie in een hardebollensysteem door gebruik te maken van drie verschillende methoden voor rare events, namelijk Umbrella Sampling, Forward Flux Sampling en moleculairedynamicasimulaties. We ontdekten dat de nucleatiesnelheden berekend met deze drie methoden heel goed met elkaar overeenstemden in eenheden van langetijdsdiffusie. De nucleatiesnelheden berekend met deze simulaties kwamen ook overeen met experimentele resultaten bij hoge oververzadiging, maar bij lage oververzadiging was er duidelijk een grote discrepantie. Daarbij ontdekten we dat de structuur van de nuclei onafhankelijk is van de simulatiemethoden, en dat ze gemiddeld aanzienlijk meer deeltjes bevatten die geordend zijn in een kubisch vlakgecentreerd (fcc) rooster dan in een hexagonale dichtste stapeling (hcp), terwijl het verschil in vrije energie per deeltje tussen fcc en hcp slechts in de ordegrootte van $10^{-3} k_B T$ is.

Omdat in experimenten de gesynthetiseerde deeltjes nooit perfecte harde bollen zijn, en er altijd een zachte repulsie tussen de deeltjes is, is er gespeculeerd dat de discrepantie tussen de nucleatiesnelheden verkregen uit experimenten en uit computersimulaties afkomstig zou kunnen zijn van deze zachte repulsies tussen de deeltjes in de experimenten. Hoofdstuk 4 onderzoeken we daarom het fasegedrag van de Weeks-Chandler-Andersonpotentiaal (WCA) met $\beta\epsilon = 40$, d.w.z. een interactie lijkend op die van harde bollen. Kristalnucleatie in dit modelsysteem is recentelijk bestudeerd door Kawasaki en Tanaka [20], die stelden dat de berekende nucleatiesnelheden goed overeenkwamen met experimenten, een ontdekking die eerdere simulatieresultaten tegensprak. Wij rapporteren hier ons uitvoerige numerieke onderzoek naar kristalisatie in het WCA-model, gebruik makend van drie geheel verschillende technieken (Brownse dynamica, Umbrella Sampling en Forward Flux Sampling). We bemerken dat al deze simulaties in wezen identieke simulatiesnelheden opleveren. Deze snelheden verschillen echter aanzienlijk van de waardes die Kawasaki en Tanaka noemen en daarom stellen wij dat de enorme discrepantie van nucleatiesnelheden tussen simulaties en experimenten voortduurt. Als we het WCA-model afbeelden op een hardebollensysteem vinden we een goede overeenstemming tussen de resultaten van huidige simulaties en die van hardebollensimulaties.

Bovendien kunnen in het lab gesynthetiseerde deeltjes niet perfect bolvormig zijn, en mogelijkerwijs zou de kleine anisotropie van de deeltjes de resulterende nucleatiesnelheden kunnen benvloeden. Daarom bestuderen we in Hoofdstuk 5 de homogene kristalnucleatie in suspensies van colloïdale harde dumbbelldeeltjes. Als de vorm van de harde dumbbells bijna bolvormig is, blijkt dat het systeem nucleëert tot een plastic kristalfase, in overeenstemming met het equilibriumfasediagram uit vrije energieberekeningen [86]. Ook blijkt dat bij lage oververzadiging de vrije energiebarrières lichtelijk hoger worden bij toenemende anisotropie van de dumbbells [96]. Als de oververzadiging groter wordt, krimpen de barrières juist bij toenemende anisotropie, hetgeen alleen maar verklaard kan worden door een verschillende drukafhankelijkheid van de oppervlaktespanning voor verschillende lengtebreedteverhoudingen van de dumbbells. Hoewel de nucleatiesnelheid voor de plastic kristalfase niet veel varieert met de afmetingsverhouding, wordt de dynamica wel aanzienlijk langzamer. We hebben ook moleculairedynamicasimulaties uitgevoerd en de nucleatiesnelheden verkregen uit spontane nucleatie-events vergeleken met die uit de Umbrella Sampling Monte-Carlosimulaties en vonden een goede overeenkomst binnen een foutenmarge van één ordegrootte. Ook hebben we de nucleatie van de aperiodieke kristalfase van harde dumbbells bestudeerd voor het geval dat de vorm van de dumbbells dichtbij die van twee elkaar rakende bollen ligt. Onze resultaten lieten zien dat bij dezelfde druk de nucleatiebarrière van de aperiodieke kristalfase van harde dumbbells met $L^* = 1.0$ lichtelijk hoger is dan die van harde bollen, hetgeen vooral komt door een klein verschil in oververzadiging.

De aandacht voor positioneel en oriantationeel geordende assemblages van anisotropische deeltjes is gedreven door hun enorme technologische potentie vanwege hun anisotropische optische eigenschappen, maar komt ook voort uit een fundamenteler interesse. De kinetische reactiepaden van de zelfassemblage van anisotropische deeltjes worden echter niet goed begrepen. Het fasediagram van harde staafjes is bijvoorbeeld al zo'n vijftien jaar bekend en laat zien dat er stabiele isotropische, nematische, smectische en kristalijne fases zijn, afhankelijk van de afmetingsverhouding [114]. Heel recent pas werd het kinetische reactiepad van de faseovergang tussen isotroop en nemaat beschreven, maar de faseovergang tussen isotroop en smectisch (ISm) en tussen isotroop en kristal (IX) is nog steeds niet ontrafeld. In Hoofdstuk 6 bestuderen we de kinetische reactiepadedn van IX en ISm in systemen van harde staafjes, en identificeren we drie dynamische regimes in oververzadigde isotrope fludums van korte harde staafjes: bij gematigde oververzadiging zien we nucleatie van gelaagde kristalijne structuren; bij hogere oververzadiging vinden we nucleatie van kleine kristallieten die zich ordenen in langlevende lokaal geprefereerde kubaat-achtige structuren die kinetisch opgesloten worden, terwijl bij nog hogere oververzadiging de dynamische opsluiting komt door de conventionele "cage-trapping" glasovergang. Voor langere staafjes ontdekken we dat de formatie van de (stabiele) smectische fase uit een oververzadigde isotrope staat sterk wordt onderdrukt door een isotropische-nematische spinodale instabiliteit, en we lieten voor het eerst zien dat de clusters voor temperingen dichtbij een spinodale divergeren in grootte.

De aandacht voor positioneel en oriantationeel geordende assemblages van anisotropische deeltjes is gedreven door hun enorme technologische potentie vanwege hun anisotropische optische eigenschappen, maar komt ook voort uit een fundamenteler interesse. De kinetische reactiepaden van de zelfassemblage van anisotropische deeltjes worden echter niet goed begrepen. Het fasediagram van harde staafjes is bijvoorbeeld al zo'n vijftien jaar bekend en laat zien dat er stabiele isotropische, nematische, smectische en kristalijne fases zijn, afhankelijk van de afmetingsverhouding [114]. Heel recent pas werd het kinetische reactiepad van de faseovergang tussen isotroop en nemaat beschreven, maar de faseovergang tussen isotroop en smectisch (ISm) en tussen isotroop en kristal (IX) is nog steeds niet ontrafeld. In Hoofdstuk 6 bestuderen we de kinetische reactiepadedn van IX en ISm in systemen van harde staafjes, en identificeren we drie dynamische regimes in oververzadigde isotrope fludums van korte harde staafjes: bij gematigde oververzadiging zien we nucleatie van gelaagde kristalijne structuren; bij hogere oververzadiging vinden we nucleatie van kleine kristallieten die zich ordenen in langlevende lokaal geprefereerde kubaat-achtige structuren die kinetisch opgesloten worden, terwijl bij nog hogere oververzadiging de dynamische opsluiting komt door de conventionele "cage-trapping" glasovergang. Voor langere staafjes ontdekken we dat de formatie van de (stabiele) smectische fase uit een oververzadigde isotrope staat sterk wordt onderdrukt door een isotropische-nematische spinodale instabiliteit, en we lieten voor het eerst zien dat de clusters voor temperingen dichtbij een spinodale divergeren in grootte.

Door uitvoerige moleculairedynamicasimulaties uit te voeren bestuderen we voorts, in Hoofdstuk 7, nucleatie in een systeem van deeltjes met interne vrijheidsgraden, namelijk colloïdale polymeren die bestaan uit flexibel verbonden harde bollen. Het is recentelijk gerapporteerd dat willekeurige pakkings van granulaire balkettingen opvallende overeenkomsten met de glastransitie in polymeren laten zien [131]. Daarom kunnen deze hardebollenkettingen dienen als een eenvoudig model voor polymeersystemen en een beter begrip van het gedrag van deze kralenkettingen zou de glastransitie en kristallisatie van polymeren kunnen ophelderen. Wij presenteren een nieuwe gebeurtenisgedreven moleculairedynamicasimulatiemethode die zowel gemakkelijk te implenteren als zeer efficint is. We ontdekken dat de nucleatiesnelheden overwegend bepaald worden door het aantal banden per bol in het systeem, in plaats van de exacte bijzonderheden van de kettingtopologie, kettinglengte en polymeersamenstelling. Onze resultaten laten dus zien dat de kristalnucleatiesnelheid van kralenkettingen verhoogd kan worden door monomeren toe te voegen aan het systeem. Bovendien ontekken we dat de resulterende kristalnuclei aanzienlijk meer fcc- dan hcp-geordende deeltjes bevatten. Verassender is dat de resulterende kristalnuclei een verscheidenheid aan kristalmorfologin bevatten, inclusief structuren met een vijfvoudige symmetrie.

Vergeleken met de nucleatie in systemen die bestaan uit monodisperse deeltjes is kristalnucleatie in systemen die verschillende typen deeltjes bevatten veel ingewikkelder. In systemen met binaire mengsels zouden dynamische heterogeniteiten bijvoorbeeld de kinetische reactiepaden van nucleatie kunnen vormen [28], of ze zouden benvloed kunnen worden door de orde-parameter die gebruikt wordt in de Umbrella Sampling-simulaties. In het laatste hoofdstuk van het eerste deel van deze scriptie, namelijk Hoofdstuk 8, bestuderen we kristalnucleatie in een binair mengsel van harde bollen en onderzoeken we de samenstelling en grootte van (niet-)kritische clusters met Monte-Carlosimulaties. We laten zien dat de keuze van orde-parameter de samenstelling van niet-kritische clusters sterk kan bevloeden vanwege de projectie van het Gibbs vrije-energielandschap in het tweedimensionale samenstellingsvlak op een éé ndimensionale orde-parameter. Anderzijds is de kritische cluster onafhankelijk van de keuze van orde-parameter, vanwege de geometrische eigenshappen van het zadelpunt in het vrije-energielandschap, dat invariant is onder cordinatentransformaties. We onderzoeken het effect van de orde-parameter op de clustersamenstelling voor nucleatie in een substitutionelevastestofoplossing in een eenvoudig model van identieke harde bollen maar gemarkeerd met verschillende kleuren, en voor nucleatie van een interstitilevastestofoplossing in een binair hardebollenmengsel met diameterverhouding $q = 0.3$. In beide gevallen vinden we dat de samenstelling van niet-kritische clusters afhangt van de keuze van orde-parameter, maar dat dat goed verklaarbaar is met behulp van de voorspellingen uit klassieke nucleatietheorie. Belangrijker is dat de eigenschappen van het kritische cluster niet afhangt van de orde-parameterkeuze.

In het tweede deel van deze scriptie bestuderen we het entropiegedreven fasegedrag van colloïdale systemen. In Hoofdstuk 9 bepaalden we met behulp van vrije-energieberekeningen in Monte-Carlosimulaties het volledige fasediagram van colloïdale harde superballen, waarvan de vorm interpoleert tussen kubussen en octaders, via bollen. We ontdekken niet alleen een stabiel fcc-plastickristalfase voor bijna-bolvormige deeltjes, maar ook een stabiel kubisch ruimtelijk gecentreerd (bcc-) plastic kristal dichtbij de octadervorm, en zelfs coxistentie van deze twee plastic kristallen met een aanzienlijke energiekloof. De plastic fcc- en bcc-kristallen zijn echter instabiel in de kubus- en octaderlimiet, hetgeen erop wijst dat de afgeronde hoeken van de superballen een belangrijke rol spelen bij het stabiliseren van de rotatorfases. Verder bemerken we een smeltfenomeen in twee stappen voor harde octaders, waarin het Minkowski-kristal smelt tot een metastabiele bcc-plastickristal voordat het smelt tot een fludumfase.

Tenslotte verwezenlijken we in Hoofdstuk 10 een type patchy deeltje door de oppervlakteruwheid op specifieke plaatsen op het deeltje te beheersen. We bestuderen in het bijzonder een systeem van deeltjes met slechts éé n attractieve patch. We ontdekken dat als het interactiebereik relatief groot is, hij goed beschreven kan worden met een Wertheimachtige theorie. In experimenten is het interactiebereik echter meestal klein, wat het zeer moeilijk maakt om evenwicht te bereiken. Directe Monte-Carlosimulaties zorgen voor clustergroottedistributies die in goede overeenkomst zijn met experimenten, hoewel beiden niet overeenkomen met de resultaten verkregen uit vrije-energieberekeningen.

# 中文概述

本论文研究了熵驱动胶体体系的相行为以及相变的动力学路径。在第一章中，我们首先概括性的介绍了胶体以及本论文用到的计算机模拟方法，即Monte Carlo和分子动力学模拟方法。论文的研究内容可以被分为两个部分：胶体体系内的成核和相行为。

在本文的第一部分中，我们研究了由各种胶体粒子组成的体系内的成核问题。在第二章中，首先以临界温度下Van der Waals流体的汽 – 液相变为例，介绍了成核的物理背景。随后我们给出了一个常用于描述成核问题的唯象理论，即经典成核理论。在接下来的章节中，我们进一步研究了不同胶体体系中的成核问题。

硬球流体可能是我们研究胶体体系所能使用的最简单的模型系统，且计算机模拟的平衡态相图和实验结果吻合得很好 [8]。 然而，采用Monte Carlo模拟方法计算得到的硬球流体的晶体成核率和实验观测到的数值相差至少六个数量级，在过去的十多年内，学术界对此巨大的差异的成因存在很大争议。 为了解决这个问题，我们在第三章中采用了三种完全不同的计算机模拟方法，即umbrella sampling、forward flux sampling及分子动力学模拟方法，系统研究了硬球流体的晶体成核问题。我们发现采用长时扩散时间为单位，三种方法得到的成核率几乎一样。 此外，我们还发现这三种模拟方法得到的成核率在高压的情况下和实验结果吻合的很好，但是在低压的情况下仍然有巨大的差别。我们进一步研究发现 形成的核的结构不受模拟方法的影响，且三种模拟方法都发现包含较多的具有面心立方（fcc）对称性的粒子，但是只有少量具有六边形（hcp）对称性的粒子。这个结果是让人意想不到的，因为fcc和hcp两种晶体的自由能差别很小，即每个粒子$10^{-3}k_BT$的数量级上。

此外，实验中得到的胶体粒子不可能是完美的硬球，且粒子之间总是有一些软的排斥作用。因此，一些学者认为也许成核率的差别是源于 这些软的排斥相互作用。所以，我们在第四章中研究了一种类硬球的软排斥粒子体系的相变，即具有$\beta\epsilon = 40$的Weeks-Chandler-Andersen（WCA）相互作用。该体系的成核已经于近期被Kawasaki和Tanaka [20]用布朗动力学方法研究过，且他们认为用计算机模拟得到的成核率和实验结果吻合的很好，但和前人的模拟结果不一样。我们采用三种完全不一样的模拟方法，即布朗动力学模拟、umbrella sampling和forward flux sampling模拟，研究该WCA体系的成核问题。我们发现三种方法得到的模拟结果几乎一致，但是与Kawasaki和Tanaka得到的结果存在差异，因此我们认为计算机模拟得到的成核率和实验观测值间的差异仍然存在。但如果将WCA模型的相边界映射到硬球体系上，我们发现模拟结果和硬球的成核率吻合的很好。

另外，实验里合成的也不可能是完美的球形粒子，也许粒子的非均向的形状也是影响成核率的一个原因。因此，在第五章，我们研究了哑铃状胶体粒子体系的均向成核。当粒子的形状接近球形的时候，我们发现体系会成核生成一种塑性晶体，该结果和通过自由能计算得到的平衡态相图吻合的很好 [86]。此外，在低压情况下，成核的自由能垒随着粒子的非球性增加而缓慢增加，我们认为这是由于界面自由能随着粒子形状发生变化导致的差异 [96]。当压强增大的时候，成核的能垒随着粒子的非球性增加而降低，这是由于界面自由能对于压强的依赖随着粒子形状的变化而变化导致的。虽然成核率随着形状的改变并没有发生太大变化，但是体系的运动速度却随着粒子的非球性的增加而明显降低。此外，我们还进行了分子动力学的模拟，并将得到的成核率和umbrella sampling模拟得到的成核率做比较，其结果吻合的很好，这也验证了我们的umbrella sampling模拟的结果是可靠的。除此以外，我们还研究了由较长哑铃状粒子

形成的非周期性晶体的成核问题。我们发现当形状因子$L^* = 1.0$时，哑铃状粒子的成核能垒稍稍低于同压强下的硬球流体，我们认为这是由于过饱和度的少量增加导致的。

非球形粒子形成具有位置和取向都有序的结构的研究重要性不仅仅由于其在光学器件上的应用，从更加基础的物理研究来看也是有重大意义的。然而，目前对于非球形粒子自组装的动力学路径中物理规律一直没有清楚的理解。例如，硬棒体系的相行为已经在十五年前就被报道了，其相图显示由于形状因子的不同，硬棒体系可以行成稳定的流体态、向列型（nematic）和层列型（smectic）液晶态及晶体态[114]。仅仅在最近几年，硬棒流体从流体到向列型液晶态（IN）的成核路径才被发现，但是流体到层列型液晶态（ISm）及流体到晶体（IX）的成核路径仍然是亟待解决的问题。在第六章中，我们研究了硬棒流体的IN和IX成核路径问题，我们在过饱和的短硬棒流体里发现了三个动力学区域：当压强较低的时候，硬棒流体会通过成核相变成晶体；在稍微高一些的压强下，硬棒流体会形成一种类cubatic局部最优结构；再高的压强会使得体系发生玻璃态转变。对于稍长一些的硬棒，我们发现稳的层列型液晶由于亚稳态的列向型液晶spinodal的微扰导致不能成核，并且我们观察到当逼近spinodal的时候相关长度的发散。

在第七章中，我们采用分子动力学的方法研究了具有内部自由度的粒子体系（即由自由链接的硬球行成的胶体高分子）的晶体成核问题。最近，研究者发现颗粒链的随机堆积问题和高分子的玻璃态转变有着惊人的相似[131]。这表明硬球链是一种研究高分子体系物理的一个很好的模型，对其的研究可以加深我们对于高分子结晶和玻璃态转变的理解。我们提出了一种高效且易于实现的事件驱动分子动力学方法。我们发现硬球链体系的成核率主要取决于链上每个节点所具有的键的数目，而其具体的拓扑结构、链长和组分结构都不是很重要。我们的模拟结果表明此种长链高分子的成核率可以通过添加单体的方式得到提升。此外，我们还发现最终得到的晶体里，绝大部分粒子都是具有fcc对称性的，而只有很少一部分粒子具有hcp对称性。更加让人吃惊的是，我们在晶体生长过程中发现了大量具有五阶对称性的粒子。

和单组分体系的成核相比，多组分体系的成核就更加复杂了。比如，在二元体系里，动力学的不均匀性可以使得成核路径完全脱离平衡态[28]，且在umbrella sampling模拟中很容易被采用的序参数影响。因此，在第一部分的最后一个章节里，即第八章，我们采用Monte Carlo方法研究了二元硬球体系的晶体成核及其非临界核的组分问题。我们发现序参数的选择可以很大程度地影响非临界核的组分。另外一个方面，我们也发现，临界核的组分不受序参数选择的影响，这是由于自由能面上鞍点的几何性质导致的。我们研究了尺寸比为$1 : 1$和$1 : 0.3$的硬球混合物，发现得到的非临界核的组成符合经典成核理论的预测。

在本文的第二部分，我们研究了熵驱动胶体体系的相行为。在第九章中，我们通过Monte Carlo自由能计算得到了胶体超级球从立方体变化成正八面体的相图。当超级球的形状接近球体的时候，我们不仅发现了一种fcc塑性晶体，也发现当超级球的形状接近正八面体的时候，超级球具有一种稳定的体心立方（bcc）的塑性晶体结构。但是这两种塑性晶体在完美多面体，即立方体和正八面体，的时候都不是热力学稳定态，这表明光滑的顶点和楞对于稳定塑性晶体起到了很重要的作用。此外，在硬正八面体的体系内，我们还发现一个两步融化过程，即闵可夫斯基晶体在融化成流体前先融化成一个亚稳的bcc塑性晶体。

最后，在本文的第十章中，我们通过控制粒子表面的粗糙度实现了一种斑块粒子。特别地，我们研究了具有一个活性位点的斑块粒子（即两面不对称粒子）的自组装。我们发现当相互作用的距离相对较长的时候，可以采用一种类Wertheim的理论去描述该体系。然而，在实验中，相互作用的距离往往非常短，导致体系很难达到平

衡。Monte Carlo模拟的结果和实验结果吻合很好，但是和平衡态自由能计算得到的结果差别较大。

# Acknowledgments

In the end of this thesis, I would like to extend my thanks to all the people who have helped me bring this thesis into being. The past four years is a great experience in my life, and it would not have been interesting, fun and satisfying without the support of those around me. First of all, I would like to extend my sincere gratitude to my supervisor, Marjolein, who always has time for answering my questions and listening to my wildest theories. Majorlein, many thanks for your help during the past four years. I learned a lot from you not only from the scientific point of view, and I can not wish for a better supervisor. I hope to keep collaborations with you in my future academic life. Also I want to thank Alfons, René and Arnout for bringing excellent experimentalists, theorists and simulators together to form such a great Soft Condensed Matter (SCM) group. Especially, I want to thank Alfons for his critical questions in the group work discussions and the critical reading of my thesis, which did have positive impacts on my research.

Moreover, I want to thank my lovely officemates, Frank, Laura and Anjan. It is very nice of sharing the same office with you, and I enjoy our many discussions and debates on statistical mechanics, simulation methods and nucleation and so on, which is the most important part of my past four years. I would also like to thank the collaborators inside and outside the SCM group in Utrecht. Alessandro, thanks you for helping me start up with umbrella sampling. Michiel and Laura, thanks for our collaboration on nucleation of hard spheres. Simone, thanks for our collaboration on nucleation of rods, and hopefully we can have more collaborations in the future! Matthieu, thanks for our discussions on hard dumbbells, nucleation and MD simulations. Bas, thank you for translating the summary of my thesis into Dutch. Douglas, thanks for our collaboration on the project of superballs with non-adsorbing polymers. Joost, thanks for our collaboration on the phase diagram of hard superballs. Your "picky" English writing style did help me quite a lot and I hope we can have more collaborations in the future. Willem Kegel and Daniela, thanks for your beautiful experiments on colloidal micelles. I would also like to thank the Chinese subdivision of our group for the lunches and dinners we had in the coffee corner: Bo（彭博）, Bing（刘冰）and Weikai（齐维开）. We actually have successfully influenced the group lunch place to move from Minneart to our coffee corner! In addition, I want to thank all the present and former members of the SCM group for the past beautiful four years we shared together: Niels, Thijs, Nina, Arjen, Jissy, Marlous, Marjolein, Kristina, Bart, Johan, Thomas, Rao, Matthew, Djamel, Linh, Teun, Stéphane, Ahmet, Anke, Peter×2, Krassimir, Esther, Jacob, Judith, Henriëtte, Thea and Patrick. Also I want to thank Henk Mos and his colleagues for keeping our computation cluster running smoothly.

Additionally, I want to thank some people from the outside of Utrecht for the help and fruitful discussions in the past four years. Prof. Bob Evans, thank you for flying all the way to attend my thesis defense, and I did enjoy very much our discussions during your stay as the Kramers Chair of Theoretical Physics in Utrecht. Prof. Wenbing Hu, thanks for your hospitality during my visit to your group in Nanjing University, and I do like the atmosphere in your group and your lovely group members. Andrea Fortini, it is very nice of meeting you in Lausanne and Vienna, and thanks for your help with the application for Rubicon fellowship, although in the end it was figured out that I was unfortunately not eligible. I would also like to thank Yang Jiao in Princeton for our fruitful discussions, and I wish you all the best with your new professorship in U.S.! Mingcheng（杨明成）, thank you and your wife for the hospitality during our trip in Jülich, and the nice time we shared in Granada. Xiaofeng Sui（隋晓锋）and Yuying Gao（高玉英）, thanks for all the fun we had in the past four years, and please also bring my greetings to your daughter baby. I also want to thank Dr. Fajun Zhang（张发军）in Universität Tübingen for the nice time we shared in Vienna. Huanyang（陈焕阳）, it is very nice of meeting you in Soochow, and thanks for flying all the way to visit me in the Netherlands. Hopefully we can have the chance to collaborate at some time in the future. Shuolei（关硕磊）, thank you and your wife for the hospitality during my trip to Beijing, and all the time we shared with football games. I wish all the best to you two in the future! Yuxi（田玉玺）, thanks for the nice time we have had during my stay in Beijing. Profs. Guangfeng Jiang（姜广峰）and Wenyan Yuan（袁文燕）, thanks for your help in these years and bringing me into the field of computational Mathematics. I want to thank Dr. Marleen Kamperman for helping me find the postdoctoral position in Wageningen. I would also like to thank Huanhuan（冯欢欢）for the nice time we shared in Han sur Lesse in 2010, and Junyou（王俊友）and his wife for the hospitality during my first visit to Wageningen. I want to thank Bernhard Reischl in Tampere University of Technology for offering me the cartoon of Boltzmann used in the back cover of this thesis. Here, I would like to thank all the Chinese friends outside of SCM but also in Utrecht for all the nice time we shared together: Zhixiang Sun（孙志祥）, Qiulan Zhang（张秋兰）, Tianhui Zhang（张天辉）, Qingyun Qian（钱庆云）, Jinbao Gao（高金宝）, Wenhao Luo（罗文豪）, Yiming Zhao（赵一鸣）, Jianxi Feng（冯建喜）, Xin Jin（金鑫）, Yinghuan Kuang（匡迎焕）, Yanchao Liu（刘彦超）, Shaoyu Yin（殷邵宇）, Yaowei Yan（阎耀威）, Xinglin Zhang（张兴林）, Xiaoman Liu（刘晓曼）, Meng Chen（陈萌）and Xiaowei Wang（王小伟）. Special thanks go to Tianhui, thanks for your hospitality during my visit to Soochow University. 另外，我希望感谢在Utrecht工作的进先生，感谢每次去您家里享用的美食以及与您愉快的聊天。I also want to thank the supervisors of my master thesis, Profs. Wenchuan Wang and Dapeng Cao in Beijing University of Chemical Technology, for bringing me into the field of computer simulations.

　　最后，我想感谢我的家庭，倪爱仓先生及徐娅女士和杨长锦先生及邓克花女士，感谢您们的养育之恩以及多年来的支持与鼓励！另外，还要感谢妹妹，杨艳女士，和弟弟，李珉先生，一直以来对于我的支持和信任。最重要的，我要感谢我的妻子，杨娟女士，感谢你对于我多年来的理解与帮助，此生有你相伴是我最大的幸福！

# List of publications

This thesis is based on the following publications:

- Laura Filion, Michiel Hermes, <u>Ran Ni</u> and Marjolein Dijkstra, *Crystal nucleation of hard spheres using molecular dynamics, umbrella sampling and forward flux sampling: A comparison of simulation techniques*, Journal of Chemical Physics **133**, 244115 (2010) (Chapter 3)

- L. Filion*, <u>R. Ni</u>*, D. Frenkel and M. Dijkstra, *Simulations of nucleation in almost hard-sphere colloids: The discrepancy between experiment and simulation persists*, Journal of Chemical Physics, **134**, 134901 (2011) (Chapter 4)

- <u>Ran Ni</u> and Marjolein Dijkstra, *Crystal nucleation of colloidal hard dumbbells*, Journal of Chemical Physics, **134**, 034501 (2011) (Chapter 5)

- <u>Ran Ni</u>, Simone Belli, René van Roij and Marjolein Dijkstra, *Glassy dynamics, spinodal fluctuations, and the kinetic limit of nucleation in suspensions of colloidal hard rods*, Physical Review Letters, **105**, 088302 (2010) (Chapter 6)

- <u>Ran Ni</u> and Marjolein Dijkstra, *Effect of bond connectivity on crystal nucleation of hard polymeric chains*, submitted to Physical Review Letters (Chapter 7)

- <u>Ran Ni</u>*, Frank Smallenburg*, Laura Filion and Marjolein Dijkstra, *Crystal Nucleation in binary hard-sphere mixtures: The effect of order parameter on the cluster composition*, Molecular Physics, **109**, 1213 (2011) (Chapter 8)

- <u>Ran Ni</u>, Anjan P. Gantapara, Joost de Graaf, René van Roij and Marjolein Dijkstra, *Phase diagram of colloidal hard superballs: from cubes via spheres to octahedra*, Soft Matter, in press, (2012) (Chapter 9)

- Daniela J. Kraft, <u>Ran Ni</u>, Frank Smallenburg, Michiel Hermes, Kisun Yoon, Dave Weitz, Alfons van Blaaderen, Jan Groenewold, Marjolein Dijkstra, Willem K. Kegel, *Surface Roughness directed Self-Assembly of Patchy Particles into Colloidal Micelles*, Proceedings of the National Academy of Sciences (USA), in press, (2012) (Chapter 10)

Other publications by the author:

- L. Filion, M. Hermes, <u>R. Ni</u>, E. C. M. Vermolen, A. Kuijk, C.G. Christova, J. Stiefelhagen, T. Vissers, A. van Blaaderen and M. Dijkstra, *Self-assembly of a colloidal interstitial solid solution with a tunable sublattice doping*, Physical Review Letters, **107**, 168302 (2011)

---

*These authors contributed equally to the corresponding work.

# Curriculum Vitae

Ran Ni was born on $10^{th}$ August, 1984 in Lujiang county, Anhui Province, P. R. China. He graduated from Fanshan Middle school in Anhui Province in 2001, after which he entered Beijing University of Chemical Technology majoring in Electronic Information Engineering as an undergraduate student. He changed his major into Mathematics in 2003, and obtained the bachelor degree in Mathematics in 2005 in Beijing University of Chemical Technology. Then he continued his master study in the Department of Chemical Engineering, and focused on the computer simulations on the properties of polyelectrolytes systems under the supervision of Prof. Wenchuan Wang and Prof. Dapeng Cao. During his study in Beijing University of Chemical Technology, he received numerous awards in mathematical modeling including the *Cup of Higher Education Press* (Champion) in National Wide Mathematical Contest in Modeling in China in 2002, the *Honorable Mention* in International Mathematical Contest in Modeling in 2004 and the *First Prize* in National Postgraduates Mathematical Contest in Modeling in China in 2005. In 2008, he obtained his master degree of Chemical Engineering (*cum laude*), then he started his PhD study on computer simulations of colloidal self-assembly under the supervision of Prof. dr. ir. Marjolein Dijkstra in the group of Soft Condensed Matter in Utrecht University, the Netherlands. He presented his work as oral and poster presentations at several national and international conferences, including DFG/FOM-CODEF Young Researchers Meeting, $10^{th}$ Dutch Soft Matter Meeting, $8^{th}$ Liquid Matter Conference, $2^{nd}$ International Soft Matter Conference, CECAM Workshop and Physics@FOM Veldhoven. He received the *Chinese Government Award for Outstanding Self-financed Students Abroad* in 2011. His research on computer simulations of entropy-driven colloidal self-assembly is described in this thesis.